%% file: quantyfyingSB_v2.tex
\documentclass[preprint,12pt]{elsarticle}

\usepackage{amssymb}
\usepackage{amsmath}
\usepackage{mathrsfs}
\usepackage{longtable}
\usepackage{tabularx}
\usepackage{url}
\usepackage[normalem]{ulem}
\usepackage{braket}
\usepackage{placeins} % for floatbarrier
\usepackage{graphicx} %for figures locations
\usepackage[percent]{overpic}
\usepackage{tocloft} % for adding the part in the table of contents
\usepackage{pgffor} %for inserting n blank pages
\usepackage[hidelinks]{hyperref} % or colorlinks=true, linkcolor=blue

\usepackage{enumitem}

\usepackage{booktabs}
\usepackage{array}
\usepackage{multirow}
\usepackage{longtable}

\makeatother
\newlength\mylen

\renewcommand\cftpartpresnum{Part~}
\usepackage{color}

\definecolor{darkgreen}{rgb}{0.0, 0.5, 0.0}

\definecolor{ethred}{rgb}{0.7176, 0.2078, 0.1765}

\definecolor{petrol}{rgb}{0.0, 0.45, 0.45}

\usepackage{tikz}
\usetikzlibrary{positioning, calc}
\usepackage{pifont} % \ding{55} cross glyph
\usepackage{ifthen} % \ifthenelse
\definecolor{ink}{HTML}{2A2A2A}
\definecolor{accent}{HTML}{B5533C}

\newif\ifarxivversion
\arxivversionfalse % Change to \arxivversiontrue for arXiv
\arxivversiontrue % Comment this line for PhysRep

\ifarxivversion
    \newcommand{\aiyappacredit}{%
        under the
        \href{https://creativecommons.org/licenses/by-nc/4.0/}
        {CC BY-NC 4.0 license}
        (\textcopyright~2024 The Authors)%
    }
\else
    \newcommand{\aiyappacredit}{with permission from AAAS}
\fi

\journal{Physics Reports}

\begin{document}

\begin{frontmatter}



\title{\centering Structural balance in complex networks: \newline from measurement to dynamics, emergence and implications
} %% Article title
\author[label1]{Giacomo Vaccario}
\author[label2]{Piotr J. G\'orski}
\author[label3]{Georges Andres}
\author[label4,label5]{Manuel S. Mariani}
\author[label2]{Janusz A. Ho{\l}yst\corref{cor1}}

\affiliation[label1]{
    organization={Chair of Ecosystem Management, ETH Zurich},
    city={Zurich},
    country={Switzerland}
}
\affiliation[label2]{
    organization={Faculty of Physics, Warsaw University of Technology},
    % addressline={Koszykowa 75},
    city={Warsaw},
    country={Poland}
}
\affiliation[label3]{
    organization={Chair of Systems Design, ETH Zurich},
    % addressline={Weinbergstrasse 58},
    city={Zurich},
    % postcode={8092},
    country={Switzerland}
}
\affiliation[label4]{
    organization={Department of Network Science, Institute of Data Analytics and
Information Systems, Corvinus University of Budapest},
    % addressline={Plattenstrasse 14},
    city={Budapest},
    country={Hungary}
}
\affiliation[label5]{
    organization={Department of Business Administration, University of Zurich},
    % addressline={Plattenstrasse 14},
    city={Zurich},
    country={Switzerland}
}

\cortext[cor1]{Corresponding author}
\ead{janusz.holyst@pw.edu.pl}

% \author[1]{Giacomo Vaccario}
% \author[2]{Piotr J. G\'orski}
% \author[1]{Georges Andres}
% \author[3]{Manuel S. Mariani\corref{cor1}}
% \author[2]{Janusz A. Ho{\l}yst}

% %% Corresponding author
% \cortext[cor1]{Corresponding author}
% \ead{Janusz A. Ho{\l}yst}   % example, replace with actual email

% %% Affiliations
% \affiliation[1]{
%     organization={Chair of Systems Design, ETH Z\"urich},
%     addressline={Weinbergstrasse 58},
%     city={Z\"urich},
%     postcode={8092},
%     country={Switzerland}
% }

% \affiliation[2]{
%     organization={Faculty of Physics, Warsaw University of Technology},
%     addressline={Koszykowa 75},
%     city={Warsaw},
%     postcode={00-662},
%     country={Poland}
% }

% \affiliation[3]{
%     organization={Department of Business Administration, University of Z\"urich},
%     addressline={Plattenstrasse 14},
%     city={Z\"urich},
%     postcode={8032},
%     country={Switzerland}
% }

%% Abstract
\begin{abstract}
Structural balance theory, introduced in social psychology to describe how friendly and hostile ties organize into consistent patterns, has been reinterpreted through statistical physics and complex networks.
In the classical formulation, a triad with positive or negative links, representing friendship and hostility, is balanced when the product of its signs is positive; a balanced network splits into at most two internally friendly, mutually hostile groups.
This review synthesizes how the triadic formulation has been extended to a statistical-ensemble perspective in which balance is encoded by immanent three-body interactions described by a proper Hamiltonian, balanced configurations correspond to low-energy states, and balance measures serve as order parameters for transitions between consensus, polarization, and disorder.
We organize the field into three sections: \textit{Measurement}, \textit{Dynamics and Emergence}, and \textit{Implications}.
In \textit{Measurement}, we compare measures of structural balance and null models, and show that conclusions for a given system depend on the chosen measure and null model.
In \textit{Dynamics and Emergence}, we review deterministic and stochastic systems, including the gradient, Hamiltonian, and local triad models, and examine how noise, topology, and competing mechanisms determine whether balance emerges.
In \textit{Implications}, we survey consequences for belief formation, opinion contagion dynamics, polarization, international relations, and markets.
Across domains, the evidence is strongest for local triadic stability and weakest for system-wide balance.
Balance is not synonymous with harmony: balanced structures can reflect cohesion, polarization, or persistent conflict.
Its empirical relevance is well supported in some reciprocal social systems, but remains less settled in biological and other complex networks.
\end{abstract}



%% Keywords

\end{frontmatter}
\pagenumbering{roman}

\tableofcontents
\clearpage
\pagenumbering{arabic}
\input{introduction}

\section{Background}

\subsection{Networks and formal definition of structural balance}

This section introduces the basic network notation used throughout the review and states the formal definitions of structural balance and several related notions.
We first define signed networks, neighborhoods, and adjacency matrices, and then use these objects to formulate balance in terms of signed cycles and equivalent partition structures.
For reference, the notation used throughout the review is summarized in \ref{app:notation}; model-specific notation is introduced locally when needed.

\subsubsection{Signed network definition}
\label{subsubsec:signed-net-def}

Networks are powerful models of social systems as they allow us to capture the complex interdependence of their components in a formal way.
Importantly, networks represent the interdependence between components through \textit{dyadic edges}.
A network $\mathcal{G}$ is then defined by the set of all nodes $V$ and the set of all edges $E$.
Henceforth, we denote the total number of nodes in $\mathcal{G}$ as $N = |V|$ and the total number of edges as $m = |E|$.
Dependent on the data and the research question, we may choose to model a system through directed or undirected networks.
In a directed network, edges represent ordered pairs $(i,j)\in E \subseteq V \times V$ of nodes.
Hence, for directed networks, $(i,j)$ and $(j,i)$ are different edges.
In communication systems, for instance, this may be a suitable representation if we think of a message as a carrier of information, going from one individual to another.
Undirected networks do not make this distinction between edges $(i,j)$ and $(j,i)$.
There, edges represent unordered pairs of nodes.

% \pgcomment{I removed vector of signs because it is not needed, and vector s is used for vector of spins in Section 3.}
A signed network $\mathcal{S}$ associates a sign $s_{ij}\in\{-1,+1\}$ with each edge $(i,j)\in E$.
The network still contains the set of nodes $V$ and edges $E$ as before.
Commonly, the sign $s_{ij}$ represents the relation between nodes $i$ and $j$, either being positive or negative.
A signed network may also be weighted, with each edge being associated with a weight $w_{ij}>0$, representing the strength of the positive or negative relation.
We denote the corresponding signed relation by
\begin{equation}
x_{ij}=s_{ij}w_{ij}.
\end{equation}
Thus, $s_{ij}$ determines the sign of the relation, $w_{ij}$ its magnitude,
and $x_{ij}$ combines both.
For an unweighted signed network, $w_{ij}=1$ and hence $x_{ij}=s_{ij}$.

Unless stated otherwise, a signed network is defined on an edge set \(E\), and the sign \(s_{ij}\in\{-1,+1\}\) is assigned only to edges \((i,j)\in E\).
In the corresponding adjacency matrix, \(A_{ij}=0\) denotes absence of an edge, whereas \(A_{ij}=\pm1\) denotes an existing positive or negative edge.
Neutral or zero-valued relations (\(x_{ij}=0\)) are considered only in models that explicitly allow neutral links; in those cases, \(x_{ij}=0\) denotes a neutral relation and not a missing edge.
Unless otherwise stated, self-loops are excluded.

For consistency, let us stress that we distinguish three related quantities.
A binary sign of an existing edge is denoted by \(s_{ij}\in\{-1,+1\}\).
A weighted signed relation or tie strength is denoted by \(x_{ij}\); in continuous models, \(x_{ij}\) may take values in a bounded or unbounded real interval (see Sec. \ref{sec:emerg}).
A signed adjacency matrix is denoted by $\mathbf{A}$, with \(A_{ij}=0\) for absent edges and \(A_{ij}=s_{ij}\) for present binary signed edges, unless a different weighted definition is stated explicitly.

\paragraph{Signed neighborhoods and adjacency matrix} The \textit{neighborhood} of a node $i \in V$ is the set of nodes adjacent to $i$ and is denoted by
\begin{equation}
\mathcal{N}(i) = \{\, j \in V \mid (i,j) \in E \,\}.
\end{equation}
In signed networks, it is often useful to distinguish between neighbors connected by positive and negative edges.
Accordingly, we define the \textit{positive neighborhood} and \textit{negative neighborhood} of node $i$ as
\begin{equation}
\begin{split}
\mathcal{N}^{+}(i) &= \{\, j \in V \mid (i,j) \in E \text{ and } s_{ij} =+1 \,\}, \\
\mathcal{N}^{-}(i) &= \{\, j \in V \mid (i,j) \in E \text{ and } s_{ij} =-1 \,\}.
\end{split}
\end{equation}
Clearly, for unsigned networks we have $\mathcal{N}(i) = \mathcal{N}^{+}(i)$.
For directed signed networks, one may further distinguish between in- and out-neighborhoods.
The \textit{positive in-neighborhood} of node $i$ is defined as
\begin{equation}
\mathcal{N}^{+}_{\mathrm{in}}(i) = \{\, j \in V \mid (j,i) \in E \text{ and } s_{ji} =+1 \,\},
\end{equation}
with analogous definitions for $\mathcal{N}^{-}_{\mathrm{in}}(i)$, $\mathcal{N}^{+}_{\mathrm{out}}(i)$, and $\mathcal{N}^{-}_{\mathrm{out}}(i)$.

Neighborhood information can equivalently be encoded in the signed adjacency matrix $\mathbf{A}$. For an unweighted signed network, $A_{ij}=s_{ij}$ if $(i,j)\in E$ and $A_{ij}=0$ otherwise.
For a weighted signed network, we use $A_{ij}=x_{ij}$ if $(i,j)\in E$ and $A_{ij}=0$ otherwise.
% whose entries are $A_{ij}\in \{-1,0,1\}$, where $A_{ij}$ denotes the sign of edge $e_{ij}$.
For undirected signed networks, $\mathbf{A}$ is symmetric. For directed signed networks, $\mathbf{A}$ need not be symmetric. The diagonal entries $A_{ii}$ encode self-loops. If self-loops are excluded, then $A_{ii}=0$ for all $i\in V$.

\paragraph{Degree and signed degree}
The \textit{degree} $k_i$ of a node $i \in V$ is defined as the number of edges incident to it.
In directed networks, we distinguish between the \textit{in-degree} $k_i^{\mathrm{in}}$, counting edges pointing toward $i$, and the \textit{out-degree} $k_i^{\mathrm{out}}$, counting edges originating from $i$.
In signed networks, it is often useful to further separate the degree into its positive and negative components: the \textit{positive degree} $k_i^{+}$ is the number of positive edges incident to $i$, and the \textit{negative degree} $k_i^{-}$ is the number of negative edges incident to $i$.
Thus, for undirected signed networks without self-loops, we have
\[
k_i = k_i^{+} + k_i^{-}.
\]
When the network is weighted, these quantities can be generalized by summing over the absolute weights of the corresponding positive or negative edges instead of simply counting their number.

\paragraph{Walks, cycles and triads}
Two important but distinct concepts in the analysis of signed networks are \textit{walks} and \textit{cycles}.
A \textit{walk} of length $k$ in a network is an ordered sequence of nodes $(v_0, v_1, \dots, v_k)$ such that $(v_{\ell-1}, v_\ell) \in E$ for all $\ell = 1,\dots,k$.
A walk is said to be \textit{closed} if $v_0 = v_k$.
Closed walks may revisit nodes or edges multiple times.

A \textit{cycle} is a closed walk in which all intermediate nodes are distinct, i.e., $v_i \neq v_j$ for all $0 < i < j < k$.
Thus, cycles are simple closed paths without self-intersections.
In signed networks, the \textit{sign} of a walk or cycle is defined as the product of the signs of its constituent edges.
A walk (or cycle) is called \textit{positive} if this product is positive, and \textit{negative} otherwise.

A \textit{triad} consists of three nodes connected by three edges.
In an undirected network, a triad forms a cycle of length three.
For an undirected network, we denote the set of all existing triads by
\begin{equation}
\mathcal{T}
=
\left\{
(i,j,k)
\;\middle|\;
i,j,k\in V,\;
i<j<k,\,
(i,j),(j,k),(k,i)\in E
\right\}.
\end{equation}
Henceforth, we denote the total number of triads as
$N_T=|\mathcal{T}|$.
Throughout this review, we use the convention that $\Delta_i$ denotes a triad with exactly $i$ negative links ($i=0,1,2,3$).

Cycles play a central role in classical structural balance theory: a signed network is balanced if and only if all its cycles are positive~\citep{cartwright_1956, harary1959_measuringSB}.
For complete graphs, requiring all cycles to be positive is equivalent to requiring all triads to be balanced~\citep{cartwright_1956, antal_2005}.
Closed walks generalize cycles and provide a computationally convenient alternative, as their enumeration can be expressed through powers of the signed adjacency matrix.

Throughout the review, “positive” and “negative” refer to the sign product of a cycle, whereas “balanced” and “unbalanced” refer to its structural-balance status.
Under classical (strong) structural balance, these notions are equivalent for cycles: a positive cycle is balanced and a negative cycle is unbalanced.
This equivalence does not hold for weak structural balance, where a cycle may be negative yet still satisfy the weak-balance criterion.

\paragraph{Semicycles}
% Finally, let us define \textit{transitive semicycles}.
In a directed network, a \textit{semicycle} is a closed sequence of nodes $(v_0, v_1, \dots, v_k)$ with no repeated nodes except $v_0=v_k$, such that each consecutive pair is connected by a directed edge in either direction; for $k=2$, both reciprocal directed edges must be present~\citep{cartwright_1956}.
Unlike a cycle, the orientation of these edges need not agree with the order in which the nodes are traversed.
Thus, both $(i,j),(j,k),(k,i)$ and $(i,j),(j,k),(i,k)$ form semicycles.
For the triadic case relevant in this review, the latter is \textit{transitive} because its directed edges satisfy transitivity~\citep{Aref2020}.
In other words, a transitive semicycle contains a two-step directed path from one node to another and its shortcut, i.e., a direct edge from the initial source to the final destination bypassing the intermediate node.
%Transitive semicycles are used later when considering structural balance in directed networks~\citep{Doreian1996,Aref2020,Gorski2025}.

\subsubsection{Formal definition of structural balance}
\label{subsubsec:sb-formal}

In many social systems, relationships between individuals are not neutral but signed: they may be positive, as in friendship or cooperation, or negative, as in hostility or competition. Structural balance theory provides a framework for understanding how such signed relations organize at the network level.
Originating in social psychology with Heider's balance theory~\citep{heider1946attitudes} and later formalized in graph-theoretic terms by Cartwright and Harary~\citep{cartwright_1956}, structural balance captures the idea that some configurations of positive and negative ties are more stable or internally consistent than others.
%For example, triads consistent with the maxims ``the friend of my friend is my friend'' and ``the enemy of my enemy is my friend'' are balanced, whereas a configuration in which ``the friend of my friend is my enemy'' is unbalanced (see Fig.~\ref{fig:intro-balance-schematic}a).

\citet{heider1946attitudes} hypothesizes that an individual $P$ tends to form a positive tie with a second person $O$ when they hold the same attitude toward a third person $Q$, or impersonal entity $X$, and a negative tie when their attitudes differ.
In this first formulation, the focus is on the cognitive process of an individual person when assessing two types of triads: interpersonal triad $P$--$O$--$Q$ or a triad $P$--$O$--$X$ involving an impersonal entity.
Then a triad is referred to be balanced ``if all three relations are positive [...] or if two are negative and one positive''\citep[p. 110]{heider1946attitudes}.

Throughout this review, we distinguish between \textit{interpersonal} and \textit{impersonal} structural balance.
Interpersonal structural balance considers triads or cycles composed only of individuals.
Impersonal structural balance refers to the case where at least one element is an impersonal entity, such as an object, issue, or opinion.
However, following common later usage in the literature, we use the triad $P$--$O$--$X$ notation generically with $X$ representing either another person or an impersonal entity.

A signed graph is said to be \textit{structurally balanced} if every cycle has positive sign product, or equivalently, if every cycle contains an even number of negative edges~\citep{cartwright_1956,harary1953notion}. Harary's structure theorem states that this condition holds if and only if the vertex set can be partitioned into two subsets such that all positive edges lie within subsets and all negative edges lie between subsets~\citep{harary1953notion}.

A signed graph is \textit{weakly balanced} if it contains no cycle with exactly one negative edge~\citep{davis1967clusteringSB}. Davis's structure theorem states that a weakly balanced signed graph can be partitioned into an arbitrary number of subsets such that positive edges lie within subsets and negative edges lie between subsets~\citep{davis1967clusteringSB}. Thus, structural balance can be seen as a special case of weak structural balance.
Structural balance restricts the network to at most two internally positive and mutually antagonistic groups, whereas weak structural balance relaxes this restriction and allows any number of such groups.
The classical notion is therefore sometimes referred to as \textit{strong structural balance}.
Considering the structural-balance maxims, weak structural balance allows both ``the enemy of my enemy is my friend'' and ``the enemy of my enemy is my enemy''.
The latter is unbalanced under strong structural balance but balanced in the weak sense (Fig.~\ref{fig:balance-variants}a).

Several further variants of the definition for directed, weighted, multilayer, relaxed and antibalanced structures are briefly presented in the next section.

\subsubsection{Variants of the definition}
\label{subsubsec:sb-variants}

\begin{figure}[!htbp]
    \centering
    \includegraphics[width=\textwidth]{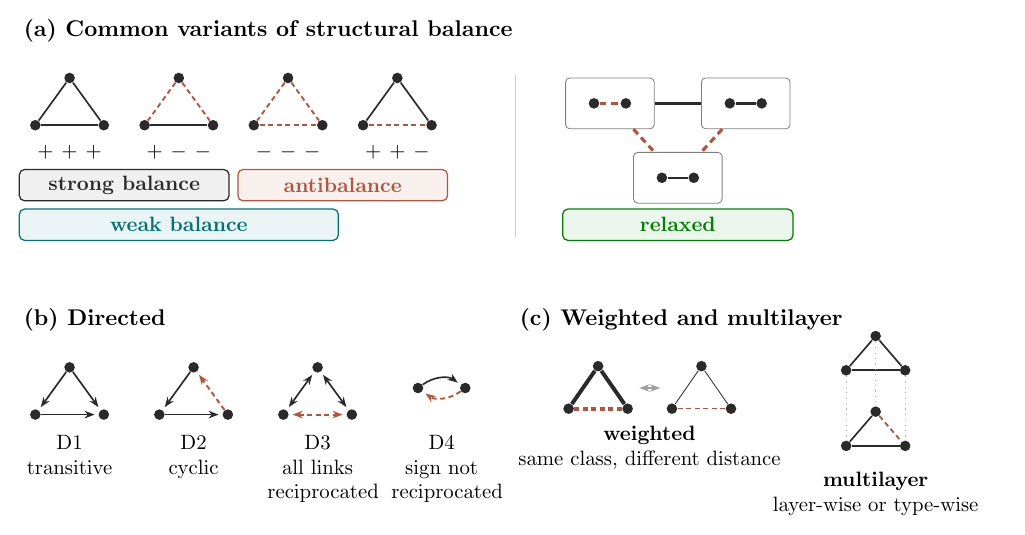}
    \caption{\textbf{Definitions of structural balance.}
Solid dark edges are positive, dashed red edges negative.
\textbf{(a)} Four different variants of structural balance: strong, weak, antibalance and relaxed.
The first three can be understood by looking at which triads they consider balanced and which not.
The relaxed version requires considering the block structure of the adjacency matrix.
\textbf{(b)} Different configurations in directed network: transitive and cyclic triads, reciprocated links in triads and reciprocated dyads.
Depending on the study and convention adopted to define structural balance, these are either considered or excluded (e.g., see Fig.~\ref{fig:triad-census-scopes}).
\textbf{(c)} Two types of extensions: weights change how far a network is from balance, and in multilayer networks, balance may be required layer by layer, type by type, or as a whole.
}
    \label{fig:balance-variants}
\end{figure}

The definitions above concern undirected, unweighted signed graphs.
Five variants recur in the literature reviewed here, each relaxing a different part.
We collect them in one place because they are not interchangeable: a statement that some network is ``balanced'' is only meaningful once the variant is named.
Fig.~\ref{fig:balance-variants} collects different definitions of structural balance and examples of triads discussed in directed, weighted, and multilayer networks.

\paragraph{Directed balance}
When edges are directed, even a set of three nodes may contain several distinct \emph{semicycles} (Sec.~\ref{subsubsec:signed-net-def}), and several inequivalent notions of balance coexist.
% Three are used in this review.
In the formulation of Cartwright and Harary for directed networks~\citep{cartwright_1956},
% First, one may disregard directions altogether and
balance is defined over {semicycles}: a semicycle is balanced when the product of the signs of its edges is positive, and a directed network is balanced when all of its semicycles are balanced.
% ; this recovers the undirected definition of structural balance.
This formulation includes semicycles of length two, formed by reciprocal \textit{dyads}, i.e., with $(i,j),(j,i)\in E$, as well as longer semicycles, and is also adopted by~\citet{talaga_polarization_2023}.

\begin{figure}[]
    \centering
    \includegraphics[width=\linewidth, clip, trim=0 0 0 0]{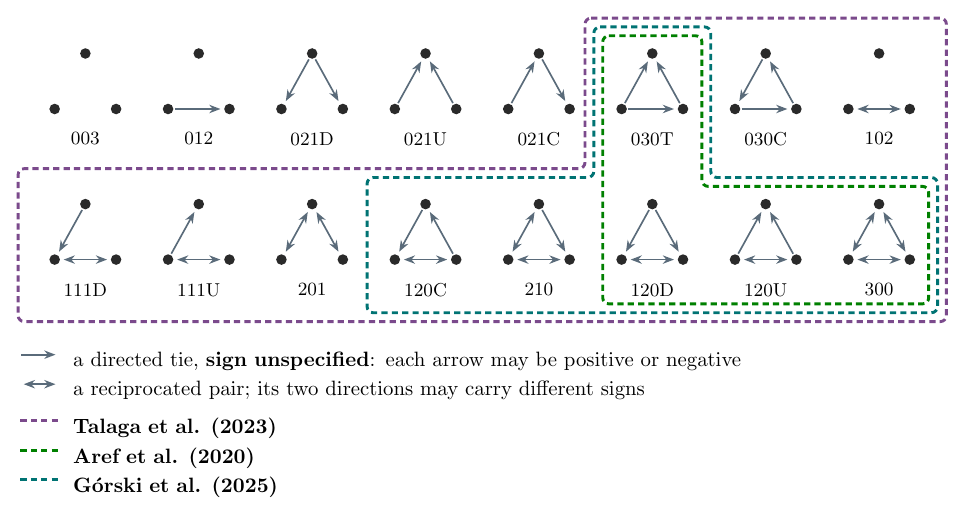}
    \vspace*{-0.7cm}
    \caption{\textbf{Which directed triads different approaches evaluate.}
The sixteen classes of directed triads, labeled as in~\citep{holland1977method}, constitute the triadic census. Each directed edge can either be positive or negative.
The dashed boxes indicate which triad classes enter the balance assessment in the three approaches shown.
% enclose the triads admitted by the conventions of three different approaches.
In the triadic census shown here, the semicycle approach adopted by~\citet{talaga_polarization_2023} corresponds to all triad classes containing at least one semicycle. Such a semicycle may be a length-two semicycle formed by a reciprocated pair or a length-three semicycle, yielding eleven of the sixteen types.
% , formed by a reciprocated pair, or of length three: eleven of the sixteen types.
% that is, whenever it closes a triangle once directions are ignored or contains a reciprocated pair: eleven of the sixteen types.
Within the transitive-semicycle approach,~\citet{Aref2020} evaluate four triad types (030T, 120D, 120U and 300), and a triad is considered balanced only if all of its transitive semicycles are balanced.
\citet{Gorski2025}, instead, consider all triads containing at least one transitive semicycle and evaluate balance separately for each transitive semicycle, allowing for partially balanced triads.
The five triads at the top left contain no semicycle and are excluded by all three approaches.
}
    \label{fig:triad-census-scopes}
\end{figure}

Other approaches restrict attention to other local configurations.
\citet{Doreian1996} consider only \emph{transitive} semicycles in triads, i.e., configurations $(i,j)$, $(j,k)$, $(i,k)$$\in E$.
% Two conventions build on this, differing both in which triads they evaluate and in how they score them.
\citet{Aref2020} evaluate four triad types and call a triad balanced only when all of its transitive semicycles are balanced, whereas \citet{Gorski2025} evaluate every triad containing at least one transitive semicycle and score each semicycle separately, so that a triad may be partially balanced (see Fig.~\ref{fig:triad-census-scopes}).

Balance can also be evaluated specifically at the \emph{dyadic} level.
% , where a reciprocated pair is balanced if $x_{ij}=x_{ji}$, so that balance reduces to sign reciprocity~\citep{mei2022structural,Gallo2025}.
\citet{Gallo2025} focus on cycles of length two and relate their balance or frustration to signed reciprocity, distinguishing strong and weak interpretations of balance.
These notions need not agree, and hence measures built on them can rank the same network differently.

\paragraph{Weighted balance}
When edges carry magnitudes, $x_{ij}=s_{ij}w_{ij}$ with $w_{ij}>0$, the \emph{binary} property is unchanged: because the weights are positive, the sign of a cycle product depends on the signs $s_{ij}$ alone, so a weighted signed graph is balanced exactly when its unweighted version is.
Weights therefore do not alter \emph{whether} a network is balanced; they alter \emph{how far} from balance it is, depending on the measure used.
% and they do so only through the measure one chooses.
% Two routes are used in practice.
Two corresponding extensions of balance measures are discussed in Sec.~\ref{subsubsec:directed-weighted}.
In weighted frustration measures, the magnitude scales the cost of a frustrated edge, so that strong relations are more expensive to violate than weak ones~\citep{aref_weightedFrustration}. In walk-based measures, edge magnitudes weight the contribution of closed walks, so that walks contaning weak ties contribute less~\citep{talaga_polarization_2023}.
Both reduce to the unweighted definitions when all $w_{ij}=1$.
Because the two approaches weight different objects, i.e., edges in the first case, walks in the second, they do not always agree on which of two networks is closer to balance.

\paragraph{Multilayer balance}
When the same actors are connected by several types of relation at once, balance may be required of each layer separately.
The layer-wise convention calls a multilayer network balanced when every layer is balanced on its own, even if the sign of a given pair differs between layers~\citep{Gorski2017}; interlayer coupling then acts as a force favoring agreement between replicas of the same link rather than as part of the balance condition (Sec.~\ref{subsubsec:multilayer}).
% An aggregate convention instead evaluates balance across levels of the same system, from triads to subgroups to the whole network, and treats these as distinct properties rather than as one quantity measured at different scales~\citep{Aref2020}.
Another possibility arises when a pair of actors is joined by exactly one link whose \emph{type} may vary; then, a triad of three positive links of two different types may count as unbalanced~\citep{Oloomi2021}.
A different kind of multiplicity appears in \citet{schweighofer_2020_2,schweighofer_2020}, where the several dimensions belong to the \emph{agents} rather than to their relations: each agent holds a vector of weighted attitudes from which the signed tie is derived, so balance is defined in an attribute space rather than across layers (Sec.~\ref{subsubsec:sb-agent-attribute}).
Balance in multilayer settings is therefore a statement about the chosen aggregation as much as about the signs.

\paragraph{Relaxed balance}
A further generalization of structural balance is \textit{relaxed structural balance}, introduced by \citet{Doreian2009partitioning}.
A signed graph has a relaxed balance structure if its nodes can be partitioned into an arbitrary number of subsets such that the links within each subset and from one subset to another have a consistent sign, i.e., they are either all positive or all negative.
Thus, unlike strong or weak structural balance, relaxed balance may allow negative relations within a subset or positive relations between subsets.
This broader formulation accommodates additional social mechanisms, such as mediation \citep{Doreian2009partitioning}.
We do not review relaxed structural balance in detail, but note its relevance because structures consistent with it have been reported in empirical networks~\citep{DiVece2025} and emerge in some of the dynamical models discussed later~\citep{Traag2013, Krawczyk2019Heider}.

\paragraph{Antibalance}
Finally, structural balance has a dual counterpart called \textit{antibalance}~\citep{harary1957structural}: ``a signed graph is antibalanced if every cycle has an even number of positive lines'', or equivalently if its nodes can be split into two subsets with all negative edges within subsets and all positive edges between them.
Signed graphs that are neither balanced nor antibalanced are called \textit{strictly unbalanced}, and this three-way classification has a spectral characterization: the spectral radius of the signed adjacency matrix is smaller than that of its unsigned counterpart if and only if the network is strictly unbalanced~\citep{tian2022spreading}.
The balanced/antibalanced/strictly unbalanced classification was developed in relation to dynamics and is based on the spectrum of the signed adjacency matrix.
The three classes produce different behavior under linear and nonlinear spreading processes~\citep{tian2022spreading}, with extensions to networks carrying complex weights~\citep{tian2024structural}.
We do not survey this classification further, but note that antibalanced networks and their properties are considerably less explored than those of balanced networks.

\FloatBarrier
\subsection{Structural balance and signed networks across domains}

Structural balance can, in principle, be defined and measured in any network whose edges carry positive or negative signs. Signed networks are common across empirical domains~\citep{diaz2025signed}.
In the literature where structural balance has been measured or discussed, we identify broad classes of signed networks: social, economic, international-relations, cognitive, and biological networks. In the following, we introduce these classes, summarize their typical structural properties, and review the extent to which structural balance is observed in each domain and what its presence or absence may imply.

\subsubsection{Social networks}

Structural balance was originally introduced in social psychology by Heider~\citep{heider1946attitudes} to describe how individuals experience cognitive tension in configurations of attitudes.
Cartwright and Harary~\citep{cartwright_1956} subsequently reformulated Heider's theory in graph-theoretic terms, thereby establishing structural balance as a theory of signed social networks.
In this formulation, nodes typically represent individuals or social actors, while edges represent positive or negative relations, such as friendship and enmity, trust and distrust, support and opposition, or cooperation and conflict.
These relations belong to the broader classes of social ties studied in social network analysis~\citep{borgatti2009network}.
Balance is first expressed at the level of triads and then generalized to the sign structure of cycles.

% \paragraph{Observation of social networks and typical properties}
Social networks are reconstructed through various techniques, including surveys, direct observation, archival records, and digital traces of interaction.
%~\cite{barbier2011data, catanese2011crawling}
They often display typical topological properties, including high clustering, short average path lengths, heavy-tailed degree distributions, and community structure~\cite{barabasi2016network}.
Related properties are observed both in face-to-face groups~\cite{atzmueller2011face, isella2011s}, local communities~\cite{chami2017diffusion,isakov2019structure,rogers2010diffusion}, and in large-scale online systems~\cite{ahn2007analysis,fu2008empirical,backstrom2012four}.
Signed social networks share these basic topological features but incorporate an additional layer of information through the sign of the links.
Empirically, positive ties are substantially more common than negative ones, leading to highly asymmetric sign distributions~\citep{leskovec2010countingTriads,ghasemian2024structure}.

Negative relations are often harder to measure than positive ones.
In surveys, asking respondents to nominate disliked, avoided, or conflictual alters can be socially sensitive, so elicited negative ties depend on the wording of the survey questions~\cite{labianca1998social}.
More generally, sensitive survey questions can produce response bias depending on privacy, wording, and the local context~\cite{krumpal2013determinants}.
Recent work has begun to systematize name-generator instruments for antagonistic ties~\cite{de2024systematic}, and to empirically map signed relations at scale in local communities~\cite{isakov2019structure,de2024systematic,ghasemian2024structure}.
In online platforms, the visibility of negative relations also depends on platform design, since some platforms provide explicit negative feedback channels, whereas others do not~\cite{Leskovec2010Predicting,leskovec2010countingTriads}.
As a consequence, signed online social networks are often reconstructed through indirect inference methods, including text-based extraction and sentiment analysis~\cite{keuchenius2021important} as well as sign prediction from network structure and behavior~\cite{Leskovec2010Predicting}.

\begin{figure}[t]
    \centering
    \includegraphics[width=0.9\linewidth]{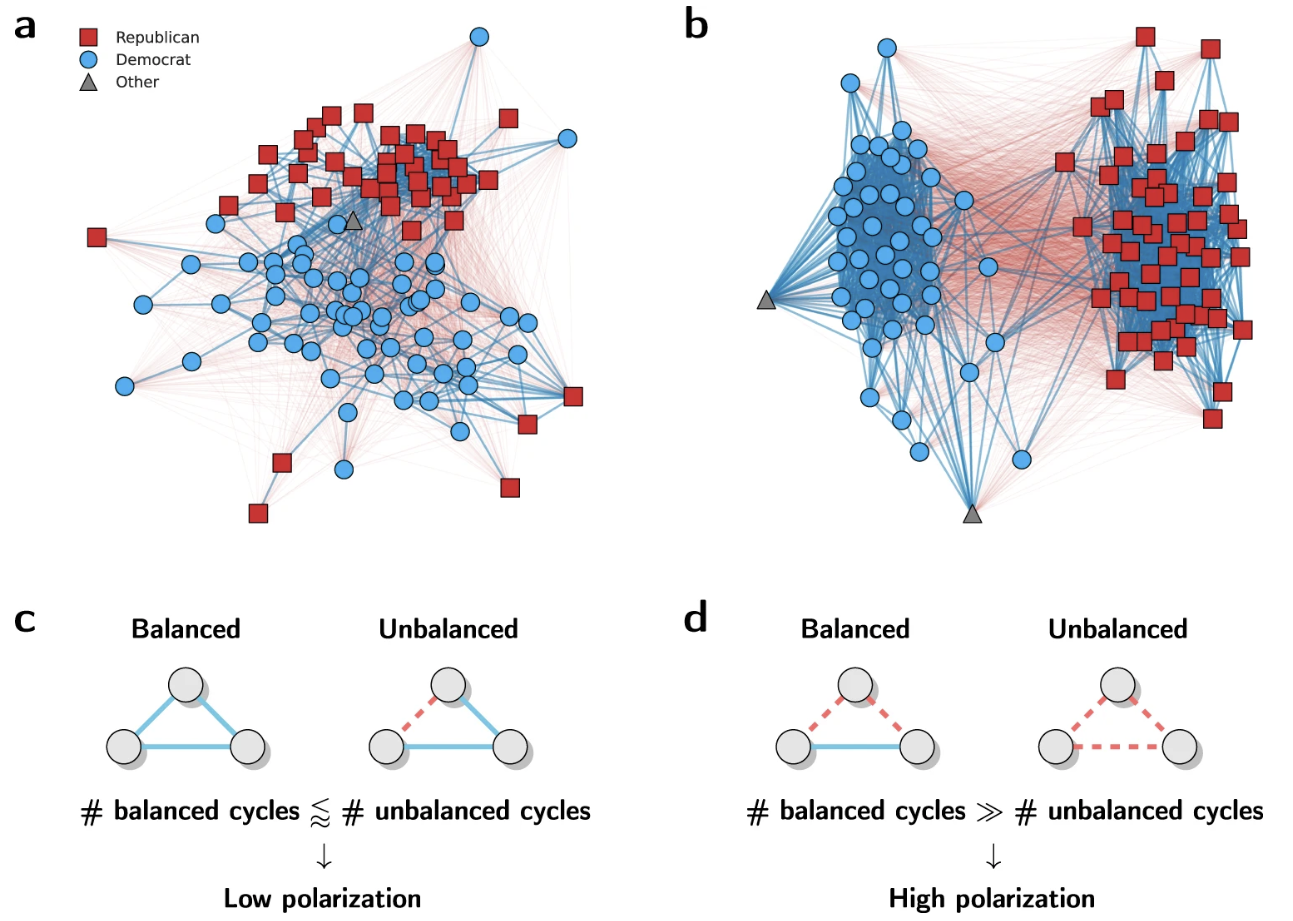}
    \caption{\textbf{Structural balance in an empirical political network.} Networks representing signed bill co-sponsorship relationships in the U.S. Senate during the 96th (1979-1981; \textbf{a}) and 114th (2015-2017; \textbf{b}) Congresses. Blue links denote positive relations, whereas red links denote negative relations. The more polarized network exhibits the characteristic balanced structure of predominantly positive within-group and negative between-group ties. Panels (\textbf{c,d}) illustrate how higher polarization is associated with a predominance of balanced over unbalanced cycles. Reproduced from~\citep{talaga_polarization_2023}; licensed under a Creative Commons Attribution 4.0 International license.}
    \label{fig:congress}
\end{figure}

The first systematic empirical investigations of structural balance were conducted primarily in the 1960s and 1970s within experimental and perceptual psychology~\cite{galesic2025experimental}.
These studies typically examined whether balanced triads were perceived as more pleasant, stable, or cognitively coherent than unbalanced ones~\cite{gerard1967recall}.
Subsequent research shifted toward observational and longitudinal studies of real social systems.
Small-group settings, such as classrooms~\cite{rambaran2015development}, religious communities~\cite{sampson1968novitiate}, and sport organizations~\cite{zachary1977information}, were analyzed using repeated surveys to track the evolution of signed relations over time.
These datasets show clear tendencies toward balanced configurations and factional splits.
Signed relations have also been inferred in political networks, most notably in the US Congress and Senate~\cite{Neal2020,talaga_polarization_2023}, using voting patterns or bill co-sponsorship behavior as proxies for positive and negative ties (see Fig.~\ref{fig:congress}).
These studies provide evidence that balance tends to increase over time and is closely linked to polarization and the emergence of cohesive factions.

In the 2000s, the availability of data from large online platforms enabled the study of structural balance at unprecedented scales.
Online platforms such as \texttt{Epinions}, \texttt{Slashdot}, and Wikipedia administrator elections allow users to explicitly express positive and negative relations, resulting in signed networks with thousands of nodes and hundreds of thousands of links~\cite{Leskovec2010Predicting,leskovec2010countingTriads}.
Related analyses have extended balance theory to multiplayer online games~\cite{szell_2010,szell2012SignedMMOGmoreInfo,belaza2017_StatPhysBalanceTheory,Belaza2019}, Bitcoin trust networks~\cite{Kumar2017}, and communication networks among traders~\cite{Askarisichani2019}.
%These data made it possible to quantify deviations from random expectations, test balance-related hypotheses statistically, and explore temporal dynamics over long observation windows.
More recently, structural balance has been investigated using large-scale longitudinal experiments.
In particular, group experiments involving several hundred participants have tracked the coevolution of dyadic relations and triadic structures over time~\cite{galesic2025experimental}.
By comparing the predictive power of competing dynamical models~\citep{MinhPham2020, pham2021balance}, experimental approaches provide evidence supporting or invalidating mechanisms that might drive the emergence of structural balance (see Section~\ref{subsec:cognitive_balance}).

\subsubsection{International-relations networks}
\label{subsubsec:ir-networks}

\begin{figure}[!htbp]
    \centering
    % \fbox{\parbox[c][42mm][c]{0.85\linewidth}{\centering\itshape Note for reviewers and editors: Proposed figure to be included after copyright clearance.\\ Intended figure: Cold War alliance network (1960), with positive and negative links shown separately. See caption for further details.}}
    \includegraphics[width=\textwidth]{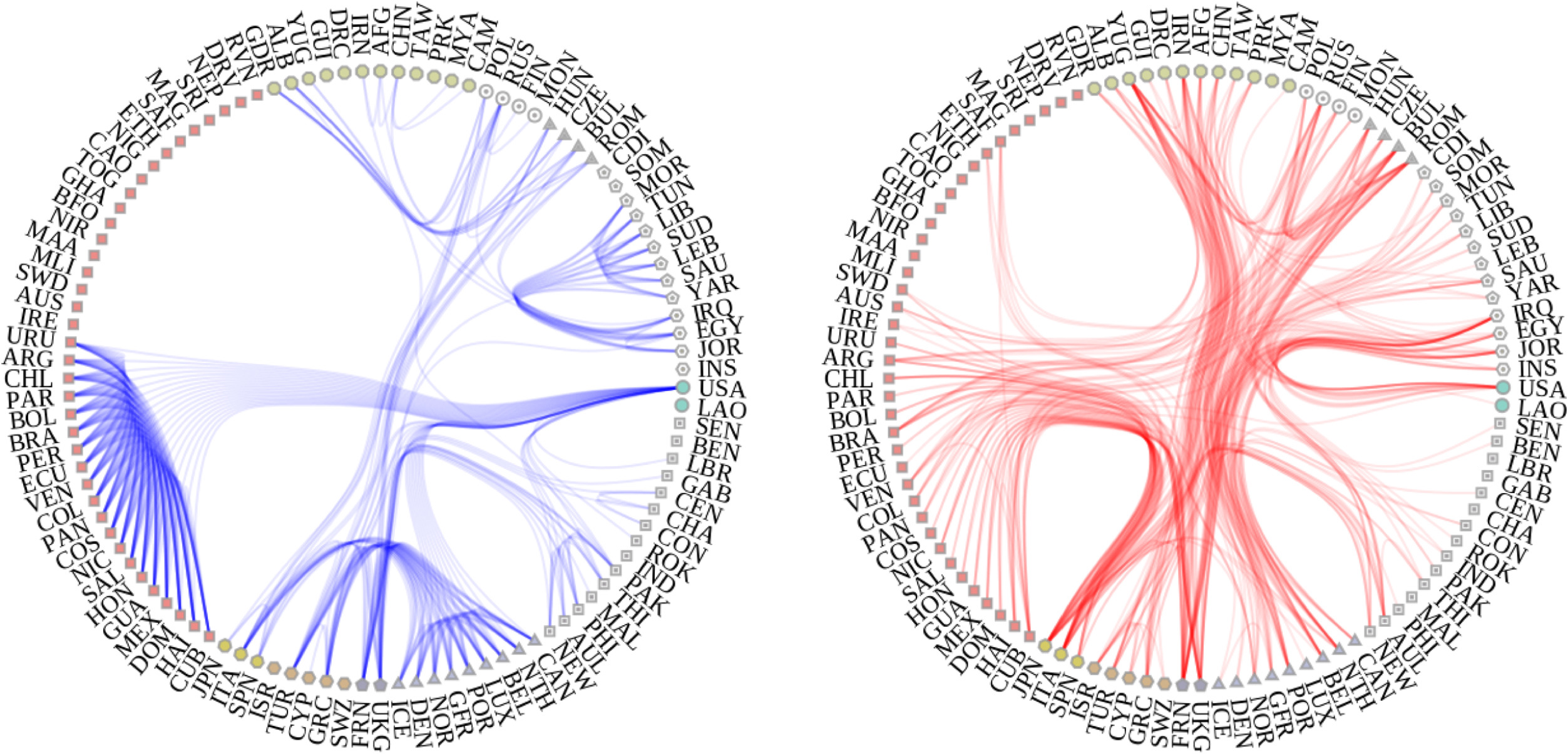}
    \caption{
    % \textbf{Note for reviewers and editors: The above figure is Fig.~1 from~\citep{Belaza2019}. Before publishing one needs to obtain copyright clearance.}
\textbf{Signed network of international relations.}
Nodes are sovereign states and links encode alliance (left) and hostility (right), drawn in two separate panels, with node colors giving the blocks recovered from the network.
A balanced configuration corresponds to a partition of the system into mutually antagonistic blocs.
Reprinted from~\citep{Belaza2019}, Copyright 2019, with permission from Elsevier.
}
    \label{fig:ir-networks}
\end{figure}

In international-relations networks, nodes most commonly represent sovereign states, although related applications also consider coalitions, militant organizations, or other political actors (see Fig.~\ref{fig:ir-networks}).
Positive edges may represent military alliances, defense-cooperation agreements, shared international institutional affiliations, or cooperative political actions, whereas negative edges may encode interstate rivalries, militarized disputes, warfare, or other hostile actions~\citep{harary1961structural,doreian,Bellutta2020,Gallo2024, Fritz2025}.
Depending on the available data, these relations may be binary or weighted, directed or undirected, and may appear, disappear, or change signs over time.
This makes international relations a natural domain for structural balance theory because bilateral cooperation and conflict are interdependent: the relation between two actors may be shaped by their respective relations with common partners or adversaries.

International-relations networks are reconstructed primarily from curated historical datasets and event data.
% [add citations to following datasets: COW, Fritz: (DCA data collected by
% Kinne (2020), Militarized Interstate Dispute (MID) Data provided by Palmer et al. (2022).].
Among the most widely used sources are the Correlates of War datasets on formal alliances~\citep{Gibler2009}, membership in intergovernmental organizations~\citep{Pevehouse2020}, and militarized interstate disputes~\citep{Palmer2022}.
Studies of structural balance use these data to infer positive relations from formal alliances and, in some cases, common membership in intergovernmental organizations, and negative relations from militarized interstate disputes or other acts of aggression~\citep{doreian,Dekker2024}.
Other studies combine formal alliance and interstate-rivalry records~\citep{Oishi2025}, use defense-cooperation agreements together with militarized disputes~\citep{Kinne2020, Fritz2025}, or construct signed and sometimes weighted relations from news-derived event data~\citep{askarisichani20201995}.
Similarity in United Nations voting records has also been used to infer continuous levels of friendliness between countries~\citep{Summers2013}.
These networks are generally sparse and strongly time-dependent: their size changes as states enter and leave the international system, while their density and sign composition vary across historical periods.
They also display pronounced coalitional and bloc structure, although the composition and number of blocs may change over time~\citep{doreian,Belaza2019,Oishi2025}.

A central methodological difficulty is the interpretation of absent or neutral relations.
In long-term interstate networks, most pairs of states have neither an observed alliance nor an overt conflict, so neutral or inactive ties can constitute the large majority of all possible relations~\citep{Belaza2019,Dekker2024}.
Treating such ties as missing information, rather than as a distinct relational state, can therefore discard an important part of the network structure.
The operationalization of positive and negative ties is similarly consequential.
For example, short-lived military confrontations and long-term interstate rivalries capture different forms of antagonism, while defense-cooperation agreements and formal alliance treaties capture cooperation on different temporal scales~\citep{Oishi2025,Fritz2025}.
Consequently, conclusions about structural balance may depend on whether the analysis distinguishes active, neutral, and absent relations and on how cooperation and conflict are defined.

Whether the international system is globally balanced is contested: longitudinal studies find no monotone drift toward balance~\citep{doreian}, the system is nonetheless closer to balance than sign-reshuffled null networks~\citep{Aref_2018}, and support varies markedly across historical periods~\citep{Oishi2025}.
Local triadic interdependence is better supported than global convergence~\citep{Belaza2019}.
The outcome-level implications of this partial drive toward balance are treated in Section~\ref{subsec:ir}.

\subsubsection{Economic networks}

\begin{figure}[]
    \centering
    % \fbox{\parbox[c][42mm][c]{0.85\linewidth}{\centering\itshape Note for reviewers and editors: Proposed figure to be included after copyright clearance.\\ Signed stock-correlation network before, at, and after a loss of balance. See caption for further details.}}
    \includegraphics[width=\textwidth, clip, trim = 0 1700 900 500]{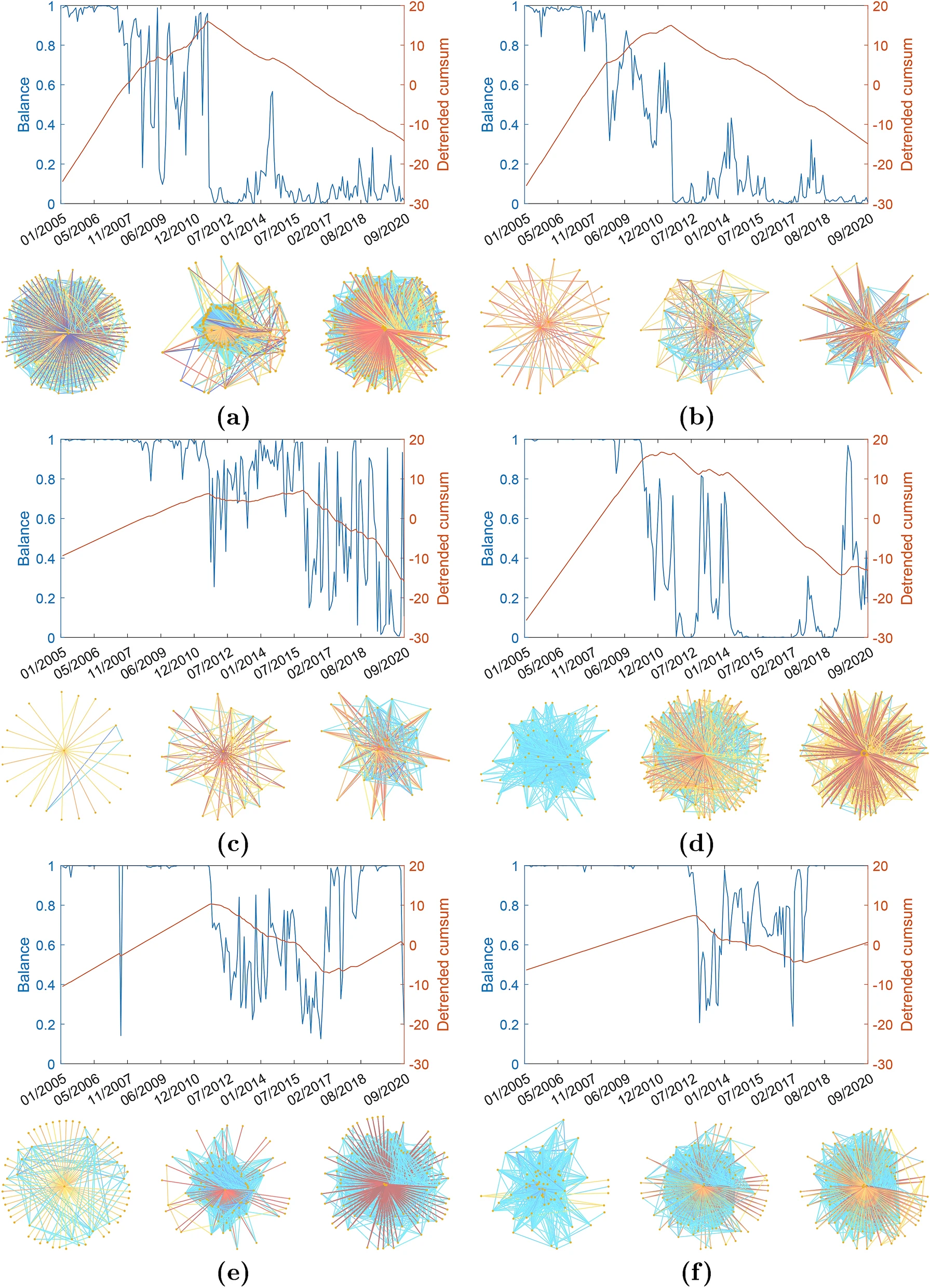}
    \caption{
    % \textbf{Note for reviewers and editors: The above figure is from panel (a) of Fig.~2 from~\citep{ferreira_loss_2021}.
    % Before publishing one needs to obtain copyright clearance.}
    %Proposed source: panel (a) of Fig.~2 from~\citep{ferreira_loss_2021}. If not, the full Fig.~2 is also acceptable.}
    \textbf{Signed network representing a stock correlation matrix.}
    Nodes are financial assets of the US stock market, and link colors encode Kendall’s tau estimates between their returns, varying continuously from red for negative values to blue for positive values.
    The three snapshots are taken before, at, and after a sudden loss of balance, so that a period of market stress can be read directly from the network as a reorganization of the negative links.
    Selected panels reproduced from~\citep{ferreira_loss_2021} under the Creative Commons Attribution 4.0 International License.}
    \label{fig:economic-networks}
\end{figure}

In economic networks nodes represent economic actors such as firms, banks, financial assets, or even countries, and links encode mutual interactions, such as collaboration, trading relationships, credit-debt relations, and correlations between prices~\cite{schweitzer2009economic, jackson2010social} (see Fig.~\ref{fig:economic-networks}).
%cooperative or competitive interactions
%These networks are particularly relevant for balance theory because they capture strategic behavior among interdependent actors whose relationships may be simultaneously cooperative and adversarial.
These interactions are typically weighted and directed, reflecting heterogeneity and asymmetry such as buying vs selling~\cite{chakraborty2024inequality}, exports versus imports~\citep{de2014network,bhattacharya2008international}, or alliances versus competition~\cite{tomasello2017data, sytch2014friendsandfoes}.
Empirical studies report characteristic structural properties including heavy-tailed degree distributions, community organization, and core--periphery structures~\citep{tomasello2017data, tomasello2017rise,
fricke2015core, Kuyyamudi2019}.

Structural balance has been studied in a variety of economic contexts.
In inter-firm networks, positive links often represent R\&D alliances~\citep{tomasello2017rise}, partnerships~\cite{sytch2014friendsandfoes}, or other forms of cooperation~\cite{Choi2009triadsinsupplynetworks}, while negative links may encode legal disputes~\citep{kito2021inter}, competitive rivalries~\cite{sytch2014friendsandfoes}, or other adversarial relations~\cite{Choi2009triadsinsupplynetworks}.
Balance theory has also been applied to financial networks, where nodes may represent stocks and links capture correlations or anti-correlations between asset returns~\citep{Kuyyamudi2019,Ehsani2020,ferreira_loss_2021,Zahedian2022}.
The level of balance in these networks has been linked with stock predictability~\citep{ferreira_loss_2021} and system risk~\citep{Kuyyamudi2019,bartesaghi2025global}.

In the microeconomic context, these networks are a natural setting for structural balance because their signs are the outcome of strategic choices by interdependent actors who can be cooperative and adversarial at the same time~\citep{sytch2014friendsandfoes,jackson2010social}.
The formation of these ties is often described using game-theoretic models: agents choose whom to befriend or fight to maximise a payoff.
Then, balanced or factional configurations arise as equilibria of the resulting game rather than as the endpoint of a relaxation dynamics~\citep{Hiller2017,Konig2017}.

\FloatBarrier
\subsubsection{Cognitive networks}
\label{subsec:cognitive}

\begin{figure}[!htbp]
    \centering
    % \fbox{\parbox[c][42mm][c]{0.85\linewidth}{\centering\itshape Note for reviewers and editors: Proposed figure to be included after copyright clearance.\\ Empirical belief networks for two contested topics. See caption for further details.}}
    \includegraphics[width=\textwidth, clip, trim= 0 180 0 0]{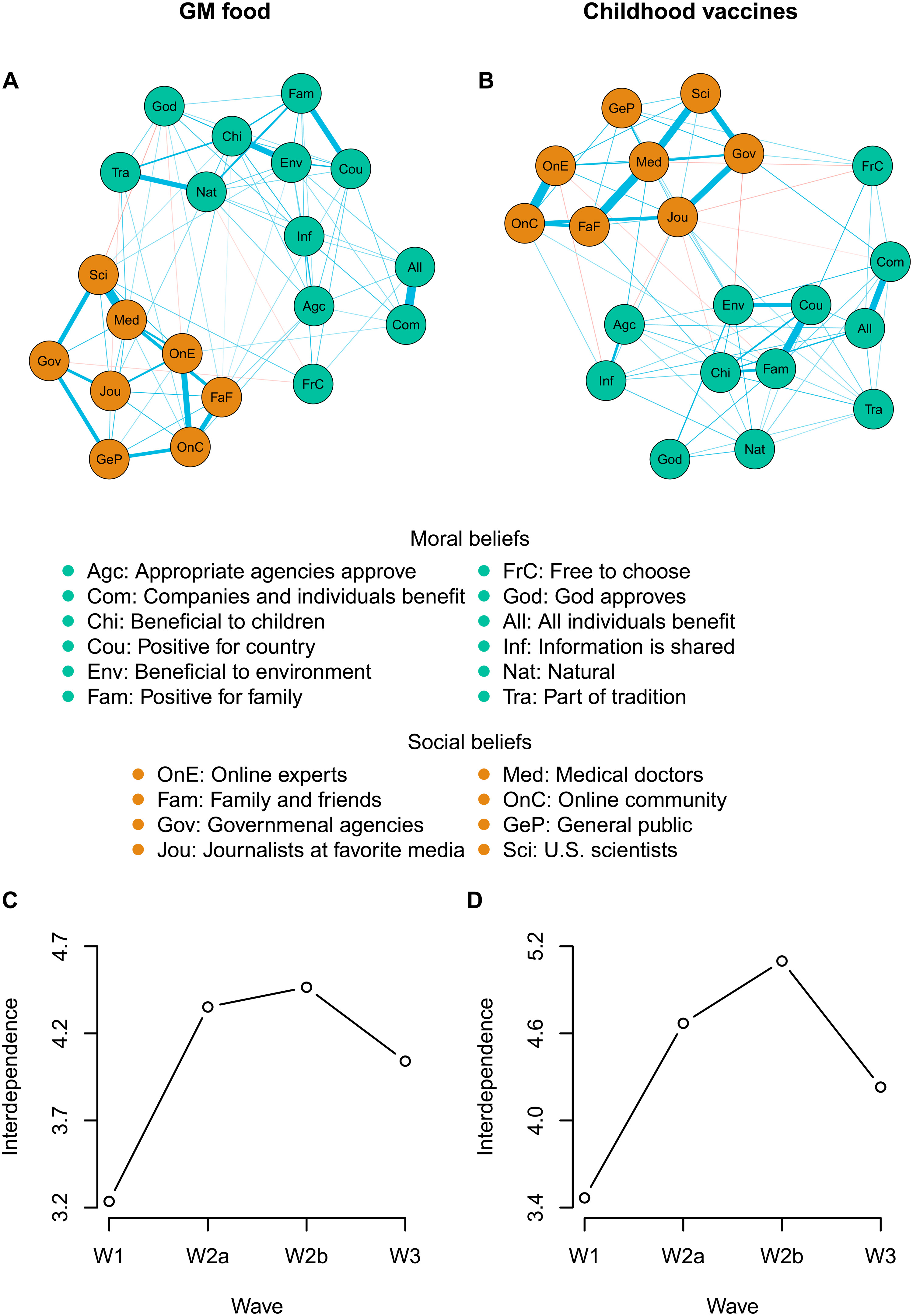}
    \caption{
    % \textbf{Proposed source: Fig.~2, panels A and B with legend, from~\citep{dalege2022using}.}
% \textbf{Note for reviewers and editors: The above figure is Fig.~2 from~\citep{dalege2022using}. Before publishing one needs to obtain copyright clearance.}
\textbf{Signed network of a belief system.}
Nodes are individual beliefs elicited in a longitudinal survey, with moral beliefs shown in green and social beliefs in orange. Links represent partial correlations between pairs of beliefs: blue links indicate positive correlations, red links negative correlations, and link width represents correlation strength.
% mutually reinforcing or mutually inconsistent pairs, so that
In this representation, structural balance can be interpreted as the internal coherence of the belief system.
The two panels show the networks reconstructed for beliefs about genetically modified food (left) and childhood vaccination (right).
% In both panels, beliefs separate into two groups: moral beliefs (green) and social beliefs (orange), which are densely and positively connected within groups and more sparsely and negatively connected between groups.
Selected panels reproduced from~\citep{dalege2022using} under the Creative Commons Attribution 4.0 International License.
}
    \label{fig:cognitive-networks}
\end{figure}

Cognitive networks model the organization of mental representations such as concepts, beliefs, opinions, and words, among others (see Fig.~\ref{fig:cognitive-networks}). They capture how cognitive elements are interconnected and how their structure affects reasoning, learning, and decision-making~\citep{vitevitch2020network,dalege2024networks}. In contrast to social networks, where nodes represent individuals, cognitive networks treat mental elements themselves as nodes and represent associative, logical, or affective relations as edges. The network approach has its roots in psycholinguistics research that developed associative and semantic network models to explain lexical access and memory retrieval. In particular, early works represented concepts as nodes linked by weighted associations~\citep{collins1975spreading,steyvers2005large}. These works established the idea that cognition is structured and relational, rather than a collection of independent propositions.
The growth of network science provides formal tools to analyze these relational structures quantitatively. Concepts, attitudes, and beliefs can be represented as interconnected systems whose topology affects cognitive dynamics~\cite{rodriguez2016collective}.

Classical work in social psychology emphasizes that people experience psychological discomfort when holding incompatible beliefs, and they are motivated to restore coherence, a process described by cognitive dissonance theory~\citep{festinger1957theory}.
In the network framework, structural balance theory becomes directly relevant: unbalanced triads in a network of beliefs can be interpreted as forms of cognitive dissonance, whereas balanced configurations correspond to internally coherent belief systems~\citep{heider1946attitudes,cartwright_1956}.
This interpretation provides a bridge between psychological theories of cognitive consistency and network science modeling~\citep{dalege2024networks}.

The specification of cognitive networks varies depending on the research theme and questions. In semantic or conceptual networks, nodes represent concepts and edges represent associative relations, often inferred from word co-occurrence, free association norms, or syntactic dependencies~\citep{motter2002topology,teixeira_revealing_2021}. Depending on the research question, edges may encode similarity, causal implication, agreement, or affective polarity. In belief or attitude networks, nodes represent propositions or evaluative statements, and edges capture perceived logical, causal, or emotional connections among them~\citep{thagard1989explanatory,abdelzaher2021belief,aiyappa2024emergence}. These networks can be empirically reconstructed using survey data, longitudinal panel studies, or text analysis.
For example,~\citet{dalege2022using} infer networks of moral and social beliefs about child vaccination and genetically modified crops, tracking their structural properties and temporal evolution. In organizational contexts, survey data have been used to map belief networks related to risk management and strategic practices~\citep{ellinas2017dynamics}.

The structure of such cognitive networks -- and their degree of structural balance -- influences belief formation and the diffusion of new belief systems~\cite{rodriguez2016collective,aiyappa2024emergence}.
Empirical studies suggest that semantic networks typically display heavy-tailed degree distributions and modular structure, with highly-connected hub concepts coexisting with more sparsely connected nodes~\citep{steyvers2005large,motter2002topology}. Reconstructions of belief networks similarly reveal a clustered organization, often corresponding to thematic or evaluative dimensions of the belief system~\citep{dalege2022using}. These networks tend to exhibit predominantly positive connectivity, which reflects coherent belief clusters, but may also contain negative ties representing perceived contradictions between propositions. For example, longitudinal survey-based reconstructions have identified structurally negative relations between beliefs grounded in different epistemic or moral frameworks~\citep{dalege2022using}.
Building on these works, this Review focuses on how principles of structural balance apply to cognitive networks. Specifically, it surveys model-based findings on when and how individuals might form balanced belief systems, and how balance-based mechanisms affect belief spreading, stability and change (see Section~\ref{subsec:cognitive_balance}).

\subsubsection{Emerging biological and ecological applications}

In several biological domains, signed network representations make it possible to compute structural-balance measures, but the theoretical interpretation of these measures is still less established than in social and cognitive networks.
Here, we briefly outline emerging biological applications in which balance can be measured, while the mechanisms expected to generate it remain domain-specific and only partly consolidated.

\paragraph{Molecular and cellular networks}
Biological networks provide a rich and diverse setting in which it is possible to investigate structural balance, as they encompass regulatory, metabolic, and signaling networks.
Many biological networks are directed, weighted, and modular. Gene regulatory and transcriptional networks display hierarchical modules and stable feedback patterns, consistent with their high balance levels~\cite{Iacono2010,IaconoAltafini2010}, whereas metabolic and signaling systems are more heterogeneous and thus more structurally unbalanced.
In gene regulatory networks,~\citet{Rizi2021} show that cancer cells tend to contain fewer unbalanced triangles than normal cells, with imbalances concentrated in the giant component.
Signed gene-gene relationships inferred from similarity matrices~\cite{Allahyari2022} support a strong version of balance, and incorporating longer cycles reveals that networks of non-essential genes are more balanced.
Broader analyses of transcriptional, signaling, and metabolic networks~\cite{Iacono2010,IaconoAltafini2010} show that transcriptional systems are highly balanced, while signaling and metabolic ones are not.
Other work links unbalancedness to monotone systems, showing that the distance of several biological networks from being monotone equals their minimum number of negative fundamental cycles~\cite{Iacono_frustrationAlgorithms}.
At the same time, the interpretation of structural balance in biological networks is less settled than in social systems, leaving room for future work to clarify what balance may explain and predict in these domains.

\begin{figure}[!htbp]
    \centering
    % \fbox{\parbox[c][42mm][c]{0.85\linewidth}{\centering\itshape Note for reviewers and editors: Proposed figure to be included after copyright clearance.\\ Construction of a signed brain network from resting-state fMRI time series. See caption for further details.}}
    \includegraphics[width=\textwidth]{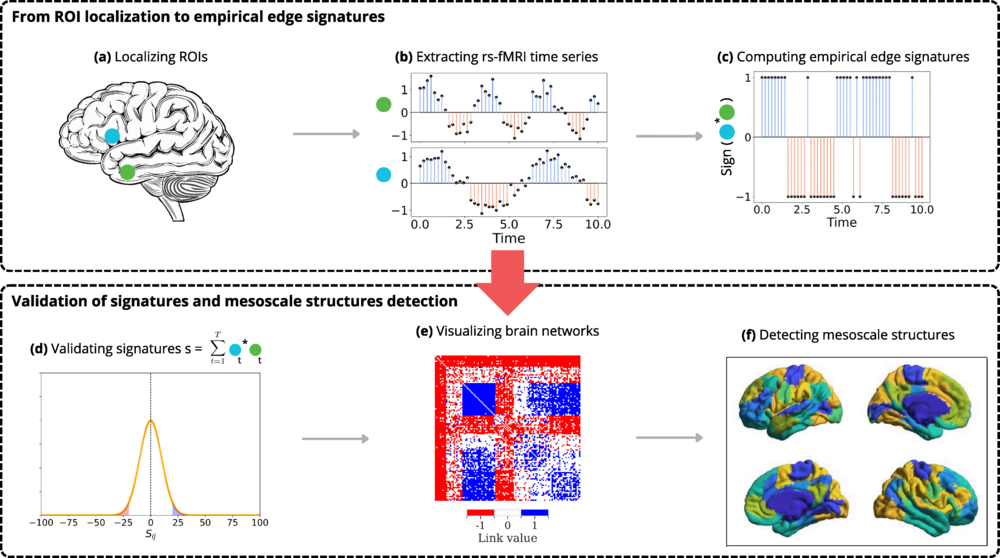}
    \caption{
    % \textbf{Proposed source: Fig.~1 from~\citep{DiVece2025}.}
    % \textbf{Note for reviewers and editors: The above figure is Fig.~2 from~\citep{dalege2022using}. Before publishing one needs to obtain copyright clearance.}
\textbf{Signed network of brain activity.}
Nodes are brain regions and the sign of a link records whether the activity of two regions is concordant or discordant significantly more often than a null model would predict. Thus, the sign is the outcome of a statistical test rather than of the raw correlation.
The panels trace the construction from the recorded time series to the validated signed projection and its communities.
    Reproduced from~\citep{DiVece2025} under the Creative Commons Attribution 4.0 International License.
}
    \label{fig:brain-networks}
\end{figure}

\paragraph{Brain networks}
In neuroscience, signed networks are constructed by looking at correlation and anti-correlation of brain regions~\citep{Moradimanesh2021, DiVece2025} (see Fig.~\ref{fig:brain-networks}).
Brain networks display typical properties, such as modularity, rich-club cores, and asymmetric signed connectivity, that modulate the distribution of signed triads~\citep{DiVece2025}.
However, evidence of structural balance is mixed.
For example, \citet{DiVece2025} find evidence aligning to a relaxed balance theory as they show a predominance of $(+,-,+)$ triads and very few $(-,-,-)$ triads.
A recent work explicitly tests structural balance theory in functional brain networks~\cite{saberi2025empirical}.
It finds that balanced triads are more temporally stable, while unbalanced triads are more transient, supporting the strong version of structural balance theory in dynamic signed brain networks.

\paragraph{Animal social networks}

Animals also organize in groups and display social behavior, including cooperation, affiliation, avoidance, aggression, and dominance~\cite{wey2008social}.
Animal social networks have been reconstructed in several systems, including mobile fish groups~\citep{rosenthal2015revealing}, baboons~\citep{Gelardi2019}, and wild rock hyraxes~\citep{Ilany2013}.
In such animal social systems, interactions may be characterized as cooperative or antagonistic, and structural balance has been studied in this context.
Structural balance appears in baboon co-location networks~\citep{Gelardi2019} and in wild rock hyraxes~\citep{Ilany2013}, where new individuals create social instability counteracting the tendency toward balance.
Animal social networks often show community structure and cooperative clustering, and empirical studies indicate that their signed ties can organize toward balanced configurations~\citep{Gelardi2019,Ilany2013}.

\paragraph{Ecological networks}

In ecology, there is a long tradition in representing interactions between species through networks.
Living organisms are grouped in ecological populations or communities, and their interactions are captured by a signed community matrix~\cite{May1973}.
The analysis of this signed matrix helps to understand the stability of ecological systems~\cite{May1973,Jeffries1974,Logofet1982}.
Researchers often examine the role of positive interactions, such as mutualistic interactions~\cite{bascompte2003nested}, or negative interactions, such as competition and parasitism~\cite{wilson1975sociobiology}, and sometimes consider both together~\cite{bastolla2009architecture,saiz2017evidence}.
A typical example of signed network in ecology is food webs, where interactions among prey and predators are represented as directed links that are either positive or negative depending on the direction considered~\cite{polis1996foodwebs}.
Signed links can also be inferred among organisms that interact indirectly because they share a prey or predator, as in apparent competition~\cite{pimm1991balance}, or to represent spatial association and co-location patterns~\cite{saiz2017evidence,saiz2018structure}.

In this literature, stability has been examined from various perspectives~\cite{mariani2019nestedness}, including linear stability analysis (that is, through the eigenvalues of the community matrix)~\cite{May1973,allesina2012stability,mccann2000diversity}, structural stability (that is, the study of how topological patterns affect the parameter regions where a stable coexistence of species is feasible)~\cite{rohr2014structural}.
Explicit applications of structural balance in ecology are more recent.
In particular, spatial ecological networks have been analyzed using structural-balance concepts, with positive and negative links inferred from spatial associations among species~\cite{saiz2017evidence,saiz2018structure}.
Thus, ecological networks can potentially provide a useful bridge between signed-matrix stability analysis and structural balance, although the connection remains less developed than in social networks.

\subsubsection{Inference of signed networks}
% another possible title: How structural balance measure contribute to network analysis
% \subsubsection{Social information extraction: Sign observation and inference}
\label{subsec:sign-obs-and-inf}
In the most general sense, a positive sign in a signed network denotes a relationship of attraction, affinity, or facilitation, while a negative sign represents repulsion, antagonism, or inhibition.
The precise meaning depends on the domain under study: in social systems, positive ties often represent friendship, trust, or alliances, while negative ties signify hostility, distrust, or rivalry; in biological networks, positive and negative edges may indicate activation or suppression between genes or neurons; in political or economic settings, they may reflect cooperative or competitive interactions among actors.

To conduct data-driven analyses of signed networks, one must first identify or infer the signs of the edges.
The most direct approach is to observe signed relations explicitly, for example through surveys or self-reported relationships on online platforms.
However, such data are often unavailable, incomplete, or difficult to collect at scale.
As a result, considerable effort has gone into developing methods to infer edge signs from various types of observed data—ranging from interaction logs to textual content or structural features of the network itself.

\paragraph{Inference from interaction data}

When direct observations of signed relationships are unavailable, signs can be inferred from patterns of repeated interactions between nodes.
Such interactions (e.g., co-authorship, co-attendance, co-sponsorship, and co-location) are often captured as multi-edge networks, where edge multiplicity reflects the frequency of interaction.
In these settings, the challenge lies in translating interaction frequency into relational sign: do repeated interactions imply affinity, and does their absence signal antagonism?

A common methodological strategy is to compare the observed interaction frequency between node pairs against what would be expected under a null model of random interactions.
Let $A_{ij}$ denote the observed number of interactions between nodes $i$ and $j$, i.e., the corresponding entry of the weighted adjacency matrix of the interaction network, and let $X_{ij}$ denote the corresponding random variable under the null model.
% Formally, one estimates the probability that the observed number of interactions $A_{ij}$ between nodes $i$ and $j$ exceeds the expected number under the null:
One may then consider the probability

\begin{equation}
    \mathbb{P}_{\text{null}}(X_{ij} < A_{ij})\;.
\end{equation}

From this statistical comparison, different authors infer signs using various approaches.
~\citet{neal_2014_backbone} and~\citet{Aref_2021} apply hypothesis testing to identify statistically significant over- or under-represented ties, assigning binary signs based on thresholded $p$-values.
~\citet{Gelardi2019} use $z$-scores from the null-model comparison to assign continuous signed weights.
Separately,~\citet{Andres2023} infer weighted signed ties by computing linear combinations of link probabilities, resulting in a fully continuous signed network.

A critical modeling choice is the null model itself, as it strongly influences which ties are deemed surprising and thus assigned a positive or negative sign.
While multi-edge configuration models are often used due to their simplicity and node-preserving properties, other alternatives such as stochastic block models (SBMs) can be employed~\citep{Andres2023}.
However, incorporating group structure into the null model introduces interpretive challenges: if community structure is already encoded in the null model, then a positive residual implies a relationship that exceeds group-level expectations.
Yet, it is precisely such community structure that often underlies the formation of positive ties.
Disentangling intrinsic group influence from genuine dyadic affinity thus becomes a subtle and non-trivial task in signed network inference.

A related but distinct line of work infers signed edges directly from interaction metadata that already encode signs.
This is the case when platforms allow users to express approval or disapproval through structured actions—such as upvotes/downvotes, helpful/unhelpful ratings, or retweet versus reply behaviors.
In such signed multi-edge settings, individual interactions already carry sign information.
~\citet{Fraxanet2024} leverage this by aggregating positive and negative feedback on platforms like Birdwatch and Twitter to infer dyadic polarity.
They use a Bayesian model to estimate whether a user pair’s interaction pattern is predominantly positive or negative, assigning signs only when the evidence is strong.

\paragraph{Inference from text}
Signed ties can also be inferred by extracting semantic cues such as sentiment, morality, or trust from textual exchanges between nodes.
For instance,~\citet{Diesner2015} apply sentiment analysis to the Enron email corpus to infer directed signed links from the emotional tone of messages, and analyze how balance evolves over time,
Expanding on this idea,~\citet{Dinh2023} and ~\citet{Rezapour2024} develop a framework that infers directed signed relations from communication data using multiple textual dimensions—sentiment, morality, and trust—across datasets such as emails and online discussions.
Their approach preserves directionality and allows for comparing different semantic interpretations of edge signs.

Textual sign inference has also been combined with machine learning to detect stance and relational polarity.
For example, ~\citet{keuchenius2021important} apply natural language processing to online debates to infer both users’ stances on controversial topics and whether specific interactions are supportive or antagonistic.
Separately, ~\citet{biyong_2023_emeddingInference} introduce a model that learns user and topic embeddings directly from signed interactions in social media, using patterns of agreement and disagreement across multiple topics.

While a variety of methods exist to infer signed links from text, it is crucial to distinguish the types of output they generate.
Most commonly, these approaches produce binary or weighted signed networks, either by analyzing the content of messages or inferring alignment from node-level attributes such as ideology.
In some cases, especially when each message is treated as an individual interaction, the result may be a signed multi-edge graph that captures repeated, valenced exchanges between pairs of nodes.
Ultimately, the chosen representation—binary, weighted, or multi-edge—depends on both the modeling assumptions and the structure of the available data.

\paragraph{Inference from signed networks}
Most methods discussed above infer signs from unsigned data sources such as text or interaction patterns.
A complementary setting starts from an already signed network and aims to infer missing, uncertain, or future edge signs or weights.
This problem is commonly referred to as \emph{link prediction} or \emph{sign prediction} in signed networks.

We do not survey this large literature, but note that it connects back to balance directly.
Many balance measures are benchmarked by their ability to predict held-out signs~\citep{kirkley2018balanceInSN,singh2017WalkSB}, and predictors in turn exploit balance-theoretic closure together with signed motifs, node features, or embeddings~\citep{Leskovec2010Predicting,Derr2019,ruiz2023triadic}, with extensions to weighted signed networks~\citep{Kumar2017}.
A recent counterpoint is that balance-driven embeddings can overstate polarization by suppressing the unbalanced ties that bridge antagonistic groups, motivating imbalance-aware methods that preserve them~\citep{du2026beyond}.

Balance-theoretic rules are also studied as decision support tools.
\citet{meng2022whom} ask how an observer should decide whether to trust an unknown user from the signs linking them to a few nodes trusted by the observer.
They derive the optimal solution to the problem defining a Bayesian rule and compare it to a balance-theoretic heuristic.
The heuristic is to follow the shortest path connecting the observer to the unknown user and multiply its signs, so that an enemy's enemy counts as a friend.
This heuristic equals the optimal solution on tree-like networks.
However, accuracy is lost when the target is reachable through several overlapping paths.

It is important to note that inference in this setting is inherently limited by the quality of the observed signed network: predicted signs are only as reliable as the signs already present.
Moreover, when incorporating additional information such as node attributes or contextual features, one must consider whether this information was already used in the original sign inference process.
Otherwise, link prediction methods may infer relations of a different nature, capturing subtly different notions of positivity and negativity than those encoded in the original signed network.

\section{Measurement of structural balance}
\label{sec:meas}
\subsection{Measures of structural balance}

Structural balance as originally formulated is a yes/no property: by the theorems of
\citet{harary1953notion} and \citet{davis1967clusteringSB} (Sec.~\ref{subsubsec:sb-formal}), a network either admits the required partition or it does not.
Real-world signed networks almost never do, which is why they are described as being in a state of \emph{partial balance}~\citep{aref2017measuring}, and measures of structural balance provide quantatitive descriptions of partial balance in observed signed networks.
% quantify how close an observed signed network is to a balanced state.
We now provide an overview of these measures; their main properties are summarized in Table~\ref{tab:bottomline_measures} at the end of Sec.~\ref{sec:meas}.
% and it is no overstatement to say that there exists a whole forest of them.
Unless stated otherwise, we first consider unweighted, undirected signed networks with binary edge signs; extensions to directed and weighted networks are discussed in Sec.~\ref{subsubsec:directed-weighted}.

\subsubsection{Balanced subgraph counting measures}

\begin{figure}[]
    \centering
    \includegraphics[width=\textwidth, clip, trim = 40 12 10 6]{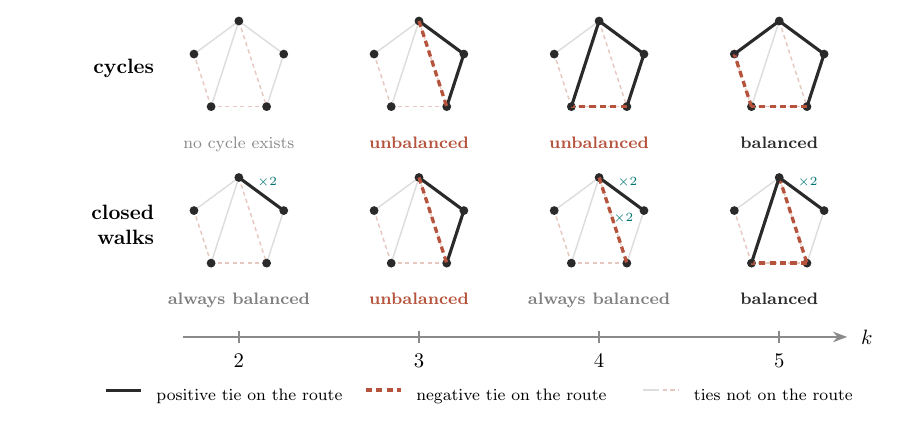}
    \caption{\textbf{Cycles vs. closed walks.}
Comparison  of cycles (top) and closed walks (bottom) on a five-node signed network with different lengths $k$. Solid lines indicate positive links on the illustrated route, dashed lines indicate negative links, and faint lines indicate links of the underlying network that are not traversed by the route. $\times 2$ indicates that the edge is traversed twice. Note that no cycle of length $k=2$ exists, whereas self-retracing walks ($k=2, 4$) are always positive.
}
    \label{fig:cycles-walks}
\end{figure}

Local aggregation measures constitute a broad class of balance measures that assess structural balance by examining smaller substructures of the network and aggregating their contribution.
Although a universal formal definition is difficult to provide, these measures typically follow the same general principle: (i) specify a type of local subgraph of interest, (ii) count how many such subgraphs are balanced in the network, and (iii) normalize this count by the total number of such subgraphs.
The most common choices for these substructures are \emph{cycles} and \emph{closed walks}.
A cycle is a closed walk in which every vertex is visited exactly once, whereas a closed walk may revisit nodes or edges (see Fig.~\ref{fig:cycles-walks}).
Then, these measures quantify structural balance through the proportion of balanced local structures.

\paragraph{Cycle-based measures}
The original cycle-based approach to measuring the balance of a network is from~\citet{cartwright_1956}.
They define a network as balanced ``\textit{if all of its cycles are positive}'' where ``\textit{a cycle is positive if it contains an even number of negative [links]}''~\citep[][p.284]{cartwright_1956}.

Cycles of size 3, i.e., \textit{triads}, have historically received the most attention and~\citet{harary1959_measuringSB} defines the first measure of balance through them; he defines the~\textit{triangle index} $T(\mathbf{A})$:

\begin{equation}
  \label{eq:triangle index}
  T(\mathbf{A}) = \frac{O^{+}_{3}(\mathbf{A})}{O_{3}(\mathbf{A})}\;,
\end{equation}
where $\mathbf{A}$ is the signed adjacency matrix of the network, $O_{3}^{+}(\mathbf{A})$ is the network's number of positive triads and $O_{3}(\mathbf{A})$ is the total number of triads in the network.
In practice $T(\mathbf{A})$ never requires enumerating triads.
To express it in matrix form, we denote by $|\mathbf{A}|$ the unsigned counterpart of $\mathbf{A}$, with entries $(|\mathbf{A}|)_{ij}=|A_{ij}|$.
Then $\operatorname{Tr}(\mathbf{A}^{3})$ and
$\operatorname{Tr}(|\mathbf{A}|^{3})$ count, respectively, the signed and unsigned closed walks of length
three, so Eq.~\eqref{eq:triangle index} can be evaluated in a single matrix expression,
\begin{equation}
  \label{eq:triangle-index-trace}
  T(\mathbf{A}) = \frac{1}{2}\left(1 + \frac{\operatorname{Tr}(\mathbf{A}^{3})}{\operatorname{Tr}(|\mathbf{A}|^{3})}\right)
       = \frac{1-U}{2}, \qquad U = -\frac{\operatorname{Tr}(\mathbf{A}^{3})}{\operatorname{Tr}(|\mathbf{A}|^{3})},
\end{equation}
where $U$ is the normalized triadic energy (also called work function) used as an order parameter in Hamiltonian models of balance~\citep{Malarz2023Thermal} (Sec.~\ref{subsec:hamil}).
For binary signed networks with homogeneous unit triadic couplings, \(U=H/N_T\), where \(H\) is the triadic Hamiltonian defined in Eq. \eqref{eq:hamiltonian_triad}.

Considering all possible cycles of different lengths,~\citet{cartwright_1956} proposes the \textit{degree of balance} measure

\begin{equation}
  \label{eq:cartwrightBalance}
  D(\mathbf{A}) = \frac{\sum_{k=3}^{\infty}O^{+}_{k}(\mathbf{A})}{\sum_{k=3}^{\infty}O_{k}(\mathbf{A})}\;,
\end{equation}
where $O_{k}^{+}(\mathbf{A})$ are the numbers of \textit{positive} cycles of length $k$ and $O_{k}(\mathbf{A})$ are the numbers of \textit{all} cycles of length $k$.
The summation of $k$ goes from 3 as this is the minimum length of a cycle in an undirected network to a maximum that depends on the network topology.
%GV{check the original paper of cartwright_1956. There you see in figure 5 a network with four nodes and six links having seven cycles: four of length 3 and three of length 4.}

These definitions assume that cycles of different lengths contribute equally to the overall structural balance of the network.
However, longer cycles are much more prevalent than shorter cycles in most networks.
Therefore, these measures that count all cycles equally will typically be dominated by the presence or absence of long positive cycles.
To solve for this problem, it was already proposed by ~\citet{cartwright_1956} to attribute higher importance to cycles of shorter length.
The earliest formalization of this concept dates back to 1972, when ~\citet{NORMAN1972} introduced a weighted adaptation of the cycle measures:

\begin{equation}
  \label{eq:normanBalance}
  D_\text{weighted}(\mathbf{A}) = \frac{\sum_{k=3}^{N} f(k) O^{+}_{k}(\mathbf{A})}{\sum_{k=3}^{N} f(k) O_{k}(\mathbf{A})}\;,
\end{equation}
where $f(k)$ is a positive function.
Choosing a monotonically decreasing function would lead to the desired effect of assigning lower weights to longer cycles.

\paragraph{Walk-based measures}

Counting the number of closed cycles is computationally expensive, and for large networks, it is not feasible.
Generalizations of such cycle-based measures therefore consider \textit{closed walks}, where backtracking along the same nodes is possible~\citep{pelino2012class}.
The benefit of employing closed walks instead of cycles is that counting closed walks can be done efficiently via the powers of the adjacency matrix.
Specifically, for a signed adjacency matrix $\mathbf{A}$, $A^{k}_{ii}$ is the difference between the number of positive and negative walks that start and end in $i$.
This practical property has been used by several authors to define computationally efficient walk-based measures of structural balance~\citep{estrada2014_RandomWalkSB,estrada2019RethinkingSB,pelino2012class, singh2017WalkSB, kirkley2018balanceInSN, talaga_polarization_2023}.

Also closed walk measures risk being skewed toward longer walks and hence, it is common to penalize longer walks, in line with Eq.~\ref{eq:normanBalance}.
The specific form of the penalty function $f(k)$ is diverse across measures.
Specifically, one may define
\begin{equation}
  \label{eq:estradaRandomWalk}
  K(\mathbf{A}) = \frac{\sum_{k=0}^{\infty}f(k)(Q^{+}_{k}(\mathbf{A}) - Q^{-}_{k}(\mathbf{A}))}{\sum_{k=0}^{\infty} f(k)Q_{k}(\mathbf{A})}\;,
\end{equation}
where $Q_{k}^{+}(\mathbf{A})$ and $Q_{k}^{-}(\mathbf{A})$ are, respectively, the network's number of positive and negative walks of length $k$ and $Q_{k}(\mathbf{A})$ is the number of its total walks of length $k$.
~\citet{estrada2014_RandomWalkSB} define the weighting function as $f(k) = \frac{1}{k!}$ and count for $N$ trivial closed positive walks at $k=0$ and zero closed walks at $k=1$.
\footnote{To be precise, \citet{estrada2014_RandomWalkSB} write $\sum_{k=1}^{\infty}\operatorname{Tr}(\mathbf{A}^{k})/k!$ and identify it with $\operatorname{Tr}\exp(\mathbf{A})$; the two agree only if one starts the summation from $k=0$, with the convention $\operatorname{Tr}(\mathbf{A}^{0})=N$.}
With this normalization and summation convention, the index $K(\mathbf{A})$ can be simplified using the eigenvalues $\pmb \lambda$, $\pmb \mu$ of $\mathbf{A}$ and its unsigned counterpart $|\mathbf{A}|$, respectively:

\begin{equation}
    \label{eq: estradaEigenvalues}
    K(\mathbf{A}) = \frac{\sum_{j=1}^{N} \exp{(\lambda_{j})}}{\sum_{j=1}^{N} \exp{(\mu_{j})}
  }\;.
\end{equation}

This latter form of the index $K(\mathbf{A})$ clearly highlights the benefits of moving from cycles to closed walks, since Eq.~\ref{eq: estradaEigenvalues} can be computed with ease.
Note that, differently to $D_\text{weighted}(\mathbf{A})$, $K(\mathbf{A})$ quantifies the \textit{difference} between the number of positive and negative walks in the numerator, while $D_\text{weighted}(\mathbf{A})$ only considers the (normalized) number of balanced cycles.
However, as noted for instance by ~\citet{aref2017measuring}, with the simple manipulation $D_\text{weighted}^{\text{walk}} = \frac{1}{2} (K(\mathbf{A}) + 1)$, the original form of $D_\text{weighted}(\mathbf{A})$ is regained.

The introduction of this measure has sparked the development of several extensions.
First,~\citet{singh2017WalkSB} replace the fixed weighting of~\citet{estrada2014_RandomWalkSB} by a parametrized version $f(k,\beta)=\beta^{k}$, with $0<\beta<1/\rho(|\mathbf{A}|)$, where $\rho(|\mathbf{A}|)$ denotes the spectral radius of $|\mathbf{A}|$.
The form is the attenuation factor of the Katz index, and the motivation for introducing it is that $f(k)=1/k!$ fixes how fast longer walks are penalized, whereas $\beta$ leaves that rate free: changing $\beta$ changes the range of walk lengths that contribute more, so the scale at which balance is assessed becomes something one can vary rather than a modeling assumption fixed in advance.
Exploiting this, they report that short cycles predict edge signs better than long ones, and they derive an approximation that is useful for practical computation.

Second, and in the same spirit,~\citet{tian2025balance} replace the exponential weighting with a Mittag-Leffler matrix function.
The resulting balance index is obtained from a nonconservative diffusion dynamic on the signed network, and it accounts for the memory of that process by diminishing the penalization that long cycles receive under $f(k)=1/k!$.
Both leave the rate of decay free, but where \citet{singh2017WalkSB} posit the penalization directly, \citet{tian2025balance} derive it from the assumed dynamics.

Third,~\citet{diazdiaz_mathematical_2024} define a local version of $K(\mathbf{A})$, which attributes the global index to individual vertices and thus measures how much a given node contributes to the balance of the network as a whole:
% \begin{equation}
%   \label{eq:estradaRandomWalk2}
%   K(\mathbf{A})_v = \frac{\sum_{k=0}^{\infty}f(k)(Q^{+}_{k}(\mathbf{A})_{vv} - Q^{-}_{k}(\mathbf{A})_{vv})}{\sum_{k=0}^{\infty} f(k)Q_{k}(\mathbf{A})_{vv}}\;.
% \end{equation}
\begin{equation}
  \label{eq:estradaRandomWalk2}
  K_v(\mathbf{A}) =
  \frac{
  \sum_{k=0}^{\infty}\frac{1}{k!}
  \left(Q^{+}_{k}(\mathbf{A})_{vv}-Q^{-}_{k}(\mathbf{A})_{vv}\right)
  }{
  \sum_{k=0}^{\infty}\frac{1}{k!}Q_{k}(\mathbf{A})_{vv}
  }\;.
\end{equation}
where $Q_{k}^{+}(\mathbf{A})_{vv}$ and $Q_{k}^{-}(\mathbf{A})_{vv}$ are, respectively, the number of positive and negative walks of length $k$ starting and ending at node $v$, and $Q_{k}(\mathbf{A})_{vv}$ is the total number walks of length $k$ starting and ending at node $v$.
The motivation for computing balance at the node level is that a single network-wide number cannot say \emph{where} the imbalance sits.

Fourth,~\citet{kirkley2018balanceInSN} define a tractable measure of imbalance $B(\mathbf{A}, z)$, which employs the weighting function $f(k,z)= 1/z^k$, with $z>1$, which corresponds to the weighting of~\citet{singh2017WalkSB} under $\beta=1/z$:

\begin{equation}
  \label{eq:kirkleybalance}
  B(\mathbf{A},z) = \sum^{\infty}_{k = 1} \frac{Q^{-}_{k}(\mathbf{A})}{z^k}=\frac{1}{4} \log \frac{\det [zI -\mathbf{A}]}{\det [zI - |\mathbf{A}|]}\;,
\end{equation}
where $I$ is the identity matrix.
% In the original formulation, the authors compute the $Q^{-}_{k}(\mathbf{A})$ using the trace of two specific adjacency matrices (see Eq.(6) of \citep{kirkley2018balanceInSN}), and hence, have an extra normalization factor of $1/2k$, where the factor $1/2$ compensates for double counting each walk in both direction and $1/k$ for counting the same walk $k$ times (described after Eq.(4) of \citep{kirkley2018balanceInSN}).
Two properties motivate this construction: i) its calculus requires the computation of two determinants (no enumeration of closed walks) and ii) the same construction is possible for weak balance (Sec.~\ref{subsubsec:weak-measures}), so strong and weak balance can be assessed with one measure.

Fifth,~\citet{talaga_polarization_2023} generalize $K(\mathbf{A})$ and $D_\text{weighted}^{\text{walk}}(\mathbf{A})$ to various scales, from node-contributions to pairwise contributions, and to directed and weighted networks, using the weighting function $f(k,\beta) = \frac{\beta^k}{k!}$.
Their stated justification for penalizing longer walks is the \emph{Locality Principle}, namely that patterns in short cycles matter more than patterns in long ones, with $\beta$ setting the resolution at which that principle is applied.
For instance, they define the analog of the node-level measure $K_v(\mathbf{A})$ but for $D_\text{weighted}^{walk}(\mathbf{A})$:

% \begin{equation}
%     \label{eq:talagaMultiscale}
%     D_\text{weighted}^{\text{walk}}(\mathbf{A})_{v} = \frac{\sum_{k=3}^{N} f(k) Q^{+}_{k}(\mathbf{A})_{vv}}{\sum_{k=3}^{N} f(k) Q_{k}(\mathbf{A})_{vv}}\;.
% \end{equation}
\begin{equation}
    \label{eq:talagaMultiscale}
    D_{\text{weighted},v}^{\text{walk}}(\mathbf{A},\beta)
    =
    \frac{
    \sum_{k=3}^{N} f(k,\beta) Q^{+}_{k}(\mathbf{A})_{vv}
    }{
    \sum_{k=3}^{N} f(k,\beta) Q_{k}(\mathbf{A})_{vv}
    }\;.
\end{equation}

\paragraph{Benefits and limitations of subgraph counting measures}
subgraph based measures provide an intuitive and interpretable framework for quantifying structural balance.
By focusing on small, well-understood configurations, such as cycles or closed walks, they connect directly to the original formulations of balance theory and offer clear sociological meaning: balanced or unbalanced relations emerge ``locally'' through triads and short feedback loops.
Walk-based extensions further enable scalable computation, leveraging matrix powers to efficiently incorporate information from longer paths, while retaining a clear connection to cycle-based intuition.

However, these approaches also face well-known limitations.
Exact enumeration of simple cycles becomes computationally prohibitive for large networks, motivating the use of closed walks and weighting schemes.
Yet, closed walks may revisit vertices and edges, causing certain patterns to be disproportionately amplified.
The necessity of a decay function to penalize longer walks introduces an additional modeling choice: different functional forms or parameter values can yield markedly different balance scores, and no universally accepted weighting scheme exists, despite recent efforts at axiomatic justification~\citep{talaga_polarization_2023, aref2017measuring, estrada2019RethinkingSB}.

\subsubsection{Frustration index}
\label{subsubsec:frustration}

A distinct approach quantifies balance by measuring the inconsistency or \textit{frustration} in edge signs.
The classic metric here is the frustration index $L(G)$ (also called the line index of imbalance in Harary’s terminology)~\citep{harary1959_measuringSB}.
$L(G)$ is defined as the minimum number of edges that need to be removed (or equivalently, whose signs need to be flipped) to make the graph balanced.
If instead of a sign, we consider the removal of a node one obtains the so called point index for balance~\citep{harary1959_measuringSB}.
In other words, the point index is the minimum number of nodes whose removal results in a balanced network.
Such measure is not often investigated in the literature.
We instead focus on the edge-based frustration index, which is more widely used.

\begin{figure}[!htbp]
    \centering
    \includegraphics[width=\linewidth]{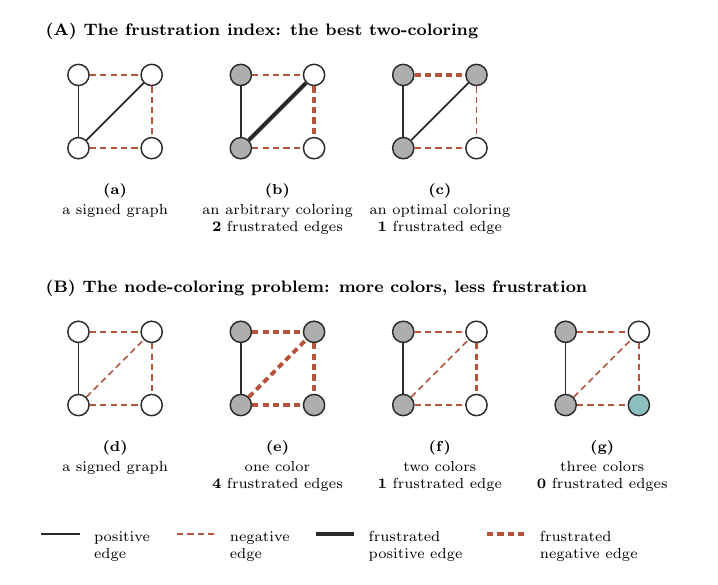}
    \caption{\textbf{The frustration index and the node-coloring problem.}
An edge is \emph{frustrated} when a positive edge joins two colors, or a negative edge joins the same color; frustrated edges are drawn thick.
\textbf{(A)} With the number of colors fixed at two, which is strong balance, the frustration index is the smallest number of frustrated edges over all colorings.
(b) is an arbitrary coloring and (c) presents the optimal coloring. One frustrated edge remains, so $L(G) = 1$.
\textbf{(B)} Allowing more colors, which is weak balance, reduces the frustration: the same graph leaves four frustrated edges with one color, one with two, and none with three, so it is clusterable.
Adapted from~\citep{aref2019thesis} with permission.
}
    \label{fig:frustration-coloring}
\end{figure}

\paragraph{Frustration as a node-coloring problem}
This definition is equivalent to the following~\citep{Doreian1996}.
Imagine we try to partition a network $G$ into two all-positive subgroups: every edge that does not ``fit'' that perfect partition (either a negative edge within a subgroup or a positive edge between subgroups) is frustrated.
$L(G)$ is the smallest count of such frustrated edges over all possible 2-group splits.
Equivalently, if we assign each node a binary label (black/white or 1/0 representing two factions), an edge is satisfied if its sign is positive and the endpoints have the same label, or its sign is negative and the endpoints have opposite labels.
Edges that fail this are frustrated.
The frustration index is the minimum number of frustrated edges attainable by the best choice of labels for all nodes.
Fig.~\ref{fig:frustration-coloring}(A) shows the construction on a small example: a poor two-coloring leaves two frustrated edges and the best one leaves a single edge, so the index is one and no coloring can do better.
This is a problem akin to community detection in signed networks (see \cite{diaz2025signed}).

Following the node-coloring formulation of~\citet{Aref_2018}, we partition the nodes $V$ into two sets, $\mathcal{X}\subseteq V$ and $\bar{\mathcal{X}}= V\setminus \mathcal{X}$.
We denote the \textit{color} of node $i$ by $c_i\in\{0,1\}$, with $c_i=1$ (\textit{black}) if $i\in \mathcal{X}$ and $c_i = 0$ (\textit{white}) if $i\in \bar{\mathcal{X}}$.
% Then, the \textit{frustration count} is the number of frustrated edges of $G$ given the partition $\mathcal{X}$:
For each edge $(i,j)\in E$, we then define the \textit{frustration indicator} $f_{ij}(\mathcal{X})$, which equals $1$ if the edge is frustrated by the coloring induced by $\mathcal{X}$ and $0$ otherwise:
\begin{equation}
f_{ij}(\mathcal{X}) =
\begin{cases}
0, & \text{if } c_i = c_j \text{ and } s_{ij}=+1,\\[4pt]
1, & \text{if } c_i = c_j \text{ and } s_{ij}=-1,\\[4pt]
0, & \text{if } c_i \neq c_j \text{ and } s_{ij}=-1,\\[4pt]
1, & \text{if } c_i \neq c_j \text{ and } s_{ij}=+1.
\end{cases}
\end{equation}
The \textit{frustration count} for the partition $\mathcal{X}$ is the total number of frustrated edges:
\begin{equation}
f_G(\mathcal{X}) = \sum_{(i,j)\in E} f_{ij}(\mathcal{X}).
\end{equation}
The frustration index is then obtained by finding the subset $\mathcal{X}^\star \subseteq V$ that minimizes the frustration count:

\begin{equation}
L(G) \;=\; \min_{\mathcal{X} \subseteq V} \, f_G(\mathcal{X}) \;.
\label{eq:frustration:strong}
\end{equation}

By definition, $L(G)=0$ if and only if the graph is balanced (no frustrated edges, a perfect split exists).
The larger L(G), the more sign conflicts must be “fixed” to eliminate all imbalance.

The optimal partition $\mathcal{X}^\star$ carries more information than the single number it produces, and balance can be assessed at more than one level, each highlighting a different property.
\citet{Aref2020} do this at three levels: the triad (micro), the subgroup (meso), and the whole network (macro).
At the macro level, $L(G)$ is rescaled to the unit interval as the \textit{normalized line index} $F(G)=1-2L(G)/m$, using the upper bound $m/2$ on the frustration count.
At the meso level, the same $\mathcal{X}^\star$ is read group by group and defines \textit{cohesiveness} and \textit{divisiveness}.
The former is the fraction of internal edges, those with both endpoints in the same set, that are positive, while the latter is the fraction of external edges, those joining the two sets, that are negative, so that intra-group solidarity and inter-group antagonism are measured separately.
At the micro level, \citet{Aref2020} compute the fraction of transitive triads that are balanced, a triad counting as transitive when all of its semicycles are (Sec.~\ref{subsubsec:sb-variants}).
Results of structural balance at different levels do not need to agree: a network can be cohesive within its groups yet barely divided between them, and \citet{Aref2020} argue that balance at the micro and macro levels is not the same property measured at different resolutions.

\paragraph{Frustration as an energy minimization problem}

Computationally, this again is equivalent to computing the ground state of a (nonplanar) Ising spin glass~\citep{Barahona_1982_spinGlass}.
Following the formalism of~\citep{facchetti2011computing}, we now attribute a spin (instead of a color) to each node $s_i\in \{-1,+1\} = \mathbb{B}_2$.
Note that this is equivalent to the previous formulation, with $s_i = 2c_i - 1$.
We can then define the energy functional:

\begin{equation}
    \label{eq:facchetti_energy}
    H_G(\mathbf{s}) = \frac{1}{2}\sum_{(i,j)\in E} (1-A_{ij}s_is_j),
\end{equation}
where $\mathbf{s}$ is the vector of spins.
Note that this is related to the so called Hamiltonian formulation of structural balance that we review in detail in Sec. \ref{sec:emerg}.

The network is then balanced when there exists a configuration $\mathbf{s}$ such that all terms in $H_G(\mathbf{s})$ are equal to zero.
Calculating the frustration index is then equivalent to finding the configuration $\mathbf{s}$ such that the energy functional is minimized:

\begin{equation}
    \label{eq: facchetti_minimization}
    L(G) \;=\; \min_{\mathbf{s} \in \mathbb{B}^N_2} \, H_G(\mathbf{s}) \;.
\end{equation}
% where $\tfrac12(N^{2}-2m)$ is a constant that depends on the number of nodes $N$ and edges $m$ in the network.

\paragraph{Computing the frustration index}
Computing the exact frustration index is challenging, as it requires identifying the optimal partition $\mathcal{X}$, the optimal spin assignment vector $\mathbf{s}$, or equivalently the optimal set of edges to delete or flip in order to minimize frustration.
This task is computationally difficult and is equivalent to solving an NP-hard problem~\citep{facchetti2011computing, Barahona_1982_spinGlass}.
Consequently, a variety of methods have been proposed to either compute the frustration index exactly or approximate it.

\citet{facchetti2011computing} introduced a nonlinear energy minimization approach for estimating the frustration index, employing a heuristic local search strategy.
Further non-exact optimization techniques have been developed~\citep{Iacono_frustrationAlgorithms}, using, for instance, memetic algorithms~\citep{Sun2014, Wang2016}.

Computing the frustration index is a task closely related to signed community detection since both aim to maximize positive ties within groups and minimize negative ties between them.
Therefore, several authors have adapted known community-detection tools to signed networks and used them to proxy frustration.
For instance,~\citet{anchuri2012spectralSB} define signed modularity:
\begin{equation}
  \label{eq:anchuri}
  Q(\mathbf{s}) = \sum_{i,j} \left(A_{ij} + \frac{k^-_{{i}}k^-_{{j}}}{2m} - \frac{k^+_{{i}}k^+_{{j}}}{2m}  \right) {s_is_j} \;,
\end{equation}
where $\mathbf{s}\in \mathbb{B}_2^N$ is the assignment vector and $k^+_i$, $k^-_i$ are the positive and negative degrees of node $i$.
For another definition of signed modularity, see~\citet{gomez2009_modularitySigned}.
Additionally, Infomap-based methods~\citep{esmailian2014mesoscopic} and block-models~\citep{Doreian1996,Doreian_Batagelj_Ferligoj_2004} have been used to approximate the optimal partitioning $\mathbf{s}$.
Some of these heuristic techniques have successfully estimated frustration values for networks with up to approximately $700{,}000$ edges.

For exact computation, binary programming formulations have been proposed by~\citet{Aref2018_frustrationUBQP}, enabling the frustration index to be computed exactly for networks with up to $100{,}000$ links~\citep{Aref_2018}.
These exact optimization models represent a substantial advance, as they allow the frustration index to be determined for large real-world signed networks where previous algorithms could only provide approximations.

\paragraph{Parametric frustration index}
The classical frustration index treats all frustrated edges equally, assigning the same penalty to a misplaced positive edge (a positive edge between groups) and a misplaced negative edge (a negative edge within a group).
However, it might be possible that these two types of inconsistencies do not generate the same level of tension.
To address this issue,~\citet{Doreian1996} introduced a \emph{parametric} generalization of the frustration index that weights the two error types differently.
Let $L^{-}(\mathcal{X})$ denote the number of negative edges within groups and $L^{+}(\mathcal{X})$ the number of positive edges between groups for a given partition $\mathcal{X}$.

The parametric version of the frustration count is then defined as:
\begin{equation}
f_G(\mathcal{X},\alpha) = \alpha\,L^{-}(\mathcal{X}) + (1-\alpha)\,L^{+}(\mathcal{X}), \qquad 0 \le \alpha \le 1,
\end{equation}
where the parameter $\alpha$ controls the relative importance of the two types of inconsistency.\footnote{Note that one can define a parametric frustration index as the minimum of this count over all partitions $L(G,\alpha) \;=\; \min_{\mathcal{X} \subseteq V} \, f_G(\mathcal{X},\alpha)$
}
When $\alpha = 1/2$, both error types are penalized equally, recovering the behavior of the traditional frustration index.
Values $\alpha < 1/2$ weight positive edges between groups more heavily, while $\alpha > 1/2$ give greater importance to negative edges within groups.

\paragraph{Benefits and limitations of the frustration index}
The frustration index offers a fundamentally global measure of balance.
By definition, it evaluates how far the entire network is from achieving perfect structural balance, identifying the minimum number of edge sign changes (or removals) required to reach a balanced configuration.
Unlike subgraph counting measures, the frustration index evaluates all edges simultaneously within a single global partition, so each edge contributes at most once to the frustration count rather than once for every unbalanced subgraph in which it appears.
This makes the frustration index conceptually appealing when the goal is to capture systemic inconsistency rather than localized tensions.

At the same time, this global perspective introduces limitations.
Because the measure aggregates imbalance across the entire network, highly localized but structurally important tensions may be treated equivalently to widespread but weak imbalances: for instance, a single frustrated edge embedded in one triad can contribute the same to the index as the same edge being frustrated across many triads.
Moreover, the frustration index takes the globally closest balanced configuration as its reference.
Interpreted dynamically, this corresponds to assuming access to a global optimum, whereas in empirical systems other social, cognitive, or structural forces may constrain the dynamics and make this optimum unreachable or irrelevant.
Computationally, determining the frustration index is NP-hard, and although recent advances in binary programming allow exact computation for networks of substantial size, scalability remains limited compared to walk-based approximations, which rely only on eigenvalue computations and therefore scale more readily to large graphs.

\subsubsection{Weak structural balance measures}
\label{subsubsec:weak-measures}

Classical structural balance theory assumes that a signed network can be partitioned into two mutually antagonistic but internally cohesive groups. Davis~\citep{davis1967clusteringSB} generalizes this idea by introducing \emph{weak structural balance}.
A signed graph is \emph{$j$-balanced} if its nodes can be partitioned into $j \ge 2$ disjoint subsets such that all edges within subsets are positive and all edges between subsets are negative.
Under this definition, triads with three negative edges are considered balanced, since each node may be thought of as belonging to its own group.

This extension broadens the applicability of balance theory in settings where more than two antagonistic factions may coexist. Consequently, it is natural to ask how measures of partial balance can be adapted to incorporate weak balance.

\paragraph{Cycle-based extensions}
Subgraph-counting measures can also be adapted to weak balance by treating some negative triads as balanced, rather than unbalanced.
In general, a network is weakly balanced if, and only if, it contains no simple cycles with exactly one negative edge \citep{davis1967clusteringSB}, meaning that any other number of negative edge larger than one is fine.
Therefore, we may define the degree of weak balance as the extent to which a network is free of cycles with exactly one negative edge.

~\citet{kirkley2018balanceInSN} use this approach to generalize their measure of structural imbalance to weakly balanced networks.
Additionally,~\citet{talaga_polarization_2023} derive a measure for weak balance of the form:
\begin{equation}
    \label{eq;talagaWeakBalance}
    W(\beta) = 1- \frac{\mu_W}{\mu^+ + \mu^-}\;,
\end{equation}
where $\mu^+$ and $\mu^-$ are weighted counts (by a function dependent on $\beta$; see Eq. \eqref{eq:talagaMultiscale} and the earlier definition of $\beta$) of positive and negative walks, while $\mu_W$ is sum of weighted counts of weakly unbalanced closed walks.
~\citet{talaga_polarization_2023} further provide a weak balance formulation for walk-based measures across multiple scales, from node-level to networks.

\paragraph{Frustration-based extensions}
The frustration index generalizes straightforwardly to weak balance by allowing partitions into $k>2$ communities.
Let $C_1, \dots, C_k$ denote the communities and let $c_i\in\{1,\ldots,k\}$ denote the community assignment of node $i$; some communities may be empty.
As in the strong-balance case, an edge is frustrated if it is negative and connects nodes in the same community, or positive and connects nodes in different communities.
Adapting the formulation of~\citet{Doreian1996,anchuri2012spectralSB} to our notation, the generalized frustration count can be written as
% \begin{equation}
%   \label{eq:doreian}
%   F(C_{1},C_{2},\ldots,C_{k})
%   =
%   \sum_{i<j} A_{ij}^- \, \delta(c_i,c_j)
%   +
%   \sum_{i<j} A_{ij}^+ \, (1 - \delta(c_i,c_j)),
% \end{equation}
\begin{equation}
\label{eq:doreian}
f_G(C_1,\ldots,C_k)
=
\sum_{\substack{(i,j)\in E\\ s_{ij}=-1}}
\delta(c_i,c_j)
+
\sum_{\substack{(i,j)\in E\\ s_{ij}=+1}}
\left[1-\delta(c_i,c_j)\right],
\end{equation}
where $\delta(c_i,c_j)=1$ if nodes $i$ and $j$ belong to the same community and $0$ otherwise.
% where $\mathbf{A}^-$ and $\mathbf{A}^+$ are the negative and positive parts of the adjacency matrices, and $\delta(c_i,c_j)$ equals $1$ if $i$ and $j$ are in the same community and $0$ otherwise.
% The first term penalizes negative edges within communities and the second penalizes positive edges across communities.
Minimizing this count over all partitions into $k$ communities gives the corresponding frustration index for weak balance:
% \begin{equation}
% \label{eq:doreian-min}
% F_{k}(G) \;=\; \min_{C_{1},\ldots,C_{k}} F(C_{1},C_{2},\ldots,C_{k}) \;,
% \end{equation}
\begin{equation}
\label{eq:doreian-min}
L_k(G)
=
\min_{C_1,\ldots,C_k}
f_G(C_1,\ldots,C_k).
\end{equation}
% where the minimizing partition is the weakly balanced one.
For $k=2$, this reduces to the strong-balance frustration index, $L_2(G)=L(G)$ [compare with Eq.~\eqref{eq:frustration:strong}].
A graph admits a weakly balanced $k$-group partition when $L_k(G)=0$.
Fig.~\ref{fig:frustration-coloring}(B) shows the effect of that relaxation: the same graph requires four frustrated edges with one community, one with two, and none with three.

Community-detection methods, such as spectral clustering~\citep{anchuri2012spectralSB} or more recent block-model approaches~\citep{Gallo2024}, provide practical heuristics for approximating this generalized frustration index.

\subsubsection{Directed and weighted measures}
\label{subsubsec:directed-weighted}

Structural balance concepts and their associated measures can be naturally extended to directed and weighted signed networks.

\paragraph{Subgraph-based measures}
As shown by~\citet{talaga_polarization_2023}, subgraph-based measures can be generalized to directed graphs and weighted adjacency matrices by adjusting the underlying algebra.
The general strategy involves replacing binary entries in the signed adjacency matrix $A_{ij}=s_{ij}$ with real-valued edge weights $A_{ij}=x_{ij}=s_{ij}w_{ij}$ and incorporating directionality into the definitions of walks or cycles.
% This allows the derivation of aggregate, node-level and pairwise balance scores for complex network topologies.

\paragraph{Frustration index}
For the frustration index,~\citet{aref_weightedFrustration} propose a direct incorporation of edge weights into the objective function.
Their formulation uses binary node colors and continuous signed edge weights.
    Adapting their notation to the notation used here, we denote the node colors by
$c_i,c_j\in\{0,1\}$ and the signed relation by $x_{ij}=s_{ij}w_{ij}\in[-1,+1]$, with $w_{ij}\in[0,1]$.
% Instead of using discrete signs $s_{ij} \in \{-1,+1\}$, they define edge weights $w_{ij} \in [-1,1]$, which contribute proportionally to frustration.
The frustration of an edge is redefined as:

\begin{equation}
    \label{eq:arefWeightedFrust}
    f_{ij} = \frac{1 - x_{ij}}{2} + x_{ij}(c_i + c_j - 2c_{i}c_j) \;.
\end{equation}
The term $c_i + c_j - 2c_ic_j$ equals 0 when the two nodes have the same color and 1 when they have different colors.
Thus, the magnitude of $x_{ij}$ determines how strongly an edge contributes to frustration, while its sign determines which coloring is favored.
Importantly, the method can be embedded into a binary optimization model that preserves the core logic of the classical frustration index while expanding its expressiveness to weighted settings.
Note that this coloring formulation using the notation $c_i$ is linked to the models for the emergence of structural balance presented in Sec.~\ref{subsubsec:sb-agent-attribute}.
There a node $i$ is assigned with a binary attribute $b_i$ that plays a role similar to the color $c_i$.
However, we kept different notations across sections to highlight the distinct origin and use of these two approaches, i.e., the coloring formulation $c_i$ to measure structural balance and the formulation with attitudes $b_i$ used to model its emergence.

\subsection{Network null models for measuring structural balance }

Structural balance can emerge for reasons beyond cognitive dissonance within cycles.
It can arise from at least two broad classes of mechanisms: i) social-cognitive mechanisms and ii) structural or random constraints.
An illustration of the former is the spin glass problem studied by ~\cite{facchetti2011computing}, which is mathematically equivalent to computing the line index.
In the spin-glass formulation, each node is assigned a spin (positive or negative), and the system's energy depends on the configuration of spins and the signed edges between them.
By flipping spins, the system seeks the minimum-energy state, and the deviation of this minimum from zero corresponds to the line index.
Translated into a social context, one could think of individuals either changing their own opinions or modifying their ties to others to reduce tension in a group.
According to structural balance, it is the change of a tie's sign that moves the network toward a balanced state.

Structural and random constraints represent a second class of mechanisms that can generate high levels of balance \textit{without invoking explicit} social-cognitive processes.
For example, a high density of positive links increases the expected proportion of balanced triads, while network topology and clustering affect the number of triads on which triadic balance is measured~\citep{gallo2024testing}.
In these cases, the apparent evidence for structural balance may arise partly from structural properties of the network rather than from social-cognitive mechanisms such as cognitive dissonance or homophily.

These observations raise a key methodological point: observed structural balance might be explained in at least two ways as the outcome of social-cognitive mechanisms (e.g., cognitive dissonance, homophily) or as a byproduct of structural constraints and random processes.
To address this ambiguity, researchers have developed null models.
A null model defines a reference ensemble of randomized networks that preserves chosen low-level features of the data while removing higher-order dependencies, allowing one to assess which patterns cannot be explained by structural constraints alone.
Thus, null models allow us to distinguish (a) whether observed structural balance exceeds what can be explained by structural or random constraints, and (b) whether it points to one social-cognitive mechanism versus another.
Note that choosing the right null model is not trivial: using an inappropriate null model for a given task can, for example, lead to false negatives, i.e., failing to detect structural balance even in networks that were explicitly constructed to contain it~\cite{gallo2024testing, Hao2024}.

\begin{figure}[!htbp]
    \centering
    \includegraphics[width=\linewidth]{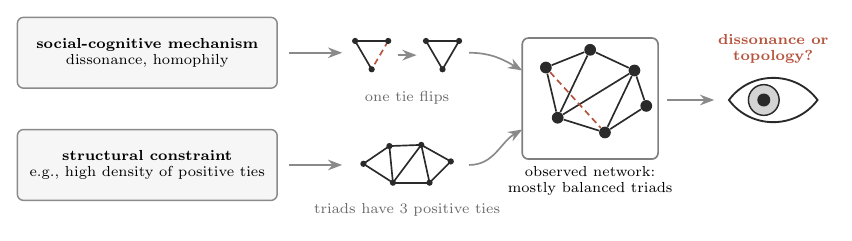}
    \caption{\textbf{Why a null model is needed.}
Many balanced triads can arise either from balance-seeking social-cognitive mechanisms, illustrated by a sign change that balances a triad, or from structural properties such as a high density of positive ties, which by itself makes most triads come out with three positive ties.
An appropriate null model is needed to distinguish these explanations.
% A high count of balanced triads can be consistent both with a social-cognitive mechanism and with a purely structural constraint; the null model is what distinguishes them.
}
    \label{fig:nullmodel-need}
\end{figure}

\begin{figure}[!htbp]
    \centering
    \includegraphics[width=\linewidth]{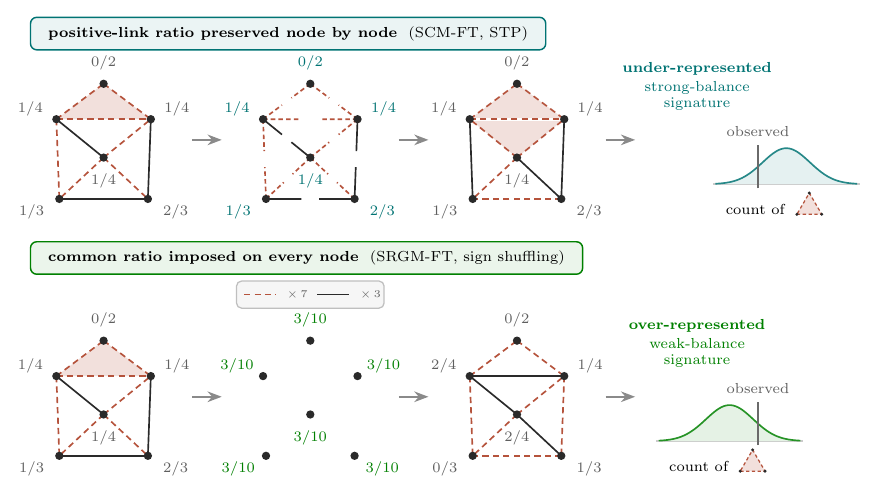}
    \caption{\textbf{How the choice of the null model can reverse the verdict.}
Two topology-preserving null models applied to the same network.
Red shading is used to indicate triads with three negative links.
Top: the positive-link ratio is preserved node by node: every edge is cut in half and the signed stubs stay at their own node.
% , so each node's ratio is unchanged by the randomization.
% The schematic shows exact node-by-node preservation;
SCM-FT and STP extend the same idea to a canonical ensemble, preserving signed degrees only in expectation rather than exactly, as drawn here.
Bottom: all nodes share a common expected positive-link ratio (e.g., SRGM-FT, sign shuffling): all ties are removed and pooled and then placed back into the network.
In both cases, the topology stays identical; only the signs are reassigned.
Under the node-by-node constraint the all-negative triad is under-represented, a strong-balance signature; under the common ratio it is over-represented, a weak-balance signature.
The shown null distributions for establishing the under- and over-representation are schematic.
}
    \label{fig:nullmodel-verdict-flip}
\end{figure}

The models reviewed below differ along four axes, and many disagreements in the literature are disagreements about which axis to choose rather than about the data.
The first is \emph{what is randomized}: the topology, the signs, or both.
The second is \emph{how tightly a constraint binds}: exactly on every realization (microcanonical ensemble), or only in expectation (canonical ensemble) \citep{cimini2019statistical}.
The third is \emph{what the test is conditional on}: fixing the observed topology makes every subsequent statement conditional on it, so a network judged balanced under a topology-preserving model is compared only with alternative sign assignments on the same topology.
The fourth is \emph{how the reference distribution is obtained}: by randomization, or from a fitted generative model.
Fig.~\ref{fig:nullmodel-need} illustrates why a null model is needed,
Table~\ref{tab:bottomline_nullmodels} summarizes key differences of each model, and Fig.~\ref{fig:nullmodel-verdict-flip} shows how much the choice of the null model can matter.

\subsubsection{Fixing density of signs and shuffling the topology: G(N,m), G(N,p) and configuration model for signed links}
\label{subsubsec:topology-reshuffling}

The simplest topology-randomizing null models build on the classical Erd\H{o}s--R\'enyi random-graph ensembles $G(N,m)$ and $G(N,p)$.
% A useful baseline for the null models considered in this subsection is provided by the classical random-graph ensembles $G(N,m)$ and $G(N,p)$.
In $G(N,m)$~\citep{ErdosRenyi1959}, a graph is drawn uniformly from all simple undirected graphs with $N$ nodes and exactly $m$ links.
In $G(N,p)$~\citep{Gilbert1959}, each of the $\binom{N}{2}$ possible links is present independently with probability $p$, so the number of links fluctuates around its expected value $p\binom{N}{2}$.
Thus, $G(N,m)$ fixes the number of links exactly, whereas $G(N,p)$ fixes it only in expectation.
In statistical-physics terminology, these correspond to microcanonical and canonical constraints, respectively~\citep{cimini2019statistical}.
Both ensembles otherwise randomize the topology, making them a natural baseline for the signed null models considered here.
% The simplest class of null models controls only for the most basic network characteristics, namely the number of nodes and the overall density of positive and negative links.
% These models randomize the signed topology while preserving the number of nodes and either the exact or expected density of positive links.
% More generally, topology reshuffling, also called rewiring, can be implemented in multiple ways and under different constraints.
% This re-shuffling of the topology (also called rewiring) can occur in multiple ways.
This reshuffling of the topology, also called rewiring, can be implemented in multiple ways and under different constraints.

% \citet{Hao2024} present two such rewiring null models, the rewire null model and the signed rewire null model.
% In the rewire null model, they randomly select two links involving four different nodes, $(A,B)$ and $(C,D)$, and rewire them to either $(A,D)$ and $(C,B)$ or $(A,C)$ and $(B,D)$.
% Starting from the original network, they perform $40 m$ link rewires.
% This procedure preserves the degree of every node while randomizing the topology.
% In the signed rewire model, the same procedure is applied only to links of the same sign, thereby preserving each node's positive and negative degrees.
% % are selected for each rewire.
% % This second approach preserves exactly the signed degree of the nodes.

The $G(N,m)$ ensemble can be extended to signed networks by fixing the number of positive and negative links separately.
% ~\cite{ErdosRenyi1959} for random networks to include signs.
This model is equivalent to randomly rewiring the positive and negative links among all possible node pairs, while preserving their exact total numbers as in the original network.
On the randomized networks, one can compute different measures of structural balance, such as the fraction of balanced triangles.
By this, the null model tests whether both the observed structure and signed patterns are consistent with structural balance theory, assuming that the absolute abundance of positive and negative links is independent of it.
In other words, the numbers of positive and negative links are constrained and generated by another unobserved underlying process, while the exact locations of these links in the network may depend on structural balance theory.

A similar approach is to use a signed version of the $G(N,p)$ model, known as $G(N,p,q)$~\cite{el2012balance}.
In this model, a positive link is created with probability $p$, a negative link with probability $q$, and no link with probability $1-p-q$.
The Signed Random Graph Model (SRGM) of~\citet{gallo2024testing} provides a maximum-entropy formulation of the same canonical idea, with global probabilities chosen so that the expected numbers of positive and negative links match their empirical values.
Thus, unlike the signed $G(N,m)$ model, these models preserve the \textit{expected} numbers of positive and negative links rather than their \textit{exact} numbers.
Consequently, all network topologies with a given number of nodes are possible, from the empty to the fully connected network, although with different probabilities.
In signed $G(N,m)$, by contrast, only networks with the prescribed numbers of positive and negative links are possible, and they are equiprobable.
This is the microcanonical--canonical distinction introduced above.
% Hence, the difference between the two is similar to the microcanonical and canonical ensembles in physics where the energy is fixed respectively to an exact value or in expectation~\cite{cimini2019statistical}.
% The correspondence is exact once one identifies the constrained quantity as a sufficient statistic of an exponential-family ensemble, in which case the associated Lagrange multiplier plays the role of an inverse temperature.
In the empirical networks analyzed by~\citet{gallo2024testing}, however, all signed triangle types are strongly over-represented under the SRGM, making it uninformative about structural balance because the result is dominated by the complete randomization of the topology.

Topology can also be reshuffled while preserving additional node-level constraints.
\citet{Hao2024} present two rewiring null models, the rewire null model and the signed rewire null model.
In the rewire null model, they randomly select two links involving four different nodes, $(A,B)$ and $(C,D)$, and rewire them to either $(A,D)$ and $(C,B)$ or $(A,C)$ and $(B,D)$.
Starting from the original network, they perform $40m$ link rewire attempts.
This procedure preserves the degree of every node while randomizing the topology.
In the signed rewire model, the same procedure is applied only to pairs of links with the same sign, thereby preserving each node's positive and negative degrees.

% A further approach replaces the global edge-number constraint of the Erdos-Renyi models with node-level degree constraints.  is to use a signed variation of the configuration model, which fixes the degree sequence rather than only the total number of links.
The same idea of preserving node-level degree information underlies configuration model approaches.
The degree sequence can be preserved either exactly, by stub-matching or edge rewiring (the microcanonical configuration model)~\citep{bollobas1980probabilistic,molloy1995critical}, or only in expectation, as in canonical, or ``soft'', configuration models, such as the Chung--Lu model~\citep{Chung2002, Chung2002a}.
For a single global constraint such as the total number of links, the microcanonical and canonical choices agree asymptotically, but when the number of constraints grows with the size of the network, as it does for a full degree sequence, they do not, and the ensembles associated with the two models remain non-equivalent in the thermodynamic limit~\citep{cimini2019statistical}.
% Signed extensions preserve the positive and negative degree sequences: in expectation for the signed Chung--Lu model~\cite{derr2018_chungLuSB,gallo2024testing}, and exactly for the microcanonical signed configuration model~\cite{saiz2017evidence}.
Signed configuration models analogously constrain the positive and negative degree sequences:
in expectation in the canonical formulation~\citep{gallo2024testing},
and exactly in the microcanonical formulation~\citep{saiz2017evidence}.

A related Chung--Lu-based approach is the Balanced Signed Chung--Lu model of~\citet{derr2018_chungLuSB}.
The model aims to reproduce the unsigned degree distribution and the global positive-to-negative link ratio, while also accounting for clustering and the proportion of balanced triangles.
Because structural balance is itself incorporated into the generative mechanism, the model is better viewed here as a related generative model rather than as a null model for testing whether structural balance is present.

All these approaches are based on fundamental models of network science, however, they have two main limitations.
First, reshuffling the topology changes the number and arrangement of triangles and cycles in the network.
For example, real-world networks often have clustering coefficients substantially larger than those expected under a $G(N,p)$ model, so topology reshuffling typically decreases the number of triads on which structural balance measures are then computed~\citep{gallo2024testing}.
% The second problem relates to the fact that reshuffling treats the positive and negative link as identical~\citep{feng2022_nullmodelSB}.
A second limitation of simple reshuffling null models, emphasized by~\citet{feng2022_nullmodelSB}, is that they may effectively treat the placement of positive and negative links as exchangeable despite their different structural organization.
Negative links, for instance in social networks, are usually less numerous than positive ones and, when considered separately, can produce networks that are comparable to those generated by $G(N,p)$ or $G(N,m)$ models.
Positive links are instead more abundant, and the networks induced by their aggregation have topologies that strongly deviate from these null models.
The differences between positive and negative links are further explored in~\citep{szell2010_paradusInDepth,szell2012SignedMMOGmoreInfo, derr2018relevance}.
To overcome these limitations, other null models have been introduced that either fully or partially fix the network topology.

\subsubsection{Fixing links topology and random signs reshuffling}
\label{subsubsec:fixed-topology}

% One way to overcome the limitations of the previous models is to fix the underlying topology.
At the opposite extreme from the topology-randomizing null models discussed above, one can keep the empirical topology fixed and reshuffle only the signs.
A very simple and clear null model is the one used in~\citep{szell_2010} and is explicitly defined in their Supplementary materials.
They keep the topology of the signed network fixed, and reshuffle the positive and negative signed links.
This null model is equivalent to a permutation test on the observed links in the network.
Then, to check the level of balance, they compare the observed number of triads with the one expected under $1000$ sign reshufflings.
For the MMOG Pardus data that they analyze, they find that balanced triads are significantly over-represented, while unbalanced are under-represented.
The significance is defined by looking at the $z$-score of the observed number of triads relative to the null model, i.e., the difference between the observed and expected values divided by the standard deviation.
% compared to mean one expected from the 1000 reshuffling.
This same null model is used in ~\citep{leskovec2010countingTriads} to test whether under- and over-represented triads are compatible with structural balance.
They also assess balance through triad counts.
They find evidence for \textit{weak balance} in two (Wikipedia and Epinions) of three datasets analyzed as only triads with one negative link are under-represented, while all other triads are over-represented.
For the third dataset, Slashdot, they have mixed evidence about structural balance: triads with three positive links are over-represented, while triads with one or more negative links are always under-represented.
To quantify how significant is the over and under representation, they define a surprise function defined as $s(N_i) = (N_i - \mathbb{E}[N_i])/\sqrt{N_T p_0(\Delta_i)(1-p_0(\Delta_i))}$ where $N_i$ is the observed number of triads with $i$ number of negative links\footnote{The original paper uses the notation that $T_i$ is the number of triads with $i$ positive links.}, $N_T$ is the total number of triads in the network, $p_0(\Delta_i)$ is the a priori probability of a triad having $i$ number of negative links given the overall fraction of positive edges\footnote{The paper never prints the exact formula. However, the denominator $\sqrt{N_T p_0(1-p_0)}$ is a binomial standard deviation, which suggests $p_0(\Delta_i) = \binom{3}{i}p^{3-i}(1-p)^{i}$, the probability of a triad being of type $i$ after the signs have been reshuffled at fixed $p$, where $p$ is the fraction of \textit{positive links}.}, and $\mathbb{E}[N_i] = N_T p_0(\Delta_i)$ is the expected number of triads of type $i$ under the null model.

There are two main criticisms raised against the approach of~\citep{szell_2010, leskovec2010countingTriads}.
First, by preserving the exact numbers of positive and negative links, the null model conditions on those counts and does not account for uncertainty in them.
% In other words, the observed edge count should not be subject to random variations, like measurement errors.
Second, by preserving the topology and reshuffling signs across links, the same expected ratio of positive links is assumed for all nodes.
In other words, there is a strong assumption that all nodes have the same propensity for positive/negative relations.
The fixed-topology counterpart of the SRGM, denoted SRGM-FT, introduced in~\citep{gallo2024testing} relaxes the first assumption and preserves the numbers of positive and negative links only in expectation.
However, because the same global sign probabilities apply to all existing links, SRGM-FT still imposes the same expected ratio of positive links on every node.
The signed configuration model with fixed topology (SCM-FT) introduced by~\citep{gallo2024testing} relaxes this second assumption as well, by preserving each node's positive and negative degrees in expectation.
In this null model, the empirical topology is still fixed, while the numbers of positive and negative links for each node are preserved only in expectation.
This allows the expected ratio of positive links to vary across nodes.
Moreover, the total numbers of positive and negative links are also preserved only in expectation.
With this null model, any signed configuration on the fixed empirical topology is possible, but configurations that deviate more from the node-specific signed-degree constraints receive lower probability.
% However, the probabilities of signed configurations decrease when the number of positive and negative links deviates from the ones observed in the data.
When applying their null model, they find evidence for structural balance in social networks and the opposite for biological networks.
Note that the same null model is introduced separately by~\cite{Hao2024}, however, the details of the numerical implementations are different.
They name their null model a signed degree and topology preserving (STP).
Using their null model, they find in all real networks analyzed evidence for structural balance as balanced triads and balanced motifs with four nodes are over-represented while unbalanced triads and unbalanced motifs with four nodes are under-represented.

\subsubsection{What we learn by comparing these studies}
\label{subsubsec:comparing-nullmodels}

Both works~\citep{gallo2024testing, Hao2024} provide an informative comparison of the effects of choosing different null models.
There are mainly five points that we can draw.
First, they show different null models can give opposite results about a triad being significantly over- or under-represented across a range of social, biological and synthetic networks.
Hence, the choice of null model is not purely a convention but has impact on the conclusions one can make.

Second, they find differences when preserving the expected ratio of positive links for each node or forcing it to be the same for all nodes.
Social networks show evidence of strong balance, i.e., the all-negative triad ($---$) is under-represented, under the SCM-FT and STP null models, which preserve node-specific signed degrees, and hence positive-link ratios, in expectation.
Under null models that impose a common expected positive-link ratio on all nodes, such as sign shuffling or SRGM-FT, the verdict reverses, and the $---$ triad becomes over-represented.
Fig.~\ref{fig:nullmodel-verdict-flip} contrasts the two randomization schemes that produce this reversal.
Recalling that the all-negative triad is expected to be under-represented under strong balance but is allowed under weak balance, this suggests that the positive-link ratio is an important characteristic for distinguishing weak structural balance from strong structural balance.

Third, the topology-reshuffling null models rewire~\citep{Hao2024} and SRGM~\citep{gallo2024testing} agree that triads with 0, 1, and 2 negative links are over-represented.
However, they disagree on the over- or under-representation of triads with three negative links.

Fourth, when comparing the results of the sign reshuffle performed by~\cite{Hao2024} with the ones coming from the sign reshuffle performed by~\cite{leskovec2010countingTriads}, we find that they agree on the Epinions data, but disagree for the Slashdot.
The two studies reach opposite verdicts on the triad with one positive link and on the all-negative triad.
Recomputing both analyses from the dataset of~\citet{leskovec2010countingTriads} shows that the disagreement is caused neither by the number of randomizations nor by the small difference in the reported fraction of positive links.
It is caused by the triad census: the $1\,508\,105$ triads of~\citet{leskovec2010countingTriads} are recovered exactly by counting a triangle once for each combination of the parallel edges of its reciprocated pairs, whereas the underlying simple graph contains $579\,565$ triangles, and under the latter count both disputed triads change sign.
A study reporting a triad census should therefore state whether reciprocated pairs were merged into a single edge, since that choice alone reverses two of the four verdicts here.

Fifth, two distinct sources of uncertainty are at work: one \emph{inferential}, the other \emph{observational}.
The former arises because a sampled null model explores its ensemble only finitely, so a $z$-score computed from $R$ randomizations carries a Monte Carlo uncertainty\footnote{Monte Carlo uncertainty typically affects the magnitude of the $z$-score, but can also change its sign when the observed value is close to the expected one.} of order $1/\sqrt{R}$; while the latter arises because the observed network is not ground truth.
The signs are themselves estimated, whether elicited by survey or inferred from behavior (Sec.~\ref{subsec:sign-obs-and-inf}), and a null model that fixes the observed sign counts exactly treats them as error-free.
Models with closed-form ensemble statistics remove the first source but say nothing about the second.
Addressing the second requires modeling the observation itself.
This can be done by either treating the signs as noisy, which shows that ignoring this inflates the apparent number of antagonistic blocks and can mimic a need for weaker notions of balance~\citep{Gallo2024}, or by treating the network as one realization of a generating process and reporting a confidence interval for the expected proportion of balanced triangles~\citep{chen2026nonparametric}.
Note that neither approach relies on a sampled null ensemble: \citet{Gallo2024} infer the block structure by minimising a Bayesian information criterion and \citet{chen2026nonparametric} obtain the confidence interval in closed form.
Hence, both approaches address the observational source of uncertainty without incurring the Monte Carlo error of the inferential one.

\subsubsection{Null models for directed signed networks}
Balance in directed signed networks is far less developed, and the few available null models disagree on what directed balance should mean.
% Extending their undirected approach.
\citet{Hao2025} define balance for directed triads and introduce a maximum-entropy null model that preserves the signed in- and out-degrees; applied to large online social networks, it reveals triad frequencies consistent with two proposed directed versions of strong balance.
\citet{Gallo2025} instead extend the exponential random graph framework to binary directed signed networks and focus on length-two cycles, where balance reduces to sign reciprocity; they find that a directed extension of balance theory does not consistently explain the observed motif frequencies unless a coarser definition of balance is adopted.
Directed structural balance therefore remains open both in its definition and in the choice of an appropriate null model.

\subsubsection{Network ensembles preserving local clustering or meso clusters (i.e., communities)}
\label{subsubsec:meso-nullmodels}

% The two families of null models discussed so far sit at opposite extremes.
% Those that reshuffle the topology (Sec.~\ref{subsubsec:topology-reshuffling}) discard the local structure of the empirical network, so that any excess of balanced triads is credited to balance by default.
% Those that fix the topology and permute the signs (Sec.~\ref{subsubsec:fixed-topology}) go to the other extreme: they preserve too much information from the empirical signed network, not allowing for exploring other plausible configurations that could have differing levels of structural balance.
% Precisely, preserving the full network topology and the exact number of positive and negative links means that any new link cannot be formed or any of the existing link cannot be removed.
% A third family sits in between.
% It leaves the topology free but constrains a structural property that is itself known to correlate with sign, so that the test asks whether balance explains anything beyond that property.
% Two such properties have been used: local clustering, measured by how many triangles a link takes part in, and mesoscale group structure.
Topology-reshuffling null models (Sec.~\ref{subsubsec:topology-reshuffling}) discard the local structure of the empirical network, whereas fixed-topology null models (Sec.~\ref{subsubsec:fixed-topology}) preserve the empirical topology and vary only the signs.
A further strategy is to preserve or explicitly model structural properties that are themselves known to correlate with sign, so that the test asks whether balance explains anything beyond those properties.
Two such properties considered here are local clustering, captured by how many triangles a link takes part in on the fixed topology, and mesoscale group structure, modeled with signed block models.

\paragraph{Preserving local clustering}
The first is the null model of~\citet{feng2022_nullmodelSB}.
The authors introduce the concept of \textit{embeddedness} of links as the number of triangles that a link belongs to.
Then, they shuffle the signs separately among edges with the same level of embeddedness.
The motivation to preserve link embeddedness derives from the fact that transitivity and homophily enhance the embeddedness of positive links.
Then, a null model preserving this quantity tests whether structural balance further favors the specific over- and under-representation of triads that could not be explained by the link embeddedness.
% They show that under some assumptions, the distribution in the network space of their model approaches a normal distribution (in the number of unbalanced triads), making it possible to sample efficiently from this space.
They show that, under some assumptions, the number of unbalanced triads is asymptotically normally distributed under both their embeddedness-preserving null model and the standard sign-permutation null model.

\paragraph{Preserving mesoscale group structure}
The second constrains the network at the mesoscale by introducing a block structure and letting link and sign probabilities depend on it.
The natural device is the signed stochastic block model, which is also the tool used to partition signed networks into blocks when balance is measured rather than tested (Sec.~\ref{subsubsec:frustration}).
% Used as a null model rather than as an estimator, it supplies the ensemble against which the observed level of balance is judged.
\citet{Gallo2024} rephrase structural balance in statistically testable terms on such a model, and show that reading the observed signs as noise-free is what makes most real graphs appear to require more mutually antagonistic blocks than they actually have (Sec.~\ref{subsubsec:comparing-nullmodels}).
On the generative side, \citet{morrison2020_communityInSignedNetworks} reveal communities in signed networks based on the stochastic block model by defining the probability of having a link within each of two groups and across the two groups.
Then they define the probability that a link is positive or negative within or across groups.
This allows them to generate networks with a community structure and derive their spectrum.
Finally, they choose deterministic, continuous time dynamics~\citep{kulakowski_2005, marvel2011_evolutionOFSBcontinuous} to see how the initial community structure affects the evolution of structural balance.
The caveat of such a procedure is that the expectations come from a model fitted to the same network, so the inferred block structure can absorb the very effect being tested.
This risk is not hypothetical as block models may be overfitted~\citep{Doreian2008}.
When this happens, several equally well-fitting partitions can be found, and no principled choice between them can be made~\citep{Doreian2008}.

\subsubsection{Model-based tests: regressions, ERGMs, and actor-oriented models}
\label{subsubsec:ergm-fitting}

Randomization null models remove the dependence between edge signs non-parametrically.
An alternative is to specify a statistical model in which that dependence is itself a parameter to be estimated, so that a drive toward balance can be weighed against competing explanations such as link density, degree heterogeneity, reciprocity, homophily, and status.
The simplest version treats each dyad as an observation and fits a logistic regression of the sign on the signed configuration of shared neighbours.
\citet{lerner2016_SBwikipedia} shows that it matters whether one models the marginal probability that two actors interact at all or the conditional probability that an existing interaction is hostile rather than friendly, and finds support for balance only in the conditional model.
Ordinary logistic regressions that treat dyads or triads as independent observations, rather than modeling network dependence, have been used in the same spirit to test triadic hypotheses in survey data~\citep{Kalmijn2003,Chiang2019,Koster2018}.
Exponential random graph models (ERGMs) keep this regression logic while accounting for the mutual dependence of ties: signed networks are handled either as two coupled layers of positive and negative edges through multiplex ERGMs~\citep{pattison1999_mutlivarERGM,wang2013_multivarERGM,Huitsing2012} or, more recently, through a signed ERGM with balance-derived network statistics such as ``the enemy of my enemy is my friend''~\citep{Fritz2025}.
When several social mechanisms are tested jointly, balance and homophily typically emerge as the strongest~\citep{Yap2015}, and some authors add contextual refinements to balance theory on this basis~\citep{Sarazin2021}.
For evolving networks the same models are estimated over time: temporal ERGMs show that distinguishing the formation from the dissolution of balanced triads improves out-of-sample tie prediction~\citep{Lee2025}, while stochastic actor-oriented models~\citep{snijders2010introduction} and relational event models~\citep{butts2008relational,lerner_2020} attribute tie changes to actor-level or event-level balance tendencies.
These approaches connect directly to the Hamiltonian models reviewed in Section~\ref{sec:emerg}: a Gibbs distribution $P(\mathcal{S})\propto e^{-H}$ over signed configurations is an exponential-family model whose sufficient statistics are counts of signed dyads, sign-homogeneous pairs, and balanced triads.
So fitting the triadic Hamiltonian of \citet{belaza2017_StatPhysBalanceTheory} by maximum likelihood (Sec.~\ref{subsubsec:hamil-multitype}) is formally equivalent to fitting an ERGM.

A recurring caveat of these models is degeneracy: triadic terms can place most probability mass on nearly empty or nearly complete graphs, so estimates are sensitive to specification and to the treatment of neutral ties.

Signed ERGMs are a relatively recent development.
\citet{Fritz2025} introduce theirs to address what they describe as an absence of inferential models tailored to signed networks.
Earlier ERGM-based tests of balance instead treated positive and negative ties as separate or coupled (multiplex) layers~\citep{pattison1999_mutlivarERGM,wang2013_multivarERGM,Huitsing2012}. Cross-study comparison therefore remains limited.

\FloatBarrier

\subsection{Bottom line: how to measure structural balance?}
\label{subsec:bottomline-measures}

Assessing structural balance requires two coupled choices: a measure, and a null model against which the measured value is judged.
Neither choice is neutral: each encodes assumptions about how edges depend on one another, how much of the system an observer can see or control, and whether positive and negative ties are treated symmetrically.
Different choices can reverse the qualitative conclusion for the same network~\citep{aref2017measuring,gallo2024testing,Hao2024}.

Structural balance is not universally expected; whether it appears depends on the nature and dynamics of the system.
In social networks it is commonly attributed to cognitive and social pressure toward consistency, and empirical support is comparatively strong at the triadic level~\citep{leskovec2010countingTriads,szell_2010}.
In international relations the evidence is period- and definition-dependent, and balance among blocs need not imply convergence to a single global configuration~\citep{doreian,Aref_2018}.
Economic and financial networks may display balance when stable cooperative blocs form, but competition and crises can disrupt it~\citep{Kuyyamudi2019,ferreira_loss_2021}.
Cognitive networks show partial balance as beliefs evolve toward internal coherence, bounded by conflicting evidence and entrenched attitudes.
Biological and ecological systems rarely exhibit full balance: regulatory networks may approximate it, whereas signaling, metabolic, brain and ecological networks typically do not~\citep{Iacono2010,Rizi2021}.
Overall, balance is most likely where relations are reciprocal and adaptive, and least likely where they are asymmetric, externally driven, or volatile.
Structural balance is thus not a default property to be confirmed but a hypothesis whose plausibility differs by domain; the measure and null model should be chosen accordingly.

The measures divide into two conceptual families.
\emph{Local aggregation measures} pick a substructure (e.g., a cycle or a closed walk) and count how many instances are balanced, and normalize~\citep{harary1959_measuringSB,cartwright_1956,estrada2014_RandomWalkSB,kirkley2018balanceInSN,talaga_polarization_2023}.
\emph{Global optimization measures} ask how far the whole network is from balance; for example, the frustration (line) index is the minimum number of edge-sign flips that make it balanced (Sec.~\ref{subsubsec:frustration})~\citep{harary1959_measuringSB,Doreian1996,facchetti2011computing,Aref_2018}.
The first practical question is scale.
Balanced-triad density and, more generally, walk-based indices reduce to traces or determinants of matrix functions and remain computable for networks with $10^{5}$--$10^{6}$ edges~\citep{estrada2014_RandomWalkSB,kirkley2018balanceInSN}.
Exhaustive enumeration of long cycles does not.
The frustration index is NP-hard: binary-programming formulations give exact values up to roughly $10^{5}$ edges, and community-detection or spectral heuristics extend approximate values to larger graphs~\citep{Aref2018_frustrationUBQP,Aref_2018,anchuri2012spectralSB}.

\begin{figure}[!htbp]
    \centering
    \includegraphics[width=\linewidth]{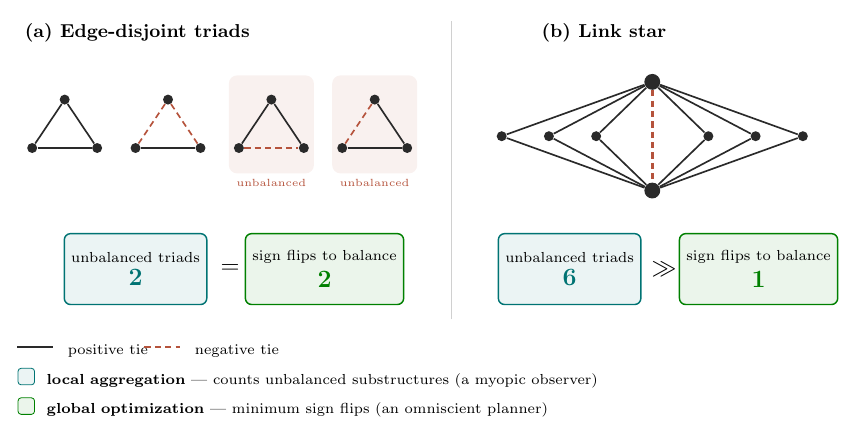}
    \caption{\textbf{Local and global balance measures can agree or disagree on the same network.}
\textbf{(a)} On edge-disjoint triads the two families coincide: each unbalanced triangle needs exactly one sign flip, and because the triangles share no edges the flips do not interact, so the number of unbalanced triads equals the frustration index.
\textbf{(b)} On a link star, six triads share a single negative edge.
Every triad is unbalanced, so a local aggregation measure reports near-complete imbalance, while the frustration index reports one flip.
Solid edges are positive, dashed edges negative; the networks are schematic.}
    \label{fig:measures-families}
\end{figure}

The two families can give very different assessments of the same network.
Cycle measures that weight all lengths equally are dominated by long cycles and drift toward uninformative mid-range values~\citep{aref2017measuring}. Weighted cycle- and walk-based indices avoid this by downweighting longer structures, but the decay function and its parameters then become modeling choices~\citep{talaga_polarization_2023}.
Comparative analyses place the same network at very different points on the $[0,1]$ scale depending on whether a cycle-, eigenvalue-, or frustration-based measure is used, and recommend frustration-based measures on axiomatic grounds and for their power to distinguish graphs~\citep{aref2017measuring}.
One reason the two families can diverge is the overlap among the network's cycles: on a set of edge-disjoint triads, the number of unbalanced triads equals the frustration index, whereas when many triads share edges the two quantities can differ substantially.
A link-star network, i.e., a network made of many triads sharing one central edge, exemplifies this point: if that central edge is negative and all others positive, all triads are unbalanced, while the frustration index reports a single flip.
Fig.~\ref{fig:measures-families} contrasts the two cases: on edge-disjoint triads the number of unbalanced triads equals the frustration index exactly, whereas on a link star the two measures differ by a factor of $k$.

Underlying the disagreement is a difference in stance.
Aggregation measures treat the counted substructures separately, even when they overlap, and describe a myopic observer sampling local configurations; the frustration index takes the globally closest balanced configuration as its reference and describes an omniscient planner able to identify the global optimum.
This matters for interpretation: the frustration index can suggest that balance is ``close'' when locally driven dynamics would never reach it (Sec.~\ref{sec:emerg}).
A second caveat concerns the asymmetry of the two signs.
Negative ties are typically rarer, less reciprocated and less clustered than positive ones, and antagonism does not propagate through a triad the way affinity does.
This asymmetry can be obscured in different ways.
Sign-permutation null models treat sign placement as exchangeable across edges and therefore disregard differences in the structural organization of positive and negative ties, while triad-based measures require a choice about which signed configurations are considered balanced. This choice is particularly important in multipolar systems: at the triad level, the distinction between strong and weak balance lies precisely in whether the all-negative triad is classified as unbalanced or balanced~\citep{davis1967clusteringSB,feng2022_nullmodelSB}.

Whether balance should be assessed on triads alone or on cycles of every length is unsettled.
The two are equivalent only on complete graphs (Sec.~\ref{subsubsec:signed-net-def}), so on a sparse network the maximum cycle length entering a measure is a modeling assumption and not an approximation.
\citet{aref2017measuring} argue for measures defined over all cycles, on the grounds that restricting attention to triads discards most of the cycle structure.
Against this, the mechanism the measures are meant to capture is local: the tension \citet{heider1946attitudes} describes arises in configurations an individual perceives, and a cycle of length six through a node is not one of them~\citep{Gorski2025}.
The choice therefore follows the system: where ties are set by agents acting on what they observe, short cycles are the appropriate unit; where the network records a process external to the nodes, there is no corresponding ground for truncation.

The many types of measures with their properties and suggestions for application are summarized in Table~\ref{tab:bottomline_measures}.

\begin{table}[!htb]
\centering
\caption{Choosing a measure of structural balance. Scale limits are
order-of-magnitude and hardware-dependent.}
\label{tab:bottomline_measures}
\footnotesize
\setlength{\tabcolsep}{4pt}
\renewcommand{\arraystretch}{1.15}
\begin{tabularx}{\linewidth}{@{}
  >{\raggedright\arraybackslash}p{2.9cm}
  >{\raggedright\arraybackslash}p{2.2cm}
  >{\raggedright\arraybackslash}X
  >{\raggedright\arraybackslash}X@{}}
\toprule
\textbf{Measure family} & \textbf{Practical scale} & \textbf{Key assumption / free choice} & \textbf{Use when} \\
\midrule
Balanced-triad density / triangle index~\citep{harary1959_measuringSB}
  & Any %($\operatorname{Tr}(\mathbf{A}^{3})$)
  & Only triads count; triads counted separately and weighted equally
  & Quick comparison; near-complete networks; longer cycles ignored
  % strong vs.\ weak via the $---$ triad
  \\
\addlinespace
Cycle-based, all lengths~\citep{cartwright_1956}
  & Small only
  & Long cycles dominate; value drifts to mid-range
  & Rarely advisable in practice \\
\addlinespace
Walk-based indices~\citep{estrada2014_RandomWalkSB,singh2017WalkSB,kirkley2018balanceInSN,talaga_polarization_2023}
  & Up to $10^{5}$--$10^{6}$ edges;
  matrix-function methods
  %eigen\slash determinant
  & Decay function $f(k)$ and its parameter ($\beta$, $z$)
  & Large networks; node- and pair-level balance; directed and weighted \\
\addlinespace
Frustration / line index (Sec.~\ref{subsubsec:frustration})~\citep{harary1959_measuringSB,Doreian1996,facchetti2011computing,Aref_2018}
  & Exact $\lesssim10^{5}$ edges; heuristics beyond
  & A global optimum is the relevant reference; generalized forms may weight $+/-$ errors differently
  & Minimum intervention to balance; systemic inconsistency; centrally controllable systems \\
\addlinespace
Weak-balance variants~\citep{davis1967clusteringSB,anchuri2012spectralSB,kirkley2018balanceInSN,talaga_polarization_2023}
  & As parent family
  & $\ge2$ antagonistic groups allowed; $---$ counts as balanced
  & Multipolar systems (blocs, factions) \\
\bottomrule
\end{tabularx}
\end{table}

A measured value of balance is interpretable only relative to a null model, because purely structural properties of the network, such as sign density, clustering and degree heterogeneity, can generate patterns that appear consistent with structural balance even in the absence of any social-cognitive mechanism~\citep{feng2022_nullmodelSB,gallo2024testing}.
Null models form a hierarchy of preserved constraints, from signed Erd\H{o}s--R\'enyi-type models that constrain only the node count and global sign densities while reshuffling the topology, through signed configuration models that additionally preserve node-level degree information, to models that keep the observed topology fixed while constraining global sign densities (SRGM-FT) or node-level signed degrees in expectation (SCM-FT, STP)~\citep{el2012balance,saiz2017evidence, szell_2010,gallo2024testing,Hao2024}.
The constraint that most changes the verdict is the per-node ratio of positive links: preserving it yields evidence of strong balance in social networks, while forcing a common ratio across nodes turns the all-negative triad $---$ from under- to over-represented, i.e., a weak-balance signature.
For directed signed networks the situation is even more complex: the few available null models disagree on what directed balance should even mean~\citep{Hao2025,Gallo2025}.

A second consideration is the tractability of the null model.
For dyad-independent canonical null models such as SRGM, SRGM-FT, the canonical signed configuration model, and SCM-FT, the moments of triad counts can be evaluated analytically~\citep{gallo2024testing}.
Other models may admit asymptotic approximations; for example, the unbalanced-triad count under the embeddedness-preserving model of~\citet{feng2022_nullmodelSB} is asymptotically normal.
When the relevant distribution is instead estimated by sampling, Monte Carlo uncertainty is introduced; examples include repeated permutation and rewiring procedures, as well as more complex model-based approaches such as signed ERGMs~\citep{Fritz2025}.
% Null models with closed-form ensemble statistics, such as the signed ERGM family (Sec.~\ref{subsubsec:ergm-fitting}), and models for which the unbalanced-triad count is asymptotically normal give exact, cheap $z$-scores and scale to large networks~\citep{feng2022_nullmodelSB,Fritz2025}.
% Sampling-based models pay a Monte Carlo error instead: given $R$ randomizations, the error scales as $1/\sqrt{R}$.
In practice, for a small network one can afford a richer, less tractable model that controls many expected properties and is sampled carefully, whereas a large network is better served by a model with analytical or asymptotic results, which saves computation time.

A third consideration lies outside the null model altogether, in how the statistic itself is computed.
\citet{leskovec2010countingTriads} and~\citet{Hao2024} apply the same sign-reshuffling null to the same Slashdot data and reach opposite verdicts on two of the four triad types, because a triangle is counted once for each combination of the parallel edges of its reciprocated pairs in the first study and once in the second (Sec.~\ref{subsubsec:comparing-nullmodels}).
Reporting the null model is therefore not sufficient on its own: the triad census has to be specified alongside it.

The many types of null models with their properties and caveats are summarized in Table~\ref{tab:bottomline_nullmodels}.

\begin{table}[!htb]
\centering
\caption{Choosing a null model for the statistical significance of balance.
\emph{Closed form} indicates whether expectations and variances of the relevant statistics are
available analytically (\checkmark), only partially or asymptotically ($\sim$),
or only by sampling (\ding{55}).}
\label{tab:bottomline_nullmodels}
\footnotesize
\setlength{\tabcolsep}{4pt}
\renewcommand{\arraystretch}{1.15}
\begin{tabularx}{\linewidth}{@{}
  >{\raggedright\arraybackslash}p{2.85cm}
  >{\raggedright\arraybackslash}p{3.25cm}
  >{\centering\arraybackslash}p{1.15cm}
  >{\raggedright\arraybackslash}X@{}}
\toprule
\textbf{Null model} & \textbf{Preserves} & \textbf{Closed form} & \textbf{Main caveat} \\
\midrule
Signed $G(N,m)$
  & $N$; exact $\#{+}$, $\#{-}$
  & $\sim$
  & Does not preserve clustering or triad counts; $+/-$ exchangeable \\
\addlinespace
Signed $G(N,p,q)$~\citep{el2012balance}
  & $N$; expected $\#{+}$, $\#{-}$
  & \checkmark
  & As above; canonical \\
\addlinespace
Microcanonical signed configuration~\citep{saiz2017evidence,Hao2024}
  & Exact signed degree of each node
  & \ding{55}
  & Still reshuffles the topology; requires sampling \\
\addlinespace
% Signed configuration~\citep{gallo2024testing,saiz2017evidence,Hao2024}
%   & Signed degree sequence, exact or expected
%   & $\sim$
%   & Still reshuffles the topology \\
Canonical signed configuration~\citep{gallo2024testing}
  & Expected signed degree of each node
  & \checkmark
  & Still reshuffles the topology \\
\addlinespace
Sign permutation on fixed topology~\citep{szell_2010,leskovec2010countingTriads,feng2022_nullmodelSB}
  & Full topology; exact $\#{+}$, $\#{-}$
  & $\sim$
  & Assumes exact global sign counts; one $+$ ratio for all nodes \\
\addlinespace
SRGM-FT~\citep{gallo2024testing}
  & Topology; expected $\#{+}$, $\#{-}$
  & \checkmark
  & One $+$ ratio for all nodes \\
\addlinespace
SCM-FT / STP~\citep{gallo2024testing,Hao2024}
  & Topology; expected signed degree per node
  & \checkmark
  &
  % Exact for SCM-FT, approximate for STP; implementation details affect the verdict
  Numerical implementations differ
  \\
\addlinespace
Embeddedness-preserving~\citep{feng2022_nullmodelSB}
  & Topology; sign counts by embeddedness
  & $\sim$
  & Asymptotically normal only; assumes embeddedness captures the confound \\
\addlinespace
Blockmodel / meso-scale~\citep{morrison2020_communityInSignedNetworks,Gallo2024}
  & Community structure
  & $\sim$
  & Expectations come from the fitted model, which may absorb the effect being tested \\
\addlinespace
% Signed ERGM row, drafted by PG and left out on 2026-10-28 (GV): an ERGM is a regression
% approach, not a null model -- balance is tested through a fitted coefficient against zero,
% not against a reference ensemble -- so it belongs in Sec. 3.2.6 and the Bottom line rather
% than in this table. Kept here in case we decide otherwise. Checked against the paper: the
% sample space is all three-state (+, -, 0) signed networks, so the topology is not fixed;
% the normalizing constant needs T*3^{N(N-1)/2} summands, "prohibitive even for small
% networks", hence MCMC-MLE and the cross; and the degeneracy caveat is the authors' own,
% worst precisely when the balance parameters are positive, which is why they use the
% geometrically weighted (alternating k-triangle) variants GWESE+/- and GWESF+/-.
% Signed ERGM (SERGM)~\citep{Fritz2025}
%   & Chosen statistics in expectation; topology not fixed
%   & \ding{55}
%   & Balance enters as a fitted parameter, not a reference ensemble; triadic terms are degeneracy-prone unless geometrically weighted \\
% \addlinespace
Directed signed~\citep{Hao2025,Gallo2025}
  & Signed degrees, or dyadic constraints
  & $\sim$
  & Directed balance itself is not agreed on \\
%   \addlinespace
% Sparse signed graphon~\citep{chen2026nonparametric}
%   & Nothing about the realized network; node exchangeability assumed
%   & \checkmark
%   & Balance must be carried by the fitted graphon, which may absorb the effect being tested; signs are drawn independently, so no triadic mechanism is represented \\
\bottomrule
\end{tabularx}
\end{table}

No measure or null model is a neutral instrument: each fixes a stance on edge interdependence, on how much the analyst knows or controls, and on whether positive and negative signs are treated symmetrically.
We therefore recommend reporting more than one measure alongside an explicitly justified null model, and reading ``this network is (not) balanced'' as conditional on both choices.

\FloatBarrier
\clearpage

\section{Dynamics and emergence of structural balance}
\label{sec:emerg}

Having established the theoretical foundations and mathematical formulations of structural balance, we now turn to its emergence and stability in dynamical systems.
This section examines how balanced states can arise spontaneously from local interactions among agents. The central questions are: (i) under what conditions do initially disordered networks evolve toward structural balance? and (ii) under what conditions do initially balanced networks lose stability and become disordered? Depending on the update rules and assumptions, these dynamics may lead not only to perfectly balanced systems but also to network structures that resemble real-world configurations.

The development of models for the emergence of structural balance followed a trajectory similar to other agent-based frameworks, becoming possible only with the rise of computational power and the spread of simulation techniques. Two seminal papers, both published in 2005, form the cornerstone of almost all later work and remain among the most widely cited. Kułakowski, Gawroński and Gronek~\citep{kulakowski_2005} introduced a model of continuous-time signed-link dynamics,
which we will refer to as \emph{gradient triad dynamics} (GTD)\footnote{This model does not have a broadly accepted name. It is sometimes called a ``continuous-time’’ model, although discrete-time variants also exist.
% Following established practice, we use the name of the first author.
We adopt the name ``gradient triad dynamics'' because the standard undirected dynamics has a gradient-flow structure with respect to a triadic Hamiltonian. Some extensions discussed below do not preserve this gradient structure, but we still refer to them as extensions of GTD.
% We also occasionally use the name ``Kułakowski--Gawroński--Gronek model'' when referring to the original model or its broader family.
% when necessary, we refer to the broader family using the names of all three authors.
}. Antal, Krapivsky and Redner~\cite{antal_2005,antal_2006} introduced local triad dynamics (LTD) and constrained triad dynamics (CTD).
These models have their unrecognized predecessors.
In 1994, Lee, Muncaster and Zinnes~\cite{Lee1994} proposed a model incorporating ``memory’’ and triadic interactions, of which GTD is a special case.
Since this work appeared in a philosophy journal, it remained largely unnoticed by physicists and network scientists studying structural balance\footnote{The first modeling paper to explicitly acknowledge this precursor is~\citet{Morrison2023} (2023).}. In 2003, \citet{Wang2003} proposed a model equivalent that reduces to a special case of LTD. These earlier contributions are important to acknowledge, but they did not have a significant impact on subsequent developments in the physics and network science literature. For this reason, we do not name later models after these predecessors.

This Section focuses on models that capture the emergence and stability of structural balance through dynamical rules inspired, sometimes loosely, by the principles of structural balance theory. It does not cover the extensive literature on optimization approaches to structural balance~\citep{facchetti2011computing}. That body of work aims to identify the optimal partition of a signed network into two antagonistic factions, often by minimizing the number of elements that need to be changed (see the frustration index, Sec. \ref{subsubsec:frustration}). These approaches include exact and heuristic algorithms~\citep{Aref2018_frustrationUBQP, Iacono_frustrationAlgorithms}, as well as methods based on genetic algorithms~\citep{Sun2014, Wang2016}, and address NP-hard computational problems. Their goal is to find efficient algorithms for partitioning, and they do not explain how societies or other systems originate and evolve. Although important in their own right, such studies fall outside our focus here.

The structure of the following sections reflects different perspectives on the dynamics of structural balance. After the introductory Sec.~\ref{subsect:concept-emerg}, Sec.~\ref{subsec:kk-model}-\ref{subsec:stoch-sign-dyn} examine the emergence and stability of structural balance solely through the dynamics of signed relations.
As link signs constitute the core variables of structural balance theory, these models are the natural starting point. Section \ref{subsec:sign-attribute-dynamics} considers models in which signed relations interact with agent attributes. These attributes may remain fixed and influence link signs, or they may coevolve with them.
This section also explores the dynamics of dyadic interactions that facilitate or destabilize structural balance. Such approaches fall under structural balance emergence because they correspond to the impersonal case of the $P$--$O$--$X$ formulation, in which $X$ denotes an impersonal entity (e.g., an object, an opinion, etc.).
Finally, section \ref{subsec:generalized} explores models of balance in more complex network structures, including multilayer, temporal, and higher-order networks.

\subsection{Conceptualizing emergence of structural balance \label{subsect:concept-emerg}}

The notion of ``emergence of structural balance'' carries a dual meaning. In the first sense, it refers to an initially unbalanced system that evolves into a balanced state. In the second, more general sense, it denotes the appearance of configurations in which the principles of structural balance theory become observable. Such states need not be perfectly balanced, but they exhibit structural features that reflect the influence of structural balance theory.

Most research in this area seeks to identify the conditions under which networks reach fully balanced states. When this does not occur, systems may become trapped in so-called \emph{jammed states} or exhibit persistent unbalanced dynamics. In models admitting an energy formulation, jammed states correspond to local minima in the associated energy landscape, although whether these minima are dynamically trapping depends on the update rule. If the measure of balance is based on triads (see Eq. \eqref{eq:triangle index}), a typical Hamiltonian\footnote{In classical mechanics, the Hamiltonian generates the equations of motion in canonical variables. In the statistical physics models considered here, it is used primarily as an effective energy function, which need not generate canonical equations of motion or be extensive. For simplicity, we use the terms \emph{Hamiltonian} and \emph{energy} interchangeably when referring to these model functions.} is given by
\begin{equation}
H = - \sum_{(i,j,k)\in\mathcal{T}} J_{ijk} x_{ij} x_{jk} x_{ki} \overset{J_{ijk}=1}{=} - \sum_{(i,j,k)\in\mathcal{T}} x_{ij} x_{jk} x_{ki},
\label{eq:hamiltonian_triad}
\end{equation}
where $\mathcal{T}$ is the set of all existing triads, $x_{ij}$ denotes the signed relation between agents $i$ and $j$, and $J_{ijk}\ge0$ specifies the strength of interaction associated with the triad $(i,j,k)$. A vanishing coefficient $J_{ijk}=0$ indicates that the corresponding triple of agents does not contribute to the balance energy, for example, because the corresponding triad is not considered meaningful or active for the interaction under study. Usually, the model is simplified by assuming homogeneous interactions $J_{ijk}=1$ for all triads.
Models can be broadly divided into \emph{continuous} and \emph{discrete}. In continuous formulations, $x_{ij} \in \mathbb{R}$ or is restricted to a bounded interval, typically $(-1,+1)$. In discrete formulations, $x_{ij}$ takes values from a finite set, most commonly $\{-1,+1\}$. % In general, $x_{ij} \in \mathbb{R}$, though specific models restrict this range, for instance to values in the range $(-1,+1)$ or in the discrete case to $\pm 1$ \msmcomment{I would probably frame slightly differently, saying there are ``continuous" models where $x_{ij} \in \mathbb{R}$ or $x\in(-1,+1)$ and discrete models where $x\in\{-1,+1\}$ or other sets, e.g., $x\in\{-1,0,+1\}$}.
For binary relations and homogeneous interactions, the global minimum of the Hamiltonian corresponds to configurations in which all existing triads are balanced, that is, $x_{ij} x_{jk} x_{ki} > 0$ for every triad $(i,j,k) \in\mathcal{T}$.
Understanding the nature of jammed states, that means ``when they arise'', ``how many exist'', and ``what forms they take'', has been a central focus of many studies.
These issues are discussed in detail in Sec.~\ref{subsec:jammed}.

When balance is achieved, a further question arises: \emph{what kind of balance emerges?} Is it the ``paradise state'', in which all links are positive? Is it a division into two antagonistic factions? Or, when considering the weak formulation of structural balance theory, does a multipolar system appear? Initial conditions may also be crucial: for some models, only a single class of stable solutions exists, whereas in others, multiple attractors coexist, so that the final state depends strongly on the starting configuration. Moreover, phase transitions are often observed when system parameters cross critical thresholds, leading to qualitative changes in the type of stability that the network exhibits.

Beyond the type of balance, another studied point is the \textit{dynamics of convergence}, i.e., ``How long does it take for a system to reach balance, and how does this time scale with system size or vary across parameter regimes?''
In some cases, convergence can be relatively fast, while in others it grows sharply with $N$ or depends on the chosen rules.
Comparing these convergence properties across related models provides valuable insights into their dynamics.

% \subsection{Ku{\l}akowski model: Gradient descent on a Hamiltonian}
% \subsection{The Ku{\l}akowski--Gawro\'nski--Gronek model and its extensions}
\subsection{Gradient triad dynamics and its extensions}
\label{subsec:kk-model}

\citet{kulakowski_2005} assume that $x_{ij}$ are real numbers and time evolution of these variables is continuous.
The most basic version of gradient triad dynamics (GTD) is defined as follows. Each link $x_{ij}\in\mathbb{R}$ connecting agents $i$ and $j$ evolves in time according to the differential equation
\begin{equation}
  \frac{dx_{ij}(t)}{dt} = \sum_{k} x_{ik}(t)\,x_{kj}(t),
  \label{eq:kulakowski}
\end{equation}
where the sum is taken over all common neighbors $k$ of $i$ and $j$.
The network is assumed to be undirected and without self-loops, so that $x_{ij}=x_{ji}$ and $x_{ii}\equiv0$.
Unless specified otherwise, we consider a complete graph, as is common in most studies, in which case the summation involves $N-2$ terms. Nowadays, the dynamics equation (\ref{eq:kulakowski}) is typically not used in its original form due to the unbounded growth of $x_{ij}$ -- see Sec.~\ref{subsubsec:kk-sat} for more details and a modified version addressing this limitation.

This dynamics directly reflects the principles of structural balance theory. To illustrate, let us examine a few representative cases:
\begin{enumerate}[label=(\arabic*)]
    \item ``Friend of my friend is my friend'': If the relations between agents $i$ and $k$, and $j$ and $k$ are positive (i.e., $x_{ik} > 0$ and $x_{kj} > 0$), then, $x_{ik}x_{kj} > 0$, which drives $x_{ij}$ to increase (become more positive).
    \item ``Friend of my enemy is my enemy'' or ``enemy of my friend is my enemy'': If one relation is positive and the other is negative (e.g., $x_{ik} > 0$ and $x_{kj} < 0$), then $x_{ik}x_{kj} < 0$, which drives $x_{ij}$ to decrease (become more negative).
    \item ``Enemy of my enemy is my friend'': If both relations are negative (i.e., $x_{ik} < 0$ and $x_{kj} < 0$), then $x_{ik}x_{kj} > 0$, which again drives $x_{ij}$ to increase (become more positive).
\end{enumerate}
% (1) the relation between agents $i$ and $k$, and $j$ and $k$

The system of differential equations minimizes the Hamiltonian (\ref{eq:hamiltonian_triad}) with $J_{ijk}=1$, as can be seen from the fact that the time derivative of $H$ is always non-positive~\citep{kulakowski2019stabilityOfSBsolutions}:

\begin{equation}
  \frac{dH}{dt} = \sum_{i<j} \frac{\partial H}{\partial x_{ij}}\, \frac{dx_{ij}}{dt} = -\sum_{i<j}\left(\sum_{k}x_{ik}\,x_{kj} \right)^{2} \le 0.
  \label{eq:kk:grad:des}
\end{equation}

Equivalently, the dynamics \eqref{eq:kulakowski} can be written in the form
\begin{equation}
   \dot{x}_{ij} = -\frac{\partial H}{\partial x_{ij}}.
\end{equation}
This expression is mathematically identical to the motion of an overdamped particle in a potential field~\citep{VANKAMPEN2007292}, or, in optimization terms, to continuous-time gradient descent~\citep{Boyd_Vandenberghe_2004}.
Consequently, the system evolves downhill on the energy landscape, although, due to the unbounded link weights, balanced or jammed sign configurations do not in general correspond to finite minima in this landscape.

\subsubsection{\label{subsubsec:kk-sat}Saturation terms}

A limitation of the dynamics (\ref{eq:kulakowski}) is that the link weights are unbounded: the absolute values of $x_{ij}$ diverge to infinity over time.
Such behavior is difficult to interpret in a sociological context.
To address this issue, saturation terms have been introduced, most notably the quadratic form $(1-x_{ij}^{2})$~\citep{kulakowski_2005}, the absolute form $(1-|x_{ij}|)$~\citep{Deng2012}, the Heaviside step function $\Theta(1-|x_{ij}|)$~\citep{Krawczyk2015, Hassanibesheli2017}, and the saturating normalization~\citep{Jia2016, Mei2019}.
Intuitively, saturation terms act to prevent relations from becoming arbitrarily strong, ensuring that the weights remain within finite bounds.
The quadratic term is by far the most widely used:
\begin{equation}
   \dot{x}_{ij} = \left(1 - x_{ij}^{2}\right)\sum_{k} x_{ik}\,x_{kj},
   \label{eq:kulak-sat}
\end{equation}
where $\dot{x}_{ij}$ denotes the time derivative. We will refer to the dynamics defined by Eq.~(\ref{eq:kulak-sat}) as standard GTD.
In some formulations, an additional normalization factor $\tfrac{1}{N-2}$ is included \citep{Krawczyk2015, Hassanibesheli2017, Gorski2017, Gorski2023}.
For the standard GTD, this factor merely rescales the solution in time, but it becomes relevant in more complex variants of the model where additional processes (e.g., reciprocity, multilayer connections, agent similarity) are incorporated, as discussed in Sec. \ref{subsubsec:kk-variants}, \ref{subsubsec:combined-interpersonal-agent-attribute}, \ref{subsubsec:multilayer}

When the initial values of $x_{ij}$ lie in the interval $(-1,1)$, the quadratic saturation term constrains all link weights to remain within this range.
The interval can be extended to a symmetric domain $(-R,R)$ with $R>0$, by modifying the term to $(1-(x_{ij}/R)^{2})$. As with the scaling normalization factor discussed above, this rescaling does not alter the qualitative character of the evolution. It changes the effective range of possible link weights and rescales time.
The quadratic form is advantageous because it is differentiable and therefore more amenable to analytical treatment, whereas the absolute form is not \citep{Gorski2017}.

Importantly, the quadratic saturation preserves the gradient-flow structure of the original dynamics. Using the Hamiltonian~\eqref{eq:hamiltonian_triad} with $J_{ijk}=1$, Eq.~\eqref{eq:kulak-sat} can be written as
\begin{equation}
\dot{x}_{ij} = -(1-x_{ij}^{2})\frac{\partial H}{\partial x_{ij}}.
\end{equation}
Consequently,
\begin{equation}
\frac{dH}{dt} = -\sum_{i<j}(1-x_{ij}^{2}) \left(\frac{\partial H}{\partial x_{ij}}\right)^2 \leq 0,
\end{equation}
since $1-x_{ij}^{2}\geq0$.
Thus, the saturated model retains a gradient-flow form with state-dependent mobility, and the Hamiltonian is non-increasing~\citep{kulakowski2019stabilityOfSBsolutions}.
%
% Following~\cite{kulakowski2019stabilityOfSBsolutions}, we note that the dynamics (\ref{eq:kulak-sat}) also monotonically decreases the Hamiltonian \eqref{eq:hamiltonian_triad}.
Since the saturation term confines the dynamics to a bounded domain, balanced configurations with $x_{ij}=\pm1$ correspond to global energy minima, while stable unbalanced configurations may correspond to local energy minima, i.e., jammed states.

Two additional approaches to limiting the growth of link weights appear in the literature. One relies on hyperbolic tangent functions~\citep{Morrison2023,Goto2024}, which do not directly bound the weights themselves but instead saturate the structural balance interaction terms; these formulations are discussed in Sec.~\ref{subsubsec:kk-variants}. The other arises in time-discrete versions of the model~\citep{Muncaster1998, Jia2016}, where saturating normalization factors are introduced to prevent unbounded growth of weights. These discrete-time formulations are discussed in more detail in Sec.~\ref{subsubsec:kk-weak} and~\ref{subsubsec:kk-discrete}.

% Another way to constrain link weights is introduced in time-discrete versions of the model~\cite{Muncaster1998, Jia2016}. In one such approach, a saturating normalization term is  used to prevent unbounded growth of links. These models are discussed in detail in Sections~\ref{subsubsec:kk-weak} and \ref{subsubsec:kk-discrete}.

\subsubsection{\label{subsubsec:kk-self}Model variant with self-loops: Marvel et al.'s approach}

The inclusion of self-loops fundamentally changes the mathematical structure of gradient triad dynamics, enabling a more tractable analytical treatment.
A seminal contribution in this direction is made by \citet{marvel2011_evolutionOFSBcontinuous} (2011), where a variant of the model with self-loops is analyzed.
In this formulation, the basic dynamics (\ref{eq:kulakowski}) can be expressed in matrix form. Specifically, the system can be written as
\begin{equation}
   \dot{\mathbf{X}} = \mathbf{X}^{2},
   \label{eq:kk-self}
\end{equation}
where $\mathbf{X}$ is the matrix of link weights.
Unlike the standard formulation, Eq.~(\ref{eq:kk-self}) also governs the diagonal entries $x_{ii}$, which evolve in time rather than being fixed at zero.
These diagonal values can be interpreted as self-appraisals~\citep{marvel2011_evolutionOFSBcontinuous,Mei2019}.
Equation~(\ref{eq:kk-self}) belongs to the family of matrix Riccati equations \citep{AbouKandil2003}, a well-studied class in matrix analysis and control theory.

\citet{marvel2011_evolutionOFSBcontinuous} exploit this structure to derive analytical results and make two major contributions.
First, assuming random initial link weights drawn independently from an absolutely continuous distribution with bounded support, they prove that a system of size $N$ reaches a balanced state with probability approaching one as $N$ grows\footnote{The proof is based on the eigen-decomposition of $\mathbf{X}(t)$, $\mathbf{X}(t) = \mathbf{Q}\,\mathbf{D}(t)\,\mathbf{Q}^{\top}$, where $\mathbf{Q}$ is the orthogonal matrix of eigenvectors of $\mathbf{X}(0)$, and $D(t)$ is the diagonal matrix with eigenvalues of $\mathbf{X}(t)$. Also $\mathbf{D}(t) = \mathrm{diag}[\ell_{1}(t), \ldots, \ell_{N}(t)]$, where $\ell_{m}(t) = \lambda_{m}/(1-\lambda_{m}t)$. The dynamics therefore have a finite \textit{blow-up time} at $t^{*} = 1/\lambda_{1}$, where $\lambda_{1}$ is the largest eigenvalue of the initial matrix $\mathbf{X}(0)$.
At $t \to t^{*}$, the leading eigenvalue $\ell_{1}(t)$ diverges, while all remaining eigenvalues vanish in comparison. Consequently, the dominant term is $\mathbf{X}(t)\sim (t^*-t)^{-1}\boldsymbol{\omega}_1\boldsymbol{\omega}_1^\top$,
% the matrix $\mathbf{X}(t)$ becomes dominated by the rank-one term associated with the leading eigenvector, yielding $\mathbf{X}(t^{*}) = \boldsymbol{\omega}_{1}\boldsymbol{\omega}_{1}^{\top}$,
where $\boldsymbol{\omega}_{1}$ is the eigenvector corresponding to $\lambda_{1}$. Since, with probability one, all components of $\boldsymbol{\omega}_{1}$ are nonzero~\citep{marvel2011_evolutionOFSBcontinuous}, the resulting sign pattern is structurally balanced: diagonal entries are positive, and nodes partition into two groups according to the sign of the components of $\boldsymbol{\omega}_{1}$, with positive (negative) relations within (between) the groups.}.
Second, the authors show that when the initial off-diagonal link weights are drawn from distributions symmetric about their common mean, a positive mean favors a paradise state, whereas a non-positive mean favors a polarized state, with some results requiring additional assumptions on the variance.
The above reasoning is further extended to directed networks~\citep{Traag2013}, as described in Section \ref{subsubsec:kk-directed}.

It is not entirely clear to what extent the conclusions from the model with undirected links \citep{marvel2011_evolutionOFSBcontinuous} with self-loops apply to the model without them. In general, self-loop weights always increase:
\begin{equation}
    \dot{x}_{ii}=\sum_k (x_{ik})^2
\end{equation}
and, when positive, they strengthen existing weights:
\begin{equation}
    \dot{x}_{ij}=\sum_k x_{ik}x_{kj}=\sum_{k\neq i,j}x_{ik} x_{kj} + x_{ij}(x_{ii}+x_{jj}).
\end{equation}
Thus, self-loops introduce an additional self-reinforcing mechanism proportional to $x_{ij}$
that can strengthen existing link signs when $x_{ii}+x_{jj}>0$.
% that may stabilize configurations which would otherwise evolve further. In particular, this effect can promote the persistence of jammed states by reinforcing local inconsistencies.
In the absence of self-loops, this reinforcing term vanishes, and the dynamics is driven solely by triadic interactions.
% Consequently, reaching jammed states should be more difficult, suggesting that convergence to structurally balanced states should be more robust.
However, whether this difference facilitates or hinders convergence to structural balance cannot be inferred from this term alone.
The standard GTD formulation considered here also differs from the formulation of Ref.~\citep{marvel2011_evolutionOFSBcontinuous} by the presence of the quadratic saturation term, which bounds the link weights and modifies the dynamics. We hypothesize that the convergence-to-balance result by~\citet{marvel2011_evolutionOFSBcontinuous} may nevertheless extend to the standard GTD without self-loops. Further investigation is required to confirm this hypothesis.

\subsubsection{\label{subsubsec:kk-anal}Analytical considerations for standard gradient triad dynamics}

Analytical investigations of the standard GTD~(\ref{eq:kulak-sat}) are typically based on Jacobian analysis in the vicinity of fixed points~\citep{Wongkaew2015}.
At balanced states, it can be shown that all link weights converge to values $\pm 1$~\citep{Wongkaew2015, kulakowski2019stabilityOfSBsolutions}; see also the supplementary material of~\citep{Gorski2017}.
For system (\ref{eq:kulak-sat}), fixed points occur at\footnote{These are not the only fixed points. There exist degenerate fixed points for which (some of the) sums $\sum_k x_{ik}\,x_{kj}$ vanish, leading to zero eigenvalue(s) of the Jacobian. The abundance of such cases depends on the parity of the number of nodes. In a complete graph, each sum $\sum_k x_{ik}\,x_{kj}$ contains $N-2$ terms.
If $N$ is even, then $N-2$ is also even, and assuming $x_{ik},x_{kj}\in\{\pm1\}$, the sum may vanish due to cancellation between positive and negative contributions.
For odd $N$, this particular cancellation mechanism is absent because $N-2$ is odd.
To limit the occurrence of such degenerate cases, many studies restrict numerical simulations to graphs with an odd number of nodes.
} $x_{ij} = \pm 1$ for all $i,j$.
To understand their stability, we outline the proof and derive the general stability condition. The linearization of the dynamics near a fixed point $\mathbf{x}^*$ can be written as
\begin{equation}
   \dot{\mathbf{x}} \approx
   \mathbf{J}(\mathbf{x}^*)\,(\mathbf{x}-\mathbf{x}^*),
\end{equation}
where $\mathbf{x}$ is the vector of all edge variables and $\mathbf{J}(\mathbf{x}^*)$ is the Jacobian matrix evaluated at the fixed point, with entries
\begin{equation}
   J_{ab} = \frac{\partial \dot{x}_{a}}{\partial x_{b}},
\end{equation}
with $x_{a}$ and $x_{b}$ denoting the $a$-th and $b$-th links, respectively.
A fixed point is locally asymptotically stable if all eigenvalues of $\mathbf{J}$ have negative real parts~\citep{glendinning1994stability}.
% In~\citep{Wongkaew2015} it was shown that paradise state is stable, and hell state is not.

As an illustration, consider the case $N=3$.
Then system~(\ref{eq:kulak-sat}) reduces to three equations:
\begin{equation}
\begin{aligned}
   \dot{x}_{12} &= \left(1 - x_{12}^{2}\right) x_{13}\,x_{23}, \\
   \dot{x}_{13} &= \left(1 - x_{13}^{2}\right) x_{12}\,x_{23}, \\
   \dot{x}_{23} &= \left(1 - x_{23}^{2}\right) x_{12}\,x_{13}.
   \label{eq:kk-N3}
\end{aligned}
\end{equation}
Assuming this triad is connected (i.e., at least two links are nonzero), the only fixed points are $x_{ij} = \pm 1$.
At these points, all off-diagonal elements of $\mathbf{J}$ vanish, while the diagonal entries equal $-2x_{12}x_{13}x_{23}$.
Thus, fixed points corresponding to balanced states are stable, while those corresponding to unbalanced states are unstable~\citep{Krawczyk2017}.

For the general case, the Jacobian elements are given by~\citep{kulakowski2019stabilityOfSBsolutions}:
\begin{equation}
\frac{\partial}{\partial x_{kl}} \dot{x}_{ij}
= \left(1 - x_{ij}^{2}\right)
   \frac{\partial}{\partial x_{kl}} \sum_{m} x_{im}\,x_{mj}
   - 2x_{ij}\left(\delta_{ik}\delta_{jl} + \delta_{il}\delta_{jk}\right)
   \sum_{m} x_{im}\,x_{mj}.
\end{equation}
where $\delta$ denotes the Kronecker delta.
If $x_{ij} \not\equiv x_{kl}$, then the second term vanishes, and the first term becomes zero at fixed points $x_{ij} = \pm 1$.
If $x_{ij} \equiv x_{kl}$, then (without self-loops) the first term vanishes and the second term equals $-2x_{ij}\sum_{m}x_{im}x_{mj}$. Therefore, stability of a fixed point requires the condition $x_{ij}\sum_{m} x_{im}x_{mj} > 0$ for each link separately.
Equivalently, in the vicinity of the fixed point, positive links must have positive derivatives, while negative links must have negative derivatives.
In other words, in the neighborhood of a fixed point, all edges must evolve toward the $\pm 1$ value closest to their current state.

When the model includes the quadratic saturation term, the link values are confined to the interval $(-1,1)$ (or more generally $(-R,R)$).
\citet{Wongkaew2015} also analyze cases outside these limits.
It is shown that if all links exceed the upper boundary, then they converge to it over time, whereas if all links are below the lower boundary, they diverge to $-\infty$.

The probability of reaching a paradise state is studied in ref.~\citep{Krawczyk2017}.
For a single-triad system with uniformly distributed initial conditions in $(-1,1)$, each of the four balanced states is equally likely, giving a $1/4$ probability of reaching paradise.
It is further hypothesized that a similar equiprobability applies in larger systems, leading to a probability of $2^{1-N}$ for paradise.
% This is a slight contradiction of a conclusion obtained for the system with self-loops, where, when the average is positive, a paradise state should also be reached.

The time required to reach a balanced state in a complete graph, with or without saturation, scales as $N^{-1/2}$~\citep{kulakowski_2005}, i.e., the larger the network, the shorter the time.
This scaling is derived under the assumption of uniformly distributed initial conditions centered at zero.
The reasoning is as follows. The number of summation terms scales as $N$, while fluctuations of their average decrease as $N^{-1/2}$, so that the typical magnitude of the sum scales as $N^{1/2}$. The combination of these effects determines the speed of link evolution, which is the inverse of the final time.
Assuming the analysis of \citet{marvel2011_evolutionOFSBcontinuous} also applies to the model without self-loops, the result is consistent with another argument: by Wigner’s semicircle law, the largest eigenvalue $\lambda_1$ of a random symmetric matrix scales like $N^{1/2}$. Hence, using the relation for the blow-up time, $1/\lambda_1$, one again obtains the $N^{-1/2}$ scaling.

\subsubsection{\label{subsubsec:kk-directed}Model extension to directed networks}

\citet{Traag2013} consider two natural ways to extend gradient triad dynamics to directed networks with evolving self-loops, where $x_{ij}$ denotes the relation from agent $i$ to agent $j$:
\begin{equation}
   \dot{x}_{ij} = \sum_{k} x_{ik}\,x_{kj},
   \label{eq:kk-dir1s}
\end{equation}
or
\begin{equation}
   \dot{x}_{ij} = \sum_{k} x_{ik}\,x_{jk}.
   \label{eq:kk-dir2s}
\end{equation}
Both update rules can also be formulated without self-loops, by keeping $x_{ii}=0$ fixed and considering the dynamics only for $i\ne j$.
The small difference in index placement leads to distinct interpretations.

In the first case, when link $i\rightarrow j$ is updated, it is as if agent $i$ asks others what they think of agent $j$, weights these assessments by $i$’s relations to those others and then adjusts the relation from $i$ to $j$ accordingly.
This interpretation motivates the name \textit{reputation model}.
In the second case, agent $i$ asks agent $j$ about $j$’s opinions on other agents, weights these assessments by $i$’s relations to those others and then adjusts the relation from $i$ to $j$ accordingly.
This is therefore referred to as the \textit{gossip model}.
In short, the reputation model focuses on how others view $j$, while the gossip model focuses on how $j$ views others. In another paper~\citep{Mei2019}, these approaches are named influence-based and homophily-based models, respectively.

An important difference between the two models is that the gossip model satisfies $\dot{x}_{ij}=\dot{x}_{ji}$, whereas this equality generally does not hold for the reputation model. Consequently, in the gossip model, $x_{ij}(t)-x_{ji}(t)$ remains constant over time.
The mechanism of the gossip model has been linked to processes such as cooperation or gossip dynamics, where it matters more what others \textit{do} than what one \textit{thinks} \citep{Traag2013}.
It can be viewed as a form of indirect reciprocity: observing agent $j$ perform a positive act toward agent $k$ increases the chance that agent $i$ responds positively toward $j$, provided that $i$ likes $k$.

\citet{Traag2013} analyze both directed variants, extending the analytical treatment of \citet{marvel2011_evolutionOFSBcontinuous} to asymmetric initial conditions and introducing the alternative gossip mechanism.
When self-loops are included and allowed to evolve according to the same update rules, the two equations~(\ref{eq:kk-dir1s}-\ref{eq:kk-dir2s}) become
\begin{equation}
   \dot{\mathbf{X}} = \mathbf{X}^{2},
   \label{eq:kk-dir1m}
\end{equation}
and
\begin{equation}
   \dot{\mathbf{X}} = \mathbf{X}\mathbf{X}^{\top},
   \label{eq:kk-dir2m}
\end{equation}
where $\mathbf{X}^{\top}$ denotes  transposition.
Equation~(\ref{eq:kk-dir1m}) corresponds to the reputation model~(\ref{eq:kk-dir1s}), while Eq.~(\ref{eq:kk-dir2m}) corresponds to the gossip model~(\ref{eq:kk-dir2s}).

For the reputation model with self-loops, \citet{Traag2013} established sufficient conditions for convergence to structural balance. These conditions require that the initial conditions satisfy
\begin{equation}
   \mathbf{X}(0)\mathbf{X}^{\top}(0) = \mathbf{X}^{\top}(0)\mathbf{X}(0),
   \label{eq:kk-dir-cond}
\end{equation}
% and the largest eigenvalue $\lambda_{1}$ of $\mathbf{X}(0)$ is real, positive, and simple.
and a real, positive, simple eigenvalue $\lambda_1$ of $\mathbf{X}(0)$ exceeds all other real eigenvalues.
This matrix condition is always fulfilled for symmetric matrices (i.e., undirected networks) and certain special cases.
Otherwise, system~(\ref{eq:kk-dir1m}) does not necessarily converge to structural balance but may instead approach a form of \textit{relaxed balance}~\citep{Doreian2009partitioning,Doreian_Batagelj_Ferligoj_2004}, where groups emerge with homogeneous patterns of relations both within and across groups, which may be positive or negative.

The gossip model behaves differently.
Here, $\mathbf{X}(t)$ can be decomposed into symmetric and skew-symmetric parts.
Since $\dot{\mathbf{X}} = \dot{\mathbf{X}}^{\top}$, the skew-symmetric part remains constant in time, while the symmetric part evolves.
For a dense set of initial conditions, as $t$ approaches a blow-up time $t^{*}$, it was shown that
\[
   \lim_{t \to t^{*}} \frac{\mathbf{X}(t)}{\|\mathbf{X}(t)\|} = \mathbf{w}\mathbf{w}^{\top},
\]
for some vector $\mathbf{w}$.
Thus, the gossip model generically converges to structural balance.

The results above are also relevant for the corresponding directed extensions of GTD without self-loops.
For the reputation-type dynamics~\eqref{eq:kk-dir1s}, structural balance is likewise not a generic outcome for asymmetric relations~\citep{Krawczyk2015,Hassanibesheli2017}.
Although for $N=3$ all stable states are balanced, jammed states already exist for $N=5$~\citep{Krawczyk2015}, and their number can substantially exceed that of balanced configurations in larger networks~\citep{Hassanibesheli2017}.

\subsubsection{\label{subsubsec:kk-variants}Extensions of gradient triad dynamics with additional dynamical mechanisms}

\paragraph{Memory effects}
As mentioned in the introduction of Section \ref{sec:emerg}, the triadic interaction term underlying GTD already appeared in the more general model proposed by \citet{Lee1994}.
They introduce the following extended dynamics to study the evolution of international relations:
\begin{equation}
\dot{x}_{ij} = \alpha_{ij}\,x_{ij} + \sum_{k} \beta_{kij}\,x_{ik}\,x_{kj},
\end{equation}
where we have adapted the original notation to avoid confusion with the adjacency matrix $\mathbf{A}$ used throughout this review. The coefficient $\alpha_{ij}$ quantifies how strongly the past history of the bilateral relationship between nations $i$ and $j$ influences its future evolution, while tensor element $\beta_{kij}$ governs the strength of structural balance interactions among three nations.
No self-loops were assumed, i.e., $x_{ii} \equiv 0$.
% \msmcomment{A and B are matrix, can we say more explicitly, matrix element $A_{ij}$ quantifies how strongly the past history of the bilateral relationship between nations $i$ and $j$ influences its future evolution, \dots (same for B)}

This general formulation allows for various dynamical behaviors, depending on the relative magnitudes of these coefficients and initial conditions. For a homogeneous three-nation system, the authors classified configurations into six categories: unipolar, bipolar, tripolar, weakly apolar, moderately apolar, and strongly apolar. They showed that almost all initially unbalanced configurations eventually become balanced, although some exceptional unbalanced configurations can persist. Hence, the model bridges structural balance with geopolitical classifications based on stability and polarity.

Building on the idea of combining dyadic memory with triadic balance interactions, \citet{Morrison2023} propose a nonlinear extension that incorporates bounded balance interactions and transient perturbations. The dynamics is given by
\begin{equation}
\dot{x}_{ij}
= -\beta\Bigl(x_{ij}-x_{{ij}}^{(D)}\Bigr)
+ L\,\tanh\!\left(\frac{1}{\gamma}\sum_{k} x_{ik}x_{kj}\right)
+ x_{{ij}}^{(T)},
\end{equation}
where the first term represents attraction toward a historical baseline relation $x_{{ij}}^{(D)}$, with $\beta$ controlling the damping strength, the second term encodes triadic structural balance interactions, and $x_{{ij}}^{(T)}$ accounts for transient external perturbations. The hyperbolic tangent naturally bounds the contribution of the balance term, eliminating the need for explicit saturation factors. For small values of $\sum_k x_{ik}x_{kj}$, the balance interaction can be linearized, recovering the triadic interaction term of GTD, $\alpha\sum_k x_{ik}x_{kj}$, with $\alpha=L/\gamma$ controlling the sensitivity to structural balance.

The authors apply the model to international relations data and investigate a peace--war bifurcation driven by increasing $\alpha$. They observe pronounced hysteresis: restoring a peaceful state requires reducing $\alpha$ to values lower than those at which conflict first emerges.

\paragraph{Reciprocity}
The lack of structural balance in the directed extension of the GTD (the reputation dynamics, Eq.~\ref{eq:kk-dir1s}) motivated the development of a variant~\citep{Krawczyk2015} incorporating two effects: a structural balance term and a drive toward reciprocity:
\begin{equation}
\dot{x}_{ij}
=
   \alpha(x_{ji} - x_{ij})
   + \frac{1 - \alpha}{N - 2}\sum_{k} x_{ik}\,x_{kj},
\label{eq:kk+recip}
\end{equation}
where $\alpha$ controls the relative influence of reciprocity and structural balance. In numerical simulations, link weights were constrained to the interval $[-1,1]$.

This model combines homophily-driven reciprocity with the triadic consistency of structural balance theory, allowing both mechanisms to interact.
It was shown that even a weak reciprocity term is sufficient to drive all pairs toward fully reciprocated relations. Consequently, the sign matrix becomes symmetric, and the system ultimately reaches a balanced state. Over a broad range of $\alpha \in (0.2, 1)$, balanced states are consistently obtained.

\citet{Hassanibesheli2017} propose a further extension of the reciprocity model~(\ref{eq:kk+recip}) by introducing the mechanism of \textit{gain and loss of esteem}. In this formulation, the rate of change of $x_{ij}$ depends explicitly on the rate of change of $x_{ji}$. In other words, if the relation $j\rightarrow i$ is becoming more friendly (i.e., $\dot{x}_{ji}>0$), this leads $i$ to become more friendly toward $j$, whereas a decreasing relation leads to a reduction of $x_{ij}$.
The resulting dynamics is
\begin{equation}
\dot{x}_{ij}
= \left(1 - x_{ij}^{2}\right)
\left[
   \alpha\left(x_{ji} - x_{ij}\right)
   + \frac{1 - \alpha}{N - 2}\sum_{k} x_{ik}\,x_{kj}
   + \gamma\,\dot{x}_{ji}
\right],
\end{equation}
where $\gamma$ is the strength of the gain-loss-of-esteem effect.

Simulations are performed for initial weights drawn from a uniform distribution $(p - 0.3,\, p + 0.3)$, with $p$ as a control parameter.
The phase diagram reveals three regimes: jammed, polarized, and paradise states.
Large $p$ leads to paradise, while small $p$ combined with large $\alpha$ (or $\gamma$) leads to polarization; small $p$ and small $\alpha$ (or $\gamma$) produce jammed states.
Increasing $\gamma$ lowers the critical value of $\alpha$ required for the transition from jammed to balanced states.
This extension shows how coupling reciprocity and esteem dynamics can drive transitions between paradise, polarized, and jammed configurations.

\paragraph{Decay of relations}
\citet{Goto2024} propose a modified version of the standard GTD in which neutral links emerge naturally.
The dynamics is given by
\begin{equation}
\dot{x}_{ij}
= -\beta\,x_{ij}
  + A\sum_{k} \tanh\!\left(\frac{x_{ik}}{\alpha}\right)
              \tanh\!\left(\frac{x_{kj}}{\alpha}\right),
\label{eq:kk-goto}
\end{equation}
where $\beta$, $A$, and $\alpha$ are positive constants.
Here, $\beta$ represents the \textit{damping} (or forgetting) strength of direct relations,
$A$ scales the contribution of the triadic term,
and $\alpha$ determines the sensitivity of weights in structural balance theory interactions.
The hyperbolic tangents limit the rate of link change, while the damping term drives weakly connected pairs toward neutrality.

Self-loops are allowed, enabling analytical treatment similar to that in \citet{marvel2011_evolutionOFSBcontinuous}.
Initial link weights are drawn independently, up to symmetry, from a distribution with zero mean and standard deviation $\sigma$.
Two regimes are observed: a bipolar balanced state and a neutral state.
A transition between these regimes occurs upon varying $\beta$, $A$, or the system size $N$ (note that $A$ is not rescaled with $N$, unlike in other models with competing processes). The system undergoes a size-dependent transition between neutrality and polarization.

An analytical estimate of the transition point can be obtained by imposing $A = \alpha^{2}$ and considering the regime $|x_{ij}|\ll\alpha$, in which Eq.~(\ref{eq:kk-goto}) can be approximated as
\begin{equation}
\dot{\mathbf{X}} \approx -\beta\,\mathbf{X} + \mathbf{X}^{2}.
\end{equation}
Following \citet{marvel2011_evolutionOFSBcontinuous}, the solution is
\begin{equation}
x_{ij}(t) = \sum_{m=1}^{N}
(\boldsymbol{\omega}_m)_i\,l_m(t)\,
(\boldsymbol{\omega}_m)_j.
\end{equation}
where $\boldsymbol{\omega}_m$ and $\lambda_m$ are eigenvectors and eigenvalues of the initial matrix $\mathbf{X}(0)$, respectively, and $(\boldsymbol{\omega}_m)_i$ denotes the $i$-th
component of $\boldsymbol{\omega}_m$, with
\begin{equation}
l_m(t) = \frac{\beta}{
1 - \frac{\lambda_m - \beta}{\lambda_m}
\exp(\beta t)}.
\end{equation}
For large $N$, the largest eigenvalue $\lambda_1 > 0$ is simple, and all components of its corresponding eigenvector $\boldsymbol{\omega}_1$ are nonzero.
% Within this approximation, when $\lambda_1 > \beta$, links diverge at a finite time and the system reaches a polarized balanced state.
Within this approximation, $\lambda_1>\beta$ leads to finite-time divergence, signaling the instability of the neutral state. In the full nonlinear model, the bounded interaction term instead allows the emergence of a polarized balanced state.
When $\lambda_1 < \beta$, all links decay to zero.
Since the largest eigenvalue scales as $\lambda_1 \approx 2\sigma\sqrt{N}$, the critical size separating these regimes is
\begin{equation}
N_c = \left(\frac{\beta}{2\sigma}\right)^{2}.
\end{equation}
The analytical solution reveals a critical system size $N_c$ that determines whether the group remains cohesive or splits.
For smaller systems ($N < N_c$), all relations decay toward neutrality, whereas for larger systems ($N > N_c$), polarization---and consequently group division---emerges.

% Another possible system outcome is jammed states. Numerically, jammed states are likely to be observed when the lambda1 approx lambda2. This is very unlikely. Jammed states, however, become very likely when a stochastic noise is added to the dynamics:
Jammed states may also appear.
They are numerically observed when $\lambda_1 \approx \lambda_2$, a rare condition that becomes more likely when stochastic perturbations are incorporated into the dynamics.
Specifically, the model can be extended by adding a noise term:
\begin{equation}
\dot{x}_{ij}
= -\beta\,x_{ij}
  + A\sum_{k} \tanh\!\left(\frac{x_{ik}}{\alpha}\right)
               \tanh\!\left(\frac{x_{kj}}{\alpha}\right)
  + I\,\xi_{ij}(t),
\label{eq:kk-goto2}
\end{equation}
where $\xi_{ij}(t)$ is Gaussian white noise with zero mean, uncorrelated in time and across links, and $I$ is the noise intensity.

Surprisingly, noise increases the likelihood of jammed states. This is counterintuitive, as stochastic perturbations are often expected to help systems escape local minima rather than become trapped in them. The authors suggest that this behavior may be explained by the much larger number of jammed configurations than balanced ones.

% With noise it is easier to end up in a jammed state. An optimal intensity leading to the highest probability of jammed states was found. This seems unexpected because jammed states relate to the local minima in the energy landscape. The noise should move the system out of local minima and not tract it into it. The presumed reason for that is the larger number of jammed states than balanced ones.

\paragraph{Delay-induced higher-order interactions}
A delayed version of GTD is explored in~\citep{Siboni2025}:
\begin{equation}
\dot{x}_{ij}(t)
= \sum_{k} x_{ik}(t - \tau)\,x_{kj}(t - \tau),
\label{eq:kk-delay}
\end{equation}
where $\tau$ is the delay parameter.
Using a Taylor expansion in $\tau$, one obtains
\begin{equation}
\dot{x}_{ij}
= \sum_{k} x_{ik}x_{kj}
  - 2\tau \sum_{k,l} x_{ik}x_{kl}x_{lj}
  + 7\tau^{2} \sum_{k,l,m} x_{ik}x_{kl}x_{lm}x_{mj}
  + \mathcal{O}(\tau^{3}),
\label{eq:kk-delay2}
\end{equation}
where the second and third terms correspond to effective four- and five-body interactions. Thus, increasing the delay generates progressively higher-order interactions in the effective dynamics~\citep{Siboni2025}.
Interestingly, for a balanced four-cycle containing $(i,j)$, the first-order correction acts against the sign of $x_{ij}$, weakening rather than reinforcing the current relation; delay may therefore hinder balance.
The authors note that low-order Taylor truncations of delayed differential equations need not reproduce the full delayed dynamics, so this correspondence is most useful for interpreting sufficiently small delays~\citep{Siboni2025}.
The connection to higher-order Hamiltonian models is discussed further in Sec.~\ref{subsubsec:hamil-variants}.
% Considering only the first-order approximation, the corresponding Hamiltonian can be written as
% \begin{equation}
% H = -\sum_{i,j,k} x_{ij}x_{jk}x_{ki}
%   + 2\tau \sum_{i,j,k,l} x_{ij}x_{jk}x_{kl}x_{li},
% \label{eq:hamil-delay}
% \end{equation}
% and the resulting dynamics are further analyzed in Section~\ref{subsubsec:hamil-variants}.

\subsubsection{\label{subsubsec:kk-weak}Considering the weak version of structural balance}
% possibly this could be joined with the previous subsubsection

In the weak version of structural balance, the rule ``enemy of my enemy is my friend'' is omitted. It is not straightforward to adapt the Kułakowski--Gawroński--Gronek model to this weaker version, because the strong version of structural balance naturally emerges from link multiplications in the dynamics. Two approaches addressing this challenge have been proposed in the literature, which are briefly discussed below.
\citet{Saperstein2004} extended the earlier model of \citet{Lee1994} as follows:
\begin{equation}
\dot{x}_{ij} = \alpha_{ij}\,x_{ij} + \beta_{ijk}\,(x_{ik} + s)(x_{kj} + s),
\label{eq:saperstein}
\end{equation}
where $\alpha_{ij}$ and $\beta_{ijk}$ denote the strengths of the direct relationship and structural-balance-type interactions, respectively.
Heterogeneous values of these parameters were allowed, and $s > 0$ is a flexibility parameter.
Ref. \citep{Saperstein2004} considered a system of a single triad. For a general network, Eq.~\eqref{eq:saperstein} should be extended by summing over all intermediate agents $k$, i.e., over all triads involving agents $i$ and $j$.
This formulation preserves the rule ``the friend of my friend is my friend'', but modifies the others, e.g., ``the enemy of my friend \textit{may be} my enemy''.
For sufficiently large $s$, the system often converges to the paradise state, while only strong negative links tend to persist. Consequently, triads with 3 negative links can still occur. Thus, this model represents an intermediate case lying between the weak and strong versions of structural balance.

A different approach was proposed by \citet{Jia2016}, who assumed that relations following a negative link are not perceived by agents, meaning that information does not propagate through negative links. To reflect this behavior, an ego-centered perspective was introduced: it is the agent who acknowledges the balance of triads, rather than a \textit{link} as in the standard GTD. The model dynamics are given by
\begin{equation}
\mathbf{X}(t+1) = \mathbf{W}(\mathbf{X}(t))\,\mathbf{X}(t),
\end{equation}
where $\mathbf{W}$ is an influence matrix with elements $W_{ij}$ proportional to $x_{ij}$ when $x_{ij} > 0$, and zero otherwise.
The corresponding single-link update rule is
% \msmcomment{shall index in denominator be different than $j$?}\pgcomment{done}
\begin{equation}
x_{ij}(t+1)
= \frac{1}{\sum_{k} \Theta(x_{ik})\,x_{ik} + \varepsilon}
  \sum_{k} \left[\Theta(x_{ik})\,x_{ik} + \delta_{ik}\varepsilon\right] x_{kj},
\end{equation}
where $\Theta$ denotes the Heaviside function and $\delta_{ik}$ is the Kronecker delta, ensuring that the self-appraisal parameter $\varepsilon$ contributes only to diagonal term.
This model incorporates several features already encountered in other variants: the network is directed and includes self-loops. For the first time, a \textit{saturating normalization} term was used to constrain link weights, using the sum of positive weights for normalization. Moreover, the dynamics evolves in discrete time.

Because only positive relations generate interpersonal influence, only two structural balance rules -- jointly expressed as \textit{``a friend or enemy of my friend is my friend or enemy''} -- remain relevant, while the others are no longer applicable.
As a result, weak structural balance becomes a possible outcome.
Another frequently observed configuration is \textit{factions-followers-outsiders} structure, characterized by internally friendly groups, their followers who support one or more factions and hold non-positive relations toward all others, and outsiders who hold non-positive relations toward all agents.

\subsubsection{\label{subsubsec:kk-discrete}Discrete-time variants}
% again this could be joined with model variants

The model of \citet{Jia2016}, discussed in the preceding subsection because of its relevance to weak structural balance, already operates in discrete time.
Here, we consider discrete-time models related to GTD more broadly. Such formulations can reproduce the main qualitative outcomes of continuous-time models while avoiding unbounded link growth.

\citet{Muncaster1998} proposed a synchronous (and therefore deterministic) discrete-time formulation, following the time-continuous model of \citet{Lee1994}.
% , where relations are updated based on: (i) relation history, (ii) other links in a triad.
The update rule is defined as
\begin{equation}
x_{ij}(t+1) \propto \frac{1}{N-2} \sum_{k} F\!\left(x_{ik}(t)\,x_{jk}(t)\right),
\label{eq:kk-mun}
\end{equation}
where \(F(u) = \operatorname{sign}(u)\sqrt{|u|}\).
This discrete-time form, combined with the scaling normalization, naturally bounds the link weights and prevents divergence.
The network can reach a balanced state even when initialized with all negative links. The authors further extended the dynamics~(\ref{eq:kk-mun}) to include dependence on relation history, similar to the model by \citet{Lee1994}, by incorporating three previous time steps. However, this modification did not lead to qualitatively new conclusions.

A different route to discrete-time dynamics was introduced in the model of \citet{Jia2016}, discussed in Section~\ref{subsubsec:kk-weak}.
Building on this idea, \citet{Mei2019} proposed a similar formulation that includes a \textit{saturating normalization} term based on the absolute values of all links.
Following~\citep{Traag2013}, two ways of constructing the model on the directed network were considered (see Section \ref{subsubsec:kk-directed}). These are the reputation and the gossip models, respectively:

\begin{equation}
\begin{aligned}
x_{ij}(t+1) &=
\frac{1}{\sum_{k} |x_{ik}(t)|}
\sum_{k} x_{ik}(t)\,x_{kj}(t),
&\quad &\text{(reputation model)}\\[6pt]
x_{ij}(t+1) &=
\frac{1}{\sum_{k} |x_{ik}(t)|}
\sum_{k} x_{ik}(t)\,x_{jk}(t),
&\quad &\text{(gossip model)}
\end{aligned}
\end{equation}

Analytical and numerical analyses of these systems revealed results consistent with those of \citet{Traag2013}.
For the gossip model (corresponding to $\mathbf{X}\mathbf{X}^{\top}$), convergence to structural balance was proven provided that the link weights remain bounded away from zero, while Monte Carlo simulations indicate that this condition holds for generic initial conditions. For the reputation model (corresponding to $\mathbf{X}^{2}$), structural balance is not generally reached for generic initial conditions. However, convergence to structural balance is guaranteed for any network size \(N\) when the initial matrix satisfies the symmetry condition identified by \citet{Mei2019} and the link weights remain bounded away from zero.

The discrete-time models discussed above typically approach their long-time behavior through successive updates, with convergence times varying across models.
In contrast, \citet{Krawczyk2022Structural} propose a dynamics with the remarkable property that, when applied to any complete network of odd size $N$, balance is achieved in a single update step.
The model uses the same triadic term as GTD, but applies a different update function to binary signed links.
The update rule is
\begin{equation}
x_{ij}(t+1) = F\!\left(\sum_k x_{ik}(t)x_{jk}(t)\right),
\end{equation}
where $F(x) = (-1)^{\frac{N-2-x}{2}}$.
Although an analytical proof is not provided, the authors verify this property for all configurations of small networks and for a very large number of networks of small size (up to $N=15$).

 \subsection{\label{subsec:hamil}Hamiltonian-based dynamics}

The Hamiltonian formulation of signed networks under structural balance interactions, expressed in Eq.~\eqref{eq:hamiltonian_triad}, was proposed by \citet{antal_2005}, and later thoroughly explained by \citet{Marvel2009}. This approach provided a direct analogy between structural balance theory and physical systems governed by energy minimization principles. A decade after the seminal work by \citet{antal_2005}, researchers began to systematically investigate such systems from a statistical-physics perspective, applying analytical and numerical techniques analogous to those used in conventional Hamiltonian-based models.
Analytical studies typically rely on mean-field approximations, while numerical investigations employ standard methods such as Metropolis, Glauber, or heat-bath dynamics (see  \ref{app:mc}). Together, these tools enable the analysis of stochastic fluctuations, equilibrium states, and phase transitions.

In the following, we review models that explicitly define a Hamiltonian (not necessarily identical to Eq.~\eqref{eq:hamiltonian_triad}) and study its minimization, possibly under the influence of noise. The models are presented in order of increasing complexity, rather than chronologically.

\subsubsection{\label{subsubsec:hamil-ctd}Constrained triad dynamics}

One of the models introduced in the seminal paper by \citet{antal_2005} is the \emph{constrained triad dynamics} (CTD). Although this model did not require defining a Hamiltonian, it naturally belongs to the class of Hamiltonian-based dynamics, since its evolution follows an energy-minimization principle corresponding to zero temperature.
At each update step, a link is selected at random, and a flip of its sign is proposed. The flip is accepted if it decreases the Hamiltonian~\eqref{eq:hamiltonian_triad}. If the Hamiltonian remains unchanged, the flip is accepted with probability $0.5$; otherwise, it is rejected. Thus, changes increasing the number of unbalanced triads are prohibited.
As a consequence, the system may reach jammed states. However, \citet{antal_2005} observed that for networks initialized with a positive fraction of friendly links, the probability of reaching a jammed state vanishes as the system size $N$ grows, see also~\citep{Chatterjee2022}. In the large-$N$ limit, the network almost always evolves toward a fully balanced configuration, with a relaxation time scaling as $\ln N$.

The topology of an absorbing, balanced state depends on the initial density of positive links $\rho_0$.
For small $\rho_0$, the network splits into two antagonistic cliques of comparable sizes, whereas for large $\rho_0$ it converges to the paradise state. The normalized clique-size asymmetry increases continuously from 0 to 1 with the density $\rho_0$. The authors used the second moment of this quantity as the order parameter. For $\rho_0>\rho_0^* \approx0.65$ (called the critical value in~\citep{antal_2005}), the system always converges to the paradise. Thus, CTD exhibits a transition between the bipolar and paradise phases, driven by the initial conditions.
An analytical approach, based on rate equations for the probabilities of switching positive and negative links, shows that under a mean-field assumption of no correlations, only negative links tend to flip on average. When correlations are considered -- e.g., when two emerging factions exhibit internal positive link densities above 0.5 and a cross-link density below 0.5 -- the bipolar state becomes stable. The analytical prediction of the critical density $\rho_0^*=0.5$ underestimates the numerical value.
% It is important to note that in the numerical simulations, the transition did not change with larger system size.

CTD has been further analyzed for small systems. \citet{Krawczyk2017} constructed a complete state-transition network for $N\le7$, where each node represents a network configuration and each directed link denotes an allowed transition between states.
For uniformly random initial sign configurations, all balanced states were found to be equally probable as final outcomes.
Extrapolating to larger systems, this result supports the observation that the paradise is always reached only for initial link densities significantly above 0.5.

\subsubsection{\label{subsubsec:hamil-nonzeroT}Mean-field solution for the complete graph}

 \citet{Malarz2022Mean} considered the Hamiltonian \eqref{eq:hamiltonian_triad} for the complete graph of size $N\gg 1$ where the system is in the heat bath at temperature $T$ that controls the level  of system's stochasticity and no thermodynamical  meaning to this parameter  is given.  Then, the link dynamics possesses a probabilistic character as follows:
\begin{subequations}
\label{eq:stoch1}
\begin{equation}
x_{ij}(t+1)=
        \begin{cases}
	+1 & \text{ with probability }p_{ij}(t),\\
	-1 & \text{ with probability }1-p_{ij}(t),
        \end{cases}
\end{equation}
with
\begin{equation}
    p_{ij}(t)=\frac{\exp(\xi_{ij}(t)/T)}{\exp(\xi_{ij}(t)/T)+\exp(-\xi_{ij}(t)/T)} \label{eq:pMF}
\end{equation}
where the local field acting on $x_{ij}$ is
\begin{equation}
	\xi_{ij}(t)=\sum_{k} x_{ik}(t)x_{kj}(t).
    \label{eq:mMF}
\end{equation}
\end{subequations}
In the limit $T\to 0^+$, the probabilities  $p_{ij}\to 1$ when $\xi_{ij}>0$ or $p_{ij}\to 0$ when $\xi_{ij}<0$  and Eq. \ref{eq:stoch1} reduces to the following deterministic update:
\begin{equation}
x_{ij}(t+1)=sign \left(\sum_k x_{ik}(t)x_{kj}(t)\right).
\label{eq:disc}
\end{equation}
In other words, in the zero-temperature limit, the variable $\xi_{ij}$ is a local field that acts on every link $x_{ij}$, and that is ordering this link in such a way that the sum of all energies related to triangles $(i,j,k)$ where the link $(i,j)$ is present is minimal. For nonzero temperatures, the equation for thermal averages reads as follows
\begin{equation}
\label{eq:cont}
\langle x_{ij}(t+1)\rangle=\tanh\left( T^{-1} \sum_{k} x_{ik}(t) x_{kj}(t) \right).
\end{equation}
In the spirit of mean-field approximation, one can assume that all links share the same average $\overline{x}(t)=\langle x_{ij}(t)\rangle$ and the link-link correlation function factorizes $\langle x_{ik}(t) x_{kj}(t)\rangle \approx \langle x_{ik}(t)\rangle \langle x_{kj}(t)\rangle$. Then the average value of all links should evolve as
\begin{equation}
\overline{x}(t+1)=\tanh[a\overline{x}^2(t)],
\label{eq:MF}
\end{equation}
where $a=(N-2)/T$. The above equation possesses a stationary solution $\overline{x}_0=0$ that is a stable fixed point of this dynamical system for any value of the parameter $a$. For $a> a_c\approx 1.716$, two additional fixed points appear, $\overline{x}_u<\overline{x}_s$ corresponding to unstable $\overline{x}_u(a)$ and stable $\overline{x}_s(a)$ solutions, respectively.
In fact, $\overline{x}_u$ serves as the separatrix between the basins of attraction of the fixed points $\overline{x}_0$ and $\overline{x}_s$. For $a \gg 1$, one obtains the approximation $\overline{x}_u \approx 1/a$ and $\overline{x}_s\approx 1$, i.e., the system is close to the paradise.  As the parameter $a$ is reduced from large values (which corresponds to increasing the temperature $T$), the fixed points $\overline{x}_u$ and $\overline{x}_s$ collide at $\overline{x}_c\approx 0.7964$ at a critical value $a = a_c$ and annihilate. Consequently, within the mean-field approximation, the system exhibits bi-stability in the low-temperature regime $T <T_c=(N-2)/a_c$, whereas for $T > T_c$ it becomes mono-stable since the only solution is $\overline{x}=0$.   It also means that when the system is heated from the paradise phase, at $T_c$, there is a sudden decrease in the mean link sign from $\overline{x}_c$ to zero. This discontinuous change of the mean link sign $\braket{x_{ij}}$ is connected to a discontinuous change in the average triad energy $H/N_T$, from a negative value $-\overline{x}_c^3$ to zero energy.
The mean-field approximation captures the loss of stability of the the paradise-like state and agrees closely with numerical simulations initialized in paradise. It does not, however, capture balanced bipolar configurations, which can have $\overline{x}\approx0$ despite containing only balanced triads. Such configurations cannot be distinguished from unbalanced ones using the mean link sign alone. Let us note that, since the Hamiltonian~\eqref{eq:hamiltonian_triad} is not normalized, each link in a complete graph belongs to $N-2$ triads, corresponding to $N-2$ terms in the local field $\xi_{ij}$~\eqref{eq:mMF}. Consequently, the critical temperature scales as $T_c\propto N-2$ with system size. When a normalization factor is present in a Hamiltonian (e.g., see \eqref{eq:hamil-norm} below), then the dependence of the critical temperature on the system size is different (see Table  \ref{tab:Tc_scalings}).

\begin{figure}[!htb]
    \centering
    \includegraphics[width=0.95\textwidth]{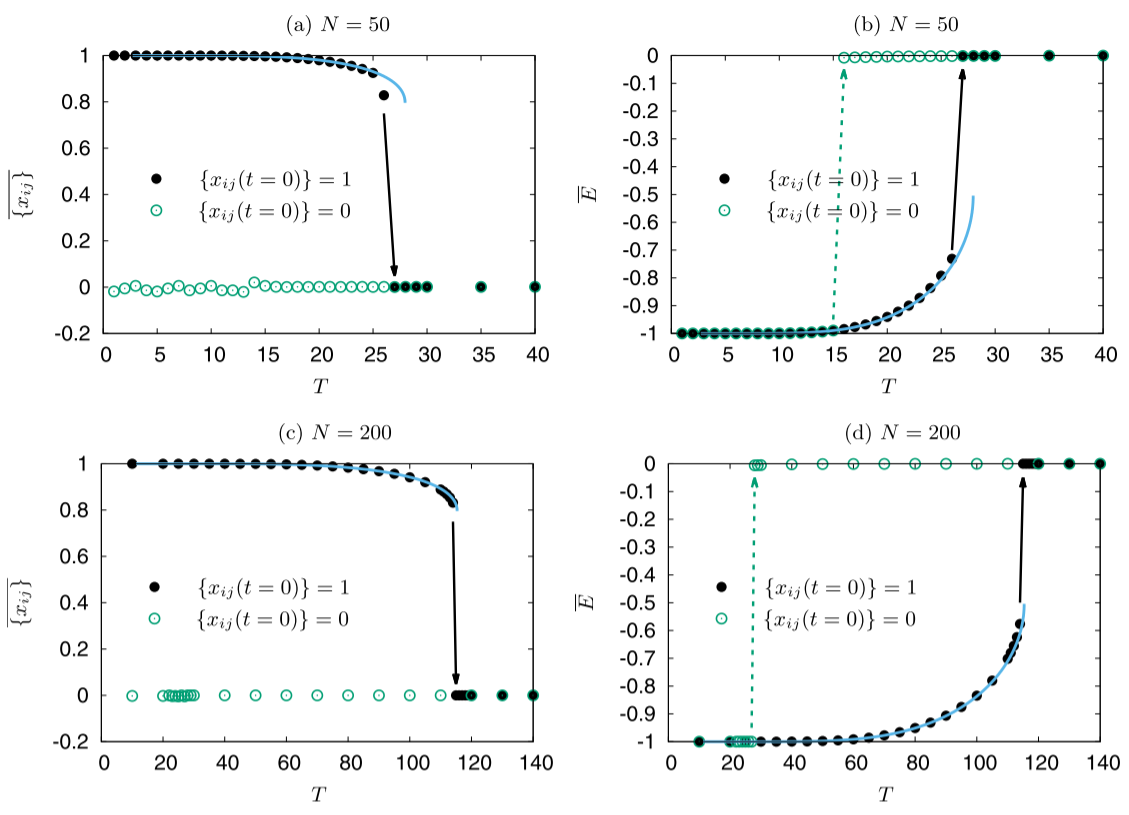}
    \caption{
    First-order transitions in Hamiltonian-based structural-balance dynamics on the complete graph.
    Panels (a,c) show the time-averaged mean link sign $\overline{x}=\braket{x_{ij}}$, while panels (b,d) show the average triad energy $\overline{E}=H/N_T$, as functions of temperature $T$ for $N=50$ and $N=200$.
    Filled markers correspond to simulations initialized in the paradise state, $\overline{x}(t=0)=1$, whereas open markers correspond to random initial conditions, $\overline{x}(t=0)=0$.
    Blue lines show the mean-field predictions, Eq.~\eqref{eq:MF}.
    The figure illustrates three regimes: paradise-like states with high $\overline{x}$ and low $\overline{E}$, polarized two-clique states with $\overline{x}\approx0$ but low $\overline{E}$, and unbalanced disordered states with $\overline{x}\approx0$ and $\overline{E}\approx0$.
    % The paradise--unbalanced and polarized--unbalanced transitions are both discontinuous, but occur at different critical temperatures.
    For these two initializations, the average triad energy changes discontinuously near \(T_c\) and \(T_d<T_c\), respectively.
    Reproduced from Ref.~\citep{Malarz2022Mean}. © 2022 American Physical Society.
    }
    \label{fig:hamiltonian-transition}
\end{figure}

% The observed first-order phase transition exhibits a specific hysteresis loop. Upon cooling from the disordered phase, the system does not return to the paradise-like solution; instead, below a much lower temperature $T_d$ it settles into a structurally balanced two-clique state.
Numerical simulations of Ref.~\citep{Malarz2022Mean} reveal different transition temperatures depending on the initial conditions. Starting from paradise, the system undergoes a discontinuous transition to an unbalanced state at $T_c$, whereas simulations initialized with random link signs produce a structurally balanced two-clique state only below a much lower temperature $T_d$.
The average triad energy is near $-1$, while the average link sign  $\braket{x_{ij}}$ is close to zero, since when two hostile cliques are of the same size (which is the most probable case because of the maximum entropy) the number of positive links inside both cliques is approximately equal to the number of negative links between them.
These two discontinuous transitions, together with the corresponding jumps in the mean link sign and average triad energy, are shown in Fig.~\ref{fig:hamiltonian-transition}. Their separation suggests hysteresis-like behavior.

\citet{Rabbani2019} likewise reported a first-order phase transition between a paradise and an unbalanced state. Using a mean-field approximation, the authors derived a coupled system of algebraic equations for the average link sign and for a correlation function describing the signs of neighboring links. By solving this system numerically, the critical temperature separating the two phases was identified, although no analysis of its scaling with system size, $T_c(N)$ was provided.

The question of how the critical temperature scales with system size has also been examined in several related works~\citep{Shojaei2019,Malarz2021Comment,Manshour2021}. \citet{Shojaei2019} investigated the system using Glauber dynamics (equivalent to the heat-bath rule for binary link variables; see Appendix~\ref{app:mc}) and a \textit{normalized} Hamiltonian,
\begin{equation}
H_{\mathrm{norm}} = -\frac{1}{N_T}\sum_{i<j<k} x_{ij}x_{jk}x_{ki},
\label{eq:hamil-norm}
\end{equation}
where $N_T$ denotes the total number of triads. Unlike \citet{Malarz2022Mean}, the authors considered random initial conditions with a prescribed initial density of positive links $\rho_0$.
Numerical simulations showed that, within this normalized formulation, the critical temperature separating balanced and unbalanced phases \textit{decreases} with increasing system size, implying that the phase transition vanishes in the thermodynamic limit ($T_c \to 0$ as $N \to \infty$).
This normalized Hamiltonian previously appeared in \citet{Marvel2009} (without temperature-driven dynamics). \citet{Manshour2021Reply} also argued that normalization is a common choice for systems with long-range interactions, and that a vanishing critical temperature may help explain the apparent absence of structural balance in large real-world networks.
In contrast, \citet{Malarz2021Comment} argued that the unnormalized Hamiltonian~\eqref{eq:hamiltonian_triad} is more natural from a sociological perspective, as social influences are expected to accumulate rather than be diluted by averaging. Using this Hamiltonian together with synchronous update dynamics, they found that the critical temperature increases with system size and that the scaling exponent $\gamma$ in $T_c \sim N^{\gamma}$ depends sensitively on the initial density of positive links $\rho_0$: specifically, $\gamma \approx 1$ for $\rho_0 \gtrsim 0.6$ or $\rho_0 \lesssim 0.4$, and $\gamma \approx 1/2$ when $\rho_0 \approx 0.5$.

It remains unclear to what extent these scaling relations, $T_c\propto N$ and $T_c\propto\sqrt{N}$, reported by \citet{Malarz2021Comment} carry over to stochastic Hamiltonian-based dynamics of the type considered in Refs.~\citep{Malarz2022Mean,Rabbani2019}, since \citet{Malarz2021Comment} employed synchronous updates in which all links are updated simultaneously. Such dynamics can differ qualitatively from asynchronous single-link updates, even when the same local update rule is employed (see also Sec.~\ref{subsubsec:hamil-network}).
\citet{Shojaei2019} also examined transitions between two-clique polarized states and paradise configurations when changing the initial density of positive links $\rho_0$. Similar transitions are known to occur for CTD~\citep{antal_2005}, as discussed in the previous section. As was shown in~\citep{Shojaei2019}, such a transition persists at nonzero temperature.
For synchronous dynamics, however, a different behavior was observed~\citep{Malarz2021Comment}: for initial densities far from $\rho_0 = 0.5$ (i.e., $\rho_0 \gtrsim 0.6$ or $\rho_0 \lesssim 0.4$), the system evolves toward paradise, whereas for $\rho_0 \approx 0.5$ it converges to a polarized two-clique state.

Taken together, these results suggest that different size-scaling behaviors of the critical temperature may be associated with distinct types of unbalanced-balanced transitions.
In particular, the scaling $T_c \propto N$ appears to be related to the paradise--unbalanced transition, whereas we conjecture that the lower temperature associated with the emergence of polarized state from random initial conditions scales as $T_d \propto \sqrt{N}$.
% may be characteristic of transitions between polarized and unbalanced states.
The shifts of both transition temperatures with increasing system size are consistent with this interpretation, as illustrated in Fig.~\ref{fig:hamiltonian-transition}.

% Overall, the scaling coefficients obtained in~\citep{Malarz2022Mean}, and those obtained for asynchronous dynamics let us hypothesize that the scaling of $T_c\propto N$ can be related to unbalanced-paradise transition and the scaling of Tc~sqrt{N} can be related to the unbalanced-polarized transition.

The mean-field approach of~\cite{Malarz2022Mean} is further extended for the observation of structural balance in Erd\"os-Renyi graphs~\cite{Mohandas2025Paradise} and at multiplexes ~\cite{Mohandas2024}. For more details, see Sec. \ref{subsubsec:hamil-network} and \ref{subsubsec:multilayer}.

\citet{Bagherikalhor2021} studied a variant of the Hamiltonian formulation in which the interaction strengths are heterogeneous. The Hamiltonian is defined as
\begin{equation}
H = -\sum_{i<j<k} J_{ijk}\, x_{ij}x_{jk}x_{ki},
\end{equation}
where coupling constants $J_{ijk}$ are drawn independently from a normal distribution with a mean $\mu$ and a variance $\sigma^{2}$.
% In this formulation, unbalanced configurations are favored in triads with $J_{ijk}<0$, introducing frustration analogous to that in spin-glass models \citep{Sherrington1975, Sherrington2025}.
In this formulation, triads with $J_{ijk}<0$ energetically favor an unbalanced sign product. Since triads share edges, heterogeneous couplings can generate competing local preferences, analogous to quenched frustration in spin-glass models \citep{Sherrington1975, Sherrington2025}.
% , then the stable triads are unbalanced ones.

For positive $\mu$, two regimes are identified. In the low-variance regime ($\mu/\sigma \ge 1$), the system behaves qualitatively similarly to the homogeneous-coupling case \citep{Malarz2022Mean, Rabbani2019}, exhibiting a first-order phase transition between balanced and disordered states.
In contrast, in the high-variance regime ($\mu/\sigma < 1$), negative couplings become common, producing competing local preferences that prevent full structural balance, with unbalanced triads persisting even at low temperatures.
Heterogeneous interaction strengths thus provide a further mechanism by which structural balance may fail to emerge in realistic systems. So they serve as an additional stochastic component.

These three-body interactions with quenched random couplings invite comparison with mean-field three-spin ($p=3$) spin-glass models~\citep{BertalanNishimori2012}. In the standard Gaussian $p=3$ model with a positive mean coupling, the phase diagram contains paramagnetic, ferromagnetic, and spin-glass phases, with a first-order transition between the paramagnetic and ferromagnetic phases and a spin-glass phase occurring at sufficiently strong disorder coupling relative to the mean coupling \citep{BertalanNishimori2012}. The low-temperature trapping observed by \citet{Bagherikalhor2021} under strong disorder is suggestive of glassy dynamics, although a spin-glass transition was not investigated. The correspondence is not exact, since here the spin variables are network edges and three-body interactions occur only among edges forming overlapping triangles.

\subsubsection{\label{subsubsec:hamil-multitype}Multiple types of interactions}

 \citet{belaza2017_StatPhysBalanceTheory} proposed a more general Hamiltonian model that simultaneously accounts for three-body, two-body, and one-body interactions between signed links. Each interaction type is weighed by a parameter reflecting its relative strength: $\alpha$, $\gamma$, and $\omega$, respectively.
% This way authors introduce a different energy value for each triad type (Delta_0, …, Delta_3). They call triads Delta_0 - strongly balanced, Delta_1 - strongly frustrated, Delta_2 - weakly balanced, and Delta_3 - weakly frustrated. They notice that triads  Delta_1 and Delta_2 are degenerated, each with 3 different micro-states.
The Hamiltonian takes the form
\begin{equation}
H = \sum_{i<j<k}
\left[
  -\alpha\, x_{ij}x_{jk}x_{ki}
  -\gamma\, (x_{ij}x_{jk} + x_{ij}x_{ki} + x_{jk}x_{ki})
  +\omega\, (x_{ij} + x_{jk} + x_{ki})
\right].
\label{eq:h:belaza2017}
\end{equation}
% H = sum_(i<j<k) [-alpha x_ij x_jk x_ki - gamma(x_ij x_jk + x_ij x_ki + x_jk x_ki) + omega (x_ij + x_jk + x_ki)]
%
Here, $\Delta_k$ denotes a triad with $k$ negative links ($k=0,1,2,3$). The parameter $\alpha$ promotes balanced triads, $\gamma$ acts as a homogenizing force favoring triads with uniformly signed edges (i.e., $\Delta_0$ and $\Delta_3$), and $\omega$ rewards the formation of negative links.
This parametrization assigns an energy $E_r$ to each of the four triad types ($\Delta_0$, $\Delta_1$, $\Delta_2$, and $\Delta_3$), with the relative ordering of their energies depending on the model parameters.
% With all the parameters positive, triads $\Delta_0$ are strongly balanced, $\Delta_1$ -- strongly frustrated, $\Delta_2$ -- weakly balanced, and $\Delta_3$ -- weakly frustrated.
Triad types $\Delta_1$ and $\Delta_2$ are each threefold degenerate, corresponding to three distinct sign configurations with the same energy, whereas $\Delta_0$ and $\Delta_3$ are nondegenerate.

Independently of the Hamiltonian, the authors used a frequentist definition of probability to estimate the occurrence of each triad type $(p_0, p_1, p_2, p_3)$ in real networks. From these probabilities, an entropy measure was defined as
\begin{equation}
S = -\sum_{r=0}^{3} p_r
    \ln\left(\frac{p_r}{g_r}\right),
\end{equation}
where $g_r$ is the geometrical degeneracy of triad type $\Delta_r$. Assuming that triads can be treated as approximately independent, the Boltzmann distribution is obtained by maximizing the entropy at fixed average triad energy $\langle E\rangle=\sum_r p_r E_r$, with the probabilities satisfying $\sum_r p_r=1$:
\begin{equation}
p_r = \frac{g_r\exp(-E_r/T)}{Z},
\label{eq:h:freq}
\end{equation}
where $Z$ is the partition function:
\begin{equation}
Z = \sum_{r=0}^3 g_r\exp(-E_r/T).
\label{eq:h:Z}
\end{equation}
Here, $T$ is an effective temperature characterizing the disorder in the distribution of triad types.

Following \eqref{eq:h:freq}, in empirical applications, relative triad energies can be estimated from observed probabilities as
\begin{equation}
E_r - E_s = \frac{1}{\beta} \ln\left( \frac{p_s}{p_r} \frac{g_r}{g_s} \right),
\end{equation}
where $\beta = 1/T$.

With the proposed Hamiltonian, at high temperatures, triad proportions approach their random (degeneracy-determined) values. At low temperatures, the lowest-energy configurations are favored. If $\Delta_0$ has the lowest triad energy, the paradise state is energetically preferred.

The model was applied to several empirical datasets by fitting its Hamiltonian parameters: two versions of the EVE online multiplayer game, records of alliances and conflicts during the Syrian civil war, and international relations data from the Cold War era. The fitted values were $\alpha \in [0.89,1.09]$, $\gamma \in [0.38,0.61]$, and $\omega \in [0.14,0.22]$. Temporal variations of triad energies and entropy were also analyzed, with interpretations linked to known geopolitical events.
% They show how triad energies and the entropy change in time. They give interpretation of behavior changes based on real situations that had happened.
% The obtained parameters were found in the following ranges: alpha in 0.89-1.02, gamma in 0.38-0.61, and omega in 0.14-0.22.

In a subsequent study, \citet{Belaza2019} extended the model to account for neutral or missing edges by introducing an additional term proportional to $x_{ij}^2$.
The modified Hamiltonian reads
\begin{equation}
H =
\sum_{i < j < k}^{N}
\left[
  -\alpha\, x_{ij}x_{jk}x_{ki}
  -\gamma\, (x_{ij}x_{jk} + x_{jk}x_{ki} + x_{ki}x_{ij})
\right]
+ \sum_{i < j}^{N}
\left[
  \omega\, x_{ij} + \mu\, x_{ij}^2
\right],
\label{eq:h:belaza2019}
\end{equation}
where $\mu$ represents a chemical potential controlling the energetic cost of creating a nonzero link. Since $x_{ij}^2=0$ for neutral or missing links and $x_{ij}^2=1$ for positive or negative links, positive $\mu$ favors link absence by penalizing existing relations.
% Since Eq.~\eqref{eq:h:belaza2017} includes the one-body term inside the triad sum, whereas Eq.~\eqref{eq:h:belaza2019} writes it as a separate pairwise sum, the same symbol $\omega$ has a different normalization in the two formulations.

Taking missing links into account leads to ten distinct triad types in undirected networks. Depending on the parameters, these triads can be ordered differently on the energy scale.
A mean-field analysis reveals that for low temperatures ($\beta^{-1} \ll \mu$) links are mostly absent; for intermediate temperatures ($\beta^{-1} \approx \mu$), a partially connected network with nonzero link polarization appears; and for high temperatures ($\beta^{-1} \gg \mu$), links become random. Parameter calibration to empirical data confirmed that the cost of creating a link (controlled by $\mu$) dominates the Hamiltonian. Transition probabilities between triad states measured in real networks were found to be consistent with theoretical predictions.

A complementary empirical application of the simpler triadic balance model was presented by \citet{Lee2017}, who employed Glauber dynamics to derive a Gibbs stationary distribution and estimate an effective ``social temperature'' from international-relations and United Nations General Assembly voting networks.

Another model involving multiple types of interactions, though not explicitly formulated within structural balance theory, was proposed in \citet{Park2005,Babaee2022}. Its Hamiltonian is given by
\begin{equation}
H = \theta \sum_{i<j} x_{ij} - \alpha \sum_{i<j<k} x_{ij}x_{jk}x_{ki}.
\end{equation}
Here, $x_{ij}\in\{0,1\}$ indicates the absence or presence of a link.
Interestingly, treating missing links as negative links leads to the transformation $s_{ij}=2x_{ij}-1$, which maps this Hamiltonian, up to an additive constant, onto the special case of Eq.~\eqref{eq:h:belaza2017}.
% If missing links are interpreted as negative edges, the same
Depending on the parameters, the global energy minimum corresponds to either an empty or a complete network, while configurations consisting of disconnected cliques may constitute local energy minima.
Under the transformation, these configurations correspond to the \emph{hell state}, the \emph{paradise state}, and weakly balanced multipolar states, respectively.
This mapping illustrates that the presence of a three-body interaction favoring structural balance does not necessarily imply that balanced configurations minimize the total energy, as other interactions may favor unbalanced states.

\citet{Burda2026Ising} studied a structural-balance Hamiltonian supplemented by an external field favoring
positive or negative relations. For complete graphs, the Hamiltonian takes the form
\begin{equation}
H
= -\frac{J}{n_t}\sum_{i<j<k}x_{ij}x_{jk}x_{ki}
-\frac{h}{n_e}\sum_{i<j}x_{ij},
\label{eq:h:burda2026}
\end{equation}
where $J>0$ controls the strength of structural-balance interactions, $h$ is the external field, and $n_t=N_T/N$ and $n_e=m/N$ denote the numbers of triads and edges per node, respectively.
The normalization ensures that both contributions to the Hamiltonian scale linearly with $N$.

In the limit $J/T\to\infty$, at fixed $h/T$, all triads satisfy $x_{ij}x_{jk}x_{ki}=1$.
On a complete graph, this condition allows the link signs to be factorized as $x_{ij}=s_i s_j$, where $s_i=\pm1$ are binary node variables. Consequently, the field acting on links becomes a pairwise interaction between node spins, yielding an Ising model without an external magnetic field.
In this mapping, the dimensionless coupling between Ising spins is $h/(n_e T)$. Using this equivalence, the authors identified a continuous field-driven phase transition on complete graphs at $h/T=1/2$.
Interpreting the Ising magnetization in terms of faction sizes, we infer that this transition separates balanced configurations with two approximately equal-sized factions from those with increasingly unequal factions as the field increases, finally approaching paradise.

More generally, the Ising correspondence applies to globally balanced signed networks independently
of their topology. Building on this result, \citet{Burda2026Lattices} investigated structural balance on triangular and square lattices, the latter discussed in Sec.~\ref{subsubsec:hamil-variants}.
The authors used the correspondence to analyze the field-driven critical behavior of these systems in the fully balanced limit.

\subsubsection{\label{subsubsec:hamil-variants}Beyond triadic Hamiltonian: higher-order couplings and weak balance}

% \msmcomment{I propose an alternative title: Beyond triadic Hamiltonians: Higher-order couplings and weak balance}

Continuing the discussion on interaction types, several extensions of Hamiltonian-based models have explored higher-order couplings and relaxed forms of structural balance.
\citet{Kargaran2020} propose a Hamiltonian that couples pairs of adjacent triads ($(i,j,k)$ and $(i,j,l)$), incorporating correlations beyond single-triad interactions:
\begin{equation}
H
= -\sum_{(i,j,k,\ell)\in\mathscr C_4}
  (x_{ij}x_{jk}x_{ki})(x_{ik}x_{k\ell}x_{\ell i})
= -\sum_{(i,j,k,\ell)\in\mathscr C_4}
  x_{ij}x_{jk}x_{k\ell}x_{\ell i},
\label{eq:hamil-quartic}
\end{equation}
where $\mathscr C_4$ denotes the set of cycles of length four, each counted once.
Assuming an undirected network ($x_{ij}=x_{ji}$) with binary links $x_{ij}\in\{\pm1\}$, this Hamiltonian evaluates balance over cycles of length four (squares). An interesting property of this formulation is that two adjacent unbalanced triads always form a balanced square, indicating that local frustration can cancel when triads share edges. However, the model neglects shorter cycles, focusing exclusively on four-body interactions.

Using a mean-field approach, the authors found that below a critical temperature $T_c$, five solutions exist, three of which are stable over a broad temperature range. Two stable solutions correspond to the paradise and hell states, while the third is characterized by an equal number of positive and negative links, $\overline{x}=0$.
As the temperature increases, the paradise and hell solutions disappear discontinuously at $T_c$, leaving the zero-average state as the only stable solution. Due to the invariance of the Hamiltonian under the inversion of all signs, the paradise and hell states are energetically equivalent.
% that above a critical temperature $T_c$ there is a single stable solution; at $T_c$ three solutions appear; and below $T_c$ five solutions exist, three of which are stable. Two of these correspond to the states of paradise and hell.
For this unnormalized Hamiltonian, the critical temperature scales with system size as $T_c \propto N^2$, in contrast to models assuming the standard triadic Hamiltonian~\eqref{eq:hamiltonian_triad}, for which $T_c$ changes linearly with $N$ in the case of paradise-unbalanced transition (see Sec. \ref{subsubsec:hamil-nonzeroT}). These results were corroborated numerically using the Metropolis algorithm for random, paradise, and hell initial conditions.

\citet{Burda2026Lattices} compared Hamiltonian-based structural-balance models on triangular and square lattices in the presence of an external field. On triangular lattices, the Hamiltonian contains three-body interactions of the form in Eq.~\eqref{eq:h:burda2026}, with the sum restricted to elementary triangles.
On square lattices, these are replaced by four-body interactions involving products of link signs around elementary squares, analogous to the quartic term in Eq.~\eqref{eq:hamil-quartic}. Both models also include the external-field term from Eq.~\eqref{eq:h:burda2026}.

Flipping all link signs changes the sign of cycles in three-body interactions but leaves four-body interactions unchanged. Consequently, the all-negative configuration is unbalanced on triangular lattices but satisfies the local balance condition on every elementary square.
A negative external field therefore competes with structural balance on triangular lattices but not on square lattices, leading to different symmetries of the field response.

We note that a related distinction between odd- and even-length cycles was previously reported for local stochastic cycle dynamics by \citet{Radicchi2007Social,Radicchi2007Universality} (see Sec.~\ref{subsubsec:ltd-extensions}).

As discussed in Sec. \ref{subsubsec:kk-variants}, \citet{Siboni2025} connected delayed extensions of gradient triad dynamics with effective higher-order interactions.
In addition to the continuous-time analysis, they compared Monte Carlo simulations of a delayed triadic-balance model with delay-free hybrid models containing triadic and higher-order interactions.
A model combining three- and four-body interactions reproduces the energy--temperature dependence of the delayed model well.
For larger delays, inclusion of five-body interactions improves the agreement.
Increasing the delay shifts the transition toward lower temperatures.
% Interestingly, unbalanced four-cycles (negative cycles or cycles with odd number of negative edges) minimize the Hamiltonian, and at large delays, they prevent the system from reaching complete balanced configurations even at low temperatures.

\citet{Manshour2021} model a weak version of structural balance by introducing a tunable parameter $\varepsilon\in[0,1]$ that interpolates between weak ($\varepsilon=0$) and strong ($\varepsilon=1$) balance. Rather than defining an explicit edge-based Hamiltonian, they assign energies to triad types and define the system energy, normalized by the number of triads, as
\begin{equation}
U = \dfrac{1}{N_T}\sum_a u_a,
\label{eq:hamil-weak}
\end{equation}
where $u_a$ denotes the energy of the $a$-th triad, taking values $0$, $1$, $0$, and $\varepsilon$ for triad types $\Delta_0$, $\Delta_1$, $\Delta_2$, and $\Delta_3$, respectively, and $N_T$ is the total number of triads. The model is analyzed using a mean-field approach and simulated numerically through Glauber dynamics.
The authors explore the phase behavior as a function of temperature and the initial positive link density $\rho_0$.

% As in \citet{Bagherikalhor2021} (see Sec. \ref{subsubsec:hamil-nonzeroT}), the energy is normalized by $N_T$, and the authors explore the phase behavior as a function of temperature and the initial positive link density $\rho_0$.
A phase transition is observed, characterized by the disappearance, at low temperatures, of $\Delta_1$ triads containing a single negative link. This signals the emergence of a weakly balanced state that may comprise multiple factions.
% Depending on the initial link density the system may end up in a paradise or polarized states. The polarized state may contain for
% Therefore, here the
Consistent with \citet{Shojaei2019} (see Sec.~\ref{subsubsec:hamil-nonzeroT}), normalization by $N_T$ leads to a critical temperature that decreases with system size and vanishes in the thermodynamic limit, $T_c\to0$ as $N\to\infty$.

The critical temperature also decreases with decreasing $\varepsilon$, implying that polarization becomes less likely as the model approaches the weak-balance limit $\varepsilon\to0$. Interestingly, the complete elimination of $\Delta_1$ triads is easier for $\varepsilon>0$ than for $\varepsilon=0$.
% This decrease of $T_c$ with decreasing $\varepsilon$ can be interpreted as a competition between energy and entropy. As $\varepsilon$ becomes smaller, the energy difference between bipolar (two-clique) and multipolar ($k>2$) weakly balanced states diminishes. However, the number of possible divisions into more than two cliques grows rapidly with $k$, increasing the entropy $S$ of multipolar configurations. Since the system minimizes the free energy $F = U - TS$ \citep{Malarz2020Expulsion}, the entropic contribution favors multipolar states at finite temperature.
In light of the conjectured scaling--transition relations discussed in Sec.~\ref{subsubsec:hamil-nonzeroT}, where $T_c\propto N$ is associated with the paradise--unbalanced transition and $T_d\propto\sqrt{N}$ with the emergence of bipolar order from random initial conditions, we further conjecture that transitions toward weakly balanced $k$-clique states with $k>2$ may occur at progressively lower temperatures as $k$ increases.

As in \citep{Bagherikalhor2021}, the final state depends on the initial density of positive links. Depending on $\rho_0$, the system converges to either a paradise or a polarized state.
Increasing $\varepsilon$ enlarges the basin of attraction of paradise. The transition occurs between paradise and a two-clique state for $\varepsilon \gtrsim 0.7$, and between paradise and a multi-clique weakly balanced state for small $\varepsilon$.
% Therefore, here the
% Another transition occurs when varying $\varepsilon$: below a critical $\varepsilon_c\approx0.67$, a multipolar (weakly balanced) phase arises, whereas above $\varepsilon_c$, the system converges to a bipolar, strongly balanced configuration.

\subsubsection{\label{subsubsec:hamil-network}Hamiltonian dynamics on non-complete graphs and alternative update schemes}
% Beyond complete graphs

Most simulations based on Hamiltonian dynamics have assumed a complete graph, which facilitates analytic treatment and reasonably models small, densely connected communities. Recent works have broadened the scope to other topologies, including regular lattices and Erd\H{o}s--R\'enyi (ER) random graphs. Some of these works consider structures beyond complete networks but also employ synchronous update schemes, in which all links are updated simultaneously.
Hamiltonian formulations have also been applied to empirical, generally incomplete signed networks to estimate model parameters, including interaction strengths and effective temperatures~\citep{Lee2017,belaza2017_StatPhysBalanceTheory,Belaza2019} (see Sec.~\ref{subsubsec:hamil-multitype}).

\paragraph{Erd\H{o}s--R\'enyi networks}

For ER graphs, each potential edge is present with probability $p$, corresponding to the $G(N,p)$ ensemble introduced in Sec.~\ref{subsubsec:topology-reshuffling}. The expected number of triads is
$
\binom{N}{3}p^{3}.
$
% and the expected number of triads per existing edge is $(N-2)p^{2}$.
To analyze the paradise--unbalanced transition and its dependence on $N$, $p$, and $T$, \citet{Mohandas2025Paradise} extended the mean-field approach of \citet{Malarz2022Mean}. Introducing a normalized temperature $\tilde T=T/(N-2)$, they obtained
$
\tilde T_c = 0.58258\, p^{2},
$
in excellent agreement with Metropolis simulations, also independent of $N$.
A similar paradise--unbalanced transition in such systems was reported by \citet{Masoumi2022}.
In both works, the dynamics was asynchronous.
The transition between balanced polarized and unbalanced states under synchronous updates was investigated numerically by \citet{Malarz2023Thermal}.
% In contrast to the asynchronous case, synchronous updates on ER graphs can lead either to a smooth crossover at low graph densities or a discontinuous phase transition between balanced and unbalanced phase.
The nature of this transition depends on graph density:
at low densities, a smooth crossover is observed, whereas
at higher densities, the transition is discontinuous.
% continous (diluted) do not lead to fully balanced states for small $p$, even at very low temperatures.
%\footnote{For instance, consider an unbalanced 4-node motif of two adjacent triads with one positive link at $T=0$. Under synchronous updates, this configuration evolves into a period-5 limit cycle.}.
A minimum density $p_{\min}\propto N^{-0.58}$ is required for a discontinuous transition to occur. Noting that for the complete graph $T_c\sim \sqrt{N}$ under the considered synchronous dynamics with random initial conditions, the authors introduced a normalized critical temperature $T_c^\star=T_c/T_c(p=1)$ to remove explicit $N$-dependence and obtained the empirical scaling $T_c^\star \propto p^{1.719}$ for $p>p_{\min}$.

Comparing the ER results highlights the distinction between the loss of balance from an initially balanced paradise state and the emergence of bipolar order from random, typically unbalanced initial configurations. For sufficiently dense graphs, both transitions are discontinuous, but their critical temperatures scale differently with graph density (see also Sec.~\ref{subsubsec:hamil-nonzeroT}).
% Both transitions are discontinuous, yet they scale differently with the increase of network density and system size.

\paragraph{Synchronous update dynamics on different network structures}

% \subsubsection{\label{subsubsec:hamil-synchronous}Synchronous update dynamics}

% In this section, we describe models based on Hamiltonian, though differently than all the models in this section, the described dynamics will be deterministic. Assuming synchronous updates of system states and

% When agents or links update the

% Having synchronous dynamics under the CTD rule (zero temperature) \msmcomment{I would rephrase, ``Under the CTD rule (zero temperature), replacing asynchronous updates by \emph{synchronous updates}"}
Under a deterministic variant of the CTD rule at zero temperature, in which energy-neutral flips are suppressed, \emph{synchronous updates} lead to a \emph{deterministic} evolution. \citet{Malarz2020Towards} examined this setting on a \textit{diluted} triangular lattice: during the initial construction, each potential link is removed with probability $f$. On the triangular lattice, every link belongs to two triads. In an update step, each link flips its sign if both adjacent triads jointly dictate the same sign. If the change is ambiguous, no action is taken.
This gives a deterministic, discrete-time, local rule.
% is the first model with deterministic dynamics, discrete signs (and time) and also with local information.
Unlike the usual CTD (which requires global knowledge), the decision depends only on the two neighboring triads.
The authors studied the average energy of the final state as a function of the link-removal probability $f$. The relationship is not monotonic. Without link removal ($f=0$), the system reaches a state in which approximately one eighth of triads are unbalanced. Increasing the probability to $f\approx 0.1$, slightly lowers the average energy. For larger $f$, however, the energy increases and reaches an observed maximum at around $f\approx 0.3$. For $f\gtrsim0.3$, the energy decreases monotonically with $f$, and for $f\approx 1$ the network becomes balanced because the remaining triads are effectively isolated.
% The resulting network is balanced for $f\approx 1$ because triads are isolated in such a case. For small $f\approx 0.1$, the authors observe a minimum of energy (but not a balanced state) \msmcomment{global/local minimum?}. For $f\approx 0.3$, they find the energy maximum \msmcomment{is it unique and global?}.
A local pattern-counting calculation reproduces the shape of the curve with a maximum near $f\approx 0.4$.
% Authors give analytical reasoning for the existence of this maximum, which is able to reproduce the shape of energy curve with maximum at f\approx 0.4. The analytics is based on the balance analysis of the triad neighboring patterns and calculating these patterns' abundance.

% Related studies were performed on the complete graph, focusing on jammed states and limit cycles~\citep{Krawczyk2021Towards}. \msmcomment{shall we add one takeaway from this paper?} In the complete graph, global knowledge is used to determine the next step. Also, unlike standard CTD, synchronous dynamics can increase the energy from one step to the next.
Related studies on the complete graph showed that this deterministic synchronous variant of CTD can lead not only to balanced or jammed states, but also to limit cycles~\citep{Krawczyk2021Towards}. In this topology, global knowledge is used to determine the next step. Also, unlike standard CTD, the energy may increase from one step to the next.
The length of the observed cycles grows with system size. Nevertheless, the probability of reaching a balanced state approaches 1 as the system size increases, similarly to asynchronous CTD \citep{antal_2005} or GTD with self-loops \citep{marvel2011_evolutionOFSBcontinuous}.

Synchronous \emph{stochastic} updates have also been explored.
\citet{Malarz2020Expulsion} studied the standard Hamiltonian~\eqref{eq:hamiltonian_triad} on the triangular lattice (heat-bath dynamics) and found a phase transition only at zero temperature: for all \(T>0\) the system relaxes to an unbalanced state, independently of initial conditions.
This behavior can be understood in terms of the competition between energy and entropy. In equilibrium, this competition is captured by the free energy $F=\langle H\rangle-TS$, where $H$ is the unnormalized triadic Hamiltonian in Eq.~\eqref{eq:hamiltonian_triad}.
For a network with $m$ signed links, there are $2^m$ possible sign configurations, giving a maximum configurational entropy $S_{\max}=m\ln 2$.
the number of links scales as $N^2$, while the number of interacting triads, and hence the characteristic energy scale, grows as $N^3$, so energy dominates entropy for large $N$. On the triangular lattice, both the number of links
and the number of triads scale linearly with system size. Thus, unlike on complete graphs, entropic contributions remain competitive with energy in the thermodynamic limit, consistent with the observed absence of a balanced ordered phase at $T>0$.
Extending this line, \citet{Malarz2025Heider} considered planar graphs composed of separated triads or chains of triads and showed that perfect structural balance cannot persist for $T>0$ on planar lattices.

The absence of a finite-temperature transition underscores how crucial neighboring triads are for sustaining balance. To probe this, \citet{Malarz2021Heider} simulated the model on a ring lattice where each node is connected to its $2r$ nearest neighbors. Increasing $r$ increases the number of triads sharing edges. They found that, as $r$ grows, the crossover between balanced and unbalanced regimes becomes sharper and shifts to higher temperatures, independently of system size.

\citet{Malarz2025Heider} applied the standard Hamiltonian \eqref{eq:hamiltonian_triad} with a heat-bath rule to update all links in parallel. Comparing synchronous and asynchronous settings across several topologies, they observed that synchronous evolution is markedly more sensitive to network structure and, overall, attains a lower level of balance than its asynchronous counterpart.

\subsubsection{Summary of critical-temperature scalings}
\label{subsubsec:hamil-scaling-summary}

The critical-temperature results reviewed above depend on the Hamiltonian
normalization, interaction order, network topology, update scheme, and
type of transition. Table~\ref{tab:Tc_scalings} summarizes representative
scaling results illustrating these dependencies.

{
\begin{table}[!htb]
\centering
\small
\setlength{\tabcolsep}{4.5pt}
\renewcommand{\arraystretch}{1.08}

\begin{tabularx}{\linewidth}{
>{\raggedright\arraybackslash}p{0.16\linewidth}
>{\raggedright\arraybackslash}p{0.27\linewidth}
>{\raggedright\arraybackslash}X
>{\raggedright\arraybackslash}p{0.16\linewidth}
}
\toprule
\textbf{Reference}
& \textbf{Model setting}
& \textbf{Critical-temperature result}
& \textbf{Basis} \\
\midrule

\citet{Malarz2022Mean}
% & Complete graph; unnormalized triadic Hamiltonian, Eq.~\eqref{eq:hamiltonian_triad}; heat-bath dynamics
& Heat-bath dynamics; paradise init. cond.
& Paradise--unbalanced transition:
\(T_c\simeq 0.58258\,(N-2)\)
& Mean field; simulations \\ \addlinespace[4pt]

\citet{Malarz2021Comment}
% & Complete graph; unnormalized triadic Hamiltonian, Eq.~\eqref{eq:hamiltonian_triad}; synchronous updates
& Synchronous heat-bath; random init. cond.
& \(T_c\propto N^{\gamma}\), with
\(\gamma\approx1\) for \(\rho_0\gtrsim0.6\) or
\(\rho_0\lesssim0.4\), and
\(\gamma\approx1/2\) for \(\rho_0\approx0.5\)
& Numerical \\ \addlinespace[4pt]

\citet{Shojaei2019}
& Hamiltonian normalized by $N_T$,
Eq.~\eqref{eq:hamil-norm}; Glauber; random init. cond.
& \(T_c\) decreases with \(N\) and vanishes in the thermodynamic limit
& Numerical \\ \addlinespace[4pt]

\citet{Kargaran2020}
& Four-body Hamiltonian, Eq.~\eqref{eq:hamil-quartic}; Metropolis
& Ordered--disordered transition:
\(T_c\propto N^2\)
& Mean field; simulations \\ \addlinespace[4pt]

\citet{Mohandas2025Paradise}
& ER graph; Metropolis; paradise init. cond.
& Paradise--unbalanced transition:
\(\tilde T_c \equiv T_c/(N-2)=0.58258\,p^2\)
& Mean field; simulations \\ \addlinespace[4pt]

\citet{Malarz2023Thermal}
& ER graph; synchronous heat-bath; random init. cond.
& Bipolar emergence:
\(T_c^\star \equiv T_c/T_c(p=1)\propto p^{1.719}\);
\(p_{\min}\propto N^{-0.58}\)
& Numerical \\

\bottomrule
\end{tabularx}

\caption{Representative critical-temperature scaling results for Hamiltonian-based structural-balance models. Unless stated otherwise, the results concern complete graphs with the unnormalized triadic Hamiltonian~\eqref{eq:hamiltonian_triad}. The results depend on Hamiltonian normalization and interaction order, network topology, update scheme, initial conditions, and the type of transition. Here, ``init. cond.'' denotes initial conditions, $\rho_0$ is the initial density of positive links, and $p$ is the edge probability in Erd\"os--R\'enyi graphs. Other notation and model details are defined in the corresponding sections.
}
\label{tab:Tc_scalings}
\end{table}
}

% As discussed in Sec.~\ref{subsubsec:hamil-nonzeroT}, we conjecture that the \(T_c\propto N\) and \(T_c\propto\sqrt{N}\) scalings are associated with paradise--unbalanced and polarized--unbalanced transitions, respectively.

\subsection{Stochastic sign dynamics}
\label{subsec:stoch-sign-dyn}

The previous two subsections described sign dynamics models based either on gradient triad dynamics or on Hamiltonian formulations of structural balance. In their standard formulations, updating a single relation in both approaches depends on global information about the current network state, either through sums over multiple triads or through system-wide energy evaluations.

In this section, we review a broad class of stochastic sign dynamics models, most of which depart from this framework of global coupling.
Such approaches usually rely primarily on local update rules, typically acting on a single triad at a time or within an ego-centered neighborhood. Yet, such models can still generate balanced or weakly balanced signed networks.

Within this group, the models reviewed differ mainly in what information enters an update and in what part of the network is allowed to change. Standard local triad dynamics (Sec.~\ref{subsubsec:ltd}) acts on a single triad, while its extensions (Sec.~\ref{subsubsec:ltd-extensions}) modify the allowed link states, interaction motifs, network topology, or the rule used to select the updated link.
Ego-based models (Sec.~\ref{subsubsec:ego-perspective-models}) further restrict balance evaluation to the perspective of a focal agent.
A central question in these approaches is whether local reductions of imbalance are sufficient to produce global structural balance or instead lead to long-lived unbalanced, weakly balanced, or jammed states.

We also discuss a smaller set of stochastic sign dynamics models (Sec.~\ref{subsubsec:other-stochastic-models}) that do not fit cleanly into the preceding categories. These models may modify earlier tension-reduction dynamics, allow the network topology itself to evolve, use nonlocal information, or employ alternative measures of imbalance.
Although some of these models are inspired by GTD or Hamiltonian-based formulations, they are discussed here because their stochastic update rules differ from the standard formulations reviewed above.
% They are grouped here because they fall outside the \kk and explicit Hamiltonian frameworks rather than because they share a common information scope.
Finally, we briefly consider strategic and game-theoretic models (Sec.~\ref{subsubsec:strategic-models}), in which balanced or low-frustration network structures emerge from agents' incentives rather than from prescribed tension-reduction rules.

% \begin{itemize}
% \item Local triad dynamics
% \item Extensions of LTD
% \item Models assuming ego perspective
% \item Combination of LTD with other models
% \item Other stochastic models
% \end{itemize}

\subsubsection{Local triad dynamics}
\label{subsubsec:ltd}

 \citet{antal_2005} introduced, alongside constrained triad dynamics discussed in Sec. \ref{subsubsec:hamil-ctd}, the local triad dynamics (LTD).
In LTD, the elementary update step is defined on a single triad: at each step, a triad is chosen uniformly at random. If the chosen triad is unbalanced, one of its links is flipped. Balanced triads are left unchanged.
Unbalanced triads are of two types: $\Delta_3$ (three negative links) and $\Delta_1$ (one negative and two positive links).
For a $\Delta_3$ triad, one of its three links is selected uniformly at random and flipped.
For a $\Delta_1$ triad, with probability $p$ the negative link is flipped, while with probability $(1-p)$ one of the two positive links is flipped at random. Hence, $p$ is the sole model parameter.

Both analysis and simulations reveal a phase transition at $p = 1/2$ between an unbalanced quasi-stationary regime and the paradise state.
The analytical treatment is based on a Master equation and a detailed balance argument.
In the stationary regime, it is assumed that the rate of creating positive links,  $pn_1+n_3$, equals the rate of creating negative links, $(1-p)n_1$, where $n_i$ denotes the density of triads $\Delta_i$ (with $i$ negative links).

\begin{figure}[!htb]
    \centering
    \includegraphics[width=\linewidth]{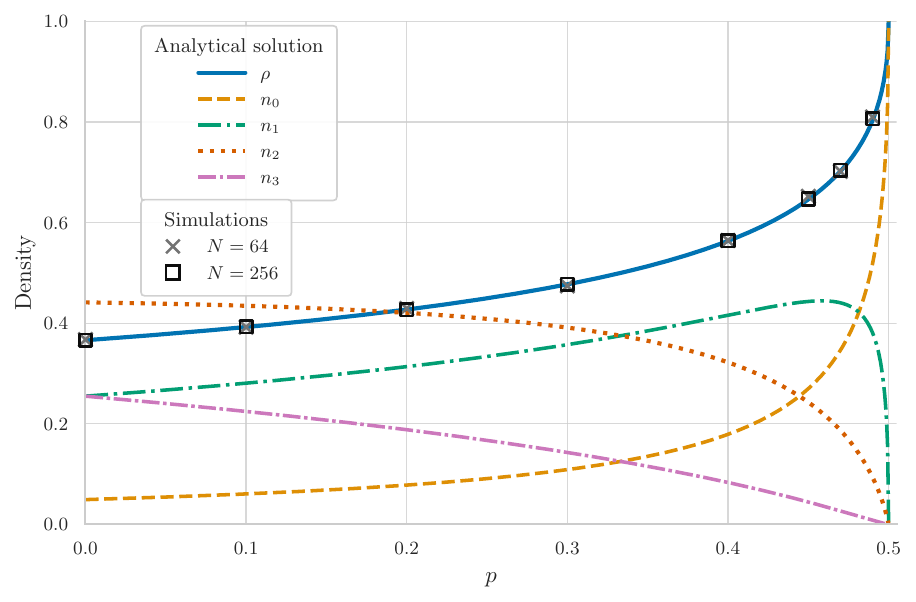}
    \caption{Phase transition in the local triad dynamics.
Lines show the analytical quasi-stationary solution derived in Ref.~\citep{antal_2005} for the density of positive links $\rho$, and the corresponding densities $n_i$ of triads with $i$ negative links.
Markers show our simulation results for complete networks with $N=64$ and $N=256$ nodes, using the system sizes considered in Ref.~\citep{antal_2005}.
For $p<1/2$, the dynamics relaxes to an unbalanced quasi-stationary state, whereas at $p=1/2$ this nontrivial solution disappears and the paradise state becomes the only stable stationary solution.
% Adapted with permission from Ref.~\citep{antal_2005}.
}
    \label{fig:ltd}
\end{figure}

Assuming further that positive and negative signs are placed at random (no correlations approximation), the triad densities follow the Bernoulli form:
\begin{equation}
  n_i = \binom{3}{i}\,\rho^{3-i}(1-\rho)^i,
  \label{eq:ltd-bernoulli}
\end{equation}
where $\rho$  is the density of positive links in the network. Inserting Eq.~\eqref{eq:ltd-bernoulli} into the detailed-balance condition, $pn_1+n_3=(1-p)n_1$, gives analytic solutions for $\rho$ \citep{antal_2005}.
For $p<1/2$, there are two solutions: a nontrivial quasi-stationary solution
\begin{equation}
  \rho = \frac{1}{\sqrt{3(1-2p)} + 1},
  \label{eq:ltd-rho-quasi}
\end{equation}
and the paradise $\rho=1$. In this regime, the quasi-stationary state is dynamically stable, while paradise is unstable, see Fig.~\ref{fig:ltd}. For $p>1/2$, the nontrivial solution disappears, and paradise becomes the only stable stationary state.

Balanced configurations (including paradise) are absorbing: once reached, the dynamics stops. The quasi-stationary state for $p<1/2$ is therefore a long-lived transient. Starting from a random initial condition, the LTD dynamics first relaxes exponentially fast to a quasi-stationary state. For finite systems, it eventually escapes to an absorbing balanced configuration. The mean time to reach balance scales with the system size $N$ as
\begin{equation}
  T(N) \sim
  \begin{cases}
    \exp\!\big(N^2\big), & p < 1/2,\\[4pt]
    N^{4/3}, & p = 1/2,\\[4pt]
    \dfrac{\ln N}{2p-1}, & p > 1/2,
  \end{cases}
  \label{eq:ltd-tau-scaling}
\end{equation}
showing an extremely slow approach to balance below the transition and a very rapid approach above it.

% While the obtained results are almost independent on the initial conditions, in~\citep{Qi2019} it is shown that in an incomplete graph, the initial conditions may slightly change the obtained

% The above results were obtained for complete graphs. At the mean-field level, the same qualitative picture also persists for sufficiently dense random graphs~\citep{Qi2019Minimum}.

\subsubsection{Variants and extensions of LTD}
\label{subsubsec:ltd-extensions}

Extensions of LTD modify several distinct ingredients of the basic model, including the allowed link states, constraints on which links may change, the motifs on which balance is evaluated, and the underlying network topology.
These modifications help identify which features of the standard LTD rule are responsible for its characteristic stationary states and phase transition.
% Over the years, many variants and extensions of LTD have been proposed.
Several studies focus on the special case $p=1/3$. For this choice, when an unbalanced triad is selected, each of its links is flipped with equal probability. In other words, the dynamics is unbiased with respect to which link is changed.

\citet{Wang2003} consider this special-case LTD on a network with neutral links, where link values can be $+1$, $-1$ or $0$. The LTD rule is modified so that, when a triad with two or three neutral links is chosen, no update is performed. Triads with exactly one neutral link are updated by adjusting that neutral link. Observing that the network always reaches a balanced configuration, the authors investigate whether it is possible to predict to which of the final groups a given agent will belong. They find that such a prediction is generally not possible from the initial configuration alone when neutral links are initially present.
% , but becomes possible once neutral links are either removed or converted during the evolution.
% prediction is not feasible in the presence of neutral links, but becomes possible once neutral links are either removed or converted during the evolution.

% \citet{Qi2019Minimum} study the same special-case LTD in a setting where each agent initially chooses one random neighbor, and the sign of the corresponding link is kept fixed throughout the dynamics. In standard LTD on a complete graph, the stationary densities of triads and the stationary density of positive links are essentially independent of the initial positive-link density. With the introduction of these frozen links, however, the initial density of positive links can significantly impact the eventual outcome.
% The topology explored is a ring lattice-like topology. The authors show that the stationary density of positive links increases with the average degree of agents.

\citet{Radicchi2007Social} extend LTD from triads to cycles of arbitrary length $k$. In the given dynamics, only cycles of fixed length $k$ are considered. When an unbalanced cycle is chosen, one of its links is flipped. If all links in the cycle are negative, any link is selected with equal probability. Otherwise, with probability $p$, a negative link is flipped, and with probability $(1-p)$, a positive link is flipped.

For complete graphs, the model behaves differently depending on whether the dynamics is driven by odd or even cycles. The cases $k = 3,5,\ldots$ are qualitatively similar to one another, but differ from $k = 4,6,\ldots$. The origin of this difference lies in the fact that flipping all signs in a balanced even cycle produces another balanced cycle. Therefore, the cycle with all links negative is balanced. This is not true for odd cycles.
For odd $k$, a phase transition at $p = 1/2$ is found. For $p > 1/2$, the system reaches paradise. For $p < 0.5$, the system remains in an unbalanced stationary regime in the thermodynamic limit, while finite systems may eventually reach balanced absorbing configurations.
For even $k$, there are two stable frozen states: paradise and hell (all links negative), and no stationary unbalanced regime exists.

Using methods similar to those of~\citep{antal_2005}, \citet{Radicchi2007Social} derive analytical predictions that agree well with numerical simulations. For $k = 5$, they obtain an exact solution that shows the phase transition. For $k = 7,9,\ldots$, the resulting equations are of too high order to be solved exactly. Under additional assumptions, however, they derive a general solution that again exhibits a transition at $p = 1/2$. In the limit $k \to \infty$, the stationary solution below the transition approaches $\rho = 0$.

\citet{Radicchi2007Social} also study LTD-type dynamics on random networks, showing that finding balanced configurations can be formulated as a three-spin optimization problem. For $p=1/3$, LTD corresponds to a stochastic local-search algorithm for this problem. This connection allows algorithms developed for related optimization problems to be adapted to structural balance, potentially reducing the time required to find balanced configurations.
% and highlight an analogy with the $k$-SAT problem~\citep{mezard2002analytic}, for which efficient algorithms to find frozen configurations are known. In particular, the special-case LTD with $p = 1/3$ can be mapped to a three-spin model.
% This connection suggests that $k$-SAT algorithms can be exploited to reduce the time required to find frozen states.

In a subsequent paper, \citet{Radicchi2007Universality} explore the LTD on a triangular lattice and their LTD modification based on cycles of length 4 on a square lattice.
In both experiments, each link belongs to exactly two elementary cycles. The authors distinguish two types of updates: diffusion, when flipping a link changes the balance/imbalance status of the two neighboring cycles, and annihilation, when the update removes the imbalance in both neighboring cycles.

Focusing on the negative link network in the triangular lattice topology, they formulate a percolation problem. Using Binder’s cumulant, they find a percolation threshold slightly below $p=1/2$ and determine critical exponents associated with this transition. For the square lattice, a transition between absorbing states is also observed, but it cannot be described by the usual percolation scaling relations.

\citet{Deng2012} consider random signed weighted networks. After choosing an unbalanced triad, the sign (not the weight) of a link with the smallest absolute weight is flipped. Therefore, the choice of the updated link is deterministic. This algorithm was further modified by \citet{Chowdhury2021}, who allow both the sign and the weight to change.
The new weight is set to the sum of the absolute weights of the other two links, with the old weight subtracted when updating the positive link.
After reducing the number of unbalanced triads, the authors use a separate heuristic to identify hypothetical coalitions among countries.

Taken together, these extensions show that LTD behavior depends not only on the bias parameter $p$, but also on what constitutes the elementary balance motif and on which links are eligible for change; modifying these ingredients can alter the absorbing states, critical behavior, and dependence on initial conditions.

\subsubsection{Models assuming ego perspective}
\label{subsubsec:ego-perspective-models}

In ego-based dynamics, it is the focal (ego) agent who experiences the tension associated with cognitive dissonance. This tension then motivates the ego to change one of its own links. Such an approach leads to a natural modification of local triad dynamics. First, in standard LTD, any link in a selected triad may be flipped, whereas in ego-based LTD, only links incident to the ego can change. Second, if agents are assumed to control their own outgoing relations, directed networks become a more appropriate framework. The following ego-based LTD, proposed in \citet{Gorski2025}, can be implemented on both directed and undirected signed networks:

\begin{enumerate}
  \item Sample an \emph{ego-based triad} with a designated ego. If this triad is balanced, repeat this step.
  \item Otherwise, flip one of the ego's two links in the triad. If both links have the same sign, choose either of them with equal probability.
  \item If the two ego links have opposite signs, then with probability $p$ the negative link is flipped to positive, and with probability $(1-p)$ the positive link is flipped to negative.
\end{enumerate}

The way ego triads are sampled can vary. \citet{Gorski2025} use a three-step procedure: first, the ego is selected; second, one neighbor of the ego is chosen from the set of agents to which the ego has an outgoing link; third, a common neighbor of this pair is chosen. The resulting triad forms a transitive semicycle. In this algorithm, only such transitive semicycle triads are ever sampled. \citet{Gorski2025} argue that, from the ego's perspective, these triads are more meaningful than purely cyclic triads: the ego needs to acquire only one piece of external information (the link between the two neighbors), while information about the other two links is assumed to be intrinsic to the ego.
% The interpretation of this sampling procedure is as follows: agent A meets B, and B talks about C, thanks to which a triad is formed.

Alternative sampling schemes are possible. \citet{Deng2021}, for instance, first sample a triad and only then choose the ego among its nodes. One could also first select an ego and then choose a pair of its common neighbors. These approaches lead to the same result for complete networks. For more general network topologies, they may result in quantitative differences.

Ego-based LTD exhibits a phase transition similar to standard LTD, but shifted to larger values of the parameter $p$. Following the analysis of \citet{antal_2005}, \citet{Gorski2025} derive analytical expressions for the stationary density of positive links. As in standard LTD, the paradise state $\rho = 1$ is always a stationary solution. In addition, for a range of $p$, there exists a nontrivial quasi-stationary state with
\begin{equation}
  \rho_{\text{ego}} = \frac{1}{1 + 2\sqrt{0.75 - p}} \,,
  \label{eq:rho-ego}
\end{equation}
which disappears only when the preference to flip a negative link exceeds $p = 0.75$.
Although the expressions for standard LTD [Eq.~\eqref{eq:ltd-rho-quasi}] and ego-based LTD, Eq.~\eqref{eq:rho-ego}, look different, they are closely related: the latter can be viewed as a stretched version of the former. In particular, both models yield the same value of $\rho$ at $p = 0$. In ego-based LTD, the special case $p = 1/2$ corresponds to an unbiased dynamics in which the ego's two links are chosen with equal probability, and one finds $\rho_{\text{triad}}(p = 1/3) = \rho_{\text{ego}}(p = 1/2)$.

\citet{Deng2012} study the special-case ego LTD on undirected networks with an additional parameter, the imbalance tolerance threshold $B_{\text{th}}$. At each step, an agent $A$ is chosen. Agent $A$ then changes the sign of one of its links in an unbalanced triad containing $A$ if and only if the fraction of balanced triads that $A$ belongs to does not exceed $B_{\text{th}}$. For $B_{\text{th}} = 1$, one recovers the standard ego-based LTD. The authors analyze the dependence on initial conditions and show that, for $B_{\text{th}} \gtrsim 0.7$, the final outcome is essentially independent of the initial positive-link density. For smaller thresholds, both low and high initial densities of positive links can be sustained in the stationary state. This leads to a counter-intuitive effect: increasing imbalance intolerance (raising $B_{\text{th}}$) may drive the system toward a lower stationary density of positive links and, consequently, a lower fraction of balanced triads.

\citet{Oishi2021} provide a different perspective with a dynamics that can be interpreted as an extension of ego LTD. In each update step, a directed pair $B \to C$ is first sampled. Then every agent $A$ that has links to both $B$ and $C$ forms a triad $(ABC)$. The interpretation is that $B$ is a donor, $C$ is a recipient, and the nodes $A$ are observers. Depending on the sign of $x_{BC}$, the donor $B$ either cooperates with or defects against $C$. Observers evaluate $B$'s action based on their existing opinions of $B$ and $C$, and update their sign toward $B$ as
\begin{equation}
  x_{AB}(t+1) = x_{AC}(t)\, x_{BC}(t) \,.
  \label{eq:oishi-update}
\end{equation}
This rule guarantees that, at time $t+1$, the triad $(ABC)$ is balanced. The proposed dynamics can thus be viewed as a modified ego-based LTD in which, during a single update, many observers may simultaneously change their links in currently unbalanced triads.

The authors investigate this dynamics on directed Erd\H{o}s--R\'enyi graphs with self-loops, with variable link density $d$. When the required connections exist, agents $B$ and $C$ can also act as observers. As the dynamics progresses, self-loops become positive and directed links tend to become reciprocated (either both positive or both negative). For very small or very large values of $d$, the time to reach an absorbing state is short. For intermediate densities, however, the relaxation time grows rapidly with system size, making it practically impossible for large systems to reach balance within realistic time scales.

A transition is observed between network structures that reach absorbing states quickly and those that remain in a long-lived stationary regime. The control parameter is $N d^{2}(1-d)$, and the transition occurs when this parameter exceeds 1. The transition point is derived analytically within a mean-field approximation that explicitly accounts for larger motifs (tetrads) containing the updated triads. This enables the authors to evaluate the impact of local update rules on the broader network structure. An interesting outcome is that ``non-observing'' neighbors—agents that do not directly participate in the update but share triads with the updated edges—tend to drive the system away from balance.

In contrast to \citet{Gorski2025}, where the dynamics is restricted to transitive semicycle triads, \citet{mei2022structural} propose a version of ego-based LTD on directed, incomplete networks that explicitly incorporates dyadic balance. In directed networks, balance can also be evaluated at the dyadic level: reciprocated links are balanced when their signs agree, and unbalanced when they disagree.
In their dynamics, an ego agent $A$ and one of its neighbors $B$ are first selected, and the directed link $x_{AB}$ is updated so as to adjust balance. With probability $p_1$, dyadic balance is enforced by setting $x_{AB} = x_{BA}$. Otherwise, the update proceeds via one of the transitive semicycle triads involving agents $A$ and $B$. With probability $p_2$, the update follows $x_{AB} = x_{AC}\,x_{CB}$, while with probability $1-p_1-p_2$ it follows $x_{AB} = x_{AC}\,x_{BC}$.
% Numerical results show that the dynamics converges to configurations in which reciprocated links have the same sign and all existing triads are balanced.
The authors prove that, when all existing links are initially reciprocal, the dynamics converges almost surely to configurations in which reciprocal links have the same sign and all existing triads are balanced.
However, on incomplete graphs, this local balance does not necessarily
imply global structural balance.
% drives all agents toward local balance, both at the dyadic level and within triads. However, global structural balance is generally not achieved in an incomplete graph. Instead, the system converges to weakly balanced states.
Therefore, ego-based update rules may promote local consistency, but they do not guarantee global balance. This topic is also analyzed in the ego-centered extensions of constrained triad dynamics proposed by \citet{Deng2021}.

\citet{Deng2021} propose two ego-based extensions of constrained triad dynamics. In the first, an ego-centered CTD is introduced: at each step, an unbalanced triad is selected, an ego within this triad is chosen, and one of the ego’s incident links is considered for update. The link is flipped only if this reduces the number of unbalanced triads involving the ego. On complete graphs, the system always reaches balance, whereas on incomplete networks, the dynamics typically becomes trapped in jammed states.

In the second modification, agents are assumed to have a limited cognitive span, an approach closely related to the model of \citet{Qi2019Minimum}. Each agent initially selects up to $N_{\mathrm{CS}}$ neighbors that define its cognitive span and this selection remains fixed throughout the dynamics.
% , and the ego-based CTD is applied only to triads formed within this set. A link is flipped if and only if this reduces the number of unbalanced triads inside the cognitive span.
The ego's decision to flip one of its links in a selected unbalanced triad depends on whether the change strictly increases the balance ratio evaluated with respect to these important neighbors. The considered triads may also involve other neighbors outside the selected set.
For complete graphs, this rule led to global balance for all tested values of $N_{\mathrm{CS}}$, even for $N_{\mathrm{CS}}=1$. For incomplete networks, global balance was not always reached. However, if the rule is relaxed to allow updates that leave the number of unbalanced triads within the cognitive span unchanged, the dynamics may fail to reach global balance for some even values of \(N_{\mathrm{CS}}\), depending also on network size and initial conditions.

% \subsubsection{Combination of LTD with other models}
% \label{subsubsec:ltd-combinations}

\subsubsection{Alternative stochastic models}
\label{subsubsec:other-stochastic-models}

The models discussed so far in this section have largely focused on local triad dynamics, its extensions, and ego-centered formulations. We now turn to stochastic sign dynamics models that do not fit naturally into these categories.
% However, local update rules leading to structural balance are not restricted to LTD-type mechanisms.
% A broad class of alternative stochastic models employs different notions of locality and balance, while still giving rise to balanced network structures.
These models employ different information scopes, update criteria, and notions of imbalance.
They include stochastic modifications of GTD and CTD, models of link formation and decay, and dynamics based explicitly on alternative imbalance measures, such as the frustration index.

\citet{Deng2012} introduce a stochastic model for weighted signed networks that applies, in discrete time, an update rule inspired by GTD. A link $(i,j)$ participating in at least one triad is first selected. Then up to $k$ neighbors of $i$ and $j$ (with $k$ as a model parameter) that form triads with $(i,j)$ are chosen at random, and the weight $x_{ij}$ is updated according to
\begin{equation}
  x_{ij}(t+1) = x_{ij}(t) + \frac{1-|x_{ij}(t)|}{k}
  \sum_{\ell \in \mathcal{N}_{ij}} x_{i\ell}(t)\,x_{\ell j}(t),
\end{equation}
where $\mathcal{N}_{ij}$ denotes the set of selected neighbors. The authors study this dynamics on random graphs and analyze how the densities of different triad types depend on the network density and on the initial density of positive links. They identify parameter ranges in which the system does not evolve toward a balanced configuration.

% \citet{Lee2017} extend CTD by introducing a temperature-like noise-control $T$. At each step, a sign change of a link $(i,j)$ is proposed and is accepted if, for each endpoint, the number of unbalanced triads $c_i(t)$ satisfies
% \begin{equation}
%   c_i(t) \leq c_i(t-1) + T\varepsilon,
% \end{equation}
% where $\varepsilon$ is a random variable drawn from a logistic distribution. Thus, $T$ controls the magnitude of stochastic fluctuations. In the limit $T\to 0^+$, one recovers CTD, where only changes that do not increase the number of unbalanced triads are allowed. For $T>0$, some changes that increase this number can also be accepted.
% % This differs from Hamiltonian (Metropolis-type) approaches, where acceptance is probabilistic: here, acceptance is deterministic once the inequality is checked.
% Depending on exceeding a certain threshold related to the value of temperature. Authors estimate the temperature for some real world networks.

\citet{Chatterjee2022} propose \emph{Best Edge Dynamics}, a modification of CTD in which, once an unbalanced triad has been drawn, the link chosen for the update is the one that participates in the largest number of unbalanced triads. At first sight, this rule appears to drive the system along the descent in the ``energy landscape''. However, this intuition is misleading: updates are still performed even when they increase the total number of unbalanced triads. The authors show that, despite these possible transient ``uphill'' moves, on the complete graph, Best Edge Dynamics eliminates jammed states and guarantees that any initial configuration eventually evolves to a balanced one.

A different class of models allows the network topology itself to evolve.
\citet{Abell2009} propose an evolving-link dynamics in which, at each step, a new link is added to the network. A parameter $\alpha$ controls the sign of the newly added link. The second parameter, $\beta$, controls imbalance intolerance, with larger values corresponding to lower tolerance. Whenever the fraction of unbalanced triads around an agent exceeds the permitted level, links incident on that agent are randomly deleted until the local imbalance falls below this level. The authors identify three regimes: sparse, semisparse, and dense networks. They derive an analytical condition separating the dense and semisparse regimes, namely
$\alpha^3 = \beta$: for $\alpha^3 > \beta$ the network becomes dense, whereas for $\alpha^3 < \beta$ cascades of deletions occur frequently. The semisparse regime exhibits signatures reminiscent of self-organized criticality: the rates of link creation and deletion are approximately proportional, triad-breakup sizes follow an approximate power-law distribution, while link-breakup sizes follow an approximately exponential distribution, degrees are positively correlated, and the degree distribution displays a long tail.

\citet{Vukainovi2014} propose a stochastic dynamical model of weighted signed networks in which links can both form and decay over time. At each step, agents create positive or negative interactions based on a combination of global popularity (measured by degree) and local triadic structure.
Some interaction signs are assigned stochastically, with the bias toward positive versus negative set by model parameters, while others are chosen to satisfy structural balance principles.
Positive interactions increase link weights, while negative interactions decrease them. Inactive links gradually evaporate, and new links may be created to close triads. The model is inspired by ant colony behavior and is evaluated against empirical social network data. The authors report that the resulting networks reproduce several empirical features commonly observed in real-world systems, including sparsity, small diameter, logarithmic scaling of average path length with system size, and relatively high clustering coefficients.

The only dynamical model in which the balance measure is based explicitly on the frustration index and an optimal partition is proposed by  \citet{Hummon2003}. The dynamics proceeds as follows:
\begin{enumerate}
  \item The system first computes an optimal partition of the network, minimizing the number of frustrated edges (see Sec.~\ref{subsubsec:frustration}); if multiple optimal partitions exist, one is chosen at random. Agents are informed about this current optimal partition.
  \item A random agent $A$ receives information from others and updates $A$’s internal image of the network.
  \item Agent $A$ then acts on its own connections. It may change a sign, create a signed link, remove a link, or do nothing. These options are ranked according to how much they reduce imbalance in $A$’s perceived network, and the move that yields the largest reduction is selected. The current optimal partition is used as a tie-breaker.
  \item Agent $A$ communicates the chosen change. Four communication schemes are considered:
    \begin{enumerate}
      \item Simple dyadic: only the recipient agent is informed;
      \item Friends: $A$ informs all its friends;
      \item Acquaintances: $A$ informs both friends and enemies;
      \item Broadcast: $A$ informs all agents in the network.
    \end{enumerate}
  \item A new optimal partition is computed on the basis of the actual network (rather than the agents’ images).
\end{enumerate}

A notable feature of this model is that the network may remain structurally unbalanced even when all agents believe it to be balanced. The longest times to reach an absorbing state occur when information is restricted to friends or acquaintances (with the friends-only case giving the longest times for initially random signs). The authors study how both group-level and individual imbalance depend on network size and on the initial density of \textit{negative} links.
For three-node networks, the average group imbalance increases with the initial density of negative links; however, for larger networks, it tends to decrease. The lowest average group imbalance is achieved under the acquaintance communication scheme, followed by the friends-only scheme. For simple dyadic and broadcast communication, almost all agents become balanced at equilibrium. The average number of clusters increases with the initial density of negative links. The differences between the simple, friends-only, and acquaintances cases can be understood by considering that agents may perceive balance even when the actual network remains unbalanced.
However, the behavior under broadcast communication is harder to interpret and may be influenced by details of the simulation protocol and stopping criterion. Another counterintuitive observation is that a higher initial density of negative links can, in some cases, lead to more balanced final states.

% For the dynamics discussed in this and previous sections, the long-time behavior typically involves convergence to either a balanced configuration, a jammed state, or an unbalanced quasi-stationary regime, with timescales varying widely across models.
% In the final model considered here,  \citet{Krawczyk2022Structural} propose a dynamics with the remarkable property that, when applied to any complete network of odd size $N$, balance is achieved in a single update step.
% The update rule is
% \begin{equation}
%   x_{ij}(t+1) = F\!\left( \sum_k x_{ik}(t)\,x_{jk}(t) \right),
% \end{equation}
% where $F(x) = (-1)^{\frac{N-2-x}{2}}$.
% Although an analytical proof is not provided, the authors verify this property for all configurations of small networks and for a very large number of networks of small size (up to $N=15$).

% There are many extensions to models proposed by Antal et al. One was suggested by  \citet{Parravano2016}, where agents possess 1-2 continuous opinions and a co-evolutionary (opinion-relation) dynamics is proposed. Opinion dynamics follows Hegselmann-Krause model and link dynamics is LTD. Agents are more likely to align their opinion to their friends'. Alignment of opinions is observed.

\subsubsection{Strategic and game-theoretic models}
\label{subsubsec:strategic-models}

The models above impose an update rule and ask what structure it produces.
A complementary approach treats structural balance as an outcome of strategic behavior. Although such models are not stochastic sign dynamics in the narrow sense, we discuss them here because their main object is the organization of the signed network or its partition, rather than the interaction of signed relations with additional agent states.
% A complementary approach specifies the agents' incentives and derives balanced structures as the equilibria of a game.

\citet{Hiller2017} studies a \emph{network-formation game} in which each agent chooses whom to befriend in order to extract resources from weaker enemies; every Nash equilibrium is either the all-positive network or a partition into internally friendly, mutually hostile groups, so weakly balanced partitions re-emerge as a strategic outcome rather than as the fixed point of a dynamical system.
% Related formulations obtain (clustering-)balanced networks as pairwise-stable or potential-game equilibria under triad-based utilities~\citep{Malekzadeh2011,CisnerosVelarde2020}.
Related game-theoretic models use triad-based utilities to study the formation and stability of signed networks~\citep{Malekzadeh2011,CisnerosVelarde2020}. In particular, \citet{CisnerosVelarde2020} show that strategic sign adjustments can lead to weakly balanced configurations.
\citet{Tan2016} let each node adaptively switch alliance to increase its own fitness, showing that the resulting evolutionary-game dynamics converges quickly to a minimum-frustration partition of arbitrary signed networks.

Thus, these models provide a complementary route to balanced or
low-frustration structures, in which network organization results from
strategic incentives rather than from a prescribed tension-reduction
rule. Game-theoretic models that additionally couple signed relations
to agents' behavioral strategies are discussed in
Sec.~\ref{subsubsec:context-dependent-updates}.

% In this game-theoretic setting, the consideration of whether triads are balanced or not changes the stability of cooperation in social dilemmas.
% In \citep{Righi2013,Righi2014}, the authors show that cooperation in a signed-network prisoner's dilemma never survives on dyadic information alone, but does once agents also weigh the balance of the enclosing triads.
% Comparable effects have been reported for public-goods games~\citep{Zhong2021} and for prisoner's-dilemma dynamics built explicitly on structural-balance rules~\citep{Song2022}.

\subsection{Jammed states}
\label{subsec:jammed}

%\GV{No paragraph or no subsection}
% \subsubsection{Definition} % or \paragraph{Definition}

The emergence of structural balance is much more difficult when the dynamics admits jammed states.
In models with a Hamiltonian formulation, jammed states are structurally unbalanced configurations corresponding to local minima of the energy defined through the Hamiltonian~\eqref{eq:hamiltonian_triad}.
For binary links with homogeneous interactions, this is equivalent to a local minimum of the number of unbalanced triads.
Depending on the specific dynamics, the system may evolve into such a state and remain trapped there.
Jammed states have been analyzed in many works; in this section, we briefly summarize the main results.

\citet{Marvel2009} consider the structural balance Hamiltonian energy landscape \eqref{eq:hamiltonian_triad} and connect network configurations that differ by the sign of a single link. They define a \emph{strict jammed state} as a configuration that is not structurally balanced and for which all neighboring configurations (differing by one link sign) have strictly higher energy. In contrast, in a \textit{weak jammed state} no neighboring configuration has lower energy, but neighboring configurations with the same energy may exist.

The energy-landscape definition does not apply directly to all models considered in this review. In particular, asymmetric directed dynamics does not have a scalar Hamiltonian whose minimization governs the evolution. In such models, the term \emph{jammed state} is also used more broadly for an unbalanced stationary or dynamically stable configuration in which the system can become trapped~\citep{Krawczyk2015,Hassanibesheli2017,Krawczyk2019Heider}. We follow this broader usage when discussing models without an energy formulation, while the distinction between strict and weak jammed states refers specifically to the energy-landscape definition above.

\subsubsection{Structure of jammed states}

% For such a setting,
Under the weaker definition, which allows neighboring states to have equal energy,  jammed states first appear for $N = 6$ and then for all $N \geq 8$ \citep{Marvel2009}, while strict jammed states first occur for $N = 9$ and then for all $N \geq 11$ \citep{antal_2005}. Jammed states cannot have energy above $0$, i.e., the number of unbalanced triads cannot exceed the number of balanced ones. Zero-energy jammed states were found only very rarely in numerical searches~\citep{Marvel2009}, possibly reflecting their lower abundance or smaller basins of attraction.
% and in fact, there are fewer jammed states at zero energy than at lower energies.
To demonstrate this,  \citet{Marvel2009} define cliques as disjoint groups obtained by removing all links that belong to unbalanced triads, and show that achieving a jammed state with higher energy requires a larger number of such cliques. This number increases as the energy approaches 0, reaching $N$ at zero energy.
In~\citep{Marvel2009}, the majority of identified jammed states contain three cliques, followed by five, six, and so on.
% Family of jammed states can be constructed using the concept of Paley graphs.

The energy can exceed $0$—that is, there can be more unbalanced than balanced triads in a jammed state—if \emph{stubborn links} (zealot links that cannot change sign) are introduced~\citep{Sheykhali2020}. A relatively high fraction of such links is required to push the energy above $0$, but if they are placed in a targeted rather than random way, this fraction can be significantly reduced.

For evolving signed networks, jammed states are expected whenever the dynamics effectively follows the energy landscape. They can therefore appear both in models such as CTD and in the standard gradient triad dynamics~\citep{antal_2005, Hassanibesheli2017, kulakowski2019stabilityOfSBsolutions}.

\begin{figure}[!ht]
    \centering

    % =========================================================
    % Top row: (a) and (b)
    % =========================================================

    \begin{minipage}[t]{0.33\textwidth}
        \centering
        \begin{overpic}[width=\linewidth]
            {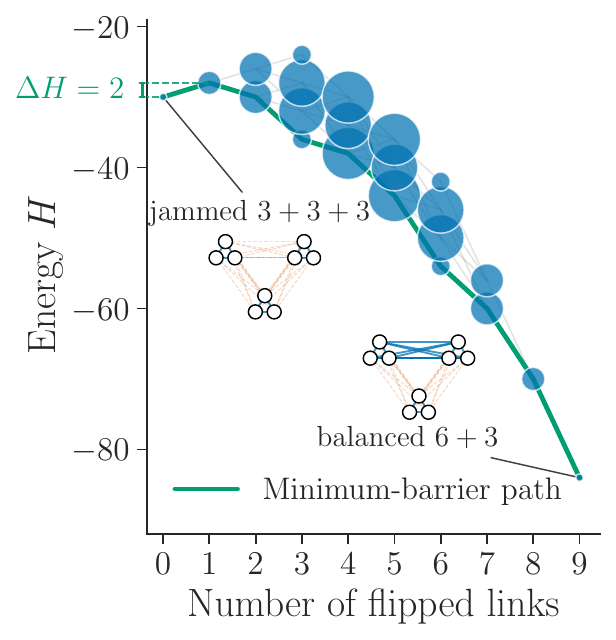}
            \put(-10,95){\textbf{(a)}}
        \end{overpic}
    \end{minipage}
    \hfill
    \begin{minipage}[t]{0.66\textwidth}
        \centering
        \begin{overpic}[width=\linewidth]
            {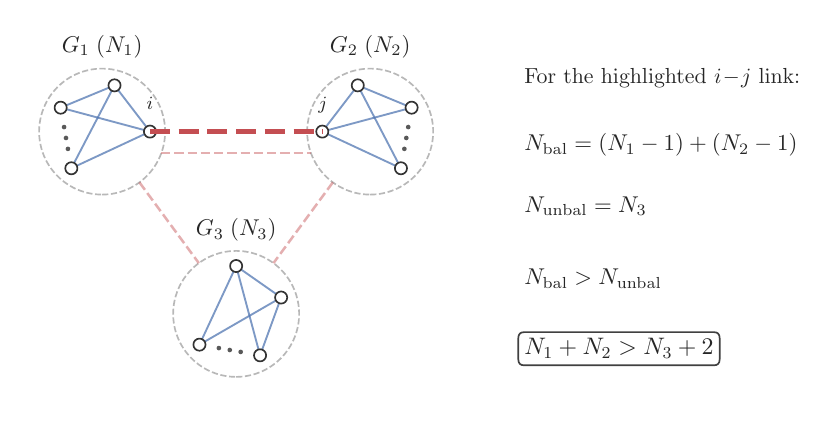}
            \put(-5,47){\textbf{(b)}}
        \end{overpic}
    \end{minipage}

    \vspace{0.6em}

    % =========================================================
    % Bottom row: (c), spanning most of the figure width
    % =========================================================

    \begin{minipage}{0.94\textwidth}
        \centering
        \begin{overpic}[width=\linewidth]
            {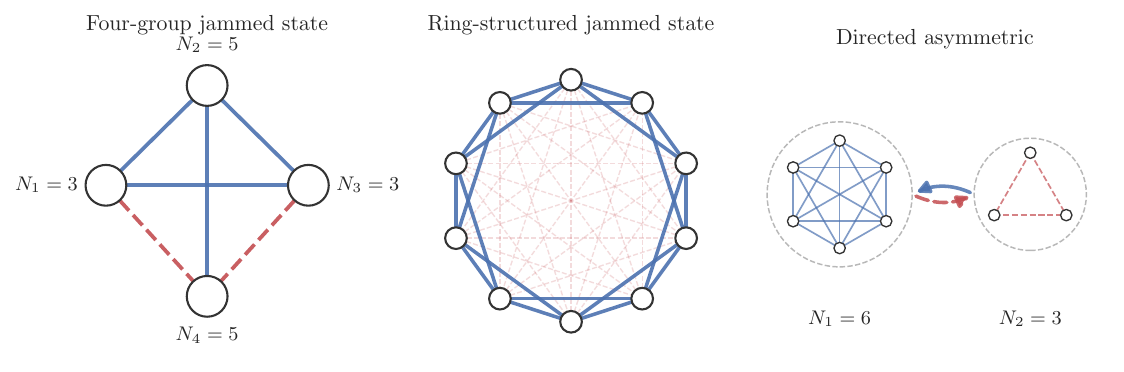}
            \put(-5.5,30){\textbf{(c)}}
        \end{overpic}
    \end{minipage}

    \caption{
    Jammed states and their structure.
    \textbf{(a)} Restricted energy landscape around the $N=9$ three-clique strict jammed state, obtained by varying the nine links between two of the cliques.
    % Starting from this configuration, only the nine links between two selected cliques are allowed to change.
    % Configurations with the same number of flipped links and the same energy $H$ are aggregated into a single point; marker size indicates their multiplicity, while faint connections indicate the existence of a single-link transition between configurations represented by the corresponding points.
    Bubble size indicates the number of configurations represented by each point. The green curve shows the energies of states along a minimum-barrier path from the jammed $3+3+3$ state to a balanced $6+3$ state.
    % The highlighted curve shows one minimum-barrier path from the jammed $3+3+3$ state, with $H=-30$, to the balanced $6+3$ state, with $H=-84$;
    Within this restricted energy landscape, the minimum energy increase required to escape the jammed state is $\Delta H=2$.
    \textbf{(b)} Generic three-clique jammed state and its stability condition~\citep{antal_2005}. The complete network is partitioned into three internally positive groups, with all links between different groups negative; the highlighted $i$--$j$ link is one of the negative links between $G_1$ and $G_2$.
    \textbf{(c)} Examples of other jammed-state architectures: a four-group state~\citep{kulakowski2019stabilityOfSBsolutions}, a ring-structured state~\citep{Chatterjee2022}, and a directed asymmetric state~\citep{Krawczyk2019Heider} (left to right). In the left and center diagrams, each circle represents a group of nodes; individual nodes are shown explicitly in the directed example with arrows indicating relations between groups.
    % Left: a stable four-group configuration of the \kk{} dynamics, with internally positive groups and the illustrated pattern of positive and negative intergroup relations; the shown group sizes satisfy the stability conditions derived in Ref.~\citep{kulakowski2019stabilityOfSBsolutions}.
    % Center: a ring-structured jammed state of Ref.~\citep{Chatterjee2022}, shown for $d=2$ and $4d+2=10$ clusters; positive relations connect clusters at ring distance at most $d$, while all remaining intercluster relations are negative.
    % Right: a directed asymmetric stationary state~\citep{Krawczyk2019Heider}.
    Blue solid and red dashed lines denote positive and negative relations, respectively.
    }
    \label{fig:jammed_states}
\end{figure}

\citet{antal_2005} analyze CTD, where updates are chosen so that states can evolve toward configurations of lower or equal energy.
As a result, the system is able to escape weak jammed states (in the sense defined above), whereas strict jammed states are absorbing.
Even for $N = 9$, there are several possible connectivity patterns of strict jammed states~\citep{antal_2005}.
The simplest consists of three cliques of size three, each internally friendly, with all links between different cliques negative.
Fig.~\ref{fig:jammed_states}a illustrates the local energy landscape around this state. In the restricted landscape shown there, only the nine links between two of the three cliques are varied. Reaching a balanced state requires crossing an energy barrier.

More generally, strict jammed states with three cliques can be constructed when their sizes satisfy the modified ``triangle inequality'' $N_a + N_b > N_c + 2$, where $N_a$, $N_b$, and $N_c$ are the sizes of the cliques.
For example, for a negative link $(i,j)$ connecting groups $1$ and $2$, there are $(N_1-1)+(N_2-1)$ balanced incident triads and $N_3$ unbalanced ones. Thus, flipping the link increases the energy when $N_1+N_2>N_3+2$, as illustrated in Fig.~\ref{fig:jammed_states}b.

Similar three-clique structures also form jammed states in standard GTD.  \citet{kulakowski2019stabilityOfSBsolutions} additionally construct stable jammed configurations with four groups and derive inequalities on group sizes that must be satisfied for these configurations to remain jammed (Fig.~\ref{fig:jammed_states}c).
\citet{Chatterjee2022} construct another family of jammed states in which approximately equal-sized groups are arranged on a ring, with positive relations between nearby groups and negative relations between more distant ones (Fig.~\ref{fig:jammed_states}c).
When asymmetric links are allowed in directed networks, jammed states appear generically in directed extensions of GTD for $N \geq 5$~\citep{Krawczyk2015}. \citet{Krawczyk2019Heider} numerically identify frequent classes of such states consisting of two, three, or four groups of nodes that behave identically toward members of their own and other groups. An example is shown in Fig.~\ref{fig:jammed_states}(c). Such jammed states can be viewed as relaxed balanced states in the sense of \citet{Doreian2009partitioning}.

For both main types of dynamics (constrained-triad and the self-loop extension of GTD) on complete graphs, the probability of ending in a jammed state decreases with system size and vanishes as $N \to \infty$~\citep{antal_2005, marvel2011_evolutionOFSBcontinuous}.
As observed by  \citet{Chatterjee2022}, for CTD this statement actually holds provided the initial density of positive links is not too small. They show that, starting from a network with a very low number of positive edges, a three-clique jammed state can still be reached with nonzero probability. Moreover, such a state is quite robust: after perturbing a substantial number of links around each agent, the dynamics typically returns to the same jammed configuration.

\subsubsection{Number of jammed states}

\citet{antal_2005} show that the number of jammed states grows much faster than the number of balanced states.
For a complete graph with $N$ labeled nodes, there are $2^{N-1}$ distinct balanced sign configurations~\citep{Gorski2017}\footnote{This result follows from a simple inductive argument. In any balanced system of $N$ agents, the agents can be partitioned into two groups, one of which may be empty in the paradise state. When a new agent is added, there are two ways to preserve balance: the new agent must form positive links with one group and negative with the other, or vice versa. Since for $N=3$ there are four balanced sign configurations, the number of balanced configurations for $N$ agents is $2^{N-1}$.}, while \citet{antal_2005} obtained an exponential lower bound of order $3^N$ for the number of jammed states.
%Yet, the probability to reach a jammed state gets smaller with system size.
A stronger lower bound, $2^{\Omega(N \log N)}$, was later established by \citet{Chatterjee2022}, implying that the number of jammed states grows super-exponentially with system size.

% \subsection{Emergence when coupling sign dynamics to node attributes -- 8 pages}
\subsection{Coupled sign--attribute dynamics}
\label{subsec:sign-attribute-dynamics}

The models described in the previous sections \ref{subsec:kk-model}-\ref{subsec:stoch-sign-dyn} focus on systems in which links carry attributes, specifically signs. In this section, we turn to models in which agents themselves possess attributes, such as opinions, traits or states, that interact with signed relationships. Such models may arise as straightforward extensions of sign-dynamics models or as synergistic combinations of structural balance mechanisms with models of attribute dynamics.

From a modeling perspective, introducing agent attributes connects structural balance theory with other well-established approaches.
Most often, it relates to homophily~\citep{mcpherson2001birds}, according to which similar agents tend to form positive relationships and sometimes dissimilar agents may form negative ones.
Homophily thus provides a natural mechanism linking attributes and signed links. Other relevant theories and modeling approaches include status theory and epidemic-spreading models.
% , where attributes may represent, for example, social status or infection state rather than fixed opinions.

A key distinction for models of emergent structural balance among agents with attributes concerns the definition of balance itself.
As noted in the introduction, in the $P$--$O$--$X$ notation used here, $X$ may represent another person or an impersonal entity (such as an object, hobby, opinion).
All models considered up to this point adopt the former, purely interpersonal interpretation, in which balance is usually defined on triads of agents. Once attributes are introduced, however, balance relations may also involve agents and attributes. In that case, attributes act as sources of agreement or disagreement that shape the signs of pairwise relationships between agents, and models may incorporate both interpersonal balance (agent--agent--agent triads) and balance relations involving attributes (agent--agent--attribute triads).

It is important to note that there is no unique way to categorize models in this area.
Many papers combine several mechanisms, and alternative organizational schemes are possible. We use two complementary levels of
classification. First, Sec.~\ref{subsubsec:agent-attr-types} classifies the attributes themselves according to their form and their homophilic effect on signed relations. The remaining subsections organize models according to the role played by attributes in the dynamics.
% Many papers combine several mechanisms, and alternative organizational schemes are possible. The categorization adopted here emphasizes the role played by attributes in shaping signs and dynamics.
Attributes may determine link signs directly (Sec.~\ref{subsubsec:attr-driven-sign-formation}), enter the definition
of balance through agent--agent--attribute relations (Sec.~\ref{subsubsec:sb-agent-attribute}), act jointly with interpersonal
structural balance (Sec.~\ref{subsubsec:combined-interpersonal-agent-attribute}), or modify how balance is evaluated and how relations are updated (Sec.~\ref{subsubsec:context-dependent-updates}).
Closely related models may therefore appear in more than one context.

The implications of some of the models described here are also discussed in Sec. \ref{subsec:social} and \ref{subsec:netpol}.

% In this section, we first categorize the possible types of agents' attributes in terms of their impact on signed relations. Next, in Sec. \ref{subsubsec:attr-driven-sign-formation} we examine models in which
% signs are determined directly from attributes and the dynamics unfolds through changes of the attributes themselves.
% % changes of attributes and signs are tightly coupled, so that one directly follows the other.
% We then turn to models that explicitly incorporate balance relations involving agents and attributes in Sec.~\ref{subsubsec:sb-agent-attribute}. Subsequently, Sec.~\ref{subsubsec:combined-interpersonal-agent-attribute} discusses models that combine interpersonal structural balance with agent--attribute relations. Finally, in Sec. \ref{subsubsec:context-dependent-updates}, we consider models in which attributes do not directly influence link signs, but instead modulate the dynamics of sign changes.

% \begin{itemize}
%   \item Types of agent attributes
%   \item Attribute-driven sign formation and attribute dynamics
%   \item Structural balance models with agent--attribute interactions
%   \item Structural balance models combining interpersonal interactions and agent--attribute relations
%   \item Attributes modulating the dynamics of sign updates
% \end{itemize}

\subsubsection{Types of agent attributes}
\label{subsubsec:agent-attr-types}

Depending on the modeling assumptions, agent attributes may represent opinions, traits, or other characteristics. Such attributes may be effectively constant over time, may evolve more slowly than link signs, or may change on comparable timescales. These distinctions enable us to distinguish between models in which fixed attributes shape the evolution of signs and models in which signs and attributes coevolve.

Building on the classification of similarity measures proposed by  \citet{Gower1971},  \citet{Gorski2023} introduces a taxonomy of agent attributes according to their homophilic impact on link signs. The classification is based on three criteria:
\begin{itemize}
    \item Ordered versus unordered attributes: attributes whose values can be meaningfully placed on a scale are called \emph{ordered}, whereas those lacking a natural ordering are \emph{unordered}.
\item Strength of dissimilarity effects: attributes may induce the same magnitude of effect for similarity and dissimilarity (denoted as \emph{negative}), or they may have no effect when agents are considered dissimilar (denoted as \emph{positive}).
    \item Number of attribute categories.
\end{itemize}

The first distinction parallels the difference between $q$-Ising and Potts-type models. The second relates directly to the strength and nature of homophily. Among unordered attributes,  \citet{Gorski2023} distinguishes between positive and negative variants, whereas for ordered attributes, only the negative case is considered. As the number of categories of an ordered attribute increases without bound, the attribute approaches a continuous one. Conversely, when the number of categories is exactly two, the ordered and unordered classifications coincide, leading to what are often referred to as \emph{binary attributes}.

Another modeling assumption is the definition of a distance in the attribute space, i.e.,  when two attributes are close or distant enough to justify the introduction of a positive or negative relation.

Most models in the literature focus on a single attribute type, typically binary or continuous attributes.
We now turn to models in which attributes determine signed relations and are themselves updated in response to structural imbalance.

\subsubsection{Attribute-driven sign formation and attribute dynamics}
\label{subsubsec:attr-driven-sign-formation}

In the models in this section, we consider a network of $N$ agents, each possessing $M$ attributes. We use $M$, rather than $G$ as in several original studies, to avoid confusion with the graph notation. Agents can thus be embedded in an $M$-dimensional attribute space.
Following  \citet{linczuk_multidimensional_2023}, the sign of a link $x_{ij}$ between agents $i$ and $j$ is defined by
\begin{equation}
    x_{ij} = f(d_{ij}),
    \label{eq:attr-sign-general}
\end{equation}
where $d_{ij}$ is a distance between their attribute vectors, and $f$ is a threshold function that assigns positive or negative relations based on similarity. This framework allows signs to be fully determined by attributes, while attributes evolve under pressure to reduce structural imbalance. In this sense, attributes are the primary dynamical variables, and the signed network is derived from them.

\paragraph{Binary attributes}

When each agent is characterized by a binary attribute vector $\mathbf{b}_i \in \{-1,+1\}^{M}$, a common rule assigns
\begin{equation}
    x_{ij} =
    \begin{cases}
        +1, & \text{if } \mathbf{b}_i \cdot \mathbf{b}_j > 0,\\[4pt]
        -1, & \text{if } \mathbf{b}_i \cdot \mathbf{b}_j < 0,
    \end{cases}
    \label{eq:binary-attr-sign}
\end{equation}
so that agents with more matching than mismatching attributes form positive links. To avoid the degenerate case $\mathbf{b}_i \cdot \mathbf{b}_j = 0$, odd values of $M$ are typically used.

\citet{Gorski2020} extend local triad dynamics and create attribute-based LTD. At each step, a triad is sampled, and if it is unbalanced, one attribute of one agent is modified in a way that may restore balance. Analogously to LTD, a parameter $p$ determines whether a positive or negative relation is targeted during updates of $\Delta_1$ triads.

Using a Fokker-Planck equation, \citet{Gorski2020} derive analytical expressions matching simulations. When the number of attributes greatly exceeds the number of nodes, the model exhibits the same phase transition as in  \citet{antal_2005}\footnote{In~\citep{Gorski2020}, it is argued that the attribute-based LTD becomes qualitatively equivalent to LTD when the number of attributes is very large.}, Eq.~\eqref{eq:ltd-rho-quasi}: for $p < 0.5$, the system enters a quasistationary state, whereas for $p > 0.5$, the system can only reach a paradise state. If $M$ and $N$ scale as $M = \mathcal{O}(N^{\gamma})$, this phase transition occurs only for $\gamma > 2$. No transition and stable unbalanced states only are observed for small numbers of attributes. Since real social systems typically involve only a limited number of salient attributes, this result helps explain why fully friendly (paradise) states are not observed in practice~\citep{Gorski2020}.

Related but not identical results are reported in  \citet{MinhPham2020,pham2021balance}, where models combine interpersonal and impersonal structural balance.
These hybrid models are discussed separately in Sec. \ref{subsubsec:combined-interpersonal-agent-attribute}.

\paragraph{Manhattan distance}

A frequently used measure of dissimilarity is the Manhattan distance between attribute vectors~\citep{linczuk_multidimensional_2023}. In this case, the sign rule becomes
\begin{equation}
    x_{ij} = f(d_{ij}) =
    \begin{cases}
        +1, & d_{ij} \le \theta,\\[4pt]
        -1, & d_{ij} > \theta,
    \end{cases}
    \label{eq:manhattan-sign}
\end{equation}
where $\theta$ represents a tolerance threshold for attribute differences. If distances are normalized so that $d_{ij} \in [0,1]$, then $\theta \in [0,1]$ as well.

 \citet{Singh2016} analyze a model where each agent carries a single three-state attribute $a_i \in \{-1,0,+1\}$ representing leftist, centrist, or rightist orientations. In this setting, agents with opposite extreme attributes form negative links, while all other pairs form positive ones—corresponding to a threshold $\theta \in (0.5,1)$.
With probability $p$, a randomly chosen agent becomes a centrist; with probability $1-p$, a random triad is inspected and updated if unbalanced. In the latter case, centrists adopt extremist views with probability $\alpha$, while extremists become centrists with probability $1-\alpha$. The parameter $p$ acts as an external field promoting moderation~\citep{Singh2016}.

Analysis of the attribute-density phase space reveals a critical value $p_{c}$ for $\alpha > 0.5$. Below $p_c$, the system exhibits a polarized quasistationary state, a saddle point, and an absorbing state. Above $p_c$ (and for $\alpha < 0.5$), only the absorbing all-centrist consensus remains.

Such a three-state attribute is also considered by \citet{linczuk_multidimensional_2023}.
Agents may possess multiple attributes, generating a richer distribution of effective tolerance thresholds.
The dynamics is a modified attribute-based LTD~\citep{Gorski2020}, incorporating additionally probabilities to change an attribute or remove a link when encountering unbalanced triads, and a link creation probability.
After calibrating the model on empirical data, the authors investigate transition probabilities between balanced and unbalanced states as functions of the tolerance $\theta$.
Empirical transitions are shown to lie between those predicted by the agent-based model and those generated by random dynamical processes, suggesting that the modeled balance mechanism captures part, but not all, of the structure present in the empirical dynamics.
Moreover, when attributes are considered separately, structural balance in empirical data are indistinguishable from those in null models.
However, when signs are inferred from multidimensional attributes and agents are assumed to have low tolerance, the effects of structural balance become detectable.

Taken together, the models described in this subsection show that, when signed relations are derived from agent attributes, the emergence of structural balance depends strongly on the representation and dimensionality of the attribute space, as well as on the rules governing attribute change.

%     \subsubsection{Coevolution of signs and attributes as separate dynamical processes}
% \label{subsubsec:coevol-separate-processes}

\subsubsection{Models with agent--attribute interactions}
\label{subsubsec:sb-agent-attribute}

As discussed in the introduction of this section, balance reasoning can be extended beyond purely interpersonal structures to the impersonal $P$--$O$--$X$ case, in which $X$ represents an impersonal entity, such as an issue, idea, or object. For simplicity, we refer to such impersonal elements generically as \textit{attributes}.
% by allowing agents to evaluate balance by considering their relationships and attitudes toward impersonal entities (such as opinions, issues, or objects), here generically referred to as {\it attributes}.
In this interpretation, balance considerations apply to triads consisting of two agents and an attribute, in the sense of Heider's $P$--$O$--$X$ formulation.

From the network perspective, the explicit interactions in such models are pairwise: the sign of a link between two agents depends on their respective attitudes toward attributes. Balance constraints are therefore implemented through dyadic interactions involving agent--agent relations, even though the underlying consistency conditions are defined on agent--agent--attribute triads.

Two closely related mechanisms are typically modeled in this class of approaches.
First, friends tend to like or dislike the same attributes.
Second, agents with similar (dissimilar) attitudes toward attributes tend to form friendly (hostile) relationships. These mechanisms closely resemble those of homophily theory.
Indeed, some works describe their dynamics in terms of homophily, while others frame them explicitly within structural balance theory.
In what follows, we do not distinguish among these perspectives and refer to all models that implement the above mechanisms as \textit{impersonal structural balance models}, regardless of the terminology used in the original papers.

\paragraph{Impersonal structural balance Hamiltonian models}

A common starting point is a Hamiltonian formulation that assumes friends (enemies) should have similar (dissimilar) attributes. A minimal form is
\begin{equation}
H = - \sum_{(i,j)\in E} x_{ij}\, b_i b_j ,
\label{eq:hamil-impersonal}
\end{equation}
where the summation is over all existing edges, $x_{ij}\in\{-1,+1\}$ denotes the signed relation between agents $i$ and $j$, and $b_i\in\{-1,+1\}$ represents a binary attribute (spin) of agent $i$, i.e., attitude of agent $i$ toward an impersonal entity.

A Metropolis study of an Ising-like model on a fixed incomplete signed network with Hamiltonian~\eqref{eq:hamil-impersonal} is carried out by \citet{Li2019Binary}, where at each step a single attribute (spin) is updated while the topology remains unchanged. A different approach is proposed in~\citep{Singh2014, Kargaran2021Heider}.
In the dynamics of \citet{Singh2014}, each Metropolis update of attributes is followed by an adaptation of weighted signed links $x_{ij}\in[-1,1]$ according to
\begin{equation}
x_{ij}(t+1) = (1-\epsilon)\,x_{ij}(t) + \epsilon\, b_i(t)b_j(t),
\end{equation}
where $\epsilon$ is an adaptation rate. The network is assumed to be complete with all links potentially nonzero. \citet{Singh2014} show that, in this setting, minimizing the impersonal Hamiltonian \eqref{eq:hamil-impersonal} is equivalent to reaching triadic balance.
% Therefore in~\citep{Singh2014}, signed links are weighted $x_{ij}\in[-1,1]$.
Defining the relaxation time $\tau$ as the characteristic time to reach balance, in the absence of noise ($T=0$), one finds $\tau \sim 1/\epsilon$. This adaptation rule resembles Hebbian learning, and the authors draw analogies with brain networks. Without noise, balance is reached rapidly, whereas noise can significantly delay convergence, which is suggested as a possible explanation for the lack of perfect balance in empirical networks.

The authors of \citep{Singh2014} further analyze the dynamics by examining trajectories in a two-dimensional space spanned by the impersonal Hamiltonian~\eqref{eq:hamil-impersonal} and the standard triadic structural balance Hamiltonian~\eqref{eq:hamiltonian_triad}. After a transient, trajectories collapse onto a single curve independent of initial conditions and subsequently converge to states balanced with respect to both dyadic and triadic criteria.

In~\citep{Kargaran2021Heider}, both attributes and links evolve on a complete signed network. Using mean-field approximations and Metropolis dynamics, the authors compare critical behavior obtained with the impersonal Hamiltonian~\eqref{eq:hamil-impersonal} to that of the standard triadic Hamiltonian studied previously \eqref{eq:hamiltonian_triad}. They find that the impersonal Hamiltonian exhibits qualitatively different behavior from the triadic Hamiltonian~\citep{Malarz2022Mean}, displaying a continuous transition without hysteresis. In the impersonal setting, suitable order parameters include not only the energy but also average agent-link correlations~\citep{Kargaran2021Heider,Noudehi2025}.

% Following the approaches of~\citep{previous papers with delay}, this model was further extended to study this Hamiltonian with time delay. (Explore earlier paper for more details.) Similarly as for the system without delays, for small delays they observe a continuous phase transition. For larger delays no transition is observed (or it exists for lower and lower temperatures).

A more detailed Hamiltonian distinguishing the influence of friends and enemies is introduced by  \citet{pham2022_SBexplainedThroughHomophily}. Following~\citep{Gorski2020}, agents possess $M$ binary attributes, and relations are defined deterministically as
$x_{ij} = \mathrm{sgn}(\mathbf{b}_i\cdot\mathbf{b}_j)$, cf. \eqref{eq:binary-attr-sign}. The Hamiltonian is written as $H=\sum_i H_i$ with
\begin{equation}
H_i = -\frac{\alpha}{M}\sum_{j\in \mathcal{F}_i}\mathbf{b}_i\cdot\mathbf{b}_j
      +\frac{1-\alpha}{M}\sum_{j\in \mathcal{E}_i}\mathbf{b}_i\cdot\mathbf{b}_j ,
\end{equation}
where $\mathcal{F}_i$ and $\mathcal{E}_i$ denote friends and enemies of agent $i$, respectively, and $\alpha\in[0,1]$ controls their relative contributions to the Hamiltonian.
Attribute updates follow Metropolis dynamics.
Studying ring lattices and empirical networks, the authors observe the emergence of structurally balanced triadic states using only dyadic interactions.
For large values of $\alpha$ and inverse temperature $\beta$, a paradise state appears, whereas for large $\beta$ and small $\alpha$, a bipolar state is reached. The model reproduces cluster-size distributions and triad statistics comparable to those observed in empirical networks.
% Critical beta_c is proportional to G and K in a power-law function with exponents nonmonotonously dependent on alpha.

This framework is further extended by \citet{Thurner2025} to weighted signed networks, with weights defined as $x_{ij}=\frac{1}{M}\mathbf{b}_i\cdot\mathbf{b}_j$. The focus there is on polarization, quantified as the variance of signed weights, rather than on structural balance itself.

A related extension is the spin-glass model of \citet{Korbel2023}, in which agents carry binary, multi-dimensional opinion vectors.
The model captures a self-assembly process in which agents form groups characterized by intra-group homophily and inter-group heterophily.
Both group sizes and the existence of links between agents are determined by local social stress.
For more than one opinion dimension, the model Hamiltonian is frustrated, meaning that no grouping can simultaneously satisfy homophily along every dimension.
Using the Pardus multiplayer-game network for comparison, the model reproduces the empirical group-size distribution for small groups ($2$--$50$ players), as well as the aggregated probability of observing a group with more than $500$ players. However, it does not reproduce higher-order motif statistics associated with structural balance.

While the models above are restricted to binary attributes, \citet{Agbanusi2018} study the coevolution of continuous attributes and weighted signed relations that may initially be asymmetric.
In their formulation, each agent possesses a single continuous attribute $b_i$, and the Hamiltonian is given by a Dirichlet-type energy
\begin{equation}
H = \frac{1}{2}\sum_{i,j} x_{ij}(b_i-b_j)^2 .
\end{equation}
Additional constraints control the magnitudes of relations and attributes, and dynamics is defined through gradient descent of a constrained free energy.
Although relations may initially be asymmetric, the steady states of the model have symmetric relations.
The only stable steady states identified analytically or observed numerically are opinion consensus, which does not necessarily imply exclusively positive links (i.e., a paradise state), and a polarized state with groups of sizes $N-1$ and $1$. The absence of more complex balanced configurations may be due to the global nature of the imposed constraints or the small system size considered in numerical simulations~\citep{Agbanusi2018,kulakowski2025heider}.

Thus, even within Hamiltonian formulations, agent--attribute consistency can be implemented in substantially different ways, ranging from coupled evolution of attributes and relations and explicit energetic terms favoring consistency between them to constrained free-energy formulations.

\paragraph{Non-Hamiltonian approaches}

The models above encode agent--attribute consistency through an explicit Hamiltonian or free-energy formulation. A second class implements the same general consistency principles directly through dynamical update rules.
% We now turn to models whose dynamics is not derived from an explicit Hamiltonian.
We again begin with binary attributes and then consider continuous ones.
\citet{Chen2014} propose a coevolving model with a three-state attribute ($+1$, $-1$, and $0$, interpreted as an ignorant state) and signed relationships. Initially, all nodes except one are ignorant; over time, ignorant nodes adopt values $\pm1$. As a result, the attribute effectively behaves as binary, with the ignorant state corresponding to its absence. Opinion updates follow balance-based rules: agents tend to align with friends and oppose enemies.
Specifically, an agent adopts opinion $+1$ with probability $\theta_i^+/(\theta_i^+ + \theta_i^-)$, where $\theta_i^+$ ($\theta_i^-$) counts friends holding opinion $+1$ ($-1$) together with enemies holding opinion $-1$ ($+1$). Relations belonging to unbalanced dyads (friends with different opinions or enemies with the same opinion) change with probability $p_q$.
Numerical simulations on Watts--Strogatz networks show that the number of opinion groups decreases with increasing average degree and approaches two for sufficiently connected networks in the studied parameter range.

\citet{Saeedian2019} propose an LTD-like model for impersonal structural balance (see Sec. \ref{subsubsec:ltd}), in which agents possess a single binary attribute, leading to balanced and unbalanced pairs of agents. When an unbalanced pair is sampled, with probability $p$ the relationship is updated and with probability $1-p$ an attribute is changed.
Mean-field analysis and simulations on Erd\H{o}s-R\'enyi networks reveal a transition from a quasistationary regime to an absorbing frozen state. The absorbing phase occurs above a critical value of $p$ or below a critical mean degree.

Models with continuous attributes often combine structural balance ideas with bounded-confidence opinion dynamics~\citep{Aghbolagh2019,aghbolagh2020balance,schweighofer_2020,Li2022Structural}. In these approaches, similar agents reinforce each other's opinions, while sufficiently dissimilar ones repel. In  \citet{Aghbolagh2019}, signs are determined by bounded-confidence thresholds for acceptance (positive) and rejection (negative), and opinions evolve according to signed weighted averaging, in which friends contribute positively, and enemies contribute negatively to an agent's opinion update.
Depending on the values of the bounded-confidence thresholds, one, two, or multiple opinion groups emerge, and the final network is analytically shown to be structurally balanced, provided that no limiting pairwise opinion distance equals either threshold.

A related line of work builds on the opinion dynamics introduced by \citet{altafini2012dynamics}, which describes the evolution of continuous opinions under positive and negative social influence. While structural balance in this model refers to the signed social network, the resulting opinion alignment in structurally balanced networks can also be interpreted in terms of impersonal balance: friends tend to agree and enemies to disagree on a common issue (see Sec.~\ref{subsec:coev}).
Building on this framework,  \citet{Liu2021} study a model in which multidimensional continuous opinions and directed signed relations coevolve sequentially. Relations are updated according to homophily-like rules,
\begin{equation}
x_{ij}(t+1) = \frac{1}{\sum_k |{b}_{ik}(t)|}\sum_k b_{ik}(t)\,b_{jk}(t),
\end{equation}
while opinions evolve following Altafini dynamics,
\begin{equation}
b_{ik}(t+1) = \frac{1}{\sum_j |x_{ij}(t+1)|}\sum_j x_{ij}(t+1)\,b_{jk}(t).
\end{equation}
The system typically converges to one or two balanced opinion groups.

Finally,  \citet{schweighofer_2020_2} study coevolving continuous attributes and signed relations using a balance measure based on signed geometric means.
The model's primary focus is polarization viewed through the lens of structural balance. Each agent possesses $M$ continuous attributes.
% Two assumptions are defined: (1) having neutral opinion about an issue (i.e., when relationship of P towards X has zero weight; or in other words when an attribute is zero), one should have a neutral relationship towards other another agent (O), and (2) the weight of the relationship (P→ O) should be between the other two relationships. These assumptions are fulfilled with following measure of balance:
% B(bid, bjd, xij) = 1-1/6(|bid - SGM(bjd, xij)| + |bjd-SGM(bid, xij)| + |xij-SGM(bid, b_jd)|),
% where SGM(b_1,…,b_n) = (multiply_i^n sign(bi)) (multiply_i=1^n | )^(1/n) is a signed geometric mean.
Signed relationships are defined in the general form:
\begin{equation}\label{eq:SGM}
x_{ij} = f\!\left(\overline{\mathrm{SGM}}(i,j)\right),
\end{equation}
where $\overline{\mathrm{SGM}}(i,j)$ denotes the average signed geometric mean of the $M$ attributes of agents $i$ and $j$, and $f$ is a chosen response function.
Using American National Election Study survey data, the authors propose
\begin{equation}
f(x) = \mathrm{sign}(x)\,|x|^{1-e},
\end{equation}
where $e \in [0,1]$ is an extremeness parameter. For $e=0$, $f(x)=x$ reduces to the identity, while for $e=1$ the function becomes a step function.

The model is studied on a complete graph. During each macro-step, agents are selected at random and updated asynchronously. In the update, agent $i$ first chooses agent $j$ to interact with. The update then consists of three stages: (i) the signed relation $x_{ij}$ is updated according to the function $f$; (ii) an ``optimally balanced'' opinion vector $\hat{\mathbf{b}}_i$ is determined using signed geometric mean of $\mathbf{b}_j$ and $x_{ij}$, and agent $i$ adjusts its attributes toward this vector as $\Delta\mathbf{b}_i = \alpha(\hat{\mathbf{b}}_i - \mathbf{b}_i)$; and (iii) additive noise, drawn from a normal distribution with zero mean and standard deviation $z$, is applied to the attributes.
The dynamics is therefore controlled by three parameters: the extremeness $e$, the opinion-adjustment rate $\alpha$, and the noise amplitude $z$.

Starting from random initial conditions, for sufficiently small noise or sufficiently large extremeness, the population splits into two clusters whose average opinions separate, forming a balanced polarized configuration. Otherwise, the system converges to a consensus state in which all attributes are close to zero. A nonzero-consensus state can also be obtained by modifying the response function $f(x)$ to allow a positive bias even for negative values of $x$.

Across both Hamiltonian and non-Hamiltonian formulations, a recurring result is that agent--attribute consistency can provide a route to interpersonal structural balance even when balance of agent--agent--agent triads is not imposed explicitly. Whether such triadic balance emerges, and what form it takes, depends on the attribute structure, noise, network topology, and details of the dynamical rules.

\subsubsection{Models combining interpersonal structural balance and agent--attribute interactions}
\label{subsubsec:combined-interpersonal-agent-attribute}
% In these models, structural balance constraints act both on interpersonal triads among agents and on triads involving agents and attributes, in the sense of Heider’s $P$--$O$--$X$formulation.

In this subsection, we review models that explicitly combine interpersonal structural balance with agent--attribute interactions. In such approaches, balance constraints may act simultaneously on triads of agents and on configurations involving agents and attributes. Some of the papers discussed here describe the attribute-related mechanism in terms of homophily rather than structural balance; as noted in the previous subsection, we do not distinguish between these perspectives. Most of the models reviewed below study the coevolution of signed relations and attributes, although in some cases, only signs are allowed to evolve. We begin with models assuming binary attributes and then turn to those with continuous ones.

\paragraph{Binary attributes}

Hamiltonian formulations are relatively rare among models combining interpersonal and agent--attribute mechanisms.  \citet{MinhPham2020} considers a signed network with each agent possessing a single, binary attribute $b_i \in \{-1,+1\}$. The Hamiltonian combines interpersonal and impersonal balance contributions,
\begin{equation}
H = -\sum_{(i,j)\in E} x_{ij} b_i b_j - g \sum_{(i,j,k)\in \mathcal{T}} x_{ij} x_{jk} x_{ki} + \frac{h}{2}\sum_{(i,j)\in E}(1-x_{ij}),
\label{eq:pham}
\end{equation}
where the parameter $g$ controls the relative strength of interpersonal structural balance, and $h\ge0$ denotes an external field which attempts to minimize negative links in the social network.

The dynamics follows a Metropolis scheme: first, $Nn$ times an agent is selected, and its attribute is updated according to the Metropolis rule; next, one Metropolis update of a link sign is performed. Here, $N$ is the number of agents, and the parameter $n$ controls the relative frequency of attribute and sign updates. The model is studied on regular ring lattices with degree $k$. As an order parameter, the authors introduce the normalized difference between numbers of balanced and unbalanced triads, equivalent to the negative of the triadic Hamiltonian in Eq.~\eqref{eq:hamiltonian_triad}, normalized by the number of triads $N_T$. This measure ranges from $-1$ to $1$. The resulting phase diagram shows that for large $k$ and low temperature $T$, the system reaches a balanced state, whereas for high $T$ and low $k$ the network becomes effectively random. Between these regimes, a narrow parameter region appears in which the balance level takes intermediate values, similar to those observed in empirical networks. The authors identify polarized configurations with balanced states and further analyze opinion polarization, which is discussed in more detail in the implications (see Sec.~\ref{subsec:coev}).

This framework is extended in~\citep{pham2021balance} by allowing each agent to possess $M$ binary attributes, with signed relations defined from attributes as in Eq.~\eqref{eq:binary-attr-sign},~\citep{Gorski2020}. At each step, an agent $i$ is selected and evaluates a Hamiltonian computed from only a fraction $q$ of the triads to which it belongs, representing limited local knowledge of the network. Using Metropolis dynamics on ring lattices, the authors observe transitions on the $(q,M)$ and $(k,M)$ parameter planes between largely unbalanced configurations and fragmented configurations with predominantly balanced triads.
The balanced states are observed for larger $q$, larger $k$, and smaller $M$. This last dependence contrasts with the attribute-based local triad dynamics studied by  \citet{Gorski2020} (see Sec. \ref{subsubsec:attr-driven-sign-formation}), where paradise states emerge only for large numbers of attributes.
We suggest that this difference may arise because Hamiltonian dynamics in~\citep{pham2021balance} promotes increasing similarity among friends and dissimilarity among enemies, a mechanism not explicitly present in attribute-based local triad dynamics. In~\citep{Gorski2020}, attribute changes are triggered only by unbalanced triads, which leaves many relations close to a sign change. In contrast, in the Hamiltonian model, balanced triads also contribute to the dynamics, driving the system toward more stable configurations. This could explain why sign updates tend to slow down in the Hamiltonian model, while in the attribute-based LTD, they remain governed by a fixed local rule.

A different approach is proposed by  \citet{Deng2016}, who consider agents with a single binary attribute and ego-centered updates. At each step, an ego agent evaluates both the level of interpersonal (triadic) balance and the level of agent--attribute (dyadic) balance with its neighbors. If both balance levels exceed respective thresholds, $B_{\mathrm{th}}$ and $H_{\mathrm{th}}$, no change occurs; otherwise, the ego modifies the sign of one incident link in a way that increases balance. The authors study two variants: independent evaluation of the two balance measures and a simultaneous rule requiring both to fall below the thresholds.

When only interpersonal balance is taken into account, a balanced network is generally not achieved. In contrast, considering only agent--attribute balance can lead to balance for sufficiently large values of $H_{\mathrm{th}}$. Under independent evaluation, very large values of $B_{\mathrm{th}}$ make the dynamics insensitive to the dyadic threshold $H_{\mathrm{th}}$, preventing the system from reaching balance. For smaller $B_{\mathrm{th}}$, the outcome depends on agent attributes, and sufficiently large $H_{\mathrm{th}}$ leads to balance. The transition threshold itself depends on the initial density of positive links: higher initial positivity requires larger values of $B_{\mathrm{th}}$ for the agent--attribute mechanism to become irrelevant.
When both balance mechanisms act simultaneously, increasing either threshold increases balance, but balanced networks are achieved only when both are sufficiently large.
Extensions to unordered negative attributes show that increasing the number of attribute values suppresses the emergence of a balanced network.

\citet{mei2022structural} extend ego-based local triad dynamics (see Sec.~\ref{subsubsec:ego-perspective-models}) by incorporating agent attributes into the dynamics. In addition to updating link signs based on interpersonal balance, the ego agent may also modify a link to satisfy balance with respect to agent attributes or adjust its own attribute based on a neighbor’s attribute and the current relation. The authors prove that the model with attributes converges to a globally balanced configuration, whereas the model without attributes guarantees only local triadic balance, which need not imply global balance on incomplete networks.

\paragraph{Continuous attributes}

Several models extend the combination of interpersonal and agent--attribute mechanisms to continuous attributes. Some of these approaches build on gradient triad dynamics (GTD)~\citep{Weinmann2022,Gorski2023}. In  \citet{Gorski2023}, attributes are fixed but influence the evolution of signed relations. The authors develop a general framework encompassing attributes of different types (binary, ordered, unordered, and continuous) and study their effectiveness in destabilizing balanced states or preventing polarization, understood here as a weakly balanced system divided into at least two groups.
The dynamics on complete networks is given by
\begin{equation}
\dot{x}_{ij} = \left( \frac{1}{N-2}\sum_{k} x_{ik}x_{kj} + \gamma g_{ij}(\mathbf{b}_i,\mathbf{b}_j) \right)\bigl(1 - x_{ij}^2\bigr),
\end{equation}
where \(\mathbf b_i\) and \(\mathbf b_j\) are the attribute vectors of agents \(i\) and \(j\), respectively, and $\gamma$ controls the relative strength of attribute effects. The function $g_{ij}$ depends on the attribute type; for a single binary attribute, it equals $+1$ or $-1$ depending on whether attributes match. Assuming more attributes of the same weight, a simple average is used.

Analytical and numerical results show that destabilization occurs once $\gamma$ exceeds the inverse of the expected single-attribute effect. For binary attributes, the expected effect is zero, so destabilization cannot be guaranteed. Among the types considered that can reduce polarization, continuous attributes are most effective, followed by  positive unordered, ordered, and binary attributes.

In the model discussed above, only link signs are allowed to evolve, while attributes remain fixed. When attributes are continuous, it becomes possible to describe their evolution through differential equations.
Such an approach is taken by  \citet{Gao2018IntJouModPhysB}, who study the coevolution of signs and attributes using a framework related to Riccati-type matrix equations, with the aim of designing control strategies that make the relation matrix converge to a prescribed structurally balanced configuration.

When signs and attributes coevolve, it is often necessary to introduce an additional parameter controlling the relative timescales of these processes. This idea is explored by \citet{Weinmann2022}, who couple GTD to Kuramoto oscillators. In their model, each agent carries a continuous phase that evolves to synchronize with friends and desynchronize with enemies, while signs evolve through a combination of interpersonal and impersonal balance mechanisms. In this model, the overall level of synchronization influences the evolution of link signs, favoring a paradise state with in-phase synchronization.
% TODO: Read new review of Kulakowski and see how he describes this paper.

Coevolution of signs and continuous attributes does not require link sign relations  themselves to be continuous. \citet{Parravano2016} combine local triad dynamics with a bounded-confidence opinion model. Agents update their opinions by moving toward the mean opinion of those friends whose opinions are sufficiently similar. Link signs evolve according to local triad dynamics; in ambiguous update situations, the link connecting the most similar agents (for $\Delta_3$ triads) or the most dissimilar agents (for $\Delta_1$ triads) is selected for updating.
For a single opinion, the system converges to either a paradise or a bipolar configuration, with convergence times significantly shorter than in standard LTD. As the confidence bound increases, opinion updates involve progressively more agents and thus rely on increasingly global information, leading to convergence-time scaling similar to that of constrained triad dynamics.

% \subsubsection{Attributes modulating the dynamics of sign updates}
% \label{subsubsec:attr-modulated-dynamics}
\subsubsection{Models with context-dependent update rules}
\label{subsubsec:context-dependent-updates}

In this final subsection, we discuss models in which agent states influence how agents evaluate balance or update relations, rather than determining link signs through a fixed attribute-based rule. In such approaches, attributes~\citep{Aguiar2015}, infection states~\citep{saeedian2017epidemic}, behavioral strategies~\citep{Righi2013,Righi2014,Zhong2021,Song2022}, or status-based considerations~\citep{Gorski2025} shape agents’ decision rules, effectively introducing heterogeneity into the dynamics of sign updates.

A prominent example is the model proposed by  \citet{Aguiar2015}, who study a signed network of agents endowed with two fixed binary attributes: an opinion ($A$ or $B$) and a behavioral trait (tolerant or intolerant). Tolerant agents update relations without considering attributes, whereas intolerant agents refuse to create or maintain friendly links with agents holding a different opinion. In this sense, intolerant agents simultaneously account for both interpersonal and agent--attribute balance, whereas tolerant agents rely solely on interpersonal balance.

The authors consider twenty possible triad configurations, which reduce to ten types upon interchanging opinions $A$ and $B$. When all agents in a triad are tolerant, or when all agents share the same opinion, the dynamics reduces to standard local triad dynamics with parameter $p$. In mixed triads involving a tolerant agent of opinion $A$, an intolerant agent of opinion $A$, and a tolerant agent of opinion $B$, an additional parameter $q$ governs which relation is updated, with $q$ interpreted as intolerance toward intolerant agents. In all remaining cases, updates lead deterministically to a balanced $\Delta_2$ configuration, in which two agents sharing the same opinion are mutually friendly.

In the numerical simulations with intolerant agents absent, or when all agents share the same opinion, the dynamics is the standard LTD. Otherwise, the system typically reaches absorbing states consisting of two factions. When all intolerant agents share the same opinion, the final group sizes depend on both $q$ and $p$: for large $q$, the intolerant group remains small, whereas for small $q$ it attracts all agents sharing the same opinion; similarly, small values of $p$ lead to groups of comparable size, while large $p$ produces a dominant group opposed by a minority of intolerant agents.

Attributes that modify agent behavior rather than balance definitions are also central to the model introduced by  \citet{saeedian2017epidemic}, who study susceptible--infected (SI) epidemic dynamics on signed networks\footnote{This work is also discussed in Sec.~\ref{subsec:contagion}, in the context of models that link spreading processes and structural balance dynamics.}. In this framework, infections do not spread through hostile links. Unstable triads include not only triads unbalanced according to interpersonal structural balance, but also such triads that contain a friendly pair connecting susceptible and infected agents.
Therefore, balance considerations are extended beyond interpersonal triads to include dyadic configurations, but only for friendly pairs, reflecting a partial incorporation of impersonal structural balance.

To formalize this, the authors extend the standard triadic Hamiltonian to penalize friendly susceptible--infected pairs and simulate a version of constrained triad dynamics. At each step, a pair of agents is selected: if it is a friendly susceptible--infected pair, the link becomes hostile with probability $1-\alpha$, or the susceptible agent becomes infected with probability $\alpha$. If the pair consists of agents of the same type, the sign of the link is flipped. Proposed changes are accepted if they decrease the Hamiltonian, or with probability 0.5 if the Hamiltonian remains unchanged.
% \pgcomment{The text below repeats what is said in Sec. \ref{subsec:contagion}. To avoid it the text below could be replaced with sth like ``The implications of this model, together with other examples linking spreading processes and structural balance dynamics, are discussed in Sec.~\ref{subsec:contagion}.'' If so, remove the footnote above.} \msmcomment{given the narrative, I prefer to keep this text here, and think that some redundancy is there but relatively small. I've referenced this section in the simple contagion section below. If you're fine with this, please remove the comment}
Starting from networks dominated by negative links, the authors show that the likelihood of reaching jammed states depends on both the infection probability $\alpha$ and the initial fraction of infected agents. Hostile links effectively act as a form of immunization by separating groups through which infection cannot spread.

Behavioral strategies provide another agent state that can interact with signed relations in game-theoretic models (compare Sec.~\ref{subsubsec:strategic-models}, where the signed network itself is the primary object of strategic choice).
In~\citep{Righi2013,Righi2014}, cooperation strategies and signed relations coevolve in a prisoner's-dilemma setting.
Cooperation does not persist when agents rely only on dyadic information, but can be stabilized when they also take the structural balance of the surrounding triads into account.
Related studies examine the role of signed relations in public-goods games~\citep{Zhong2021}, while prisoner's-dilemma dynamics explicitly coupled to structural-balance rules have also been investigated~\citep{Song2022}.

Finally,  \citet{Gorski2025} introduce a model in which agent attributes are not explicitly assigned, but their implicit existence influences the dynamics. The authors combine structural balance theory with status theory, another prominent framework for understanding signed relations~\citep{Yap2015}. In status theory, positive links point from lower- to higher-status agents, while negative links point in the opposite direction. The authors identify triadic configurations that are inconsistent with status theory, regardless of how agent statuses are assigned. Thus, explicit knowledge of statuses is not needed. Building on this observation, they propose a dynamics that alternates between balance-based and status-based update rules.

Specifically, an ego-based triad is selected (see Sec.~\ref{subsubsec:ego-perspective-models}). With probability $q$, the ego evaluates the triad according to status theory, and with probability $1-q$ according to structural balance theory. The ego then updates one of its outgoing links using LTD, with separate probabilities $p_{{ST}}$ and $p_{{SBT}}$ for status-based and balance-based updates, respectively. Treating one of the model parameters as a control variable, the authors identify a transition between quasistationary unbalanced states and a paradise state. In contrast to standard LTD and its ego-based variants, the inclusion of status theory can introduce a discontinuous transition, characterized by a separatrix between basins of attraction.

The model is calibrated on empirical datasets, enabling the authors to assess the relative importance of status theory and structural balance across various systems.
They find that networks dominated by strong relations are more strongly shaped by structural balance, whereas in systems with pronounced status effects, agents tend to reach the top of the hierarchy.
Building on this model, \citet{gorski2026more} examined how the interplay between structural balance and status affects network polarization. They found that increasing the relative importance of status-driven updates enhances local structural polarization when status dominates. The corresponding polarization measure and empirical findings relating weak ties, hierarchical triads, and polarization are discussed in Sec.~\ref{subsec:netpol}.

% \subsection{Growing networks}
% \InsertBlankPages{8}

\subsection{Dynamics in generalized network structures}
\label{subsec:generalized}

In this section, we consider models in which structural balance dynamics is defined on generalized network structures, including multilayer, temporal, and higher-order networks. Such frameworks allow one to represent multiple types of interactions, contexts, or dynamical processes acting simultaneously on the same set of agents.

\subsubsection{Multilayer networks}
\label{subsubsec:multilayer}

Most models on multilayer topologies in practice consider two layers, although extensions to more layers are sometimes proposed. Such models can be broadly divided into those in which structural balance dynamics operates in all layers and those in which structural balance is coupled to other types of dynamics, with distinct processes assigned to different layers.

\paragraph{Structural balance dynamics in multiple layers}

One approach to structural balance in multilayer networks considers link multiplex structures~\citep{Gorski2017,Mohandas2024,Mohandas2025Paradise}. In such topologies, the same set of agents exists in all layers, and interlayer coupling acts between corresponding links of different layers. Within each layer, signed links evolve according to principles of structural balance theory, while interlayer coupling can favor either alignment or opposition between replicas of the same link. Positive interlayer coupling captures correlations between different kinds of relations involving the same pair of agents; for example, friendship between two individuals at work may increase the likelihood of friendship in a leisure context.

\citet{Gorski2017} extend GTD by coupling two signed-network layers, with coupling strengths $\beta_1$ and $\beta_2$. The dynamics is given by
\begin{equation}
\dot{x}_{ij}^{(\alpha)} = \bigl(1-(x_{ij}^{(\alpha)})^2\bigr)
\left( \frac{1}{N-2}\sum_{k=1}^N x_{ik}^{(\alpha)}x_{kj}^{(\alpha)}
+ \beta_{\alpha} x_{ij}^{(\alpha')} \right),
\end{equation}
where $\alpha,\alpha'\in\{1,2\}$ and $\alpha\neq\alpha'$. Each link in layer $\alpha$ is coupled to its replica in the other layer $\alpha'$. The network is considered balanced if both layers are balanced, even though the signs of corresponding links may differ across layers.

The authors solve the dynamics analytically for a triadic network (N=3) and present numerical results for larger systems. They identify several distinct regimes: reaching balance independently in uncoupled layers for weak interlayer coupling; dominance of one layer that drives balance in both layers or induces imbalance in the other; regimes with very strong interlayer interactions, where balanced configurations become extremely unlikely in large networks; and oscillatory behavior arising when both couplings are strong but one of them is negative.

The Hamiltonian-based approach introduced in \citet{Malarz2022Mean} (see Sec.~\ref{subsubsec:hamil-nonzeroT}) is further extended to multiplex networks in~\citep{Mohandas2024} and to bilayer Erd\H{o}s--R\'enyi graphs in~\citep{Mohandas2025Paradise}.
For a system consisting of $L$ layers, the Hamiltonian contribution associated with a pair of nodes $(i, j)$ is given by
\begin{equation}
H_{ij} = -A \sum_{k=1}^{N}\sum_{\alpha=1}^L
x_{ij}^{(\alpha)}x_{jk}^{(\alpha)}x_{ki}^{(\alpha)}
- K\sum_{\alpha>\beta} x_{ij}^{(\alpha)} x_{ij}^{(\beta)},
\label{eq:energy_raw_multi-ch2}
\end{equation}
where $A$ controls intralayer triadic interactions, while $K>0$ represents interlayer coupling between replicas of the same link. The presence of multiplexity increases the critical temperature at which a system initialized in paradise states across all layers loses structural balance and gives rise to a richer phase structure due to possible mismatches between ordering patterns across layers.
In general, mismatched clique partitions across layers lose stability at temperatures lower than the critical temperature of matching ordered configurations.

For a given existing link in a monolayer Erd\H{o}s--R\'enyi graph, the mean-field contribution associated with triadic interactions scales as $p^2$, since two additional edges must be present to form a triad~\citep{Mohandas2025Paradise}. Consequently, the normalized critical temperature $\tilde T_c=T_c/(N-2)$ scales as $\tilde T_c \propto p^2$.
In coupled layers, however, the situation differs. The bilayer graph is constructed such that each link in each layer exists independently with probability $p$, so that the two layers are independent Erd\H{o}s--R\'enyi realizations. While intralayer triadic interactions still scale with $p^2$, the probability that an existing link in one layer also exists in the other is $p$. Consequently, the effective interlayer coupling scales linearly as $Kp$.
For bilayer networks with diluted links, the dependence $T_c(p)$ becomes significantly more complex than in the monolayer case.
To recover behavior analogous to that observed in monolayer complete graphs, the interaction strengths must therefore be rescaled with $p$: the intralayer Heider interaction strength scales as $A \propto p^{-2}$, whereas the interlayer coupling scales as $K \propto p^{-1}$ for random graphs.

While the above models consider general multiplex structures, a more application-driven multilayer formulation is studied by  \citet{rodriguez2016collective}. The authors propose a bilayer model in which one layer represents agents’ belief systems, modeled as signed networks subject to triadic balance, while the other layer represents social interactions governed by dyadic balance. Agents communicate beliefs to neighbors who may adopt them following Metropolis dynamics. The social network is modeled as an Erd\H{o}s--R\'enyi graph of average degree five, while each belief network is fully connected with five nodes. Further details of the Hamiltonian (energy) and implications of this model are discussed in Sec.~\ref{subsubsec:tipping-points}.

\paragraph{Multiple types of links}

An alternative generalization considers networks with multiple types of links rather than multiple layers.  \citet{Oloomi2021} study a system in which each pair of agents is connected by exactly one link, which can be of type $a$ or type $b$. Although this model could be classified differently, the resulting structure is similar to a bilayer network, where a link between two nodes is either in the layer $a$ or layer $b$ (i.e., zero interlayer overlap).
% to consider:
% add interpretation of links
% The link types are in fact competin
% Different link types compete with each other. They are assumed to represent discourses that are not independent but are in conflict of interest. Being positive on one topic  (positive or negative) agents have in a

To analyze balance, the authors represent links using complex numbers: type-$a$ links take values $\pm1$, while type-$b$ links take values $\pm \mathrm{i}$ (with $-\mathrm{i}$ denoting friendliness).
Standard balance rules are assumed in triads composed of links of single type (i.e., triads with three friendly links or one friendly and two hostile are balanced for both types). This leads
% Energies are defined for all triads, including mixed-type triads. Balanced triads contribute $-1$ or $-i$ to the total energy,
% leading
to nonstandard balance conditions in the case of mixed-links triads, for example, a triad with three positive links, one of one type and two of another, is considered unbalanced.

Dynamics follows constrained triad dynamics: a randomly chosen link is proposed to change its type or friendship/enmity status, and the change is accepted if the number of unbalanced triads does not increase. In stationary states, one link type typically dominates and the system consists of balanced triads only~\citep{Oloomi2021}. Introducing an external field leads to hysteresis effects~\citep{Oloomi2023}.

\paragraph{Structural balance interacting with additional dynamical layers}

Structural balance has also been studied in conjunction with other dynamical processes that can be viewed as coupled network layers.
\citet{Traag2011} propose a model combining structural balance with game-theoretic interactions and information spreading. In this interpretation, one layer represents signed links (reputations), while another represents a cooperation--defection decisions. Agents cooperate with positively connected neighbors and defect against negatively connected ones. After interactions, information is communicated, and reputations are updated accordingly.
Cooperation and defection are not attributes in the usual sense, since an agent may cooperate with some neighbors and defect with others. For symmetric cooperation, the resulting networks at fixed points are weakly balanced.

A different bilayer formulation is introduced by \citet{Shang2020}, who extend the unsaturated GTD equations. One layer represents agents’ true sentiments $\mathbf{X}$, while the second represents expressed or perceived sentiments $\hat{\mathbf{X}}$. The dynamics is given by
% \begin{align}
% \dot{\mathbf{X}}(t) &= \mathbf{X}(t)\hat{\mathbf{X}}(t), \\
% \hat{\mathbf{X}}(t) &= \phi\,\mathbf{X}(t) + (1-\phi)\mathbf{Y}(t),
% \end{align}
\begin{equation}
\begin{cases}
\dot{\mathbf{X}}(t) = \mathbf{X}(t)\hat{\mathbf{X}}(t), \\[4pt]
\hat{\mathbf{X}}(t) = \phi\,\mathbf{X}(t) + (1-\phi)\mathbf{Y}(t),
\end{cases}
\label{eq:bilayer-true-expressed}
\end{equation}
where $\mathbf{Y}(t)$ is an external influence and $\phi\in[0,1]$ measures agents’ resistance to influence. Analytical results show that when $\mathbf{Y}(t)=\mathbf{X}(0)$, the system converges to a weakly-balanced configuration consisting of four clusters.

\subsubsection{Temporal and higher-order networks}
\label{subsubsec:beyond-temp-higher}

\citet{Nishi2014} introduce temporal effects into structural balance dynamics by modifying the update protocol. At each step, a link is selected at random and allowed to evolve according to the GTD interaction rule for a fixed duration $\tau$, after which another link is selected at random, possibly the same one. The authors study how the time required to reach balance depends on the temporal parameter $\tau$ and on the initial average link weight $\mu$. They find that the convergence time decreases as $\tau$ becomes smaller and as $\mu$ increases.

Three-body interactions constitute the core of most structural balance models.
As discussed in Sec.~\ref{subsubsec:hamil-multitype}, some Hamiltonian formulations additionally introduce one-body and two-body terms to capture supplementary effects.
Higher-order interactions can also arise naturally. Such interactions appear, for example, when correlations between adjacent triads are introduced or when delays are incorporated into link dynamics, with successively higher-order interaction terms required to reproduce the dynamics as the delay increases (see Secs.~\ref{subsubsec:kk-variants} and~\ref{subsubsec:hamil-variants};~\citep{Kargaran2020,Siboni2025}).

An explicit treatment of higher-order interactions is proposed by  \citet{Hakimi2022}, who introduce a Hamiltonian combining three-body and four-body terms,
\begin{equation}
H = -\sum_{i<j<k} x_{ij}x_{jk}x_{ki}
    - g\sum_{(i,j,k,\ell)\in \mathscr C_4} x_{ij}x_{jk}x_{k\ell}x_{\ell i},
\end{equation}
where $\mathscr C_4$ denotes the set of cycles of length four, each counted once, and $g$ controls the relative strength of four-body interactions. The authors analyze the system's stable states and identify a critical value $g_c\propto N^{-1}$,
% in which the system either
% demonstrate that for small $g$, the triadic term dominates,
% resulting in
% reaches configurations
% without unbalanced triads or states in which all 4-cycles are balanced.
above which the four-body term dominates the triadic contribution.
% For sufficiently large $g$, however,
In this regime, unbalanced triads can persist in stationary states.
For the unnormalized Hamiltonian above, at fixed $g$ and sufficiently large $N$, the critical temperature scales as $T_c\propto N^2$, in contrast to the three-body case, where $T_c\propto N$.
% This mechanism provides an additional explanation for the absence of complete structural balance in empirical networks.

\subsection{Bottom Line: Why does structural balance emerge in complex networks?}

Across most models reviewed above, the emergence and stability of structural balance can be traced to mechanisms that reduce tension in signed triads. These mechanisms may rely either on global or on local information about the network.
% Below we compare the three main modeling approaches and then discuss attribute-based (impersonal) formulations and hybrid models.

Within this perspective, we distinguish three main modeling approaches (see Sec. \ref{subsec:kk-model}-\ref{subsec:stoch-sign-dyn}).
The first two, the gradient triad dynamics and Hamiltonian-based models, rely on global information.
In GTD, link weights evolve according to coupled differential equations.
In its standard formulation, this dynamics can be interpreted as a gradient-like descent in a global energy landscape defined via a Hamiltonian.
% Therefore both of these approaches seem to relate on the definition of Hamiltonian. However, such a connection is misleading because ...
Hamiltonian-based models make this idea explicit by defining a system energy and studying its evolution using tools from statistical physics, such as introducing a temperature, applying mean-field approximations in analytical calculations and using update rules like Metropolis algorithm in numerical simulations.

The third approach comprises a broad class of stochastic models that do not fit the previous two categories. Most of these rely only on local information and build on the first such model proposed, local triad dynamics (LTD) \citep{antal_2005}. They are formulated as agent-based models, in which agents perceive dissonance (imbalance) in their immediate neighborhood, which then drives the system's evolution. Despite relying on local updates only, such dynamics can still produce globally balanced structures.

% In the models discussed so far, structural balance has been defined in terms of triadic relations between agents. However, the concept can be extended beyond purely interpersonal interactions and applied to agent--attribute relationships.

In models where agents interact with attributes (opinions, topics, or activities), balance can be defined for agent--agent--attribute configurations.
% We refer to this formulation as \emph{impersonal structural balance}, see Sec.~\ref{subsec:sign-attribute-dynamics}.
In this case, structural balance reflects homophily: balanced motifs consist of friends sharing similar attributes or enemies holding different ones.
Models based on this impersonal formulation can likewise be classified according to whether the dynamics relies on global or local information, following the same three modeling approaches discussed above.
This suggests that the proposed classification is robust.

Some models combine both mechanisms, allowing interpersonal and impersonal balance to coevolve.
Interestingly, triadic (interpersonal) balance may emerge even when the dynamics acts solely on reducing tensions in agent--agent--attribute structures.
This relation can be understood by noting that a balanced network partitioned into hostile groups is equivalent to a polarized system in which different groups adopt different opinions.
% maybe the above sentence should be stated more clearly.
This equivalence has been used in measures of structural balance (see Sec.~\ref{subsubsec:frustration}), where identifying hostile groups corresponds to solving a node-coloring or frustration-minimization problem.

A complementary route to balanced or low-frustration structures is provided by strategic and game-theoretic models, in which network organization emerges from agents' incentives rather than from a prescribed tension-reduction rule.

\subsubsection{Comparison of typical structural balance models}

Table~\ref{tab:sign_dynamics_comparison} compares the three most widely studied modeling approaches to structural balance dynamics. It summarizes the representative dynamical structure of each approach, characteristic outcomes reported across the corresponding model families, commonly studied extensions, and network topologies considered.

In its standard formulation, gradient triad dynamics relies on global information and assumes continuous time, with all link variables evolving simultaneously under a deterministic flow.
Signed links are represented by continuous weights that evolve according to coupled differential equations. As a consequence, the dynamics is deterministic and randomness may only arise from random initial conditions.

Hamiltonian-based dynamics also relies on global information, but the evolution is defined through a system energy (Hamiltonian). Binary signed links evolve through stochastic updates, typically following Metropolis or Glauber/heat-bath rules. The temperature controls the level of randomness, including the probability of energy-increasing changes.

Local stochastic models rely on local information and are formulated using agent-based dynamics.
Links are binary and evolve in discrete time according to simple update rules acting on triads.
Asynchronous updates introduce intrinsic stochasticity through the random selection of the element to be updated.

% In all cases the systems may reach balanced states, either paradise or polarized.

% All basic formulations of models consider undirected networks.

{

\small
\setlength{\tabcolsep}{4.5pt}
\renewcommand{\arraystretch}{1.08}

\begin{longtable}{
>{\raggedright\arraybackslash}p{0.31\textwidth}
>{\raggedright\arraybackslash}p{0.22\textwidth}
>{\raggedright\arraybackslash}p{0.23\textwidth}
>{\raggedright\arraybackslash}p{0.23\textwidth}
}
\caption{Comparison of three major modeling approaches to sign-dynamics models.
Part~I describes the representative standard formulation of each approach. Part~II summarizes typical dynamical outcomes reported across the corresponding model families. Part~III highlights major extensions. Part~IV summarizes network topologies considered. The listed outcomes should not be interpreted as universal properties of all variants within a given model family.
% Part~I summarizes the most common formulation. Part~II compares typical dynamical outcomes. Part~III highlights major extensions.  Part~IV summarizes network topologies considered.
% Part~I and Part~II describe the representative standard formulation of each approach; individual variants reviewed in Sec.~\ref{sec:emerg} may depart from them.
% \msmcomment{@Piotr, in the end, here we're offering here our perspective on what the literature has focused on. This is very helpful to guide future works. I'm wondering here if we should replace yes/no with "Never/rarely/often/always", using never/always only when something is mathematically impossible/certain, as we will always miss some works in each cell. }
% \pgcomment{ If we worry that we might miss a model considering the extension that we claim "there are NO models", we could add a limitation sentence, something like 'Part~III summarizes extensions identified in the literature surveyed in this review.'
% Would it be useful to have instead of `Yes` proper Refs. (eg, Ref. [34,67,98])? My worry was that this could be too long making the table unreadable.
% I am fine to changing Yes/No to something else. We could have the "Never/rarely/often/always" but slightly differently from the proposed "Never (math impossible)/rarely/often/always (math certain)":
% Never - no paper we considered considered this extention.
% Rarely - there are 1-2 papers considering this extension.
% Often - there are a number of papers (does not mean majority) considering this extensions.
% Always}
}
\label{tab:sign_dynamics_comparison}\\

\toprule
 & \textbf{Gradient triad dynamics} & \textbf{Hamiltonian-based dynamics} & \textbf{Local stochastic (LTD-type)} \\
\midrule
\endfirsthead

\toprule
 & \textbf{Gradient triad dynamics} & \textbf{Hamiltonian-based dynamics} & \textbf{Local stochastic (LTD-type)} \\
\midrule
\endhead

\midrule
\multicolumn{4}{r}{\emph{Continued on next page}}\\
\bottomrule
\endfoot

\bottomrule
\endlastfoot

% -----------------------------
\multicolumn{4}{l}{\textbf{I. Dynamical structure and state space}} \\
\midrule
Time formulation
  & Continuous
  & Discrete
  & Discrete \\

Update timing
  & Synchronous
  & Asynchronous
  & Asynchronous \\

Update mechanism
  & Differential equations
  & Energy-based acceptance
  & Rule-based sign flip \\

Information scope
  & Global
  & Global
  & Local  \\

Deterministic vs. stochastic updates
  & Deterministic
  & Stochastic (thermal noise)
  & Stochastic (intrinsic randomness) \\

Link values
  & Continuous
  & Binary
  & Binary \\

% -----------------------------
\midrule
\multicolumn{4}{l}{\textbf{II. Dynamical outcomes}} \\
\midrule
Phase transition
  & Not in standard model; present in extensions
  & Yes (1st-order in standard thermal model)
  & Yes \\

Jammed states
  & Yes
  & Yes (at $T=0$)
  & No \\

Typical attractors / long-time behavior
  & fixed points; limit cycles in non-gradient extensions
  & Stable/metastable states; hysteresis possible
  & Absorbing balanced states; long-lived transients \\

% -----------------------------
\midrule
\multicolumn{4}{l}{\textbf{III. Extensions in the literature}} \\
\midrule
Directed networks
  & Common
  & Rare % impersonal SBT, Agbanusi
  & Common \\

Self-loops
  & Common
  & Not found
  & Rare  \\

Neutral links
  & Rare
  & Rare
  & Rare \\

Longer cycles beyond triads
  & Rare
  & Rare
  & Rare \\

Explicit weak balance
  & Rare
  & Rare
  & Not found  \\

Synchronous updates
  & By construction
  & Rare
  & Not found \\

Asynchronous updates
  & Rare
  & By default
  & By default  \\

Time delay
  & Rare
  & Rare
  & Not found \\

Ego perspective
  & Rare
  & Not found
  & Common \\

% -----------------------------
\midrule
\multicolumn{4}{l}{\textbf{IV. Network topologies}} \\
\midrule
Typical topology
  & Complete graph
  & Complete graph
  & Complete graph \\

Other topologies
  & multilayer; empirical
  & ER; lattices (triangular, ring); multilayer; empirical
  & Lattices (ring, triangular, square); ER; empirical \\

\end{longtable}
}

\paragraph{General conditions to reach balance}

Balanced states are common outcomes of undirected GTD-based models. For complete graphs with self-loops, \citet{marvel2011_evolutionOFSBcontinuous} proved that, starting from random initial conditions, the probability of reaching a balanced state approaches one in the large-system limit (see Sec. \ref{subsubsec:kk-self}). Because GTD can be related to an energy landscape, jammed states may nevertheless occur (see Sec. \ref{subsec:jammed}). In typical model formulations, such states cannot contain more unbalanced than balanced triads \citep{Marvel2009}.
For directed networks, the long-time behavior depends on the specific update rule.
Depending on whether agents update their relation to a neighbor based on others' opinions about that neighbor or on the neighbor's opinions about others, the system converges generically either to a jammed or balanced state, respectively (see Sec. \ref{subsubsec:kk-directed}).
When balance is reached, the final configuration is either a paradise or a polarized state, with the outcome in the self-loop variant \citep{marvel2011_evolutionOFSBcontinuous} determined by the initial mean link weight. A positive initial bias favors paradise.

In Hamiltonian-based dynamics, balanced states are stable in the low-temperature regime. Jammed states can occur as absorbing states only at $T=0$, typically when a majority of links are initially negative (Sec. \ref{subsubsec:hamil-ctd}). For unbiased random initial conditions (equal probability of positive and negative links), the system tends to reach polarized states. As the temperature increases, a balanced paradise state loses stability at a critical temperature $T_c$, whereas bipolar order emerging from random initial conditions is observed only below a lower temperature $T_d$. Importantly, the reviewed results suggest that these two characteristic temperatures may scale differently with system size.
With the interaction constant that is not scaled with system size, $T_c \sim N$ is associated with the paradise--unbalanced transition, whereas we conjecture that $T_d \sim \sqrt{N}$ characterizes the emergence of bipolar order from random initial conditions
% may be characteristic of the polarized--unbalanced transition
(Sec. \ref{subsubsec:hamil-nonzeroT}).
These scalings are not universal and depend on the normalization of the Hamiltonian and on the update scheme. In particular, when the Hamiltonian is normalized by the number of triads, the critical temperature may instead decrease with system size and vanish in the thermodynamic limit.

For local stochastic dynamics, the conditions leading to balance depend on the specific model. In local triad dynamics-type models, polarized states are not dynamically stable but can persist only because such configurations are absorbing. Paradise becomes stable only when the probability of creating positive links exceeds that of creating negative ones (Sec. \ref{subsubsec:ltd}). In these models, balance typically results from a competition between local tension reduction and stochastic fluctuations.

\subsubsection{Why does structural balance fail to emerge?}

Although the models reviewed above incorporate mechanisms promoting structural balance, balanced configurations do not always emerge. Several factors may prevent the system from reaching or maintaining balance:

% * Possibility of jammed states
% * Too high randomness
% * topology: too dense or too sparse topology depending on the model,
% % may boost randomness, too sparse topology lowers triads' influence on each other
% and the possibility of other links following other mechanisms
% * interplay of competing mechanisms
\begin{enumerate}
\item the presence of jammed states,
\item excessive randomness,
\item network topology,
\item competition with other social mechanisms.
\end{enumerate}

First, jammed states may arise in systems where tensions are evaluated globally. In models with a Hamiltonian, the energy landscape may contain local minima corresponding to configurations that are not balanced, yet any single update would increase the overall tension (see Sec. \ref{subsec:jammed}). The system cannot escape such states unless moves that temporarily increase imbalance are allowed. Introducing noise through temperature or using local update rules may help the system escape such states.

Second, randomness itself may prevent balance.
Moderate noise can be helpful, but when it becomes too strong the system may fail to converge to a balanced configuration and instead remain in a \emph{random}, unbalanced state. In temperature-based dynamics, this phenomenon typically manifests as a phase transition separating balanced (ordered) states from random ones (Sec.~\ref{subsubsec:hamil-nonzeroT}). In models based on local information, local updates can also have effectively random consequences at the network level, because resolving imbalance in one triad may inadvertently create imbalance in adjacent ones \citep{antal_2006}. Model parameters therefore determine whether the system converges to balance.

Third, network topology plays an important role. Changes in topology alter the number and arrangement of adjacent triads, thereby affecting how local updates propagate through the network. The consequences depend on the modeling approach.
% Topology potentially affects all systems.
In Hamiltonian-based models, denser networks generally stabilize balanced states (for instance, in Erd\H{o}s--R\'enyi graphs the normalized critical temperature scales as $\tilde{T}_c \propto p^2$, Sec~\ref{subsubsec:hamil-network}).
In contrast, in local stochastic models, higher link density may amplify competing updates and effectively increase randomness.
Moreover, networks with weakly connected modules may evolve toward a balanced system consisting of multiple factions.
In multilayer networks, coupling between layers may also lead to regimes in which maintaining interlayer consistency dominates over the drive toward the whole-system balance, Sec.~\ref{subsubsec:multilayer}.

Finally, in real social systems, the mechanisms associated with structural balance theory may coexist with other processes affecting the system's evolution. Models often incorporate additional social mechanisms such as homophily, status, reciprocity, reputation, or cooperation processes. Other extensions introduce behavioral effects such as memory or self-appraisal, or couple structural balance with dynamical processes including opinion formation, epidemic spreading, or game-theoretic interactions. Some of these mechanisms, such as reciprocity (Sec.~\ref{subsubsec:kk-variants}), tend to reinforce structural balance, whereas others may compete with it. Consequently, another reported reason for not observing structural balance is the interplay of competing mechanisms.

Homophily provides a prominent example of such interacting mechanisms. In many models, it is implemented through the coevolution of signed relations and agent attributes. These formulations connect structural balance dynamics with models of opinion formation and allow the study of phenomena such as consensus or polarization, which are closely related to, but not identical with, structural balance.
These connections are examined in more detail in the next chapter, which discusses the broader implications of structural balance.

% Overall, the reviewed models demonstrate that structural balance can arise from both global and local tension-reduction mechanisms. The specific outcome depends on the interaction between these mechanisms and other factors such as randomness, network topology, and competing social processes. Explaining when these forces reinforce or counteract each other remains a central challenge in modeling the evolution of signed social networks.

The discussion in the previous two subsections highlights that the mechanisms promoting or hindering structural balance are closely interrelated.
Table~\ref{tab:sb_emergence_summary} summarizes these mechanisms.
% Fig.~\ref{fig:emergence_summary} summarizes these mechanisms and illustrates their interactions.

{

\begin{table}[!ht]
\centering
\small
\setlength{\tabcolsep}{5pt}
\renewcommand{\arraystretch}{1.18}

\begin{tabularx}{\linewidth}{
>{\raggedright\arraybackslash}p{0.25\linewidth}
>{\raggedright\arraybackslash}X
>{\raggedright\arraybackslash}X
}
\toprule
\multirow{2}{*}{\textbf{Factor}}
& \multicolumn{2}{c}{\textbf{Why does structural balance}} \\
\cmidrule(lr){2-3}
& \textbf{emerge?}
& \textbf{fail to emerge?} \\
\midrule

Energy minimization / gradient-like descent
& Reduces tensions.
& May lead to jammed states. \\

Noise
& Moderate noise may help the system escape jammed states.
& Excessive noise prevents ordering and favors random states. \\

Local rules
& Reduce tensions, leading to self-organization.
& Local updates may destabilize adjacent triads and prevent global ordering. \\

\specialrule{1.1pt}{2pt}{2pt}
\multicolumn{3}{l}{\textbf{Role of topology is model-dependent}} \\
\midrule

More adjacent triads
& In global-rule models, balanced states become more stable.
& In local-rule models, conflicting updates become more frequent. \\

Sparse or modular topology
& Supports multi-faction balanced states.
& May prevent convergence to global balance. \\

\specialrule{1.1pt}{2pt}{2pt}
% \multicolumn{3}{l}{\textbf{Other social mechanisms}} \\
% \midrule

Other social mechanisms & Interpreting SB through agent--agent--attribute interactions, homophily leads to balance.
& Competing mechanisms (homophily, status, etc.) may favor relations inconsistent with triadic balance. \\

\bottomrule
\end{tabularx}

\caption{Conceptual summary of mechanisms that may promote or hinder the emergence of structural balance (SB). Some effects depend strongly on the modeling framework and network topology.}
\label{tab:sb_emergence_summary}
\end{table}
}

The reviewed models show that mechanisms promoting structural balance can also hinder it under different conditions. Global tension reduction can lead to trapping in local minima, noise can either facilitate escape or destroy ordering, and increased connectivity can stabilize balance in globally coupled models while intensifying conflicts between local updates. Thus, the emergence of structural balance depends not only on whether a balance-seeking mechanism is present, but also on how it interacts with the update rule, network topology, and competing constraints.

\clearpage

\section{Implications of structural balance}
\label{sec:implications}

This Section explores a range of themes concerning how structural balance influences the dynamics of social, political, and economic systems.
%In principle, structural balance -- and more generally, the structure of signed networks -- affects the drivers of outcomes for various actors and at various scales.
%Individual and collective beliefs affect indeed political behaviors such as voting in elections and parliamentary settings (see Section~\ref{subsec:cdm}) as well as diplomatic decision-making in international relations (see Section~\ref{subsec:ir}).
We begin by reviewing how cognitive balance, i.e., the structural balance of suitably defined cognitive networks, affects the internal consistency and social spreading of belief systems (Section~\ref{subsec:cognitive_balance}).
We subsequently review how social balance, i.e., the structural balance of social networks, affects opinion dynamics, collective decision-making, the spread of information and behavioral change (Section~\ref{subsec:social}) as well as connections between structural balance and network polarization (Section~\ref{subsec:netpol}).
Further, we review the implications of structural balance in international relations (Section~\ref{subsec:ir}) as well as organizations and markets (Section~\ref{subsec:eco}).

%Political outcomes emerge from individual and collective beliefs (covered in Section~\ref{subsec:cognitive_balance}), how representatives act in parliamentary settings (Section~\ref{subsec:parliament}), and how governments make decisions within international relation networks (Section~\ref{subsec:ir}).

\subsection{Cognitive balance on individual and collective beliefs}
\label{subsec:cognitive_balance}

The structure and dynamics of signed networks affect how we form beliefs as well as how beliefs spread in social systems.
At the individual level, individuals form and change beliefs on interrelated topics under the influence of both cognitive factors (e.g., cognitive dissonance minimization; individual study on a topic) and social cues.
At the collective level, the interplay between cognitive and social factors leads certain belief systems to spread with more or less resistance by a population~\cite{aiyappa2024emergence}.
Multiple topics and/or beliefs are typically interrelated through signed networks of varying degrees of structural balance, which directly influences the level of internal coherence in one's beliefs as well as the polarization of societal belief systems.
How do cognitive and social factors shape up the formation and spreading of beliefs, and what role does cognitive balance play for individual and collective belief properties? This Section aims to provide insights into the answer.

\subsubsection{Network representations of belief systems}
\label{subsec:net_belief}

The interconnections between one's beliefs can be captured through a signed network, whose level of structural balance we refer to as \textit{cognitive balance}~\cite{heider1946attitudes}.
There is not a unique definition of belief\footnote{\label{foot:beliefs} In this section, we use the term ``belief'' rather than ``opinion''. Both terms have been used to label spin-like variables in dynamic models, with ``opinion'' being traditionally more popular in sociophysics~\cite{castellano2009statistical,caldarelli2025physics}. Our choice aligns with most of the works reviewed in this Section (e.g., ~\cite{dalege2022using,aiyappa2024emergence}), which focus on individuals’ internal cognitive states without necessarily being visible to and influencing others.
% interpreted as mental representations held in memory and linked through a map~\cite{sep-belief} (represented as a network here)
This contrasts with the traditional sociophysics focus on ``opinions'', where each agent’s state is typically visible to others and influences their state updates~\cite{castellano2009statistical}.}.
We adopt here \citet{galesic2021integrating}'s inclusive definition which encompasses assumptions about states of the world, political and moral views, preferences, and subjective judgments or evaluative stances (see Section~\ref{subsec:cognitive}).
% The core assumption of network representations of belief systems is that when and how an individual's belief changes is affected by others beliefs she holds~\cite{dalege2024networks}.
% Two possible ways to specify this assumption have been examined in the literature, which differ in whether the focal individual forms new beliefs about the nodes or the edges of their cognitive network.
The core assumption of network representations of belief systems is that when and how an individual's belief changes depends on the other beliefs and associations embedded in their own cognitive system~\cite{dalege2024networks}.
The literature has formalized this idea in two related ways.
Some models encode beliefs as states of nodes, with signed edges representing perceived compatibility, opposition, or implication between issues or propositions.
Other models encode beliefs and attitudes directly as signed edges between concepts, actors, objects, or the self.

\paragraph{Beliefs as the states of nodes of a signed network} In this paradigm, each node $\alpha$ represents a concept in a given cognitive network, and it is endowed with a belief state (continuous, $b_\alpha\in[-1,1]$, binary, $b_\alpha\in\{-1,1\}$, or categorical), and the edges between the nodes represent the relationships between beliefs. For example, the nodes of such networks -- whose signed adjacency matrix we denote as $\mathbf{B}$ -- could correspond to propositions capturing moral beliefs\footnote{For example, distinct nodes may represent propositions concerning childhood vaccines, such as whether they are safe, beneficial to children or society, approved by authorities, or whether vaccination is perceived as a matter of free choice rather than coercion. A positive belief state would indicate agreement with a proposition, whereas a negative belief state would indicate disagreement~\cite{dalege2022using}.} or social beliefs\footnote{For example, distinct nodes may encode trust in different sources of information on a health-related issue, such as online health influencers, medical doctors, or scientists. A positive belief state may indicate trust in the corresponding source, whereas a negative belief state may indicate distrust~\cite{dalege2022using}.}. In this case, individual $i$'s set of beliefs about a set of $C$ concepts of interest is termed as $i$'s \textit{belief system}, and it is captured by vector
$\mathbf{b}_i=(b_{i1}, \dots, b_{iC})$, where $C$ is the number of nodes. Depending on the model, each element $b_{i\alpha}$ can be binary, categorical (ordered or unordered), or continuous~\cite{starnini2025opinion}.

This paradigm has been used in recent empirical measures of the structure and dynamics of cognitive networks from longitudinal survey data~\cite{dalege2022using}.
Some works explore how individuals form or update beliefs about each node of the cognitive network according to simple heuristics based on the network's structure -- e.g., averaging and majority rules~\cite{medo2021fragility}, dissonance minimization~\cite{dalege2022using}. For example in a network of international relations, a subject attempting to form an opinion on a set of national governments might be influenced by the sign of their perceived relation between the governments, e.g., inferred from their political interactions~\cite{askarisichani20201995}.

\paragraph{Beliefs as the edges of a signed network} This representation has its roots in the social knowledge structure from psychology research~\cite{greenwald2002unified}. The nodes represent concepts, and signed edges between them represent beliefs on the relationships between pairs of them~\cite{greenwald2002unified}.
For example, consider a scenario where an individual supports a given soccer team, and the team has been associated with match fixing (example adapted from ref.~\cite{rodriguez2016collective}). The soccer team, match fixing, and themselves are three distinct concepts (i.e., nodes) in the individual's belief network. The cognitive network features a positive link between themselves and the team they support; a positive link between the team and match fixing; and (assuming the individual values honesty) a negative link between themselves and match fixing.
Because of internal coherence mechanisms (see Section~\ref{subsec:mechanisms}), such an unbalanced configuration is likely to be solved by the individual withdrawing support for the team, which would result in the link between themselves and the team becoming negative.
With this network representation, the belief system is captured by the $C\times C$ cognitive-network adjacency matrix $\mathbf{B}$ that encompasses all the relations between the concepts, where $C$ denotes the number of concepts. When considering multiple individuals, we denote the cognitive-network matrix of individual $i$ by $\mathbf B^{(i)}$, with $B_{\alpha\beta}^{(i)}$ representing the perceived relation between concepts $\alpha$ and $\beta$. Equivalently, we denote by $\mathbf r_i=(r_{i1},\ldots,r_{iL})$ the vector containing the link values of $\mathbf B^{(i)}$, arranged in the same order for all agents, where $L$ is the number of links in the cognitive network. Moreover, relationships between concepts and the self encode the individual's beliefs about these concepts, which in the previous representation are captured by node states. This paradigm has been used in recent theoretical works on how new belief systems spread~\cite{rodriguez2016collective,aiyappa2024emergence} (see Section~\ref{subsec:spread}).

\subsubsection{Cognitive and social mechanisms of belief change}
\label{subsec:mechanisms}

Given a suitable definition of a belief system, the core question is how beliefs are formed or updated.
The literature identifies two key classes of mechanisms. One can be broadly termed as \textit{internal consistency} mechanisms: People prefer to change beliefs in order to reduce the inconsistency among their existing beliefs. This tendency is well-documented in the psychology literature, holding its roots in Festinger's cognitive dissonance theory~\cite{festinger1957theory}.
According to this class of mechanisms, inconsistency among one's beliefs leads to personal dissonance; to reduce it,
people feel the need to alter their beliefs~\cite{dalege2024networks}.
Cognitive dissonance can be captured by the presence of unbalanced triads in the cognitive networks described above.
% For example, an individual who supports a sport team (represented by a positive link between themselves and the team in their cognitive network) that has been involved in match fixing (positive link between the team and match fixing) will have an unbalanced triad in their cognitive network if they value honesty (negative link between themselves and match fixing). As a result, they are likely to change their belief about one of these links, e.g., by stopping supporting the team.
In the soccer-team example of Sec.~\ref{subsec:net_belief}, cognitive dissonance can be reduced by changing one of the three relationships, for instance by withdrawing support for the team.

Another key factor affecting belief update is \textit{social contagion}: when people make decisions, they care about their perception of what others think, say, do, and approve~\cite{bicchieri2016norms,gavrilets2024modelling}.
The connection between individuals can be captured by the adjacency matrix of the relevant social network, which we denote as $\mathbf{A}$; this network can be potentially signed\footnote{The social network itself has a certain level of structural balance, which can affect belief dynamics -- see Section~\ref{subsec:social}.} and multilayer\footnote{In this case, each layer can represent the network of interaction pertaining a given object. With this representation, one can incorporate internal coherence effects by assuming that beliefs are influenced by an individual-dependent external field~\cite{battiston2016interplay}.}.
Social contagion effects arise from diverse social and psychological mechanisms, including discomfort in deviating from existing social norms and belief systems~\cite{bicchieri2016norms}, social learning~\cite{cai2009observational}, and network externalities~\cite{katz1985network}.
Their net effect is that people tend to align their beliefs and behaviors with those of their social network contacts\footnote{Exceptions include various manifestations of counternormative behaviors rooted in psychological theory -- e.g., reactance theory~\cite{brehm1966theory} and the need for uniqueness~\cite{snyder1977abnormality}. Various works included counternormative behaviors to social dynamics models, e.g., refs.~\cite{moldovan2004cellular,mittal2024anticonformists,alipour2024enough}.
Understanding their impact on the reviewed insights on signed networks is a potential direction for future studies.}, which can be empirically described through statistical physics models~\cite{durlauf1999can,galesic2019statistical}.

\begin{figure}[ht!]
    \centering
    \includegraphics[width=0.6\linewidth]{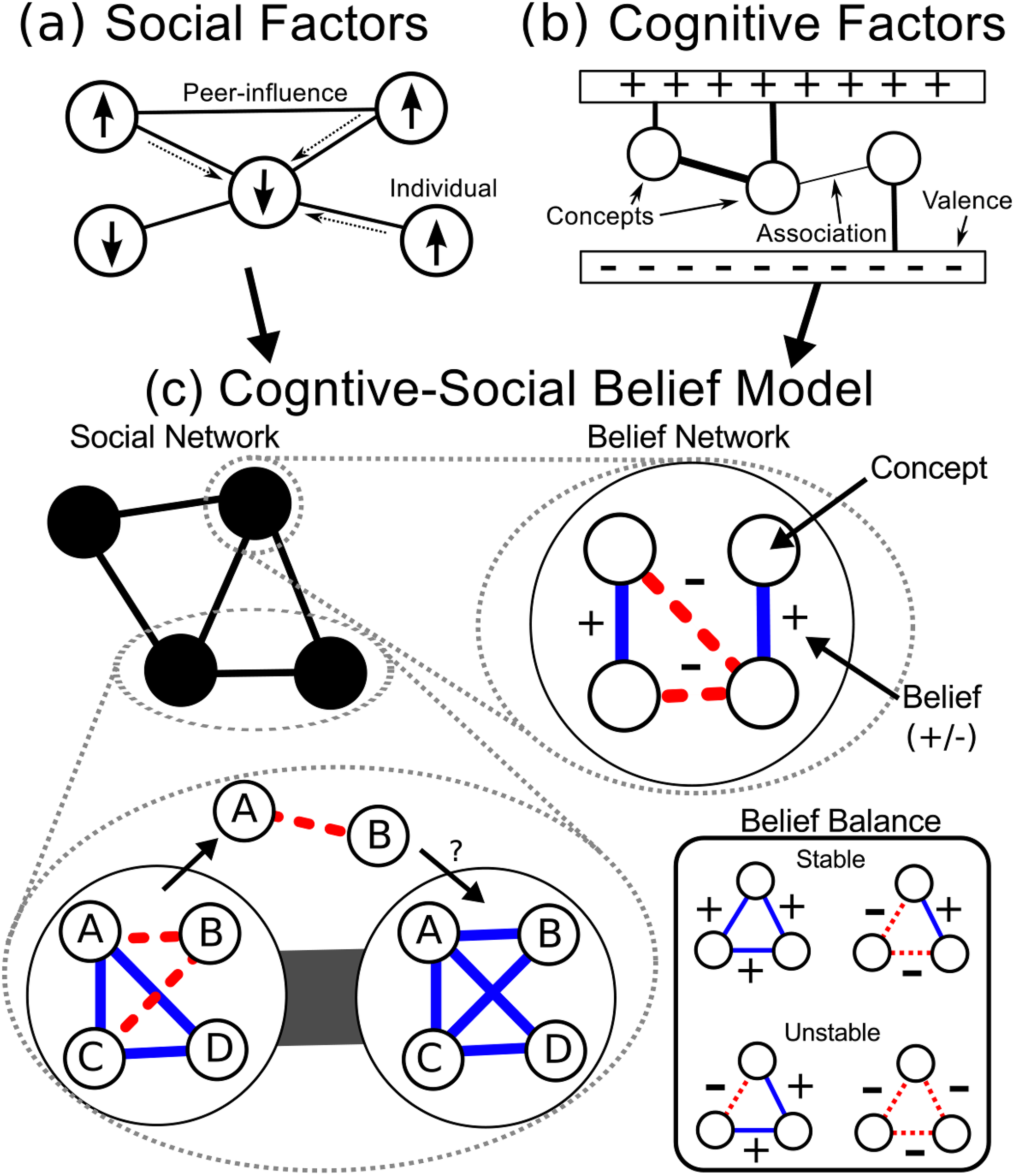}
    \caption{\textbf{An illustration of cognitive-social belief models.} People update their beliefs both due to peer influence (a) and cognitive factors (b). In models that integrate both mechanisms (c), individuals are nodes in a relevant social network and, at the same time, their belief system is represented as a belief network where concepts are related through signed links. In this modeling paradigm, individuals transmit their belief to their peers (e.g., in the figure, the negative link between A and B). The recipient's decision to accept the new belief might be influenced by both social alignment~\cite{rodriguez2016collective} and the potential impact of the new belief on the structural balance of their belief system~\cite{rodriguez2016collective,aiyappa2024emergence}.
    Reproduced from~\cite{rodriguez2016collective} under the Creative Commons Attribution License.}
    \label{fig:belief}
\end{figure}

In sum, both belief network structures (capturing the potentially signed interdependence between concepts) and social network structures (capturing the potentially signed and multilayer interdependence between belief-forming subjects) can affect beliefs' emergent properties at the individual and collective levels (see Fig.~\ref{fig:belief} for an illustration), which is the focus of this and the following Section.
%Models on the formation and spread of beliefs would deserve a review in its own, and it would be unfeasible to extensively cover all of them here. Instead, we highlight here a few theoretical results that point to the role of structural balance in shaping individual and collective belief properties, and refer the reader to refs.~\cite{castellano2009statistical,galesic2019statistical,starnini2025opinion,caldarelli2025physics} for more extensive reviews and perspectives on models of belief formation and dynamics.
We emphasize that the literature on individual belief network dynamics is much broader. Recent advances allow cognitive network structures to be inferred directly from behavioral or survey data, and use dissonance reduction mechanisms to perform out-of-sample predictions of individual belief evolution~\cite{dalege2022using}. We refer the interested reader to ~\cite{galesic2021integrating,dalege2022using,rabb2022cognitive,dalege2024networks} for related results and perspectives. Further, we refer the reader to refs.~\cite{castellano2009statistical,galesic2019statistical,starnini2025opinion,caldarelli2025physics} for more extensive reviews and perspectives on physics-inspired models of belief formation and social dynamics.
% \msmcomment{Piotr, the revised version of this paragraph should resolve the ambiguity you noticed. the footnote was a bit missplaced in my view, moved its text here and rephrased.}

\subsubsection{Formation and internal consistency of belief systems}
\label{subsec:formation}

% \pgcomment{Also, we assume "absence of other belief-forming individuals", yet later we use an example involving a “network of Twitter accounts” representing scientists vs. science deniers. Is this a contradiction?}

% \msmcomment{The assumption here is that the individual merely observes, without influencing the (assumed static) relationships. see below if my rephrasing resolves the ambiguity.}

% \pgcomment{I have one more issue with this example. First, in the paragraph we discuss beliefs, consistency etc. When thinking about beliefs, naturally, I consider impersonal issues (taxes, death penalty, view on cheating). However, in the example the described concepts are views on real people. (Maybe it is only me!) I proposed a minor change below.}

% \msmcomment{agree, replaced twitter accounts with beliefs endorsed by scientific consensus vs science deniers.}

% \pgcomment{Another potentially relevant paper, I accidentally found:
% \url{https://www.nature.com/articles/s41598-025-31643-5}
% ``The desire to avoid cognitive dissonance drives community formation in a social network model``.}

% \msmcomment{added below}

We review now recent theoretical findings on the impact of cognitive balance on the consistency of newly-formed belief systems.
We begin by examining a \textit{fixed} cognitive network
capturing the association among a set of concepts of interest as perceived by the individual (see Section~\ref{subsec:net_belief}).
A natural question concerns the impact of the network's structural balance on the consistency of the beliefs\footnote{Differently from the original paper~\cite{medo2021fragility}, we term here the spin variables as ``beliefs'' and not ``opinions'' for the reasons outlined in the beginning of Section~\ref{subsec:net_belief}.} formed by the individual about the \textit{nodes} of the network.
To address this question, \citet{medo2021fragility} analyze a simple process where the individual starts from a small set $\mathcal{C}_0$ of concepts, with $C_0=|\mathcal{C}_0|$, on which they form an independent belief; the individual eventually leverages its cognitive network $\mathbf{B}$ to form a belief about all the concepts. Whenever formed, individual belief about concept $\alpha$ is denoted as $b_\alpha=\pm 1$.
They further assume, for simplicity, that the concepts belong to one among two camps: Pairs $(\alpha,\beta)$ of concepts within the same camp are more likely to have a positive relation ($B_{\alpha\beta}=1$), whereas the opposite ($B_{\alpha\beta}=-1$) holds true for pairs of concepts belonging to different camps. Under these assumptions, belief networks $\mathbf{B}$ with different levels of structural balance can be generated.
For example, the two camps could represent two sets of health-related claims: claims endorsed by the scientific consensus and claims promoted by science-denial narratives. The individual is assumed to observe fixed perceived relations among these claims, without interacting with other belief-forming individuals.

In a perfectly consistent belief-formation process, an individual who initially accepts a small set of $C_0$ scientifically-endorsed claims should end up accepting claims from the scientific-consensus camp and rejecting claims from the science-denial camp.
Formally, denoting by $T_\alpha=\pm 1$ a variable that flags whether concept $\alpha$ belongs to camp 1, say the scientific-consensus camp ($T_\alpha=+1$), or camp 2, say the science-denial camp ($T_\alpha=-1$), a perfectly consistent belief-formation process ends up with $b_\alpha=1$ if $T_\alpha=1$ and $b_\alpha=-1$ if $T_\alpha=-1$.
It is helpful to define the degree to which the final beliefs are consistent as~\cite{medo2021fragility}
\begin{equation}
    K=\frac{1}{C-C_0}\sum_{\alpha\notin \mathcal{C}_0} b_\alpha\,T_\alpha,
\end{equation}
where the sum runs over all the beliefs that do not belong to seed set $\mathcal{C}_0$.
In what conditions does a belief formation process converge to a consistent set of beliefs?

It depends on the structural balance of the network. A perfectly-balanced network $\mathbf{B}$ is one where all the concepts of one camp are connected via positive (negative) relationships to all the concepts of the same (opposite) camp; deviations from this idealized configuration result in lower degrees of structural balance.
The dependence of belief consistency on the belief network balance can be unpacked by studying various heuristic belief-formation rules adopted by the individual.
In the simplest case, to form a belief about concept $\alpha$, the individual considers one of $\alpha$'s neighbors in the cognitive network, $\beta$, about which they already hold a belief $b_\beta$. The individual adopts the same belief on $\alpha$ when the perceived relation between $\alpha$ and $\beta$ is positive, and the opposite belief when this relation is negative.
Formally, $b_\alpha=B_{\alpha\beta}\,b_\beta$. In a perfectly balanced network, this simple rule leads to perfectly coherent beliefs, $K=1$. As soon as the network's topology deviates from perfect balance, the internal consistency $K$ of the belief system drops rapidly~\cite{medo2021fragility}. This drop is significantly faster for larger networks of concepts: $K$ decreases as a power-law of the number of concepts $C$, pointing to the fragility of belief formation in a more complex world~\cite{medo2021fragility}.

The lack of belief consistency even for mildly-unbalanced cognitive networks calls into question the reliability of individual beliefs formed through simple heuristics. Even a few perceived positive associations between scientifically-endorsed claims and claims rejected by the scientific consensus can lead an individual who initially accepts the former to accept some of the latter.
It is natural to wonder how to mitigate this problem. A natural solution for the individual would be to invest more resources into forming an independent belief about more concepts, minimizing reliance on simple heuristics. In this modeling representation, this strategy would require the individual to expand their seed belief set $\mathcal{C}_0$.
One finds indeed that the power-law decrease of $K$ as a function of network size $C$ can be prevented if the individual increases the size of their seed set linearly with $C$~\cite{medo2021fragility}.
However in complex contexts, independently forming beliefs on a large number of concepts is costly. If the budget for independent formation of beliefs about concepts is limited, an alternative way to improve belief consistency is to allocate it strategically -- either toward a cluster of close nodes or toward disperse nodes, conditional on the topology and degree of structural balance of the network~\cite{wang2025combining}.
The above modeling framework can be generalized to study more sophisticated belief formation heuristics~\cite{medo2021fragility,meng2022whom} and provide theoretical insights into the emergence of fragmentation in growing signed networks~\cite{fenoaltea2023phase,meng2023disagreement}.

\subsubsection{Simple vs. complex contagion of new belief systems}
\label{subsec:spread}

\begin{figure}[!htbp]
    \centering
    \includegraphics[width=0.5\linewidth]{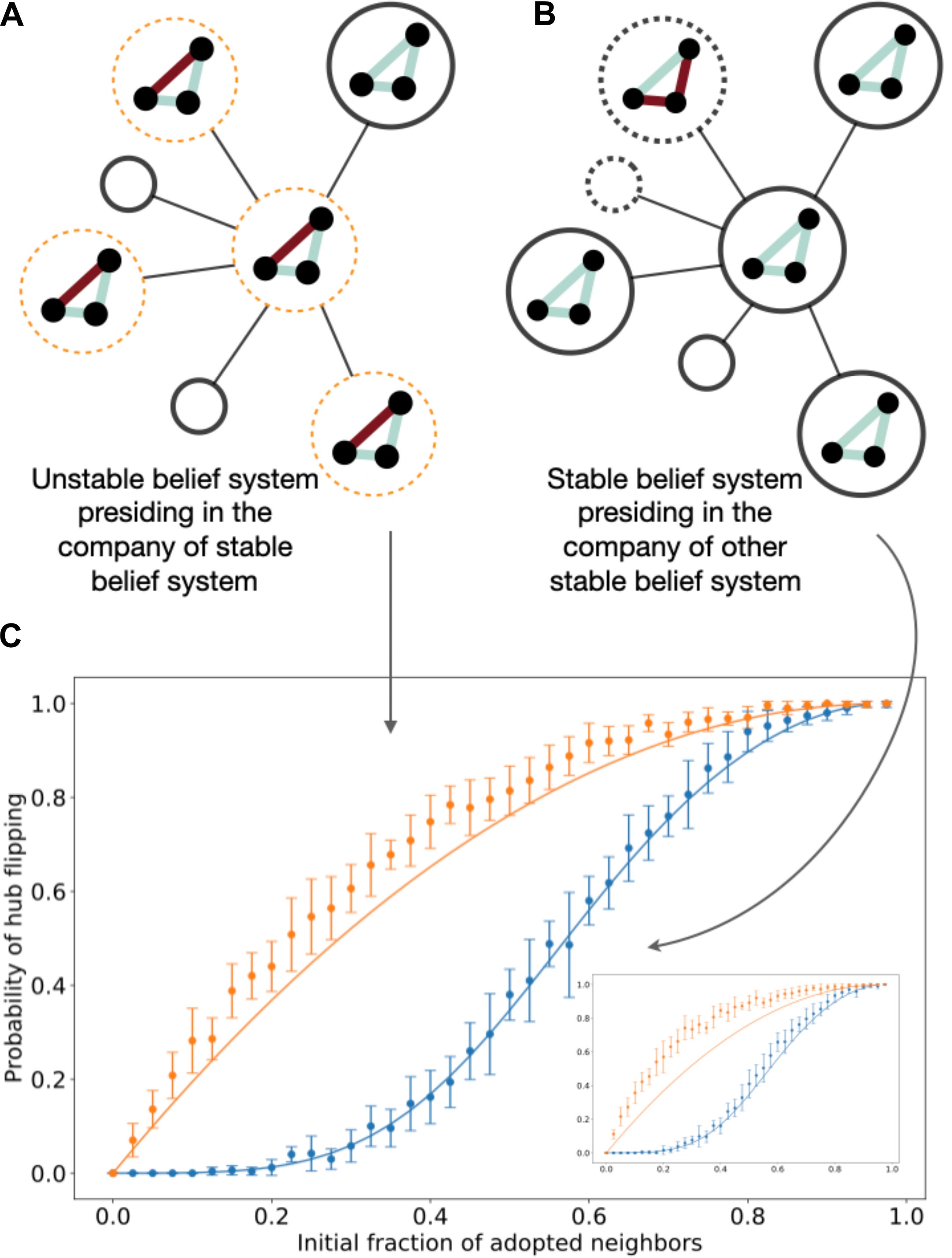}
    \caption{\textbf{New belief systems might spread as simple or complex contagions}. \textbf{(A,B)} Within an individual, concepts are related through a weighted and signed belief networks; Individuals can transmit beliefs through social ties. In this example, a hub node has six neighbors, and she holds either an unstable (A) or a stable (B) belief system. \textbf{(C)} The probability the hub holding the unstable system will adopt the stable one as a function of the number of neighbors holding the stable system follows a simple contagion (orange line); the probability the hub holding the stable system will adopt a competing stable one as a function of the number of neighbors holding the competing system follows a complex contagion (blue line). The inset in (C) illustrates a variant with dynamically evolving beliefs of some neighbors; see Ref.~\cite{aiyappa2024emergence} for details. Selected panels (A--C) reproduced from Ref.~\citep{aiyappa2024emergence} \aiyappacredit.}
    \label{fig:hub}
    % To generate permission text for arxiv or not for axiv change/comment-out lines with \arxivversionfalse or \arxivversiontrue in the preamble.
\end{figure}

The previous modeling approach focused on the effect of the structure of belief matrix $\mathbf{B}$ on individual beliefs, overlooking social contagion effects among belief-forming individuals. By combining the effects of both belief and social networks into a single model, we can understand how entirely new belief systems can spread throughout a population.
We now consider cognitive-network matrices $\mathbf{B}^{(i)}$ associated with individual agents.
% Therefore in this representation, $\mathbf{B}^{(i)}$ can be interpreted as $i$'s emergent belief system.
Unlike the models in Section~\ref{subsec:formation}, $\mathbf{B}^{(i)}$ is not fixed, but it evolves dynamically.

%\pgcomment{A comment about equations 69 and 71 on Di, and f(bjx,B(t))
%Equation Di is essentially a Hamiltonian. Eq. 71, the derivative of Di over bix, closely resembles equation 21 in Sec \ref{subsec:kk-model}, where Kułakowski model is introduced and it is shown that this model dynamics is equivalent to overdamped particle and the system goes downhill on the energy landscape.
%This analogy seems strong and may be worth explicitly highlighting, perhaps in both sections.
%} \msmcomment{nice. will study soon.}

In line with the social contagion mechanisms outlined in Section~\ref{subsec:mechanisms}, the belief-forming individuals can interact and update their beliefs based on information they receive from their neighbors.
Specifically,
individuals update their beliefs due to both a cognitive and a social contagion mechanism.
The cognitive mechanism is the minimization of the incoherence (or dissonance) of the individual's belief system, which is operationalized through the structural balance function~\cite{Marvel2009,rodriguez2016collective,aiyappa2024emergence}
\begin{equation}
    E^{(D)}_i=-\frac{1}{N_T}\sum_{(\alpha,\beta,\gamma)} B^{(i)}_{\alpha\beta} \, B^{(i)}_{\beta\gamma} \, B^{(i)}_{\gamma\alpha}=-\frac{1}{N_T}\sum_{(x,y,z)} r_{ix} \, r_{iy} \, r_{iz}
    \label{dissonance}
\end{equation}
where the first sum runs over all triads $(\alpha,\beta,\gamma)$ present in $i$'s belief network; $N_T$ denotes the total number of triads; $x,y,z$ represent indices of the elements of $\mathbf r_i$ corresponding to links $(\alpha,\beta)$, $(\beta,\gamma)$, $(\gamma,\alpha)$ in matrix $\mathbf{B}^{(i)}$. Note that Eq.~\eqref{dissonance} is the cognitive-network analogue of the structural-balance Hamiltonian in Eq.~\eqref{eq:hamil-norm}. Accordingly, the dissonance-minimization term in the update rule below can be interpreted as a force driving the belief system downhill on this energy landscape, analogously to the overdamped dynamics discussed in Section~\ref{subsec:kk-model}.

The social contagion mechanism is specified by assuming that at each step in the simulation, a randomly-selected node can transmit a randomly-selected belief to a randomly-selected neighbor; the transmitted belief influences linearly the neighbor's belief on the same link. Specifically, when $j$ transmits their belief $r_{jx}(t)$ to their neighbor $i$, $i$'s belief on $x$ evolves as~\cite{aiyappa2024emergence}
\begin{equation}
    r_{ix}(t+1)=r_{ix}(t)+f(r_{jx}(t), \mathbf{B}^{(i)}(t)),
\end{equation}
where $f$ depends on both the transmitted belief and $i$'s whole belief system at time $t$.
Function $f$ contains two terms that capture social influence and cognitive dissonance minimization, respectively~\cite{aiyappa2024emergence}:
\begin{equation}
 f(r_{jx}(t), \mathbf{B}^{(i)}(t))\sim\alpha_S\,r_{jx}(t) -\alpha_D\,\frac{\partial E^{(D)}_i(t)}{\partial r_{ix}(t)},
\end{equation}
where $\alpha_S$ and $\alpha_D$ capture the relative importance of social contagion and internal coherence, respectively; partial derivative $\partial E^{(D)}_i(t)/\partial r_{ix}(t)$ captures the direction toward which a belief should change for the internal dissonance to decrease.

Intuition about the contagion of a new belief system can be built up by considering a simple star network, where the central hub is susceptible to influence, whereas all the leaf nodes hold their belief system fixed (see Fig.~\ref{fig:hub}).
We assume that the hub has initial belief system $\mathbf{B}$; a fraction of its neighbors hold $\mathbf{B}$, whereas the others hold a different belief system $\mathbf{B}'$.
One can then study how the probability that the hub changes their belief system from $\mathbf{B}$ to $\mathbf{B}'$ depends on the fraction of their neighbors who hold $\mathbf{B}'$. If $\mathbf{B}$ is unstable and $\mathbf{B}'$ is stable, one says that the contagion that leads to belief change $\mathbf{B}\to \mathbf{B}'$ is stabilizing; on the other hand, if $\mathbf{B}$ is stable and $\mathbf{B}'$ is a competing stable system, belief change $\mathbf{B}\to \mathbf{B}'$ is destabilizing.

A core finding is that stabilizing contagion behaves like a simple contagion, meaning that the probability that the hub changes belief system is non-zero even when only a small fraction of its neighbors hold a different belief system.
By contrast, destabilizing contagion behaves like a complex contagion, meaning that the probability that the hub changes belief system remains low until sufficient social reinforcement is provided, after which it increases sharply~\cite{aiyappa2024emergence}.
Prior works argued that threshold-based, complex contagions better describe social influence in contexts where the decision-maker faces substantial levels of risk or uncertainty -- see ~\cite{guilbeault2018complex,mariani2024collective,hebert2025one,starnini2025opinion,caldarelli2025physics} for reviews.
The complexity of destabilizing contagion points to a potential cognitive pathway to complex contagion: an established stable belief system resists incompatible social influence, so stronger social reinforcement is required to destabilize it and induce a transition to a competing stable belief system.
As a consequence of this result, in agreement with the complex contagion theory~\cite{centola2007complex,nematzadeh2014optimal},
stabilizing contagion spreads more effectively in random than in highly clustered networks; for small initial seed fractions, the opposite holds for destabilizing contagion~\cite{aiyappa2024emergence}. Further, there is an optimal regime of modularity that favors the spread of a competing stable belief system in the destabilizing-contagion regime~\cite{aiyappa2024emergence}. The results~\cite{aiyappa2024emergence} suggest also a possible cognitive origin of individual heterogeneity in persuasiveness
and supportiveness in Latan{\'e}'s social impact theory~\cite{nowak1990private}. Because weakly coherent belief systems are
more susceptible to social influence, their holders may constitute less persistent sources of persuasion and support, resulting in lower effective social strength during repeated interactions.

\subsubsection{Tipping points for collective belief change}
\label{subsubsec:tipping-points}

%\pgcomment{
%I was planning to mention this paper as an example of structural balance in multilayer networks. Two possible approaches:
%(a) I briefly mention it in my section and refer readers to 4.1.5, or
%(b) I describe its dynamics in more detail in the Emergence section, allowing us to remove parts such as the adoption probability* and temperature explanation from 4.1.5.
%*Right now, when someone used Metropolis/Glauber etc., I wrote they employed Metropolis/Glauber dynamics assuming it will be obvious for the reader. Or do you think, these methods should be briefly explained the first time they are mentioned?} \msmcomment{I'd leave to you the final decision on where to present this paper. We could either explain them the first time, and then refer the reader to that section whenever we mention it again, or (better in my view) create an appendix.}
%\pgcomment{I went with idea (a) 'I briefly mention it in my section and refer readers to this section'.
%About Metropolis: TODO: create appendix; next on my list. }

%That internally coherent belief systems face less resistance against spreading than internally incoherent ones echoes the theoretical results from a previous study~\cite{rodriguez2016collective} that examined the joint effect of internal coherence and social alignment on collective belief change.^
The different spreading mechanisms of coherent versus incoherent belief systems manifest themselves in phase transitions separating regimes of homogeneous and fragmented collective opinions, and in tipping points at which a sufficiently large minority can overturn the prevailing majority belief.
To uncover these emergent phenomena, \citet{rodriguez2016collective} examined the joint effect of internal coherence and social alignment on collective belief change. They considered an individual-level energy function composed of the individual's dissonance (akin to Eq.~\eqref{dissonance}) and the degree of alignment between their beliefs and those of their peers:
\begin{equation}
    E_i=\alpha_D \, E^{(D)}_i+\alpha_S\,E^{(S)}_i,
\end{equation}
where $E^{(D)}_i$ has a form similar to Eq.~\eqref{dissonance}; $\alpha_D$ and $\alpha_S$ denote the importance of internal coherence and peer influence, respectively; $E^{(S)}_i$ denotes the social alignment perceived by $i$, defined as
\begin{equation}
    E^{(S)}_i\propto -\sum_j A_{ij}\,\mathbf{r}_i\cdot\mathbf{r}_j,
\end{equation}
where $\mathbf{r}_i$ is the vector containing the link values of $\mathbf{B}^{(i)}$ and $\mathbf{A}$ denotes the adjacency matrix of the social graph. The social alignment term can also be interpreted as structural social dissonance in triads composed of two individuals and opinions;  see also Sec. \ref{subsubsec:sb-agent-attribute} and Eq. \eqref{eq:hamil-impersonal}.
Similar to ref.~\cite{aiyappa2024emergence},~\citet{rodriguez2016collective} studied a process where throughout the dynamics, pairs of individuals are selected at random, and one individual communicates one of their beliefs to the other. The receiver accepts the belief according to the Metropolis update rule (see~\ref{app:mc}) with temperature $T$.
% if it decreases their energy; if it increases their energy, they will accept it with probability given by the corresponding Boltzmann factor, $\exp{(-\Delta E/T)}$.
The temperature plays a dual role: higher $T$ facilitates social contagion by making it easier for people to accept received beliefs that contradict their own belief system; at the same time, by facilitating social contagion, higher $T$ also helps the collective beliefs to align. In this sense, $T$ plays the role of a temperature for individual belief coherence, but a role of an inverse temperature for the system-level alignment~\cite{rodriguez2016collective}.

The model is suited to link the emergence of consensus (specified as the relative size of the population that adheres to the most popular belief system) to the micro-level parameters of belief update.
There are several key takeaways~\cite{rodriguez2016collective}. First, as expected, consensus benefits from higher temperature, lower importance of internal coherence, and higher peer influence, all of which facilitate social contagion and alignment. The effects of each of the parameters are highly nonlinear, with critical points separating regions of the phase space where consensus is vs. is not achieved (see Fig.~1 in ref.~\cite{rodriguez2016collective}). Second, the critical mass of a committed minority required to overturn a mainstream belief system is smaller the more coherent the new belief system. Finally, ``cults'' (interpreted as minority belief systems that contradict the mainstream) can survive even in non-fragmented networks, as long as people place higher importance on internal coherence than social alignment.
These findings suggest that the level of polarization in a society is an emergent phenomenon critically determined by the micro-level processes through which individuals form opinions. Dysfunctional or inaccurate belief systems might persist when they are internally coherent, particularly if individuals place greater value on maintaining internal coherence than on external accuracy.

The tipping behavior induced by committed minorities~\cite{rodriguez2016collective} is closely related to the discontinuous transitions found in social-impact models with a strong leader and social noise~\citep{kacperski1996phase,holyst2000phase}. In the strong-leader model, a localized cluster of followers coexists with a homogeneous leader-dominated state, and an increase in the leader strength or
social temperature may produce an avalanche-like transition between these states. In the model of~\citet{rodriguez2016collective}, belief
coherence plays a different but analogous role: it does not directly increase the interpersonal impact exerted by the zealots, but reduces the rate at which converted agents abandon the transmitted belief system, thereby lowering the critical density of the committed
minority and reinforcing the contagion process. The two models thus suggest a common nucleation-and-positive-feedback
mechanism. The analogy between the two temperature parameters is however  limited:
susceptibility $T$ in the cognitive--social belief model facilitates the
acceptance of socially transmitted beliefs and hence collective alignment,
whereas the social temperature in the strong-leader model represents
independent behavioral noise, which may first trigger an avalanche but
ultimately destroys collective order at sufficiently large values.

\subsection{Impact of social balance on opinion and spreading dynamics}
\label{subsec:social}

% \pgcomment{What is your approach here? I understood the distinction is that beliefs are internal states, while opinions are how others perceive our state, how this state is visible in the network. Most opinion/polarization models work with the latter. Shouldn't one use opinion here?} \msmcomment{My approach was: In 4.1, we need to use beliefs (internal states). In 4.2, both belief and opinions might work, as beliefs are indeed perceived by others. Probably opinions is more precise and I would replace belief with opinion, as you suggest.}

So far, we have focused on cognitive balance, that is, the structural balance of internal belief networks that shape how individuals form and update their beliefs. In this subsection, we shift attention to social balance, namely the structural balance of the social networks through which information, opinions, and behaviors spread. Whereas cognitive balance concerns the internal coherence of an individual's belief system, social balance concerns the pattern of positive and negative relations among interacting individuals and groups, and how this pattern constrains or amplifies collective dynamics. The literature on opinion dynamics and contagion processes in signed networks is already extensive, and a comprehensive review would go beyond the scope of this article. Instead, we focus on a few illustrative families of models describing the coevolution of network structure and opinion polarization (Section~\ref{subsec:coev}), the spread of simple and complex contagions (Section~\ref{subsec:contagion}), and collective decision-making (Section~\ref{subsec:cdm}).

\subsubsection{Impact of social balance on opinion polarization}
\label{subsec:coev}

The models reviewed in Section~\ref{subsec:cognitive_balance} that incorporate social contagion mechanisms assume that social networks offer an underlying structure throughout which new belief systems can propagate. They focus on the impact of cognitive balance for collective belief outcomes. At the same time, the balance of the underlying social network plays a key role for collective outcomes as well.
A class of models examines the impact of the structural balance of a fixed social network structure on the polarization of the resulting opinions\footnote{To align with the typical terminology in the sociophysics literature, we refer to the spin variables entering the models reviewed in this Section as ``opinions'', interpreted as beliefs that are visible to others. See also Footnote~\ref{foot:beliefs} at the beginning of Section~\ref{subsec:cognitive_balance}.}~\cite{altafini2012dynamics,altafini2012consensus,shi2019dynamics,Li2019Binary,zhai2021dynamic}. \citet{altafini2012dynamics} considers a dynamical system of the general form:
\begin{equation}
    \mathbf{\dot{b}}=f(\mathbf{b}),
\end{equation}
where $\mathbf{b}\in \mathbb{R}^N$ is an $N$-dimensional vector composed of real-valued opinions of $N$ social network's nodes. In this class of models, the influence of $j$ over $i$ is described by
\begin{equation}
    F_{ij}=\frac{\partial f_i(\mathbf{b})}{\partial b_j}.
\end{equation}
Positive and negative influences correspond to $F_{ij}>0$ and $F_{ij}<0$, respectively. One assumes that the sign of $j$'s influence on $i$ is independent of the specific opinion $\mathbf{b}$ being considered, a condition referred to as sign stability:
\begin{equation}
\text{sign}(F_{ij}(\mathbf{b}))=\text{sign}(F_{ij}(\widetilde{\mathbf{b}}))
\end{equation}
for any \(\mathbf b,\widetilde{\mathbf b}\in\mathbb R^N\); one can define thus
\begin{equation}
    A_{ij}=\text{sign}(F_{ij}(\mathbf{b}))
\end{equation}
at an arbitrary point $\mathbf{b}$, and assume the same sign patterns hold across the entire opinion space.
As the matrix $\mathbf{A}$ captures the influence of each individual on another, it can be interpreted as the ``sociomatrix'', i.e., ``the adjacency matrix of the signed graph describing the social network''~\citep[p.1]{altafini2012dynamics}.

Equipped with this framework, one can examine how the structural balance of the signed sociomatrix affects the long-term behavior of opinion dynamics.
A key theoretical result~\citep{altafini2012dynamics} is that a structurally balanced signed network can be transformed into an equivalent network with only positive interactions.
This transformation, known as a \emph{switching transformation} in signed-graph theory, corresponds to a $\mathbb{Z}_2$ \emph{gauge transformation} in statistical physics~\citep{Marvel2009,altafini2012dynamics}.
Assigning each node a factor $s_i\in\{-1,+1\}$ and defining
$\mathbf{D}_s=\operatorname{diag}(s_1,\ldots,s_N)$,
the signed adjacency matrix transforms as
\begin{equation}
    \mathbf{A}'=\mathbf{D}_s\mathbf{A}\mathbf{D}_s,
\end{equation}
or, equivalently, $A'_{ij}=s_i A_{ij}s_j$.
The sign product along every cycle is invariant under this transformation, since each node factor appears twice in the product.
For a structurally balanced network, the factors $\{s_i\}$ can be chosen such that all existing links in $\mathbf{A}'$ are positive~\citep{altafini2012dynamics,shi2019dynamics}.

Applying the same transformation to the opinion variables,
$\mathbf{b}'=\mathbf{D}_s\mathbf{b}$, transforms the influence matrix as
$\mathbf{F}'=\mathbf{D}_s\mathbf{F}\mathbf{D}_s$.
For a structurally balanced sociomatrix, all nonzero interpersonal influences in the transformed system are therefore positive, so the dynamics is equivalent to that of a purely cooperative system~\citep{altafini2012dynamics}.
This equivalence implies a predictable dynamics.
In a purely cooperative system where all influences are positive, if a single individual has an opinion at time $t=0$ ($b_i(0)\neq 0$), the whole system aligns in sign.
For a structurally balanced sociomatrix, the result above implies that a polarized equilibrium emerges in which nodes in the same faction align and nodes in opposite factions converge to values of opposite sign compared to the initial seed's opinion~\citep{altafini2012dynamics}.
Generalizations of this modeling approach include state-dependent susceptibility to influence and nonlinear dynamics~\cite{zhai2021dynamic}, time-varying networks~\cite{meng2016behaviors},
and more (see ref.~\cite{shi2019dynamics} for a review).

A similar in-group-agreement /out-group-rejection mechanism has been studied in economics with agents that are strategic and follow a Bayesian updating process by~\citet{DellaLena2026affective}.
They embed this mechanism in a social-learning model with a possibly biased media outlet and characterize when affective polarization produces ideological polarization, persistent disagreement, or a failure to learn the truth.

The models above typically treat the topology of the network as not coevolving with the nodes' opinions. This assumption is relaxed by studies that investigate the coevolution of structure and opinions~\cite{Gorski2020,MinhPham2020,Liu2021,pham2022_SBexplainedThroughHomophily,aghbolagh2020balance,mei2022structural,kovacs2025desire}.
In these models, the sign of a social link $(i,j)$ is dictated by the alignment between $i$'s and $j$'s opinions, e.g., as $A_{ij}=\text{sign}(\mathbf{b}_i \cdot \mathbf{b}_j)$~\cite{Gorski2020}.
As a result, different opinion configurations lead to different signed network structures, which lead to different levels of social tension captured by a suitable energy function $H$.
People update their opinions (and as a result, the signs of their social connections) with the objective of minimizing their perceived social tension.
Models differ in whether they assume that social tension is perceived at the dyadic level (meaning that individuals perceive tensions when their opinions misalign with those of their social contacts; see the models reviewed in Section~\ref{subsubsec:sb-agent-attribute}) or at the triadic level (meaning that individuals perceive tensions when their opinions cause them to be part of unbalanced triads; see the models reviewed in Section~\ref{subsubsec:combined-interpersonal-agent-attribute}).

These models are typically used to explain the emergence of structural balance, which is why we reviewed them and their implications in earlier sections.
We emphasize here that simultaneously, they hold implications for the collective properties of opinions, which is the focus of this Section.
A paradigmatic example is \citet{MinhPham2020}'s model, introduced in Section \ref{subsubsec:combined-interpersonal-agent-attribute}. In this model each agent (individual) possesses a binary spin variable (opinion), and social tension is described by a Hamiltonian function, see Eq. \eqref{eq:pham}.
In this energy function, the first term captures the tendency of individuals to connect positively with those holding similar opinions and negatively with those holding different opinions (i.e., \textit{homophily}, or, in the terminology of Section \ref{subsec:sign-attribute-dynamics}, \textit{impersonal structural balance}).
The second term captures the structural balance of the social network, and the third term represents external influence that penalizes negative links. Another recent work shows that cognitive dissonance avoidance alone can generate a wide range of emergent community structures, from consensus to polarization and fragmentation, even when agents start from identical internal belief systems~\cite{kovacs2025desire}.

Apart from analyzing the emergence of balanced states (which can be treated as polarized configurations, see Sections \ref{subsubsec:combined-interpersonal-agent-attribute} and \ref{subsec:netpol}), this framework also allows
% One can
the study of observables that depend on the individuals' opinions. Examples include the absolute value of the ground-truth energy, which captures the social tension in the system, and opinion diversity, characterized by
$m_{\mathrm{od}}=N^{-1}|\sum_i b_i|\in [0,1]$, where $m_{\mathrm{od}}=0$ and $m_{\mathrm{od}}=1$ correspond to a polarized opinion state and perfect consensus, respectively.
Even in already balanced networks, tuning external field $h$ can still alter opinion diversity, causing a sharp transition from high to low diversity at a critical value of external influence $h_c$~\cite{MinhPham2020}. These findings suggest that under strong exogenous influences, a society might abruptly transition from a fragmented configuration toward a global consensus.
Observables related to the collective properties of opinions have also been examined in dyadic models~\cite{pham2022_SBexplainedThroughHomophily}.
Taken together, these results suggest that the macroscopic organization of opinions emerges from the coupled dynamics of social balance and individual belief updating, rather than from either mechanism alone. Future research could further develop models that integrate social and cognitive balance, and develop experimental designs~\cite{galesic2025experimental} to empirically disentangle different mechanisms and their implications.

\subsubsection{Simple and complex contagions on signed social networks}
\label{subsec:contagion}

%\pgcomment{After hearing about this signed networks review, I had a question whether we should adjust what we write to what is there. If yes, then this section could have many overlaps. But maybe it is totally fine. } \msmcomment{Good point. My sense is that as our review is technical, we go much more in detail on how the models are specified and the exact results. Their book is informative but seems to stay at the high level, without entering into too much details. So I think it’s fine. I’ll double check to make sure we agree with the main conclusions. Relevant sections in their book are 5.1.3 and 5.1.4}

The structural balance of social networks also affects how information, products, and behaviors spread.
These processes are typically described by simple or complex contagion mechanisms.
In simple contagion models, one single exposure to a new idea or behavior is enough for an individual to reshare or adopt it. This class of models describes well how information spreads (especially when in relation to non-controversial, low-risk topics~\cite{notarmuzi2022universality}), but it has been found to be inadequate to describe the spreading of items that require significant levels of personal investment in the form of risk, cost, or uncertainty~\cite{guilbeault2018complex,mariani2024collective}.
These processes are better described by complex contagion models, which capture the idea that before adopting, individuals require a sufficient level of social reinforcement from their peers~\cite{granovetter1978threshold,centola2007complex}.
While studies on the effect of network structure on simple or complex contagions have traditionally focused on networks composed of positive links, a growing stream of studies explores the impact of signed ties on spreading.

\paragraph{Simple contagion} Appreciating the impact of structural balance on simple vs. complex contagion requires specifying specific spreading models on signed networks.
Simple contagions are typically described as epidemic spreading processes, where each node who has received the information has a certain transmission rate affecting their probability to transmit it to their neighbors.
In a simple susceptible-infectious-susceptible (SIS) model, infected individuals can infect their susceptible peers at a given transmission rate $\beta$, and infected individuals can recover and become susceptible again at a given recovery rate $\mu$~\cite{barabasi2016network}. These models can be modified to account for negative interactions in various ways~\cite{fan2012analysis,Li2015,Li2021,saeedian2017epidemic,ya2021information}, typically reflecting that the transmission rate between two nodes is affected by both the sign of the link and the structural balance of the triads they participate in.

Different epidemic models can be examined.
For example, \citet{Li2021} modified the classical SIS model to account for the signed structure of social networks. They posited that the transmission rate is different between positive and negative links.
For a positive link between two given nodes $i$ and $j$, the transmission rate $\beta_{ij}^{+}$ is higher the higher the number of balanced triads the link is part of.
Conversely, for a negative link between two given nodes $i$ and $j$, the transmission rate $\beta_{ij}^{-}$ is higher the higher the number of unbalanced triads the link is part of.
Hence, the links' transmission rates depend on the signed topology of the graph.
They find that the epidemic thresholds of signed networks are smaller than those of unsigned networks, suggesting that the signed structure of the networks favors the spreading.
In an earlier work, \citet{Li2015} studied a higher-order spreading model where, when activated node $i$ aims to influence their non-activated neighbor $j$, they select a discussion partner $k$ at random; the transmission probability from $i$ to $j$ is larger when there exists a balanced triad $(i,j,k)$ in the social network compared with cases in which no such triad exists or an unbalanced triad is present.
Through this simple assumption, they identify theoretical conditions under which negative links might actually increase individuals' ability to influence their peers~\cite{Li2015}.
Beyond epidemic spreading, random-walk diffusion offers another class of physics-inspired models of simple contagion on signed networks~\cite{tian2022spreading,babul2025strong}.
In these models, a single encounter suffices for the walker to move along a link, and unbalanced networks dampen diffusion compared to balanced or unsigned ones~\cite{tian2022spreading}.

%RW. \citet{tian2022spreading} studied a random-walk process on signed networks, finding that unbalanced networks dampen diffusion compared to balanced or unsigned ones. \msmcomment{mention? \url{https://www.nature.com/articles/s44260-025-00027-1}}

Various assumptions of the above works can be relaxed, e.g., to account for the interplay between opinion and epidemic spreading~\cite{she2022networked,feng2024heterogeneous}, competing cascades~\cite{srivastava2015social}, higher-order dynamics~\cite{ma2025social}, the coevolution of spreading and signed network structure~\cite{saeedian2017epidemic,mohandas2025structurally}. Such extensions can lead to qualitatively new insights. For example, \citet{saeedian2017epidemic} considered a variant of the SI model where no transmission through negative links is possible. They model the coevolution of network structure and epidemic spreading, finding that when starting from an unbalanced network topology, jammed states (see Section~\ref{subsec:jammed} for the definition) might emerge. These states are characterized by the coexistence of multiple clusters of susceptible and infected nodes, and in some cases they might protect the system from large-scale outbreaks~\cite{saeedian2017epidemic}. We refer to Section~\ref{subsubsec:context-dependent-updates} for a more detailed description of their model.

Together, these studies suggest that the effect of structural balance on simple contagion hinges on the assumptions about the spreading mechanism. On fixed signed networks, balanced structure generally promotes diffusion by amplifying propagation and lowering epidemic thresholds, whereas in coevolving dynamics with non-transmitting negative ties, the drive toward balance can fragment the system and limit global outbreaks.

\paragraph{Complex contagion} Differently from simple contagion models, threshold-based spreading models reflect the idea that in the presence of risk or uncertainty, individuals require social reinforcement before adopting a new product or behavior~\cite{granovetter1978threshold,centola2007complex}.
These models align with the complex contagion theory~\cite{centola2007complex,guilbeault2018complex}, which has demonstrated that the existence of non-trivial thresholds makes the resulting spreading processes -- referred to as complex contagion -- fundamentally different from traditional epidemic spreading processes.
In a typical threshold-based process, an individual $i$ adopts a new product or behavior (i.e., they are ``activated'') if and only if at least a threshold fraction $\tau_i$ of their neighbors is already active~\cite{watts2002simple,centola2007complex}.
These works typically assume a reference social network only composed of positive links. How would a complex contagion unfold on a signed network?

Threshold-based models on signed networks typically assume that negative links inhibit diffusion.
\citet{Lee2023} posit that a given inactive node $i$ is activated only if two conditions are true: (1) similar to the original threshold model~\cite{watts2002simple}, the ratio between the number of active positive neighbors and the total number of positive neighbors exceeds the node's threshold, $\tau_i$; (2) none of the node's negative neighbors is active. Hence in order to adopt, a node needs not only social reinforcement, but also absence of adoption among their negative connections. They find that compared to networks only formed by positive links, negative links shrink or prevent the emergence of global cascades~\cite{Lee2023}. At the same time, given a fixed proportion of negative links, decreasing the degree of structural balance of empirical networks (by reshuffling the signs of the links) shrinks the global cascades~\cite{Lee2023}.
This result echoes prior theoretical findings showing that structural balance can increase the speed and magnitude of cascades generated by threshold-based processes, while amplifying polarization~\cite{he2019information}.
These findings suggest that given the number of edges and the sign distribution, the sign structure observed in real networks favors the spreading of complex contagions, with the level of structural balance playing a positive role for spreading.

A more nuanced picture on the impact of structural balance on complex contagions emerges from a more recent study~\cite{de2024complex}. Instead of studying threshold-based spreading on a simple network, \citet{de2024complex} considered a generalized epidemic spreading process on second-order simplicial networks.
This higher-order epidemic-spreading model has a complex contagion component as the rate at which a given node changes its state is affected by the exposure to the decisions made by different neighbors~\cite{iacopini2019simplicial}.
In particular, state change processes can occur at the node level, at the link level, or at the triad level -- the latter possibility capturing social reinforcement effects~\cite{iacopini2019simplicial}. At the node level, they assumed that the nodes can recover at any time with a given rate $\mu$. At the link level, susceptible individuals can be infected with rate $\beta_1$ by interacting with a trusted infected individual; on the other hand, an infected individual can become susceptible again with rate $\beta_2$ by interacting with a distrusted infected individual. Similar to ~\cite{saeedian2017epidemic,Lee2023}, this assumption reflects the tendency of individuals to not adopt the same behaviors as distrusted individuals. At the triad level, they consider five possible processes, including three infection processes and two recovery processes. For example, if the three agents in the triad all trust each other, then susceptible individual $i$ becomes infected with rate $\gamma_1$; conversely, if an infected individual $i$ distrusts both interaction partners in the triad, they will become susceptible again with rate $\gamma_4$ -- we refer to the original paper for all the interaction types~\cite{de2024complex}.

The effect of structural balance on the epidemic threshold of the resulting model is nuanced.
A higher proportion of balanced triangles composed of three positive links favors the emergence of a bistable regime where an endemic state is possible even below the epidemic threshold of the first-order model; by contrast, a higher proportion of balanced triangles with two negative links and one positive link shrinks the spreading size~\cite{de2024complex}.
These findings suggest that the effect of structural balance on contagion can be positive or negative conditional on the relative prevalence of the two different types of balanced triads.

Finally, a recent study~\cite{vendeville2025modeling} used both simple and complex contagion models to link the structural balance of the social network, measured with the triadic-influence metric of \citet{ruiz2023triadic}, with the presence of echo chambers, revealed by a metric which captures the correlation between node state and group membership.
They found that structural balance tends to correlate with the presence of echo chambers.
Yet exceptions exist: echo chambers can also arise in antibalanced networks, which suggests that structural balance and the emergence of echo chambers are not in a one-to-one correspondence~\cite{vendeville2025modeling}.

Taken together, the reviewed findings suggest that whether social balance amplifies or suppresses emergent properties of spreading processes (such as cascade size, speed, and echo chamber effects) depends critically not only on network topology, but also on the detailed micro-level spreading mechanisms. While this general conclusion echoes similar conclusions for social spreading in unsigned networks~\cite{mariani2024collective}, the mechanisms affecting collective outcomes in signed networks require a more nuanced understanding of how negative ties and unbalanced triads affect peer-to-peer influence.

\subsubsection{Collective decision-making}
\label{subsec:cdm}

Collective decision-making refers to processes whereby groups make decisions or solve problems by updating and aggregating individual opinions~\cite{bang2017making}.
These processes manifest themselves in political systems, both at the national and international levels, which is why understanding them helps address questions at the heart of the functioning of democratic processes.
Crucially, collective decision-making is often mediated and influenced by the structure of the underlying communication and interpersonal relationship networks within the group~\cite{centola2022network}.
In this spirit, structural balance can play a key role for collective decision-making by affecting whether groups can reach consensus and how long this takes (see refs.~\cite{altafini2012consensus,altafini2014predictable,fontan2021role} for theoretical results and ref.~\cite{shi2019dynamics} for a review), with implications for political outcomes~\cite{fontan2021signed} and other forms of collective action and cooperation~\cite{he2018evolution}.

An application concerns government formation, where structural balance has been used to explain the duration of government negotiation talks after national general elections~\cite{fontan2021signed}. This duration has implications for political stability, policymaking, and public trust.
For example, after the 2010 general election in Belgium, it took 541 days to form a government, whereas in Norway in 2021 the process was completed in only 31 days.
What can explain these differences?

Patterns in the parliamentary discussions can be captured by a signed network where each node represents a member of the parliament~\cite{fontan2021signed,Gallo2024}.
Members of the parliament belonging to the same party are connected by a positive edge ($+1$), while members of different parties are connected through links whose sign is influenced by their ideological spectrum and pre-electoral coalition~\cite{fontan2021signed}. From the resulting signed adjacency matrix, one can derive an energy function of the individual votes whose minimum $\zeta$ over all possible partitions of political parties into two groups quantifies the frustration of the system, i.e., its distance from a structurally balanced state~\cite{fontan2021signed,Gallo2024}.
%  \msmcomment{piotr's comment: This should have a reference to some formula for frustration measures. If such formula are not there, they should be added. @georges}
A relevant question is: Does the level of frustration correlate with the duration of the negotiation talks?

Fontan and Altafini~\cite{fontan2021signed} find empirically that in European parliaments, such correlation is positive and significant.
%How to explain it?
Through a simple linear opinion dynamics model, they show theoretically that one can expect $\tau\sim(1-\zeta)^{-1}$, where $\tau$ denotes the duration of the government negotiation phase and $\zeta$ denotes the frustration of the underlying interaction network. This finding can be obtained by assuming that opinions evolve according to a nonlinear differential equation where the change of an agent's opinion depends on two terms~\cite{fontan2021role}: (1) a dampening term; (2) a nonlinear social-influence term which captures the influence among network neighbors, weighted by a parameter $\pi$ that captures the strength of social commitment among the agents. This model has a bifurcation at $\pi_1=(1-\lambda_1(\textbf{L}))^{-1}$, where $\lambda_1(\textbf{L})$ denotes the smallest eigenvalue of the normalized signed Laplacian associated to the dynamics, and one finds numerically~\cite{fontan2021role} that $\tau\sim \pi_1$ and $\lambda_1(\textbf{L})\sim \zeta$: The higher the frustration $\zeta$, the higher the distance of the network from a balanced configuration, the longer the time needed for consensus. Taken together, these results illustrate how structural balance, captured by frustration $\zeta$, provides a bridge between the microscopic structure of signed interaction networks and the macroscopic timescales of collective decision-making.

\subsection{Structural balance and network polarization}
\label{subsec:netpol}

% Structure of this subsection:
% - what do we mean by polarization in this section
% - relation between network polarization and strcutural balance
% - polarization measures based directly on balance measures
% - popular application of those measures: US parliament
% - Lerner approach (regression)
% - weak polarization - weak balance
% - should all balanced triads be included in polarization measures?

Having reviewed how social balance affects opinion dynamics, contagion, and collective decision-making, we now turn to a related but distinct question: how structural balance has been used to define and measure network polarization.
There are many definitions of social polarization. A common distinction separates \emph{opinion polarization} from \emph{network} (or group, identity, affective) polarization \citep{Rawlings2022, YoungPedersen2021}. The impact of structural balance on opinion polarization has already been discussed in Sec. \ref{subsec:coev}. Here, we focus on the relationship between balance and network polarization (see \citep{Interian2022} for a review of polarization measures).

Network polarization refers to the tendency of agents to form groups such that relations within groups are predominantly positive, while relations between groups are predominantly negative. In signed networks, this corresponds to a partition into internally friendly and mutually hostile groups.
Such group structures are often associated with differences in opinions, with groups tending to support certain viewpoints while opposing those held by others~\citep{Rawlings2022}. In such a case, the distinction between opinion polarization and network polarization becomes blurred.
However, this alignment is not necessary: network polarization can emerge purely from dynamics of social ties, without explicitly modeling agents’ opinions~\citep{Hummon2003}.

The notion of groups internally friendly and hostile with other group(s) resembles structural balance. In particular, strongly or weakly balanced networks naturally correspond to polarized configurations with two or more antagonistic groups.
As a result, many studies equate polarization with balance and employ structural balance measures as measures of polarization. Examples include the density of balanced triads \citep{Neal2020}, the negative of the normalized triadic Hamiltonian \citep{MinhPham2020}, measures based on longer cycles with or without length-dependent weighting \citep{YoungPedersen2021}, walk-based measures \citep{talaga_polarization_2023}, and the frustration index \citep{aref2020detecting, YoungPedersen2021}.
%citet{Neal2020 } uses density of balanced triads. \citet{Pham2020} use triadic normalized Hamiltonian, Eq (xx).
%\citet{Young Pedersen} use density of longer cycles and balance measure taking into account longer triads with decreasing weight, a measure defined in [Kirkwell, Cantwell, Newman], see Sec. YY.
%They also employ frustration index. Frustration index was first applied as a polarization measure by  \citet{Aref2020Detecting}.
These measures have been applied in empirical studies. In particular, analyses of signed networks derived from interactions among members of the U.S. Congress and Senate indicate a steady increase in polarization in the period 1979-2016, as reflected consistently across different balance-based metrics \citep{Neal2020, aref2020detecting, talaga_polarization_2023}.

% Analysis of signed networks extracted for US congress and senate revealed an increase of polarization (i.e., more triads became balanced) in the period 1979-2016. Apart from density of balanced triads, \citet{Aref2020Detecting} employs also frustration index for the same dataset, showing that both metrics show similar outcomes leading to similar conclusions concerning polarization increase.

Structural balance has also been incorporated into regression-based approaches. \citet{Lerner2019Network, Lerner2019Team} study team performance using explanatory variables related to the formation of balanced triads. Although they do not define polarization explicitly using a balance measure, their framework implicitly links higher levels of balance with stronger polarization. Their findings are also discussed in Sec.~\ref{subsubsec:teams}.

\citet{YoungPedersen2021} differentiate between \textit{strong} and \textit{weak} polarization, in analogy to strong and weak structural balance. They introduce corresponding metrics that capture these two regimes, allowing for the identification of polarization in systems with more than two antagonistic groups.
Interpreting structural balance measures as indicators of polarization is natural in systems far from global consensus or in contexts where distinct groups are expected and the key question concerns the extent of their separation (e.g., the parliamentary networks). However, despite its intuitive appeal, this interpretation has important limitations. In particular, a network dominated by positive links may be close to a structurally balanced (paradise) state while clearly not being polarized. As a result, balance-based measures can overestimate polarization in systems near consensus. Although \citet{YoungPedersen2021} acknowledge this issue, their proposed metrics do not explicitly exclude such cases.

An alternative approach addressing this limitation was proposed by \citet{Gorski2023}, who distinguish between local and global polarization. Local polarization is defined as the density of triads containing two or three negative edges. From the local (i.e., triadic) perspective, such triads can be interpreted as consisting of two or three antagonistic groups.
Triads with no negative edges are excluded as non-polarized.
This definition aligns naturally with the notion of weak polarization, in which systems may consist of multiple mutually hostile groups.
% and are closely related to weak structural balance.
% , thereby excluding fully positive triads. From the local (i.e., triad) perspective, triads with no negative edges are not polarized. Triads with 2 or 3 negative edges can be drawn as comprising of 2 or 3 hostile groups. Thus, this polarization metric is related to weak polarization because weakly balanced systems consisting of multiple hostile groups are considered as polarized.
Global polarization, in turn, is defined as the existence of a partition into $K>1$ groups, i.e., a weakly balanced but non-paradise state.
% A system is globally polarized when it can be split into K>1 groups, meaning that the system is weakly balanced but not in paradise.
Using this framework, \citet{Gorski2023} analyze a dynamical model (see Sec.~\ref{subsubsec:combined-interpersonal-agent-attribute} for details) and examine how different types of node attributes influence the emergence and stability of polarization. They find that continuous attributes are most effective at destabilizing polarized states or preventing their formation, whereas binary attributes are among the least effective.

The same measure of local polarization was subsequently applied by \citet{gorski2026more} to empirical signed school networks. The authors found that networks with a higher proportion of weak ties tend to have a higher density of hierarchical triads, and that a higher density of such triads is associated with greater local polarization. These empirical associations, together with the model results discussed in Sec.~\ref{subsubsec:context-dependent-updates}, suggest a mechanism through which weak ties may promote polarization by increasing the relative importance of status-driven dynamics.

These polarization measures still rely primarily on triad densities and do not distinguish between different numbers of groups. Systems with two groups and those with $K>2$ groups are assigned the same level of polarization. This suggests that further refinement of polarization measures, particularly those capable of capturing the richness of group structure, remains an open direction for future research.

\subsection{International relations}
\label{subsec:ir}

International relations is the oldest field of application of structural balance.
Friend-and-enemy reasoning was applied to interstate alignments from the outset~\citep{harary1961structural,moore1978international,moore1979structural,healy1973balance}.
The Background section~\ref{subsubsec:ir-networks} described how these signed networks are built.
It also asked whether the international system is balanced.
The answer there was mixed: support depends on the period, the definition of ties, and the measure.
% Local triadic interdependence was better supported than global convergence.
% These findings support balance
At the local triadic level, balance can be supported without implying that the entire international system must converge to the strict global configuration of one cohesive group or two mutually hostile blocs \citep{Belaza2019, askarisichani20201995}.
Here we ask a different question.
Given a partial drive toward balance, what does it imply for interstate outcomes?
We focus on four: the persistence of alliances and rivalries, the dynamics of conflict, systemic reorganization, and the prediction of coalitions.
One caveat runs through all of them.
Interstate signs are not a priori given but inferred from various data, e.g., alliances and militarized disputes, United Nations voting, or news event data.
Every implication below therefore inherits those coding choices (see Section~\ref{subsec:sign-obs-and-inf}).

\subsubsection{Persistence of alliances and rivalries}

Balance is most useful here as a diagnostic of which ties are durable.
It is less useful as a law that imbalance gets repaired.
Across long-run interstate data, balanced triads are more stable than unbalanced triads particularly after World War II~\citep{Oishi2025,askarisichani20201995}.
Open triads that close tend to close in a balanced form, although this tendency varies across historical periods~\citep{Oishi2025}.
%The converse is weaker.
However, unbalanced triads show little systematic tendency to dissolve or to flip toward balance~\citep{Oishi2025,Bellutta2020}.
The all-positive triad is the most stable, and the triad with two positive and one negative edge the least~\citep{belaza2017_StatPhysBalanceTheory}.
This hierarchy is persistent rather than transient~\citep{belaza2017_StatPhysBalanceTheory}.
The implication is that the balance status of a triad tells us how durable its ties are likely to be.
Balanced alignments tend to persist.
Specific unbalanced triads mark fragile or contested relationships.
Non-state actors sharpen the point.
In a 28-year network of Al-Qaeda and Islamic State factions, a change in one factional tie helps predict the later evolution of adjacent ties~\citep{Bellutta2020}.
Balanced triangles there are stable on short and medium horizons, but not on long ones~\citep{Bellutta2020}.
A plausible mechanism is reputation.
States carry forward the record of past cooperation and conflict, which makes balanced configurations self-reinforcing~\citep{crescenzi2007reputation}.
The inferential evidence that states cooperate more when they share cooperative partners or common enemies is discussed in Section~\ref{subsubsec:ergm-fitting}~\citep{Fritz2025}.
Here we take that tendency as given and read off its effect on alliance structure.

\subsubsection{Conflict: alignment rules can escalate as well as resolve}

The friend-and-enemy rules are not a peace mechanism.
The same rules that close a triad cooperatively can also drive coordinated hostility.
The earliest dynamical treatments wrote the four maxims as coupled nonlinear differential equations for interstate friendliness~\citep{Lee1994}, see Section \ref{subsubsec:kk-variants} for model description.
The system runs to unipolar or bipolar configurations.
Both cooperative and hostile ties intensify over time~\citep{Lee1994}.
Saperstein made the escalation reading explicit~\citep{Saperstein2004}.
Applied mechanically, ``the enemy of my enemy is my friend'' is a war-provoking rule.
It builds coalitions against a common target and can lock a three-actor system into conflict~\citep{Saperstein2004}.
A relaxed variant, with a flexibility parameter, softens this behavior and is described in Section~\ref{subsubsec:kk-weak}~\citep{Saperstein2004}.

Regional cases show the same logic.
A balance reading of the Syria conflict, covering the United States, Russia, Turkey, Syria, and Iran, identifies which bilateral shifts move the region toward or away from balance~\citep{Gner2018}.
It identifies Turkey--Syria relations as crucial for the structural balance of regional alignments~\citep{Gner2018}.
The historical precedents are the balance-of-power tradition~\citep{healy1973balance} and Harary's 1956 Middle East analysis~\citep{harary1961structural}.
Both were reasoning tools rather than predictive theories, a limitation their authors stressed.
The takeaway is simple.
Balance constrains which alignments are internally consistent.
It does not make the consistent state peaceful.

A parallel analysis comes from the economics of conflict.
Here the alliance-enmity network is held fixed and fighting groups choose how hard to fight.
\citet{Konig2017} derive a closed-form Nash equilibrium for fighting effort on such a network.
Own effort rises with the strength of one's enemies and falls with that of one's allies.
``The enemy of my enemy'' acts through equilibrium incentives, not a triad-repair rule.
Fitted to the Second Congo War, the model reproduces proxies of observed violence and the effect of counterfactual changes to the alliance structure~\citep{Konig2017}.
This econometric model holds the alliance network fixed.
When the network can instead change with new alliances forming and old ones dissolving, \citet{Jackson2015alliances} show that no alliance network is both stable and war-free unless interstate trade raises the cost of attacking a partner.
For a broader picture on the economics of conflict on networks, see~\citet{DziubinskiGoyalVigier2016}.

\subsubsection{Systemic reorganization: peace--war transitions}

At the system level, the strength of structural-balance interactions can act as a control parameter for collective transitions.
Morrison and colleagues model interstate ties as a signed-network dynamical system~\citep{Morrison2023}.
The model has a structural-balance force and a dyad-specific force.
As the strength of structural balance interactions increases, the system shows a sharp bifurcation from peace to systemic war~\citep{Morrison2023}.
There is a bistable regime in which peace and war equilibria coexist~\citep{Morrison2023}.
The reverse transition, from war back to peace, occurs at a lower interaction strength~\citep{Morrison2023}.
This hysteresis has a clear reading: de-escalation is harder than escalation.
The picture mirrors the first-order balanced--unbalanced transitions and hysteresis loops found for Hamiltonian balance dynamics on dense graphs (Section~\ref{subsubsec:hamil-nonzeroT}).
The long-run empirical record is consistent with reorganization rather than drift.

Over 1946--1999 the level of imbalance in the international system rises and falls, with no monotone trend toward global balance~\citep{doreian}.
For the Cold War, the relative frequency of balanced triads likewise shows no general increase over time~\citep{belaza2017_StatPhysBalanceTheory}.
A reanalysis of the same networks nonetheless places the system much closer to balance than sign-reshuffled null networks~\citep{Aref_2018}.

The evidence for balance is period-dependent at longer range: for alliance and rivalry networks over 1816--2009 it is stronger before 1867 and after 1942, and markedly weaker in between~\citep{Oishi2025}.
The prevalence of different balance mechanisms also shifts across 192 years of war and peace~\citep{Dekker2024}, with neutral ties playing a central role in that record.
Reorganization episodes also involve the activation of previously latent ties, not only sign changes on existing ones~\citep{Belaza2019}.

There is, however, an important limitation of saying that the system of international relations undergoes phase transitions.
The considered networks are low-$N$ and strongly time-dependent.
``Phase transition'' here is a modeling analogy supported by suggestive data, not a measured critical point.

\subsubsection{Anticipating and forming coalitions}

Balance-based partitioning is a coarse but usable tool for describing blocs and reasoning about mediation.
Frustration-minimizing and blockmodel partitions of interstate networks recover historical bloc structure~\citep{doreian,Aref_2018,estrada2019RethinkingSB}.
That structure persists even as membership changes~\citep{doreian}.
Node-level balance identifies which states contribute most to global imbalance~\citep{diazdiaz_mathematical_2024}.
These are the actors whose realignment might substantially affect the balance of the entire system~\citep{diazdiaz_mathematical_2024}.
Balance-based partitioning has also been used to predict hypothetical coalitions in a future conflict from historical data~\citep{Chowdhury2021}.
How far this generalizes out of sample is untested.
Balance dynamics also suggests a stylized model of coalition engineering.
In the dynamical balance model, a single actor can, in principle, drive the whole network to any chosen balanced partition by setting its own ties~\citep{Summers2013}.
Calibration adds a further use.
Fitting a ``social temperature'' to interstate data quantifies how strongly the system is balance-seeking rather than noise-driven~\citep{Lee2017}.
It also tracks how that tendency changes over time~\citep{Lee2017,askarisichani20201995}.

\subsection{Implications for organizations and markets}
\label{subsec:eco}

Economic actors and entities, such as firms, banks, traders, and assets, form signed networks in which positive links denote cooperation, alliance, or co-movement and negative links denote competition, litigation, or anti-correlation~\citep{schweitzer2009economic,jackson2010social}.
Structural balance matters at two levels: the tension of unbalanced configurations shapes coordination and performance within organizations, and the aggregate balance of a market network tracks systemic stress and predictability.

\subsubsection{Balance and performance in firms and teams}
\label{subsubsec:teams}

\citet{Morrissette1967} induce balanced and unbalanced three-person groups and find that the balanced, low-tension groups solved problems better, the classical reading of balance as strain reduction.
\citet{Askarisichani2019} recover this in a trading company: employees outperformed their own monthly baseline when embedded in more balanced local structure, and balanced triads were more stable over time.
The opposite holds where balance encodes factional bias: in Wikipedia editor networks (\emph{redo}/\emph{undo} as positive/negative ties), stronger adherence to balance predicts \emph{lower} article quality, because a balanced split into supportive and hostile camps leads editors to judge contributions by authorship rather than merit, an effect moderated by team topical diversity~\citep{Lerner2019Network,Lerner2019Team}.
\citet{Marineau2016} show that the amount of balance is too coarse: indirect negative ties help or hurt depending on whether the enclosing triad is open or closed.
\citet{Park2016} cast the assignment of employees to organizational units as an optimization problem: an employee is ``unhappy'' when placed with someone they dislike or separated from someone they like, and minimizing total unhappiness over all partitions is the weak-balance frustration problem, equivalent to a $q$-state Potts ground state.
Balance-theoretic reasoning has also been used outside the firm, for instance to predict how consumers re-evaluate a brand after it partners with a charitable cause~\citep{Yun2019}.
The balance--performance link is thus context-dependent: beneficial when balance reflects lower tension, detrimental when it reflects polarization in evaluation.

\subsubsection{Balance and the stability of financial markets}
\label{subsubsec:market-stability}

Financial signed networks are built from the co-movement of asset returns, with edge signs given by return correlations~\citep{Kuyyamudi2019,Ehsani2020,ferreira_loss_2021,Zahedian2022}.
Using random-matrix-based spectral filtering, \citet{Kuyyamudi2019} removed the global market mode and random noise from the return correlation matrix before constructing signed networks. Over nearly a century of New York Stock Exchange data, they found that frustration emerged mainly around the 1929--33 and 2007--09 crises, suggesting that its emergence may signal systemic risk.
% Over about a century of New York Stock Exchange data, frustration emerges only before and during the 1929 and 2007--08 crises, so that ``the emergence of frustration signals systemic risk''~\citep{Kuyyamudi2019}. Here, the signed adjacency matrix  was obtained by spectral filtering of the cross-correlation matrix.
Markets are near-balanced in calm periods, and the distance from balance has been proposed as a crisis-forecasting index~\citep{Ehsani2020}.
\citet{bartesaghi2025global} formalize this as a \emph{global balance index} derived from an information-diffusion process and link its numerical stability to the structural predictability of the network and to established market-risk indicators.
The transition is itself informative: after the August 2011 turmoil, six of nine national markets undergo a balance-to-unbalance transition driven by low-capitalization non-financial stocks turning market-wide anti-correlated, which maps onto a drop in stock predictability~\citep{ferreira_loss_2021}.
% In non-zero-temperature balance theory, a crisis is an ordered, collective phase, with a first-order transition whose critical temperature grows with crisis severity~\citep{Zahedian2022}.
% The picture contains a paradox: a crisis drives the network toward an all-positive, ``paradise''-like state in which everything co-moves, yet frustration rises in the residual structure.
Using unfiltered return correlations to define a triadic Hamiltonian, \citet{Zahedian2022} introduce an auxiliary statistical mechanical model in which the signs of empirical correlations are allowed to fluctuate with a model temperature. They find that crisis-period networks are more resistant to such imposed disorder than off-crisis networks.
The model-based result should be distinguished from the increased frustration reported by \citet{Kuyyamudi2019}, which concerns the residual network obtained after removing the global market mode and random noise. During crises, unfiltered correlations become more strongly positive, indicating greater market-wide co-movement and reduced diversification opportunities, even as the residual network can become more frustrated.
% arises from the different network representations: during crises, the unfiltered correlation network approaches an all-positive, ``paradise''-like state, while frustration increases in the residual network obtained after removing the global market mode and noise.
% This increase in market-wide positive correlations implies a loss of diversification and greater systemic risk.
At the mesoscale, crisis imbalance in the S\&P~500 network comes mainly from inter-sectoral interactions and tracks supply-chain and inflation shocks~\citep{dahake2026sectoral}, while assets whose local balance departs from the global value are candidates for concentrated exposure when diversification fails~\citep{bartesaghi2025local}.
Aggregate balance thus acts as an indicator for systemic stress: promising for early warning, but so far correlational and sensitive to the correlation window and filtering choices.

\subsubsection{Coalitions, alliances, and disputes among firms}
\label{subsubsec:interfirm}

In inter-firm networks, positive links represent R\&D alliances or partnerships~\citep{tomasello2017rise}, and negative links represent patent opposition, antitrust litigation, or competitive rivalry~\citep{kito2021inter,sytch2014friendsandfoes}.
For biotechnology and pharmaceutical firms in 1996--2006, existing collaboration or conflict breeds more ties of the same kind and crowds out the opposite.
This suggests that network formation is driven by a pull \emph{away from} unbalanced triads rather than toward balanced ones, and that the network segregates into collaborative and conflictual segments of firms~\citep{sytch2014friendsandfoes}.
In supply networks, balance theory with the structural-hole concept yields nine buyer--supplier--supplier triadic archetypes~\citep{Choi2009triadsinsupplynetworks}.
Structural balance thus helps characterize patterns of collaboration and conflict among firms, but explicit measurement against null models remains rare outside finance.

\subsection{What structural balance impacts in real-world complex networks}

Across the domains discussed in this Section, structural balance is best viewed as a lens rather than a law. And a balanced configuration is not necessarily a desirable one.

The tendency toward triadic consistency takes different forms across domains: coherence within a belief system, polarization in a social network, factional alignment in an organization, two persistently hostile blocs between states, or market-wide co-movement during a crisis.

In cognitive balance, the main implication is fragility.
When beliefs are formed through local, balance-consistent inference on a cognitive network, the consistency of the resulting belief system falls as a power law with the number of concepts.
In a sufficiently complex world, even relatively small amounts of imbalance can therefore disrupt coherent belief formation~\citep{medo2021fragility}.
Coherent belief systems may nevertheless have an advantage in competition with other systems: they spread more readily and require a smaller committed minority to displace a mainstream view.
This creates a route by which self-consistent, but dysfunctional, worldviews can persist~\citep{rodriguez2016collective,aiyappa2024emergence}.

The consequences of social balance for opinion and spreading dynamics are less uniform.
In a balanced sociomatrix, the dynamics lead to a predictable polarized equilibrium~\citep{altafini2012dynamics}, and balance can accelerate and increase the extent of complex contagion~\citep{he2019information,Lee2023}.
But this is not a general effect of negative ties.
When negative ties do not transmit contagion, balance can instead fragment the network into weakly connected groups, limiting the spread of outbreaks across those groups~\citep{saeedian2017epidemic}.
Higher-order models point to a similar dependence on structure: the outcome can change according to which types of balanced triads are most prevalent~\citep{de2024complex}.

At larger scales, balance can also provide information about the timing of collective changes. In parliamentary networks, for example, frustration in the signed network correlates with how long it takes to form a government~\citep{fontan2021signed}.
A related idea appears in financial markets, where changes in network structure can act as early-warning signals of systemic instability\cite{Kuyyamudi2019, Ehsani2020}.

Balance measures are also widely used as indices of polarization. They capture broad trends, including the increase in US congressional polarization since 1979, but they tend to exaggerate polarization close to consensus and cannot distinguish how many antagonistic groups are present~\citep{Neal2020,YoungPedersen2021,Gorski2023}.

Two limitations cut across these applications. First, most analyses rely on binary and undirected notions of balance, even though the underlying relations are often weighted and directed. Second, much of the empirical evidence remains correlational.

Taken together, these results suggest that balance is useful for describing structure and, in some settings, macroscopic timescales, but less so for making strong causal or predictive claims. A more decisive test will require closer integration with controlled behavioral experiments, both to constrain the dynamical models and to identify interventions that can actually change the resulting dynamics~\citep{tanase2024integrating}.

\clearpage

\section{Perspectives and open questions}
\label{sec:perspectives}

Structural balance is a deliberately minimal theory.
Heider's local rule, later expressed as the requirement that the signs of a triad multiply to a positive value, was recast by Cartwright and Harary as a condition on every cycle of a signed graph.
It follows that a fully balanced network partitions into internally friendly, mutually hostile groups: at most two in Harary's theorem, and any number once the all-negative triad is admitted (Davis's weak balance).
% Its appeal to physicists is that this is exactly the structure of a frustrated spin system, so that energy, ground states, metastability, and phase transitions all carry over.
For physicists, the key correspondence is with frustration in spin systems. When a link belongs to both balanced and unbalanced triads, flipping its sign balances the latter but unbalances the former.
This competition resembles the frustration of spins encountered for example in a triangular-lattice antiferromagnet, where flipping a spin may satisfy some interactions at the expense of others.
% It follows the  link is frustrated similarly as a spin in an antiferromagnet at a triangular lattice.
There is also a precise correspondence with Ising spin glasses: when signed links are treated as fixed couplings between node spins, an unbalanced cycle corresponds to a frustrated loop, whereas a strongly balanced network is frustration-free~\citep{facchetti2011computing}.
%an unbalanced cycle corresponds to a frustrated loop, whereas a fully balanced signed network is frustration-free.

This correspondence makes concepts such as energy, ground states, metastability, and phase transitions natural tools for studying structural balance.
It is fair to ask whether this correspondence delivers more than a shared vocabulary.
Reviewing the literature, we find that it does, but unevenly.
The concrete contributions are methodological: principled null models and significance tests for balance under measurement noise (Section~\ref{subsec:bottomline-measures})~\citep{gallo2024testing,feng2022_nullmodelSB}; the identification of first-order transitions and hysteresis in models of structural balance dynamics and systemic conflict, alongside abrupt balance--unbalance transitions in financial networks (Sections~\ref{subsubsec:hamil-nonzeroT}, \ref{subsec:ir} and~\ref{subsec:eco})~\citep{Malarz2022Mean,Morrison2023,ferreira_loss_2021}; and the recognition that a Gibbs measure over signed configurations is an exponential-family model, which puts Hamiltonian and inferential (ERGM) approaches on the same footing (Section~\ref{subsubsec:ergm-fitting})~\citep{belaza2017_StatPhysBalanceTheory,Fritz2025}.
Set against this successful conceptual transfer from social psychology to physics and other domains, the review also finds places where this transfer is still incomplete.
We group them under five headings.
%The open problems below are, in effect, the places where this transfer is still incomplete.
\subsection{What are the open questions?}

\paragraph{Making the measurement instrument-independent}
% The structural balance of a network is not an observable but the output of two coupled choices: a measure and a null model.
The empirical assessment of structural balance in networks that are not fully balanced depends on two coupled choices: a measure and a null model.
Different choices of these two instruments can reverse the qualitative conclusion for the same data (Section~\ref{subsec:bottomline-measures})~\citep{aref2017measuring,gallo2024testing,Hao2024}.
Local aggregation measures describe a myopic observer counting local configurations, whereas the frustration index takes the globally closest balanced configuration as its reference and describes an omniscient planner able to identify the global optimum.
% who can reach a global optimum that locally driven dynamics almost never would (Section~\ref{subsubsec:frustration}).
Locally driven dynamics, however, may never reach that state (Section~\ref{subsec:bottomline-measures}).
Also, there is no consensus on which decay function to use for walk-based measures~\citep{estrada2014_RandomWalkSB,kirkley2018balanceInSN,tian2025balance}. For directed signed networks, even the definition of balance remains unsettled~\citep{Hao2025,Gallo2025}. No common reporting standard has emerged either.
The near-term fix that we propose is procedural: report several measures (at least one local and one global), explicitly justify the null model, and, where several plausible null models exist, report the robustness of the conclusion across them.
% However, the long-term goal is to unify these different perspectives and identify a set of instrument-independent statements about whether a network is balanced.
However, the longer-term goal is to identify which conclusions about structural balance are robust across reasonable choices of measure and null model, and which are inherently instrument-dependent.

\paragraph{Taking the asymmetry of the two signs seriously}
Most measures and null models do not fully account for the empirical asymmetry between positive and negative ties. Negative ties are typically rarer, less reciprocated, less clustered, harder to elicit, and antagonism does not propagate through a triad the way affinity does (Sections~\ref{subsec:sign-obs-and-inf} and~\ref{subsec:bottomline-measures})~\citep{feng2022_nullmodelSB}.
% Note (asymmetry, strategic concealment): the under-measurement of negative ties is not only a data-collection artefact but can be strategic.
% In competitive or adversarial settings, actors may deliberately hide that they regard another as an enemy and keep an outwardly friendly tie -- ``keep your friends close, and your enemies closer''.
% Measured signed networks may therefore over-represent positive ties and mislabel latent antagonism as friendship, a bias with no positive-tie counterpart.
This asymmetry enters the measurement problem in different ways.
Sign-permutation null models treat sign placement as exchangeable across edges and therefore disregard differences in the structural organization of positive and negative ties, whereas triad-based measures require a choice about which signed configurations are considered balanced. Under strong balance, the all-negative triad is classified as unbalanced, just like a triad with exactly one negative edge.
% treated as structurally equivalent to the other unbalanced configurations.
Weak balance instead permits the all-negative configuration, reflecting the idea that three mutually antagonistic actors need not be regarded as an unstable state~\citep{davis1967clusteringSB}.
The distinction is therefore not merely formal: it determines whether a balanced network is interpreted as necessarily bipolar or whether multiple antagonistic groups are allowed.
% Hence, instruments that symmetrize the signs cannot, for example, separate a bipolar world from a multipolar one.
More generally, the empirical asymmetry of the two signs should be reflected consistently in how balance is measured, tested, and modeled.
Measures, null models, and dynamical rules that explicitly account for this asymmetry, together with better empirical capture of negative ties, are needed.

\paragraph{Separating balance from polarization and from welfare}
Whether a drive toward balance is necessary for polarization is unresolved: homophily on a few attributes can produce polarized structure with no explicit balance mechanism~\citep{pham2022_SBexplainedThroughHomophily,Thurner2025}.
Status may also drive polarization, especially in the presence of weak ties~\citep{Gorski2025, gorski2026more}.
Balance-based polarization indices have another limitation: the paradise state. Because an all-positive network is fully balanced, indices that equate balance with polarization can report high polarization even for a near-consensus network (Section~\ref{subsec:netpol})~\citep{Neal2020,YoungPedersen2021,Gorski2023}.
Balance is also not normatively neutral: a balanced international system can consist of two hostile blocs~\citep{Saperstein2004,Lee1994} and organizational balance improves performance when it reflects low interpersonal tension but degrades it when it reflects factional evaluation~\citep{Askarisichani2019,Lerner2019Network}.
These examples point to a broader distinction: structural balance is a property of relational structure, not a measure of polarization or welfare.
Greater balance can therefore mean greater cohesion, but it can equally mean sharper division between internally cohesive factions.
Treating balance as inherently desirable risks confusing structural consistency with social welfare.

\paragraph{From correlation to mechanism}
Almost all empirical support for the \emph{implications} of balance is correlational (Section~\ref{sec:implications}).
The dynamical models that would supply the mechanism are still mostly formulated on complete graphs, with binary, undirected link states and static topologies, whereas the systems of interest are sparse, weighted, directed, temporal, multilayer, and higher-order.
Closing this gap needs both dynamics on realistic topologies~\citep{Gorski2025} and controlled longitudinal experiments that can constrain the update rules and test whether candidate interventions actually change the outcome~\citep{tanase2024integrating,galesic2025experimental}.

\paragraph{Balance outside the social domain}
Signed representations of gene-regulatory, metabolic, neural, ecological, and financial networks allow for computing measures of structural balance.
For example, transcriptional networks come close to balance, signaling and metabolic ones do not (the domain-by-domain summary in Section~\ref{subsec:bottomline-measures})~\citep{Iacono2010,Rizi2021,saberi2025empirical}.
What is missing is an account, in many domains, of the generative process that would or would not produce balance, comparable to cognitive dissonance in the social case.
Without it, a balanced metabolic network and a balanced friendship network may share only a matrix property.

\subsection{Conclusions}

Karl Popper once wrote: ``In so far as a scientific statement speaks about reality, it must be falsifiable; and in so far as it is not falsifiable, it does not speak about reality''~\citep{Popper1959}.
Structural balance theory makes falsifiable predictions about the organization and evolution of signed networks.
Hence, its value lies especially where its predictions fail.
The mixed evidence on polarization in social systems, the period dependence of balance in international relations, and its domain dependence in biology and financial markets are therefore findings to be explained, and not nuisances to be dismissed.
The agenda that follows from this review is to make empirical conclusions about balance more robust to measurement choices, its dynamics more topologically realistic, and tests of its underlying mechanisms more causally informative.

\section*{CRediT authorship contribution statement}

\noindent
\textbf{Giacomo Vaccario:}
Conceptualization, Formal analysis, Investigation,
Methodology, Project administration, Software,
Validation, Visualization, Writing -- original draft,
Writing -- review \& editing.

\noindent
\textbf{Piotr J. G\'orski:}
Conceptualization, Formal analysis, Investigation,
Methodology, Project administration, Software,
Validation, Visualization, Writing -- original draft,
Writing -- review \& editing.

\noindent
\textbf{Georges Andres:}
Conceptualization, Investigation, Methodology,
Validation, Writing -- original draft.

\noindent
\textbf{Manuel S. Mariani:}
Conceptualization, Investigation, Methodology,
Validation, Writing -- original draft,
Writing -- review \& editing.

\noindent
\textbf{Janusz A. Ho{\l}yst:}
Conceptualization, Investigation, Project administration,
Validation, Writing -- original draft,
Writing -- review \& editing.

\section*{Declaration of competing interest}
The authors declare that they have no known competing financial interests
or personal relationships that could have appeared to influence the work
reported in this paper.

\section*{Acknowledgments}

The authors are grateful to Krzysztof Kułakowski for generously reviewing part of this manuscript and providing invaluable feedback.
The authors thank Frank Schweitzer for his participation in early discussions related to this work and for his support during its development.
The authors additionally thank the participants of the ETH Workshop ``Signed Relations and Structural Balance in Complex Systems: From Data to Models'', held in Zurich in May 2024, for their valuable comments and contributions, which helped inform the literature review.

G.V. and G.A. acknowledge financial support from Swiss National Science Foundation (SNSF) under grant no. 192746, "Signed Relations and Structural Balance in Complex Systems: From Data to Models".
P.J.G. acknowledges financial support from the National Science Centre, Poland under Alphorn Grant No. 2019/01/Y/ST2/00058.
M.S.M. acknowledges financial support from the SNSF under grant no. 100013-207888.
J.A.H. acknowledges financial support from the European Union under the Horizon Europe grant OMINO (grant no. 101086321). Views and opinions expressed are however those of the authors only and do not necessarily reflect those of the European Union or the European Research Executive Agency. Neither the European Union nor European Research Executive Agency can be held responsible for them. J.A.H. was also co-financed with funds from the Ministry of Science and Higher Education, Poland under the program entitled International Co-Financed Projects.

\section*{Declaration of generative AI and AI-assisted technologies in the manuscript preparation process}

During the preparation of this review, the authors used ChatGPT (OpenAI), Claude (Anthropic), and Gemini (Google) to assist with proofreading and editing, literature search and systematization, checking descriptions of published studies, identifying potential limitations, and suggesting improvements to the manuscript. These tools were also used to assist in writing Python code for implementing models described in the literature, reproducing published numerical results, and generating TikZ diagrams and figures based on computational results rather than generative image synthesis.

The authors reviewed and edited the AI-assisted outputs as needed. The computational code was reviewed and tested, and the reproduced results were checked against the original publications. The authors take full responsibility for the content of the published article.

% The authors declare the use of generative AI in the research and writing process. According to the GAIDeT taxonomy (Suchikova et al., 2026), the following tasks were partially delegated to generative AI tools under human supervision:
% \begin{itemize}
% \item Proofreading and editing
% \item  Visualization
% \item  Code generation
% \item  Quality assessment
% \item  Identification of limitations
% \item  Recommendations
% \item  Literature search and systematization
% \item  Validation
% \end{itemize}

% In particular, generative AI tools assisted in writing Python code for implementing models described in the literature, reproducing published numerical results, and generating figures based on computational results rather than generative image synthesis.

% The tools used were ChatGPT (OpenAI), Claude (Anthropic), and Gemini (Google).

% All AI-assisted outputs were critically evaluated by the authors. The computational code was reviewed and tested, and the reproduced results were checked against the original publications. The authors take full responsibility for the content of the published article.

\clearpage

\appendix

\section{Notation}
\label{app:notation}

Table~\ref{tab:notation_symbols} summarizes the notation used throughout the review.
Symbols that are specific to particular models, datasets, or empirical applications are introduced locally in the corresponding sections.

\begin{longtable}{p{2.8cm} p{3.2cm} p{0.55\linewidth}}
\caption{Symbols and notation used throughout the review.}
\label{tab:notation_symbols} \\

\textbf{Symbol} & \textbf{Name} & \textbf{Description} \\
\hline
\endfirsthead

\textbf{Symbol} & \textbf{Name} & \textbf{Description} \\
\hline
\endhead

\hline
\endfoot

% -----------------------------
\multicolumn{3}{l}{\textbf{Networks and matrices}}\\

$G=(V,E)$ & Network & Network with node set $V$ and link set $E$. \\
$N$ & Number of nodes & Cardinality of the node set, $N=|V|$. \\
$m$ & Number of links & Cardinality of the link set, $m=|E|$. \\
$\mathcal{T}$ & Triad set & Set of all existing triads in the network.\\
$N_T$ & Number of triads & Cardinality of the triad set, $N_T=|\mathcal{T}|$. \\
$i,j,k$ & Node indices & Indices referring to nodes in $V$. \\
$v_i$ & Node & A specific node in the network. \\
$(i,j)$ & Link & Link between nodes $i$ and $j$ (ordered for directed networks). \\
$(i,j,k)$ & Triad & Triad formed by nodes $i$, $j$, and $k$; $(i,j,k)\in\mathcal{T}$. \\
$s_{ij}$ & Link sign & Sign of link $(i,j)$, $s_{ij}\in\{-1,+1\}$. \\
$w_{ij}$ & Link weight & Non-negative magnitude (strength) of the relation. \\
$x_{ij}$ & Signed relation &
Signed link value, $x_{ij}=s_{ij}w_{ij}$; for an unweighted signed
network, $x_{ij}=s_{ij}$. \\
$\mathbf{X}$ & Signed-relation matrix &
Matrix with entries $x_{ij}$. \\
$\mathbf{A}$ & Adjacency matrix & Signed adjacency matrix with entries $A_{ij}$. \\
$A_{ij}$ & Matrix entry & For unweighted signed networks, $A_{ij}=s_{ij}$ on existing edges
and $A_{ij}=0$ otherwise; weighted formulations may use $A_{ij}=x_{ij}$. \\
$\mathbf B$ & Cognitive-network adjacency matrix &
Matrix representing relationships between concepts;
$\mathbf B^{(i)}$ denotes the cognitive-network
matrix of agent $i$. \\

\multicolumn{3}{l}{}\\
% -----------------------------
\multicolumn{3}{l}{\textbf{Node states and attributes}}\\

$M$ & Number of attributes &
Number of attributes associated with each agent. \\
$b_i$, $b_\alpha$ & Node/agent state &
Scalar opinion, belief or attribute associated with an agent or concept. \\
$\mathbf{b}_i$ & Agent attribute/belief vector &
Vector of attributes or concept-level beliefs associated with agent $i$. \\
$\mathbf r_i$ & Link-value vector &
Vector containing the link values of agent $i$'s
cognitive-network matrix $\mathbf B^{(i)}$,
with one component per link. \\

\multicolumn{3}{l}{}\\
% -----------------------------
\multicolumn{3}{l}{\textbf{Degrees and neighborhoods}}\\

$k_i$ & Degree & Total degree of node $i$. \\
$k_i^{\mathrm{in}}, k_i^{\mathrm{out}}$ & In-/out-degree & Incoming and outgoing degree in directed networks. \\
$k_i^{+}, k_i^{-}$ & Signed degree & Positive and negative degree of node $i$. \\
$\mathcal{N}(i)$ & Neighborhood & Set of nodes adjacent to node $i$. \\
$\mathcal{N}^{+}(i), \mathcal{N}^{-}(i)$ & Signed \mbox{neighborhoods} & Positive and negative neighbors of node $i$. \\
$\mathcal{N}^{\pm}_{\mathrm{in}}(i)$ & Signed \mbox{in-neighborhood} & Positive/negative neighbors pointing to $i$. \\
$\mathcal{N}^{\pm}_{\mathrm{out}}(i)$ & Signed \mbox{out-neighborhood} & Positive/negative neighbors pointed by $i$. \\

\multicolumn{3}{l}{}\\
% -----------------------------
\multicolumn{3}{l}{\textbf{Walks, paths, and cycles}}\\

$(v_0,\dots,v_k)$ & Walk & Walk of length $k$ in the network. \\
$Q_k$ & Walk count & Number of (possibly signed or weighted) walks of length $k$. \\
$\mathscr C_k$ & Cycle set & Set of cycles of length $k$. \\
$O_k$ & Cycle count & Number of cycles of length $k$, $O_k=|\mathscr C_k|$. \\
$O_k^{+},O_k^{-}$ & Positive/negative cycles & Number of positive/negative cycles of length $k$. \\

\multicolumn{3}{l}{}\\
% -----------------------------
\multicolumn{3}{l}{\textbf{Structural balance}}\\

$T(\mathbf{A})$ & Triangle index & Fraction of positive triads. \\
$D(\mathbf{A})$ & Degree of balance & Fraction of positive cycles among all cycles. \\
$K(\mathbf{A})$ & Walk-based balance index & Balance index based on weighted closed walks. \\
% $S$ & Balance measure & Generic balance index. \\
% $S_l$ & Balance at length $l$ & Balance associated with paths or walks of length $l$. \\
% $\hat{S}$ & Observed balance & Balance measured on empirical data. \\
% $\langle S \rangle$ & Expected balance & Balance averaged over a null-model ensemble. \\
% $\Delta S$ & Balance deviation & Difference between observed and expected balance. \\
% $T_{+++}, T_{+--}$ & Triad types & Triads labeled by sign configuration. \\
$\Delta_i$ & Triad types & Triad with exactly $i$ negative links.\\
$N_i$ & Triad count & Number of triads of type $i$. \\
$n_i$ & Triad density & Normalized frequency of triads of type $i$. \\
% $N_B, n_B$ & Balanced triads & Absolute and normalized count of balanced triads. \\
$P$--$O$--$X$ & $P$--$O$--$X$ triad & Three-element balance configuration; in this review, $X$ may denote another
person or an impersonal entity.\\
$H$ & Hamiltonian & a defined Hamiltonian in a given system. \\

\multicolumn{3}{l}{}\\
% -----------------------------
\multicolumn{3}{l}{\textbf{Frustration and partitions}}\\

$L(G)$ & Frustration index & Minimum number of frustrated links in $G$. \\
$s_i$ & Partition label / spin &
Binary node label, $s_i\in\{-1,+1\}$, used in frustration formulations. Spin in Ising model \\
$\mathcal{X}$ & Partition & Node partition used in frustration-based measures. \\

\multicolumn{3}{l}{}\\
% -----------------------------
\multicolumn{3}{l}{\textbf{Probability and statistics}}\\

$P(\cdot)$ & Probability & Discrete probability measure. \\
$P(x), p(x)$ & Distribution & Discrete or continuous probability distribution. \\
$\mathbb{E}[X]$ & Expectation & Expected value of a random variable. \\

\multicolumn{3}{l}{}\\
% -----------------------------
\multicolumn{3}{l}{\textbf{Vectors and functions}}\\

$\mathbf{v}$ & Vector & Vector of attributes or features. \\
% $S(t)$ & Time-dependent balance & Balance as a function of time. \\
% $\mathbb{I}[\cdot]$ & Indicator function & Logical indicator (1 if true, 0 otherwise). \\
$\Theta(x)$ & Heaviside \mbox{function} & Equals 1 if $x>0$, 0 otherwise.\\
$\delta_{ij}$ & Kronecker delta & Equals 1 if $i=j$, 0 otherwise. \\

\multicolumn{3}{l}{}\\
% -----------------------------
% \multicolumn{3}{l}{\textbf{Null models and ensembles}}\\

% $\mathcal{M}$ & Null model & Ensemble defining a random signed network model. \\
% $G^{\mathcal{M}}$ & Realization & A network sampled from $\mathcal{M}$. \\
% $S^{\mathcal{M}}$ & Realization balance & Balance measured on a realization. \\

\end{longtable}

\section{Monte Carlo dynamics}
\label{app:mc}

Assuming a system described by a Hamiltonian $H$ and evolving at temperature $T$, its dynamics may be implemented using one of several Monte Carlo algorithms. Among the most commonly used are Metropolis, Glauber, and heat-bath dynamics. The purpose of this appendix is to briefly summarize these algorithms. For a more detailed discussion of their properties and applicability, we refer the reader to, e.g., \cite{newman1999monte}.
Throughout this appendix and the whole review, we use units in which $k_{\mathrm B}=1$.

\paragraph{Accept--reject algorithms}

Metropolis and Glauber dynamics follow a common accept--reject scheme:

\begin{enumerate}
    \item Calculate the current system energy $H$.
    \item Propose a change in the system (e.g., flipping a spin, changing a link sign, or modifying an attribute).
    \item Compute the new energy $H'$ after the proposed change.
    \item Accept or reject the proposed change according to a specified transition probability.
\end{enumerate}

Depending on the convention, a single Monte Carlo step may consist of multiple such update attempts, often chosen such that each degree of freedom (e.g., node, edge, or other elementary variable) is selected once on average. Metropolis and Glauber dynamics differ in the transition (acceptance) probability used in step 4.

In the Metropolis algorithm, the proposed change is accepted with probability
\begin{equation}
    P_{\mathrm{acc}}^{\mathrm{M}} =
    \min\left(1, \exp\left(-\frac{\Delta H}{T}\right)\right),
\end{equation}
where $\Delta H = H' - H$. Thus, all energy-decreasing moves are accepted, while energy-increasing moves are accepted with a probability that decreases exponentially with $\Delta H$.

In Glauber dynamics, the proposed change is accepted with probability
\begin{equation}
    P_{\mathrm{acc}}^{\mathrm{G}} = \frac{1}{1 + \exp\left(\frac{\Delta H}{T}\right)}.
\end{equation}
In contrast to the Metropolis rule, this probability is always strictly between $0$ and $1$, so even energetically favorable moves may be rejected.

\paragraph{Heat-bath dynamics}

Unlike the accept--reject algorithms, which only require energy differences, heat-bath dynamics relies on explicit knowledge of the conditional distribution of a single system element. Instead of proposing a change followed by an acceptance step, the selected variable is directly resampled from its local equilibrium distribution.
The procedure is as follows:
\begin{itemize}
    \item Choose a variable $x$ (e.g., spin, link sign, or attribute).
    \item Assuming $x \in \{-1, +1\}$, assign its new value according to
    \begin{equation}
        x(t+1) =
        \begin{cases}
            +1 & \text{with probability } p_x, \\
            -1 & \text{with probability } 1 - p_x,
        \end{cases}
    \end{equation}
    where the probability $p_x$ is given by the conditional Boltzmann distribution
    \begin{equation}
        p_x = \frac{\exp\left(-\frac{H(x=+1)}{T}\right)}{\exp\left(-\frac{H(x=+1)}{T}\right) + \exp\left(-\frac{H(x=-1)}{T}\right)}.
    \end{equation}
\end{itemize}

In this approach, the variable $x$ is always updated (although it may retain its previous value with nonzero probability), and no explicit rejection step is involved.
For a binary variable, the standard Glauber and heat-bath rules are mathematically equivalent.
Let $x'$ denote the new value of $x$, and let $\Delta H=H(-x)-H(x)$, with all other variables held fixed. The probability that a heat-bath update reverses the sign of $x$ is
\begin{equation}
P(x'=-x\mid x)=\frac{\exp[-H(-x)/T]}{\exp[-H(x)/T]+\exp[-H(-x)/T]}=\frac{1}{1+\exp(\Delta H/T)}=P_{\mathrm{acc}}^{\mathrm{G}}.
\end{equation}
Thus, both methods define identical single-variable transition probabilities for binary variables. They yield the same stochastic dynamics when the Hamiltonian, temperature, and variable selection and update protocols are also identical. This equivalence does not necessarily extend to multistate variables or other variants of Glauber dynamics.

\paragraph{Update schemes}
The local update rule should be distinguished from the update scheme. In sequential (asynchronous) dynamics, variables are updated one at a time, and each update uses the configuration resulting from preceding updates. In synchronous dynamics, all variables are updated in parallel using the configuration at the previous time step. These two schemes can produce different collective dynamics even when the same local update rule is used. Therefore, for a given Hamiltonian and temperature, Glauber and heat-bath dynamics define the same stochastic process only when their variable-selection procedures and update schemes are also identical.

\clearpage

\bibliographystyle{elasrticle-num-names-nourl-nonotes}
\bibliography{references}

\end{document}

%% file: introduction.tex
% \addcontentsline{toc}{section}{Introduction}

\section{Introduction }
\subsection{How statistical physics contributes to the study of structural balance}

% \addcontentsline{toc}{subsubsection}{Structural balance in complex networks}
\subsubsection{What is structural balance}
% Network science provides powerful means to comprehend complex systems, representing their elements and interactions through nodes and links. 
% This formalism not only facilitates the examination of systems across social, economic, and biological domains but also invites a comparative exploration across disciplinary boundaries. 

This review focuses on \textit{structural balance}, an emergent network property.
Its origins, however, lie in social psychology rather than network science.
The concept was first formulated by Fritz Heider in 1946, who hypothesized that a person forms a positive tie (e.g., a friendship) with a second person when they share the same attitude toward a third entity.
When their attitudes differ, they instead form a negative tie (e.g., an enmity).
Heider described these attitude configurations as \textit{balanced}~\citep{heider1946attitudes}.
%This original formulation focused on how people seek consistency among their attitudes toward other individuals and external entities.

\citet{cartwright_1956} reformulate structural balance using graph theory.
The basic object of interest becomes a \textit{signed triad}, that is, three entities connected by positive or negative relations. %
Then, a signed triad is balanced if the product of its signs is positive (see Fig.~\ref{fig:intro-balance-schematic}a).
They also extend the notion of balance from triads to signed networks.
Signed networks are a representation of social organizations, where nodes represent individuals and signed links their relations.
They posit that a signed network is balanced if and only if the products of the signs along all its closed paths are positive~\citep{cartwright_1956}.

Figure~\ref{fig:intro-balance-schematic} gives an informal picture of these ideas.
At the level of a single triad, the friend-and-enemy colloquial expressions: ``the friend of my friend is my friend'' and ``the enemy of my enemy is my friend'' capture balanced triads; whereas ``the friend of my friend is my enemy'' and ``the enemy of my enemy is my enemy'' capture unbalanced triads (Fig.~\ref{fig:intro-balance-schematic}a).
At the level of a whole network, requiring every cycle to be balanced is a strong constraint: it forces one of only two global structures, a single group in which all ties are positive, or a split into internally friendly but mutually hostile factions (Fig.~\ref{fig:intro-balance-schematic}b,c).
The second structure already shows that balance is a statement about consistency rather than about harmony, since a perfectly balanced society can also be a polarized one.

\begin{figure}[t]
\centering
\begin{tikzpicture}[
  node distance = 1cm,
  every node/.style = {font=\small},
  dot/.style = {circle, fill=ink, draw=none, inner sep=1.6pt},
  posedge/.style = {ink, thick},                              % positive edge = solid ink
  negedge/.style = {accent, thick, dash pattern=on 3pt off 2pt}, % negative edge = dashed accent
  tag/.style = {font=\scriptsize, align=center},
]

% ---------- Panel (a): four elementary triads ----------
\foreach \x/\eA/\eB/\eC/\status/\maximA/\maximB in {
  0/posedge/posedge/posedge/{Balanced}/{``friend of my friend}/{is my friend''},
  3/posedge/negedge/negedge/{Balanced}/{``enemy of my enemy}/{is my friend''},
  6/posedge/posedge/negedge/{Unbalanced}/{``friend of my friend}/{is my enemy''},
  9/negedge/negedge/negedge/{Unbalanced}/{``enemy of my enemy}/{is my enemy''}
}{
  \coordinate (T) at ($(\x,0)+(90:0.55)$);
  \coordinate (L) at ($(\x,0)+(210:0.55)$);
  \coordinate (R) at ($(\x,0)+(330:0.55)$);

  \draw[\eA] (T) -- (L);
  \draw[\eB] (L) -- (R);
  \draw[\eC] (R) -- (T);

  \node[dot] at (T) {};
  \node[dot] at (L) {};
  \node[dot] at (R) {};

  \node[tag] at ($(\x,0)+(0,1.0)$) {%
    \ifthenelse{\equal{\status}{Balanced}}{\textcolor{ink}{\checkmark}}{\textcolor{accent}{\ding{55}}}%
    \ \status
  };

  \node[tag] at ($(\x,0)+(0,-1.0)$) {\maximA\\ \maximB};

  \ifnum\x=0
    \node[tag] at (-1.8,0) {\textbf{(a)}};
  \fi
}

% ---------- Panels (b) + (c): the two balanced macro-states ----------
\begin{scope}[yshift=-4.0cm,
    dot/.style = {circle, fill=ink, draw=none, inner sep=1.4pt},
    hub/.style = {circle, fill=ink, draw=none, inner sep=2.3pt},
  ]

  % Panel (b): paradise state -- a single all-positive group
  \node[tag] at (1.45,1.85) {\textbf{(b)}~~\emph{paradise state}};

  \draw[ink, opacity=0.25, rounded corners=12pt] (-0.2,-1.15) rectangle (3.1,1.15);

  \node[dot] (P1) at (0.30, 0.60) {};
  \node[dot] (P2) at (1.45, 0.90) {};
  \node[dot] (P3) at (2.60, 0.55) {};
  \node[dot] (P4) at (2.60,-0.55) {};
  \node[dot] (P5) at (1.45,-0.90) {};
  \node[dot] (P6) at (0.30,-0.60) {};
  \node[dot] (P7) at (1.45, 0.02) {};

  \foreach \a/\b in {P1/P2, P2/P3, P3/P4, P4/P5, P5/P6, P6/P1,
                     P7/P1, P7/P2, P7/P3, P7/P4, P7/P5, P7/P6}
    \draw[posedge] (\a) -- (\b);

  \node[tag, align=center] at (1.45,-1.65) {one group,\\ all ties positive};

  % Panel (c): bipolar state -- two internally friendly, mutually hostile factions
  \node[tag] at (6.8,1.85) {\textbf{(c)}~~\emph{bipolar state}};

  \fill[ink, opacity=0.05] plot [smooth cycle, tension=0.7] coordinates
    {(3.95,0.5) (4.5,1.5) (5.5,1.55) (6.3,1.1) (6.6,0.0) (6.3,-1.15)
     (5.4,-1.35) (4.4,-1.15) (3.85,-0.2)};
  \fill[ink, opacity=0.05] plot [smooth cycle, tension=0.7] coordinates
    {(7.05,0.4) (7.6,1.4) (8.5,1.5) (9.35,1.1) (9.75,0.0) (9.4,-1.15)
     (8.5,-1.35) (7.6,-1.1) (7.0,-0.3)};

  \node[dot] (L1)  at (4.42, 0.86) {};
  \node[dot] (L2)  at (4.98, 1.17) {};
  \node[dot] (L3)  at (5.53, 0.95) {};
  \node[dot] (L4)  at (6.03, 0.45) {};
  \node[dot] (L5)  at (6.15,-0.32) {};
  \node[hub] (L6)  at (5.01, 0.32) {};
  \node[hub] (L7)  at (5.53,-0.05) {};
  \node[dot] (L8)  at (5.78,-0.86) {};
  \node[dot] (L9)  at (5.16,-0.95) {};
  \node[dot] (L10) at (4.54,-0.77) {};
  \node[dot] (L11) at (4.17, 0.05) {};
  \node[dot] (L12) at (4.67, 0.05) {};

  \foreach \a/\b in {%
    L6/L1, L6/L2, L6/L3, L6/L7, L6/L10, L6/L11, L6/L12,
    L7/L3, L7/L4, L7/L5, L7/L8, L7/L9, L7/L12,
    L1/L2, L2/L3, L3/L4, L4/L5, L8/L9, L9/L10, L10/L11,
    L2/L10}
    \draw[posedge] (\a) -- (\b);

  \node[dot] (R1)  at (7.40, 0.50) {};
  \node[dot] (R2)  at (7.83, 1.04) {};
  \node[dot] (R3)  at (8.45, 1.04) {};
  \node[dot] (R4)  at (9.07, 0.68) {};
  \node[dot] (R5)  at (9.45,-0.05) {};
  \node[dot] (R6)  at (9.14,-0.77) {};
  \node[dot] (R7)  at (8.52,-0.95) {};
  \node[dot] (R8)  at (7.90,-0.68) {};
  \node[hub] (R9)  at (8.08, 0.23) {};
  \node[hub] (R10) at (8.76, 0.14) {};

  \foreach \a/\b in {%
    R9/R1, R9/R2, R9/R3, R9/R7, R9/R8, R9/R10,
    R10/R3, R10/R4, R10/R5, R10/R6,
    R1/R2, R2/R3, R4/R5, R6/R7, R7/R8, R8/R1,
    R2/R4}
    \draw[posedge] (\a) -- (\b);

  % four negative ties, faction-1 facing nodes paired with faction-2 facing nodes
  % in vertical order so the links fan out without crossing
  \foreach \a/\b in {L3/R2, L4/R1, L5/R8, L8/R7}
    \draw[negedge] (\a) -- (\b);

  \node[tag] at (5.15,-1.65) {faction 1};
  \node[tag] at (8.45,-1.65) {faction 2};

\end{scope}

\end{tikzpicture}
\caption{\textbf{An informal picture of structural balance.}
\textbf{(a)} The four elementary signed triads: two are \emph{balanced} (sign product $+1$) and two are \emph{unbalanced} (sign product $-1$), recovering the classic friend/enemy maxims.
Solid ink edges are positive ties, dashed terracotta edges are negative ties.
\textbf{(b, c)} Harary's structure theorem says a signed network is balanced exactly when it takes one of two forms: \textbf{(b)} a single group in which every tie is positive (the \emph{paradise} state), or \textbf{(c)} a split into two internally friendly, mutually hostile factions: positive edges within a faction, negative edges between them.
Configuration (c) is perfectly balanced yet fully polarized: balance does not imply the absence of conflict.}
\label{fig:intro-balance-schematic}
\end{figure}

Early applications of structural balance addressed ethnocentrism~\citep{berry1979reciprocity}, polarization~\citep{davis1963structural,abell1970structural,hage1976structural} and international conflict resolution~\citep{harary1961structural, moore1978international, moore1979structural, healy1973balance} -- topics that remain highly relevant today.
Structural balance is thus rooted in psychology and the social sciences. 
What can statistical physics contribute to it? 
To answer this question, we provide a brief historical overview of the Ising model and its relation to structural balance.

\subsubsection{Structural balance and the Ising model}
\label{subsec:balance-ising}

There are multiple connections between various models describing the concept of structural balance and the Ising model (and its relatives) developed for studies of magnetic systems \citep{Ising_1925_Beitrag}. 
Both frameworks describe how local binary interactions can generate macroscopic order, multiple stable configurations, and abrupt collective transitions, and they have a rich spectrum of multidisciplinary applications. 
However, while the historical roots of structural balance lie in social science and the concept was then studied using tools of other disciplines including statistical physics, the roots of the Ising model are in statistical physics and the concept was then adopted in other disciplines including social science. 

In fact, the Ising model (sometimes called Ising-Lenz model) was  developed in Ernst Ising's Ph.D. thesis in 1924 (see \citep{Ising_1925_Beitrag}) prepared under the supervision of Wilhelm Lenz to explain the emergent property of spontaneous magnetization observed below a critical temperature in ferromagnetic materials. 
Following Lenz's hypothesis \citep{Lenz_1920_Beitraege}, Ising assumed that the macroscopic magnetization is a collective effect resulting only from local interactions of neighboring magnetic moments of atoms or molecules and that these moments are quantized, i.e. a magnetic spin can be only up or down. Lenz used this quantization already in the paper \citep{Lenz_1920_Beitraege} to explain the behavior of magnetic susceptibility of paramagnetic crystals. 

Ising solved the problem posed by Lenz for a one-dimensional system, i.e. for a spin chain in which every spin interacts with its two nearest neighbors, and he proved that there is no ferromagnetic order in this model even at very low temperatures.
The solution provided by Ising is exact, but he also tried to extend his study to a three-dimensional periodic lattice. 
Unfortunately he used invalid approximations and came to an erroneous conclusion that the ferromagnetic order does not exist also in such a system. 
This negative conclusion contributed to the model receiving relatively little attention for some time. In 1936, Rudolf Peierls \citep{Peierls_1936_Ising} demonstrated that spontaneous order does occur in a two-dimensional lattice at low temperatures. 
The exact solution for the transition temperature in this system was found by Lars Onsager only in 1944 \citep{Onsager_1944_Crystal}. 

Since then the Ising model and its extensions, e.g. XY-model (where spins are not binary objects but two dimensional vectors of a constant length) or Heisenberg model (where spins are three-dimensional vectors), have been extensively explored on different lattices to understand the nature of phase transitions and, in particular, the universality of critical phenomena~\citep{Stanley1971,Fisher_1974_Renormalization,Amit_2005_Field}. 
Its later interdisciplinary use rests on the same idea: simple local interactions between binary variables may generate macroscopic order, metastability, and phase transitions~\citep{Macy_2024_IsingCentenary}.

 It is convenient to describe a large class of Ising models using the following Hamiltonian 
\begin{equation}
    H_{\rm Ising} = - \sum_{(i,j)\in E} J_{ij}s_i s_j - \sum_i h_i s_i ,
    \label{eq:ising_hamiltonian}
\end{equation}
where $s_i=\pm 1$ are binary spins placed on nodes $i$ of a network $G$, $J_{ij}$ denotes the interaction constant between neighboring spins placed at nodes $i$ and $j$ and $h_i$ is an external magnetic field acting on a spin $s_i$. For $J_{ij}>0$, the interaction is ferromagnetic and the energy is minimized when adjacent spins $s_i$ and $s_j$ are aligned, i.e.
\begin{equation}
    s_i s_j=+1   
    \label{eq:ising_spins}
\end{equation}
 and one can expect that at low temperatures majority of neighboring spins will be ordered in this way. On the other hand, for $J_{ij}<0$, the interaction is antiferromagnetic and the locally preferred state is anti-alignment. 

As noted above, structural balance theory starts from a different microscopic variable.
Instead of assigning binary states to nodes, it assigns signs to relations. A signed edge $x_{ij}=+1$ between nodes $i$ and $j$ can represent friendship, alliance, trust, or agreement, whereas $x_{ij}=-1$ can represent enmity, antagonism, distrust, or disagreement. 
In Heider's original formulation, cognitive tension arises when the signs in a triadic relation are mutually inconsistent~\citep{heider1946attitudes}. Cartwright and Harary subsequently generalized this idea to signed graphs, showing that a complete signed network is balanced if and only if it can be partitioned into two groups with positive ties inside groups and negative ties between groups~\citep{cartwright_1956}. At the local level, a triad $(i,j,k)$ is balanced when
\begin{equation}
    x_{ij}x_{jk}x_{ki}=+1,
    \label{eq:balanced_triad}
\end{equation}
and imbalanced when this product is negative. This condition encodes the familiar rules that ``the friend of my friend is my friend'', ``the enemy of my friend is my enemy'', and ``the enemy of my enemy is my friend''. All these rules define a local order using a triple of sign links.  The Eq. \eqref{eq:balanced_triad} corresponds to the Eq. \eqref{eq:ising_spins} defining equilibrium for the Ising model. 

The relation to the Ising model becomes explicit when structural balance is formulated as an energy minimization problem on signed links. For a complete graph, a minimal Hamiltonian for balance is
\begin{equation}
    H_{\rm SB} = - J\sum_{i<j<k} x_{ij}x_{jk}x_{ki},
    \label{eq:balance_hamiltonian}
\end{equation}
where a positive constant $J$ describes the strength of interactions leading to the local order between all three links in the triad $i,j,k$. In contrast to the pairwise Ising Hamiltonian in Eq.~\eqref{eq:ising_hamiltonian}, Eq.~\eqref{eq:balance_hamiltonian} contains three-body interactions among links. The contribution of a relation $x_{ij}$ cannot be evaluated independently of the third node $k$ that closes the triad. This multi-body character is one of the main reasons why structural balance naturally belongs to statistical physics: balanced triads lower the energy, imbalanced triads raise it, and the global dynamics can be studied as relaxation on a rugged energy landscape~\citep{Marvel2009,belaza2017_StatPhysBalanceTheory}.

% The concept of the frustration that is frequently used in spin glasses and even for an antiferromagnetic Ising model on a triangle

The concept of frustration provides an important connection between structural balance and the Ising model. In an antiferromagnetic Ising model on a triangle, not all pairwise interactions can be simultaneously satisfied: after two anti-alignments are imposed, the third edge creates an incompatible constraint. More generally, in an Ising spin glass, frustration occurs when the product of the signs of the couplings around a cycle is negative. The same mathematical condition characterizes an imbalanced signed triad. For example, if \(i\) likes \(j\), \(j\) likes \(k\), but \(i\) dislikes \(k\), the three relations cannot all be satisfied by a partition into two mutually hostile groups. In an isolated triad such an imbalance can always be removed by changing
the sign of one relation. In a larger signed network, however, a single
link generally belongs to several triads, and different triads may impose
incompatible requirements on its sign. Changing the sign of such a link
may therefore balance some triads while simultaneously unbalancing others,
providing a closer analogue of frustration arising from competing
constraints in spin systems.
%The same underlying incompatibility is described as geometrical frustration in statistical physics and structural imbalance in signed network theory, and is closely related to cognitive dissonance in social psychology. 

This underlying incompatibility is described as geometrical frustration in the case of the triangular antiferromagnet, more generally as frustration in spin glasses, and as structural imbalance in signed network theory; it is also closely related to cognitive dissonance in social psychology.
There is, however, a fundamental difference: in this frustration-based analogy, the signs of social relations correspond to the couplings in the Ising model, rather than to the spins themselves. In standard Ising and spin-glass models, the couplings are fixed (quenched), so changing spin orientations cannot eliminate frustration, whereas in structural balance models, the signs of relations may evolve, allowing the
system to reduce or eventually eliminate structural imbalance. %an imbalanced triad to become balanced. 
Thus, the mathematical correspondence concerns the incompatibility of pairwise constraints, rather than identical dynamical mechanisms. Nevertheless, this correspondence allows concepts and tools developed for frustrated magnetic systems and spin glasses to be adapted to social systems with positive and negative relations.
% However, an important distinction concerns the dynamics: in the standard Ising model, couplings are fixed, and changing spin orientations cannot eliminate frustration, whereas in structural balance models, the signs of social relations may evolve, allowing an imbalanced triad to become balanced. Thus, the correspondence concerns the incompatibility of pairwise constraints rather than identical dynamical mechanisms.

%The most direct conceptual bridge is frustration. In an antiferromagnetic Ising model on a triangle, not all pairwise preferences can be simultaneously satisfied: after two anti-alignments are imposed, the third edge creates an incompatible constraint. The same obstruction appears in an imbalanced signed triad. If $i$ likes $j$, $j$ likes $k$, but $i$ dislikes $k$, then the triad has negative sign product and at least one relation is inconsistent with the other two. The same mathematical structure is thus described as geometrical frustration in statistical physics, cognitive dissonance in social psychology, and structural imbalance in signed network theory. This correspondence is particularly important because it allows tools developed for frustrated magnetic systems and spin glasses to be transferred to social systems with positive and negative relations.

%There are also important differences. In the original Ising model, the couplings $J_{ij}$ are usually fixed and the node variables $s_i$ evolve. In many structural balance models, the node set is fixed but the signs of the relations $x_{ij}$ evolve. 

A more direct connection emerges when global structural balance is enforced.
The structure theorem then allows each link sign to be represented as $x_{ij}=s_i s_j$, where $s_i=\pm1$ are auxiliary node spins. Using this representation, \citet{Burda2026Ising} showed that, in the presence of an external field $h$ acting on links, the statistical ensemble of fully balanced networks   (i.e., the statistical ensemble restricted to configurations corresponding
to the degenerate ground states of the balance Hamiltonian~\eqref{eq:balance_hamiltonian}) becomes equivalent to the statistical ensemble of an Ising model without an external magnetic field at finite temperature. In this correspondence, the field $h$ acting on links plays the role of the interaction coupling between Ising spins (see Sec.~\ref{subsubsec:hamil-multitype}).

Beyond these equilibrium correspondences, social models
may also allow node states and link signs to co-evolve: opinions can change, relations can be rewired, and positive ties can become negative. This flexibility is particularly important in social systems. A disagreement between two friends can be resolved by changing an opinion, weakening a relation, or transforming friendship into hostility. Such adaptive dynamics are absent from the simplest equilibrium Ising model, but they are natural in Ising-inspired models with signed couplings, adaptive networks, or Hopfield-type interactions~\citep{Hopfield_1982_Neural,Gorski2020}. Hopfield's model is especially relevant because it combines binary node states with signed interaction weights, thereby making attraction and repulsion part of the same energy-based formalism~\citep{Hopfield_1982_Neural}.

Note that the claim is not that social ties or individuals are literally Ising spins. 
Rather, the Ising model supplies a minimal reference framework for understanding how and when binary variables, signed interactions, topology, noise, and frustration jointly produce collective order.
This reference framework has been widely extended in statistical physics over the past century, providing many new connections with structural balance theory and unique contributions to how we understand and validate this theory.

\subsubsection{Five main contributions of the statistical physics formulation}

The statistical physics formulation of structural balance offers five main contributions, which explain why the topic has recently attracted renewed attention from the physics community~\citep{antal_2005,kulakowski_2005, Marvel2009, szell_2010, estrada2014_RandomWalkSB, belaza2017_StatPhysBalanceTheory, kirkley2018balanceInSN, Gorski2020, schweighofer_2020, pham2022_SBexplainedThroughHomophily}. 

First, it provides the Hamiltonian language in which balanced and imbalanced configurations can be treated as low- and high-energy states.
The interactions in structural balance are characterized as multi-body interactions, which can be modeled using a Hamiltonian formulation incorporating link polarizations~\citep{Marvel2009, belaza2017_StatPhysBalanceTheory, Malarz2022Mean}. 
In this sense, the Ising model should be regarded as a reference model for the physics of structural balance: it contributes the vocabulary of spins, couplings, energy, temperature, frustration, metastability, and phase transitions.
Structural balance enriches this vocabulary by moving from pairwise spin alignment to signed relational consistency, from two-body to triadic interactions, and from fixed interaction networks to evolving social ties. 
This connection explains why methods originally developed for magnets, spin glasses, and interacting particle systems have become central tools for studying polarization, coalition formation, conflict, and the emergence of order in signed complex networks.

Second, it allows for a mathematical understanding of abrupt, endogenously triggered changes in the structural balance of a system.
The concept of phase transitions shows that small changes in noise, density of negative ties, or update rules may lead to abrupt transition, which can be either continuous or discontinuous, between consensus, polarization, fragmentation, and jammed or glassy states~\citep{antal_2005,belaza2017_StatPhysBalanceTheory,Rabbani2019,Malarz2022Mean, Singh2016,Chen2014,Hassanibesheli2017,Gorski2025}.
In other words, statistical physics provides a toolkit, including Hamiltonian dynamics~\citep{Marvel2009, Malarz2022Mean}, glassy transitions~\citep{Korbel2023}, and agent-based modeling~\citep{antal_2005, schweighofer_2020, schweighofer_2020_2}, which offers theoretical insights into the dynamics of polarization~\citep{rodriguez2016collective,MinhPham2020,Macy2021,Malarz2022Mean} and a foundation for mitigating its negative consequences in social organizations~\citep{Levorato2016, Gorski2023, Fraxanet2024}.

Third, statistical physics asks not only whether a transition exists between balanced and unbalanced states, but also how it depends on system size.
Comparing models with different Hamiltonian normalizations, update schemes, and transition types~\citep{Shojaei2019,Malarz2021Comment,Manshour2021,Bagherikalhor2021} reveals different scaling behaviors for the critical model parameter separating the balanced and unbalanced states as a function of the number of nodes (individuals) in the network.
This size dependence provides a physical diagnostic for determining whether an observed transition reflects a collective phenomenon or is merely a finite-size effect.

Fourth, it clarifies why simple microscopic assumptions may have robust macroscopic consequences. 
This robustness is what allows structural balance principles to travel well beyond their original social-psychological setting: they have been used to infer unobserved ties from co-behavior data~\citep{neal_2014_backbone}, to characterize the structure of financial markets~\citep{Ehsani2020}, to explain performance in teams~\citep{Askarisichani2019}, and to analyze the stability of international relations~\citep{moore1978international,crescenzi2007reputation,askarisichani20201995,doreian}.

Fifth, the combination of statistical physics with complex network theory provides a new way to measure balance. 
The introduction of network ensembles, which are probability distributions over a set of possible networks sharing a given set of properties, turns the vague question ``Is this network balanced?'' into several questions with statistical meaning, such as ``Is the count of balanced triads higher than what a null model predicts, once basic network properties are held fixed?'' 
Exponential random graph models and other network ensembles allow for asking and answering these questions, and hence for testing the presence of structural balance~\citep{wang2013_multivarERGM,Huitsing2012,feng2022_nullmodelSB,gallo2024testing}.

By synthesizing the work leading to these five contributions, the review aims to provide a physics-oriented critical picture of structural balance and its implications in complex networks.

\subsection{Review objective}

Our primary objective is to present the usefulness of statistical physics in modeling and understanding the emergence of structural balance. % across various systems.
Specifically, we emphasize how structural balance measures, originally defined in social psychology, may be formulated in the language of statistical ensemble theories and how this change of perspective explores the relationships between network topology and balance measures.
Moreover, we survey models explaining structural balance emergence and discuss their implications, highlighting their relevance in addressing, for example, polarization~\citep{Gorski2023} and radicalization~\citep{schweighofer_2020}.
By doing so, we offer a comprehensive understanding of structural balance, connecting its theoretical foundation in social psychology \citep{heider1946attitudes, cartwright_1956}, its formalization using concepts from physics \citep{kulakowski_2005, antal_2005, belaza2017_StatPhysBalanceTheory}, and its practical implications in social network analysis \citep{szell_2010, Leskovec2010Predicting, Li2019Binary, Gorski2023, linczuk_multidimensional_2023} and beyond~\citep{Adejumo2008, kirkley2018balanceInSN, Aref2020, Moradimanesh2021, Rizi2021, Fraxanet2024}. 

While this review is broad in scope, it is not intended to be a systematic review. 
Instead, it should be understood as an authoritative synthesis of the field, grounded in expert knowledge and the accumulated experience of the authors. 
Our goal is not to catalog every publication related to structural balance, but rather to organize, interpret, and critically connect the most influential theoretical, empirical, and methodological contributions. 
To provide transparency about our coverage, Fig.~\ref{fig:curated-map} illustrates the subset of papers analyzed in this review relative to the broader body of available work on structural balance and signed networks; the construction of this map and what it reveals about our scope are detailed in Section~\ref{subsec:curated-selection}.

\subsubsection{Structure of the review}
The review is divided into three main sections: Measurement, Dynamics and Emergence, and Implications.

\paragraph{Measurement -- Sec.~\ref{sec:meas}} 
We introduce structural balance and its statistical measures at both local and global levels. 
Local measures primarily focus on identifying structural balance within networks by examining the presence of (un)balanced triangles \citep{harary1959_measuringSB}, while global measures extend their scope to encompass cycles of arbitrary length \citep{pelino2012class, estrada2019RethinkingSB, estrada2014_RandomWalkSB, singh2017WalkSB, kirkley2018balanceInSN}. 
Beyond the definition of these measures, we emphasize how the interpretation of structural balance can vary significantly depending on the specific measure used~\citep{estrada2019RethinkingSB,aref2017measuring} and the network's topology under investigation~\citep{feng2022_nullmodelSB, Gelardi2019, Doreian2019, Zvereva2017, saiz2017evidence, leskovec2010countingTriads,szell_2010}. 
To address these challenges, 
we draw upon concepts bridging statistical physics and complex network theory, such as network ensembles \citep{wang2013_multivarERGM, Andres2023, Huitsing2012} and entropy maximization principles \citep{Rizi2021}.

\paragraph{Dynamics and Emergence -- Sec.~\ref{sec:emerg}} 
The review focuses on dynamics driven by (and leading to) structural balance in complex systems. 
By adopting the Hamiltonian formalism~\citep{kulakowski_2005, belaza2017_StatPhysBalanceTheory, kulakowski2019stabilityOfSBsolutions, marvel2011_evolutionOFSBcontinuous}, we illustrate how structural balance contributes to systemic properties instrumental in understanding the stability and evolution of networks.
These properties include physical ones, such as energy~\citep{belaza2017_StatPhysBalanceTheory, Marvel2009} and frustration \citep{van2014micro}, and social ones, such as fragmentation~\citep{pham2021balance,marvel2011_evolutionOFSBcontinuous} and polarization \citep{DU2016, MinhPham2020}.
Also, we review the role of structural balance in triggering drastic changes in these properties through phase transitions~\citep{Gorski2020, Malarz2022Mean}. 
Moreover, we comprehensively survey how structural balance impacts dynamics on and of networks 
in diffusion \citep{ehsani_balanced_2014, saeedian2017epidemic, Li2021}, and opinion dynamics models~\citep{altafini2012dynamics,medo2021fragility,pham2021balance,fenoaltea2023phase,meng2022whom}.
Finally, we review how different model dynamics either facilitate or hinder the emergence of structural balance in generalized network structures, such as temporal~\citep{Aref2020}, multi-layer~\citep{Gorski2017}, and higher-order networks~\citep{Hakimi2022}.

\paragraph{Implications -- Sec.~\ref{sec:implications}}
We turn from how structural balance is measured and how it emerges to what it implies for real-world systems, moving from the individual mind to societies, states, and markets.
We begin with \textit{cognitive balance} addressing the structural balance of the signed networks representing an individual's belief system~\citep{heider1946attitudes, greenwald2002unified,
galesic2021integrating}, and review how consistency-seeking mechanisms rooted in cognitive dissonance~\citep{festinger1957theory, dalege2024networks} shape the formation and internal coherence of beliefs~\citep{dalege2022using, medo2021fragility}, and determine whether new belief systems spread as simple or complex contagions~\citep{rodriguez2016collective, aiyappa2024emergence}.
We then shift to \textit{social balance}, the balance of the social networks along which opinions and behaviors propagate, surveying its impact on opinion polarization~\citep{altafini2012dynamics, shi2019dynamics}, on epidemic and complex contagion
processes~\citep{saeedian2017epidemic, Li2021, Lee2023, de2024complex}, and on collective decision making, where the frustration of a parliamentary network is predictive of the time needed to form a government~\citep{fontan2021role, fontan2021signed}.
Here, we also examine the widespread practice of using balance measures as indices of \textit{polarization}~\citep{Neal2020, MinhPham2020, aref2020detecting, talaga_polarization_2023}.
Third, we revisit \textit{international relations}, one of the oldest fields of application of the theory~\citep{harary1961structural, healy1973balance, moore1979structural}, and review what drive toward balance implies for the persistence of alliances and rivalries~\citep{belaza2017_StatPhysBalanceTheory,
Oishi2025}, for the escalation rather than the resolution of conflict~\citep{Lee1994, Saperstein2004, Konig2017}, for peace-war transitions~\citep{Morrison2023,Lee2017}, and for the identification of blocs and key states~\citep{Aref_2018,diazdiaz_mathematical_2024}.
Finally, we review implications for \textit{organizations and markets}, where balance is alternately beneficial or detrimental to the performance of teams and firms~\citep{Morrissette1967, Marineau2016, Askarisichani2019,
Lerner2019Team}, shapes alliance and conflict formation among firms~\citep{sytch2014friendsandfoes}, and acts as an order parameter for systemic stress in financial correlation networks~\citep{Kuyyamudi2019, Ehsani2020, ferreira_loss_2021, Zahedian2022}.
Across these domains, we argue that structural balance is best understood as a lens rather than a law, that a balanced configuration is not necessarily a desirable one, and that most of the available evidence remains correlational.

\subsubsection{A curated selection within the signed-network landscape}
\label{subsec:curated-selection}

\begin{figure}[hbt!]
    \centering
    \includegraphics[width=0.99\textwidth, clip, trim=45 0 0 50]{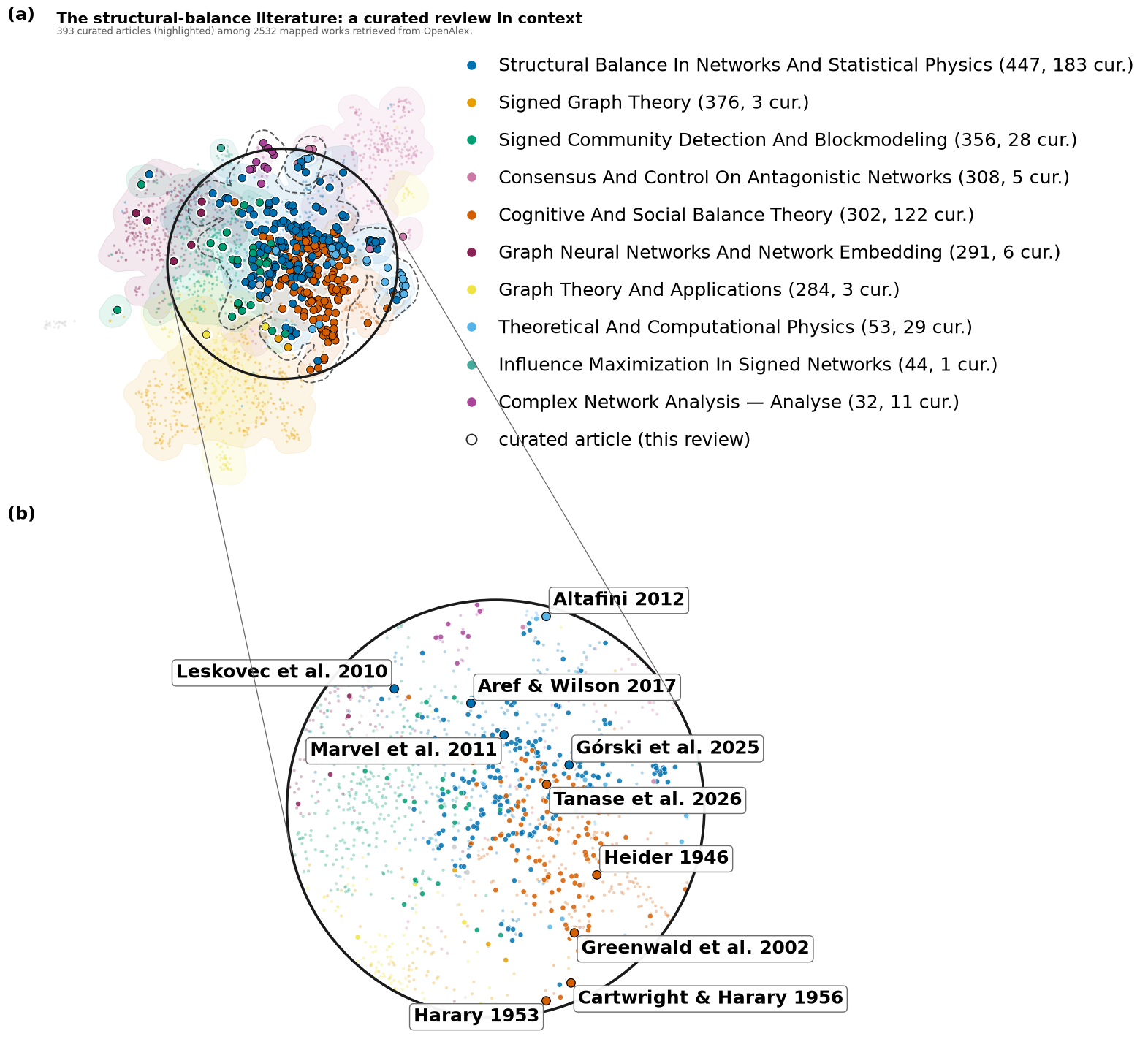}
    \vspace{-0.5cm}
    \caption{\textbf{The curated review in the context of the signed-network literature.}
    \textbf{(Top)} Each point represents a publication: the works in the reference list are shown as large outlined markers among works retrieved from OpenAlex using structural-balance search terms.
    Positions are based on a UMAP projection combining text embeddings, co-authorship, and bibliographic coupling; colors and labels indicate ten of the twelve identified Leiden communities.
    Two communities are omitted for clarity from the map (see text for details).
    The dashed line outlines the region occupied by the curated articles.
    \textbf{(Bottom)} Zoomed view of the core region, with landmark papers labeled.}
    \label{fig:curated-map}
\end{figure}

This review is a \emph{curated} synthesis: we have selected articles on the basis of expert judgment about their relevance to structural balance, rather than through an exhaustive keyword harvest. 
To make the boundaries of this selection explicit, we have placed our reference list ($393$ works) in the context of the wider literature about signed networks and structural balance. 
Using the OpenAlex bibliographic database~\citep{priem2022openalex}, we retrieved a total of $2{,}532$ works with the phrases \emph{structural balance}, \emph{signed network}, \emph{signed graph}, \emph{signed social network} and \emph{Heider balance} (title/abstract matches, restricted to items whose text explicitly concerns signed or balance-related structure).
For every work (retrieved and curated), we have combined a sentence-embedding of its title and abstract with an embedding of its co-authorship and of its bibliographic coupling; the resulting space is projected to two dimensions with UMAP~\citep{mcinnes2018umap} and partitioned into twelve communities by the Leiden algorithm~\citep{traag2019leiden}, each labeled by its dominant research topics (Fig.~\ref{fig:curated-map}).

\FloatBarrier

The curated corpus is markedly cohesive.
More than three-quarters of the selected references ($305/393$) fall in two of the largest communities, \emph{structural balance in networks and statistical physics} and \emph{cognitive and social balance theory}, with secondary contributions from \emph{theoretical and computational physics} ($29/393$) and \emph{signed community detection and blockmodeling} ($28/393$).
By contrast, the review deliberately samples only lightly from several adjacent bodies of work that are clustered separately: signed graph theory ($376$ works retrieved on OpenAlex, of which $3$ are curated), consensus and control on antagonistic networks ($308$ retrieved, $5$ curated), graph neural networks and network embedding ($291$ retrieved, $6$ curated), graph theory and applications ($284$ retrieved, $3$ curated), influence maximization in signed networks ($44$ retrieved, $1$ curated), and complex network analysis ($32$ retrieved, $11$ curated).
Two communities are omitted for clarity from the map.
One has small size as it contains only 14 works, two of which are curated; the other groups 25 works (17 having unique titles) about correlation-clustering and EU voting data. 
They were authored by 7 researchers, none of which are curated.
The dashed outline in Fig.~\ref{fig:curated-map} encloses the region occupied by the curated articles and shows that our coverage, while dense at the core of the structural-balance literature, does not extend uniformly across the full signed-network landscape. 
This figure should therefore be read as a statement of scope: the present work is an interpretive, physics-oriented review of a coherent research thread, not a systematic bibliometric survey of every study that uses signed networks.

It is also worth noting that the keyword search alone on OpenAlex recovers less than 50\% of the curated references; the missed works are often foundational or cross-disciplinary, and their titles and abstracts do not use signed-network terminology.
These works enter the corpus of curated articles because we have manually included them, underscoring the limits of keyword-based retrieval and the value of expert curation.

% \addcontentsline{toc}{subsubsection}{Existing reviews}
\subsubsection{Existing reviews}

Several previous works have reviewed aspects of structural balance theory and its implications in complex networks.

\citet{Traag2014}, in a book chapter, summarizes the most prominent models of structural balance, distinguishing between discrete-time models (e.g., local triad dynamics, LTD; constrained triad dynamics, CTD) and continuous-time models originating from the work of Ku{\l}akowski, Gawro\'nski, and Gronek~\citep{kulakowski_2005} in both undirected and directed forms.
This contribution was among the first to systematically connect theoretical formulations with the dynamical mechanisms through which balance evolves, providing a conceptual bridge between the classical sociological foundations of structural balance and modern dynamical systems approaches.

\citet{shi2019dynamics} offer a mathematically oriented review of dynamics over signed networks, concentrating on the fundamental convergence properties in deterministic and random networks. 
Their work provides a unified treatment grounded in generalized Perron-Frobenius theory, graph-theoretic tools, and elementary algebraic recursions. 
The review focuses on a small set of core results that characterize when and how signed dynamical systems converge to structurally balanced states. 

Similarly to \citep{Traag2014, shi2019dynamics}, \citet{kulakowski2025heider} provides a mini-review of dynamics based on Heider's balance theory with a special focus on continuous equations having the quadratic terms $x_{ij} x_{jk}$ where $x_{ij}\in(-1,1)$ represents the sign of the relation between $i$ and $j$.
In their emphasis on dynamical evolution rather than on empirical measures of balance, the scopes of \citep{Traag2014, shi2019dynamics,kulakowski2025heider} align with each other as all three concentrate on the mechanisms by which signed networks evolve toward balance, leaving out the problem of cataloging measures of structural balance.

\citet{Zheng2015} analyze balance at different levels (i.e., nodes, triads, and entire networks) and hence review balance measures, from the density of balanced triads to spectral-based indices, and clustering algorithms for detecting factions.
They further discuss empirical studies, both static and dynamic models of balance, and identify key methodological challenges, such as comparing measures across domains.

\citet{YoungPedersen2021} approach the topic from a logical and formal standpoint, reformulating structural balance theory within the language of propositional logic.
While their main goal is theoretical, they also review existing balance measures and perform small-scale comparative analyses on synthetic networks.
They highlight the connection between balance, logical consistency, and social polarization, emphasizing how imbalanced structures may correspond to incompatible belief systems.

\citet{aref2017measuring} focus specifically on how different measures of structural balance ``behave better than others in terms of axioms and ability to differentiate between graphs''. 
They find that the levels of partial balance vary substantially depending on whether cycle-based, eigenvalue-based, or frustration-based measures are applied, and they offer several recommendations regarding which measures should be used in future research.

\citet{diaz2025signed} provides a broad overview of signed networks as a general modeling framework that synthesizes in signed edges qualitative differences of relations, such as cooperative vs. antagonistic, supportive vs. conflicting, excitatory vs. inhibitory.
It highlights the unifying power of signed graphs for studying diverse systems from ecology to international politics, economics, and cognitive science.
While comprehensive in scope, its emphasis lies on the conceptual foundations and applications of signed networks as a whole, leaving the detailed methodological and empirical nuances of structural balance largely unaddressed.

While all the above previous reviews have addressed the concept of structural balance~\citep{Traag2014, shi2019dynamics, Zheng2015, YoungPedersen2021, aref2017measuring, diaz2025signed}, they focus mainly on models and tools used for the analysis of social systems or on theoretical results related to structural balance.
Methods provided by equilibrium or non-equilibrium statistical physics of complex networks have not been widely reviewed.  
This omission is significant, as the structural balance measures inherently depend on noisy processes running on time-dependent network topologies. 
Neglecting this crucial element can lead to a loss of meaning and accuracy in assessing structural balance, for instance, when estimating its statistical significance~\citep{gallo2024testing}, especially when a system approaches a critical point and magnitudes of fluctuations diverge \citep{Stanley1971}. 
Researchers from diverse fields are now increasingly acknowledging this limitation~\citep{feng2022_nullmodelSB, leskovec2010countingTriads,szell_2010} and physicists are providing prominent contributions to overcome it~\citep{gallo2024testing}.
Moreover, there are mixed results regarding whether the drive toward structural balance is necessary for the emergence of polarization~\citep{pham2022_SBexplainedThroughHomophily} and whether this drive positively or negatively affects the function of social organizations~\citep{Askarisichani2019}.
This mixed evidence has yet to be reviewed, and the influence of structural balance in networks beyond psychological and social domains remains largely unexplored.